\documentclass[letterpaper,12pt,titlepage,oneside,final]{book}
 
\newcommand{\href}[1]{#1} 

\usepackage{ifthen}
\newboolean{PrintVersion}
\setboolean{PrintVersion}{false}

\usepackage{amsmath,amssymb,amstext} 
\usepackage[pdftex]{graphicx} 
\usepackage{multirow}
\usepackage{enumitem} 
\usepackage{booktabs}

\usepackage{tabularx}

\usepackage[pdftex,pagebackref=false]{hyperref} 
\hypersetup{
    plainpages=false,       
    unicode=false,          
    pdftoolbar=true,        
    pdfmenubar=true,        
    pdffitwindow=false,     
    pdfstartview={FitH},    
    pdfnewwindow=true,      
    colorlinks=true,        
    linkcolor=blue,         
    citecolor=green,        
    filecolor=magenta,      
    urlcolor=cyan           
}
\ifthenelse{\boolean{PrintVersion}}{   
\hypersetup{	
    citecolor=black,%
    filecolor=black,%
    linkcolor=black,%
    urlcolor=black}
}{} 

\usepackage[automake,toc,abbreviations]{glossaries-extra} 

\let\origdoublepage\cleardoublepage
\newcommand{\clearemptydoublepage}{%
  \clearpage{\pagestyle{empty}\origdoublepage}}
\let\cleardoublepage\clearemptydoublepage

\makeglossaries

\usepackage{caption}
\usepackage{subcaption}
\usepackage[normalem]{ulem} 
\usepackage{longtable}
\usepackage{xspace}
\usepackage[english]{babel} 
\usepackage{rotating}
\usepackage{makecell}

\usepackage[sorting=none]{biblatex}
\usepackage{xcolor}

\newcommand{\unit}[1]{\ensuremath{\mathrm{\,#1}}\xspace}

\newcommand{\e}{\unit{e^{-}}}
\newcommand{\pix}{\unit{pix}}
\newcommand{\epix}{\unit{\e / \pix}}
\newcommand{\epixdia}{\unit{\e / \pix/day}}

\begin{document}
\renewcommand{\tablename}{Table}
\pagestyle{empty}
\pagenumbering{roman}

\renewcommand{\listfigurename}{List of Figures}
\renewcommand{\listtablename}{List of Tables}
\begin{titlepage}
        \begin{center}
        \includegraphics[width=4cm]{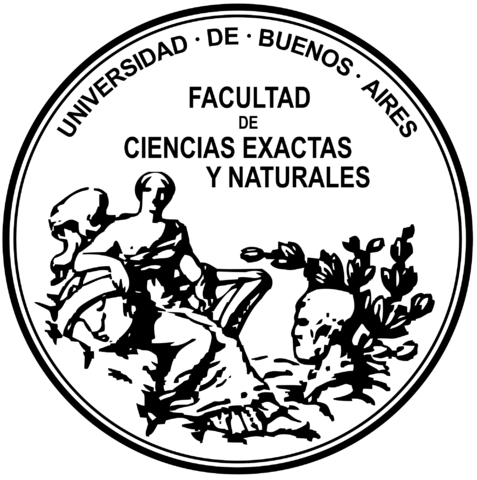}
        \vspace*{0.5cm}
        {\par UNIVERSIDAD DE BUENOS AIRES \par}
        { Facultad de Ciencias Exactas y Naturales \par}
        { Departamento de F\'isica \par}
        \vspace*{1.0cm}
        \Huge
        {\bf Origin and characterization of single-electron events in Skipper-CCDs for the search of light dark matter}

        \vspace*{0.25cm}

        \normalsize

        \Large
        Lic. Mariano Ruben Cababie \\

        \vspace*{0.5cm}

        \normalsize
        \textbf{Traducción al inglés} de la Tesis presentada para optar al título de  \\
        Doctor de la Universidad de Buenos Aires en el área Ciencias Físicas 
        \vspace*{0.5cm}

        \normalsize
        \textbf{English translation} of the Thesis presented for the degree of\\
        Doctor of Philosophy in Physics at the University of Buenos Aires
        \vspace*{0.5cm}

        \end{center}
\end{titlepage}

\newpage

\begin{titlepage}
          {\large Director : Javier Tiffenberg \par}
          \vspace*{0.1cm}
          {\large Codirector : Dario Rodrigues Ferreira Maltez \par}
          \vspace*{0.1cm}
          {\large Consejero de Estudios : Carlos Acha \par}
          \vspace*{0.1cm}
          {\large Lugar de Trabajo : Instituto de Física de Buenos Aires \par}
        \vspace*{0.25cm}
\end{titlepage}

\textbf{Important note / Nota importante:}

El presente documento es una traducción al inglés de la Tesis de Doctorado ``Origen y caracterización de eventos de un electrón en Skipper-CCDs para la búsqueda de materia oscura liviana'', cuya versión original fue escrita en español y se encuentra disponible para su consulta en el archivo digital de Tesis de Doctorado de la Universidad de Buenos Aires (Argentina). Se trata, pues, de una traducción no oficial intentada por el propio autor de la Tesis con el único propósito de facilitar su difusión y su amplio acceso para la comunidad científica. 
Sin embargo, en caso de suscitar mayor interés o bien ante cualquier duda o consulta que pueda surgir de esta traducción, se sugiere amablemente consultar la versión original a fin de respetar la rigurosidad exigida y cumplimentada en la defensa de dicha Tesis de Doctorado.

This document is an English translation of the Ph.D. thesis titled "Origin and Characterization of Single-Electron Events in Skipper-CCDs for Light Dark Matter Search." The original version of this thesis was written in Spanish and is available for reference online in the digital archive of Ph.D. theses at the University of Buenos Aires, Argentina. This translation is unofficial and has been attempted by the author of the thesis solely for the purpose of facilitating its diffusion and wider accessibility to the scientific community.
However, in case of further interest or for any questions or inquiries that may arise from this translation, it is kindly suggested to refer to the original version to ensure compliance with the rigorous standards adhered to during the defense of the Ph.D. thesis.

\textbf{Thesis link (September, 2023) / Hipervínculo de la Tesis (Septiembre, 2023):}

\url{https://bibliotecadigital.exactas.uba.ar/collection/tesis/document/tesis_n7320_Cababie}

\newpage
\pagestyle{plain}

\addcontentsline{toc}{chapter}{Resumen}
\begin{center}\textbf{Resumen}\end{center}

Desde su invención en el año 1969 los Dispositivos de Carga Acoplada (CCD, por sus siglas en ingles) han tenido un papel clave en la toma de imágenes de alta calidad tanto para fines comerciales como científicos.
A principios de la década pasada, se los comenzó a utilizar en la búsqueda de candidatos a materia oscura tipo \textit{WIMP} en el rango de 1 a 10 GeV.
No fue sino hasta el año 2017 que, gracias a la implementación de la tecnología \textit{Skipper}, que permite conocer con exactitud la carga colectada por el detector, se los empezó a utilizar para la búsqueda de materia oscura liviana.

El presente manuscrito resume el trabajo realizado en el marco de la colaboración SENSEI (\textit{Sub-Electron-Noise Skipper-CCD Experimental Instrument}), 
un esfuerzo enfocado en la búsqueda de materia oscura liviana con Skipper-CCDs.
Se presenta, en primer lugar,
el protocolo de adquisición y procesamiento de datos desarrollado para establecer un criterio de selección de eventos compatible con la señal de materia oscura. 
Dicho protocolo habilitó asimismo la estructura y las herramientas que fueron empleadas para el estudio y la caracterización de fenómenos que impactan en la calidad de las mediciones que se realizan para la búsqueda de materia oscura (medición del factor de Fano, determinación del fondo Compton y caracterización del fondo de alta energía en superficie) y que fueron llevados a cabo en el transcurso de este trabajo.
En segundo lugar, se discute el estudio exhaustivo realizado sobre el origen de los eventos de 1 electrón en Skipper-CCDs el cual permitió seleccionar los parámetros de operación del detector y mejorar la sensibilidad de detección en la búsqueda de materia oscura. Se destaca la caracterización de la correlación espacial de eventos de 1 electrón con eventos de alta energía, entre otros criterios de selección de eventos desarrollados, y la optimización del dispositivo de salida para disminuir fuentes de luminiscencia durante su operación. 
Como resultado de este estudio se reportan los valores más bajos de corriente oscura y carga espuria jamás medidos en un CCD.
Finalmente, se presentan los últimos resultados publicados desde SENSEI.
Los límites obtenidos para la dispersión de un mediador liviano son los mejores reportados a la fecha en todo el rango de masas investigado mientras que para el mediador pesado, lo son para masas menores a 10~MeV.
Mientras que, para la absorción de materia oscura, se reportan los límites más restringentes debajo de 10~eV. 
Estos resultados, logrados con una porción muy pequeña de la exposición total proyectada para SENSEI, colocaron a los SCCDs en una posición líder en la búsqueda de materia oscura liviana a nivel mundial.

\newpage
\addcontentsline{toc}{chapter}{Abstract}
\begin{center}\textbf{Abstract}\end{center}
\begin{center}\textbf{Origin and characterization of single-electron events in Skipper-CCDs for the search of light dark matter}\end{center}

Since their invention in 1969, Charge-Coupled Devices (CCDs) have played a key role in capturing high-quality images for both commercial and scientific purposes. In the early part of the last decade, they began to be used in the search for Weakly Interacting Massive Particles (WIMPs) as candidates for dark matter in the range from 1 to 10 GeV. However, it wasn't until 2017, with the implementation of the Skipper technology, which allows for an accurate measurement of the collected charge by the detector, that they started to be used in the search for light dark matter.

This manuscript summarizes the work carried out within the SENSEI (Sub-Electron-Noise Skipper-CCD Experimental Instrument) collaboration, an effort focused on the search for light dark matter using Skipper-CCDs. Firstly, it presents the data acquisition and processing protocol developed to establish a selection criteria for events compatible with the dark matter signal. This protocol also enabled the structure and tools used for the study and characterization of phenomena that impact the quality of measurements made in the search for dark matter (measurement of the Fano factor, determination of the Compton background, and characterization of the high-energy surface background), which were carried out during this work. Secondly, it discusses the comprehensive study conducted on the origin of single-electron events in Skipper-CCDs, which allowed for the selection of detector operating parameters and improved the detection sensitivity in the search for dark matter. It is highlighted the characterization of the spatial correlation between single-electron events and high-energy events, among other developed event selection criteria, and the optimization of the output device to reduce sources of luminescence during its operation. As a result of this study, the lowest levels of dark current and spurious charge ever measured in a CCD are reported. Finally, the latest published results from SENSEI are presented. The obtained limits for the scattering of a light mediator are the best reported to date across the entire investigated mass range, while for the heavy mediator, they are the best for masses below 10 MeV. In terms of dark matter absorption, the most restringent limits below 10 eV are reported. These results, achieved with only a very small fraction of the total projected exposure for SENSEI, position Skipper-CCDs as leaders in the search for light dark matter worldwide.

\cleardoublepage
\phantomsection    

\addcontentsline{toc}{chapter}{Agradecimientos}
\begin{center}\textbf{Agradecimientos}\end{center}

Quisiera comenzar con lo mismo que dije al finalizar mi Licenciatura y es que un Doctorado no es un logro individual sino colectivo, del cual forman parte un grupo muchas veces invisible de personas que aportan a la formación científica, académica, profesional y personal a lo largo de su duración. En primer lugar, quisiera agradecer a la Universidad de Buenos Aires y al CONICET por la oportunidad que brindan año a año de formar parte del sistema científico nacional y los esfuerzos realizados en su conservación, así como la preservación de una educación pública, gratuita y de calidad. 

En segundo lugar, quisiera agradecer a mis directores, Javier Tiffenberg y Darío Rodrigues, así como también a mi ex-director Ricardo Piegaia, por la oportunidad de formar parte de este proyecto que, 5 años luego de su comienzo, no sólo me formó como científico sino que cambió mi vida radicalmente en todos los sentidos. A Ricardo quisiera agradecerle su paciencia, acompañamiento y disponibilidad incluso después de haber dejado de ser mi director, hace ya varios años. De Darío solo puedo destacar su excelente calidad humana y científica, su capacidad de hacer de cualquier ambiente de trabajo un lugar ameno y producto y su infinita paciencia al momento de explicar cualquier tema ó problema, viejo ó nuevo y fuera ó dentro de su área de experiencia. En particular, quisiera agradecerle por haber sido oído y consejero de los tantos (tantísimos) lamentos, dilemas y dudas que fueron y son una parte natural de cualquier doctorado. A Javier quisiera agradecerle todas las oportunidades brindadas, por haberse preocupado siempre por mi crecimiento, presente y futuro y por todas las charlas compartidas, en persona o por zoom, de física, estadística o cualquier otro tema. Y, respecto a las oportunidades mencionadas, quisiera agradecerle en particular haberme dado la posibilidad de haber viajado al FNAL, donde conocí a colegas y amigos que hoy son parte de mi día a día.

De la gente que conocí allí, quisiera agradecerle a Guillermo, Carla, Miguel, Juan, Gustavo y Leo, así como también a sus parejas, familia y amigos por habernos hecho sentir (a mí a Daniela) tan cómodos en tan poco tiempo (y en particular a Juan y a su esposa por todos los asados compartidos). De este grupo quisiera destacar también toda la ayuda y formación que recibí, así como también al resto del grupo de \textit{SiDet}, Sho Uemura, Andrew Lathrop y Kevin Kuk. De la misma manera no quiero dejar de mencionar a todo el grupo hispanoparlante que nos acompaño durante toda nuestra visita y con el que hoy seguimos en contacto: Israel, Susana, Manuel, Carlos y Barbara, así como también a Matias, Kevin y Botti que fueron parte de este viaje.

A la colaboración SENSEI (en la cual forman parte de varios los ya mencionados) quisiera agradecerles la oportunidad no solo de haber formado parte de los esfuerzos realizados, sino de la paciencia en mi formación y adaptación así como también la chance de tener uno de los roles protagónicos en el experimento pese a mi corta experiencia en el área. Gracias por la confianza y por las oportunidades. Quisiera agradecerle a Tien-Tien, Daniel, Sravan, Luke, Prakruth, Itay, Yaron y Aviv por todas las discusiones y el trabajo realizado, así como también especialmente a Ansh, Kelly, Nate y Sho, con quienes compartí gran parte del trabajo realizado en los últimos meses. Finalmente quisiera agradecerle a Rouven, Tomer, Erez y Liron por sus consejos y paciencia, en particular a Rouven, y a Tomer, Erez y Liron por haberme invitado a conocer su maravilloso lugar de trabajo en Israel.

Volviendo a Buenos Aires, no quiero dejar de mencionar a los y las estudiantes y becarios del grupo, André, Eliana y Santiago, así como becarios de otros grupos como Nacho, Colo, Maru, Juanma, Fede, Rami, Javi entre muchos otros que formaron parte de infinitos almuerzos y una cantidad no menor de asados (gracias a Juanma). Ya saliendo del área laboral, quisiera mencionar a mi grupo de amigos de toda la vida, que saben perfectamente referirse a este mensaje, y que siguen siendo, después de tantos años, una de las partes más bellas e importantes de mi vida. Muchxs de ellxs supieron escucharme en mis momentos más complicados y festejar conmigo cuando las cosas salían bien. Crecimos juntos y seguiremos creciendo juntos. 

Llegando al final, quisiera mencionar a mi familia, cercana, un poco más lejana y a las recientes incorporaciones, por todos los momentos vividos, por el acompañamiento, por escucharme, por aconsejarme, por ser incondicionales, por todo. Y, finalmente, a Daniela, mi familia más más cercana, mi compañera de viajes, paseos y aventuras, quien supo levantarme en los momentos más difíciles y elevarme aún más durante mis logros desde hace ya casi una década que caminamos juntos. Gracias por acompañarme, gracias por ser parte y dejarme ser parte de tu vida. Este trabajo es para ustedes.

\cleardoublepage
\phantomsection    

\renewcommand\contentsname{Table of Contents}
\tableofcontents
\cleardoublepage
\phantomsection    

\addcontentsline{toc}{chapter}{List of Figures}
\listoffigures
\cleardoublepage
\phantomsection		

\addcontentsline{toc}{chapter}{List of Tables}
\listoftables
\cleardoublepage
\phantomsection		

\pagenumbering{arabic}

\chapter{Dark Matter}
\label{cap:1}

This chapter introduces the concept of dark matter (DM) and the evidence for its existence, followed by its properties, production mechanisms, and the most well-motivated and studied dark matter candidates. Subsequently, it discusses dark matter detection techniques, with a special emphasis on direct detection, and theoretical models for the scattering and absorption of certain dark matter candidates with electrons.

\section{Evidence for Dark Matter}
\label{sec:evidence}

Dark matter makes up more than 80$\%$ of the mass of the universe and 27$\%$ of the total universe, second only to dark energy with 68$\%$ \cite{aghanim2020planck} abundance. The remaining matter consists of ordinary particles described in the Standard Model or SM. The evidence for dark matter is extensive, vast, and well-known. In this section, we will present some historical examples that motivate its search.

\subsection{Fritz Zwicky and the Coma Cluster}
\label{sec:zwicky}

The first evidence of dark matter was presented in 1933 by the Swiss astronomer Fritz Zwicky \cite{zwicky1933rotverschiebung}. Although the evidence was not conclusive enough to inspire an organized search for dark matter by the scientific community, Zwicky was the first to use the term "dark matter" \cite{zwicky1933rotverschiebung}.

The concept introduced by Zwicky was as follows: he observed the motion of galaxies in the Coma Cluster over an extended period, measured their velocities, and indirectly estimated their mass. At the same time, he inferred the mass of galaxies based on their brightness, i.e., their electromagnetic radiation, and surprisingly, the two estimations did not match.

A quick way to see this, assuming a roughly spherically distributed cluster of galaxies with approximately the same velocity, is to use the Virial theorem, which states that the expected velocity of clusters should be approximately:

\begin{equation}
    v=\sqrt{\frac{5 \ G \ M_{T}}{3 \ R_{T}}}
    \label{eq.virial}
\end{equation}

where G is the gravitational constant, $M_{T}$ is the total mass of the cluster, and $R_{T}$ is the radius of the cluster. This expected velocity turned out to be much lower (about 10 times) than the measured velocity. The galaxies were moving too fast for the given \textit{visible} mass. There had to be additional mass, and it had to be invisible or \textit{dark}. The mystery lingered without significant developments for many years until the 1970s.

\subsection{Galactic Rotation Curves}
\label{sec:rotation}

While studying the rotation of galaxies, Vera Rubin and her team concluded that for a given group of galaxies \cite{rubin1970rotation,rubin1980rotational}, and quoting from \cite{rubin1983dark}, "...the rotation speed (...) either remains constant with increasing distance from the center or increases slightly..." when it is expected to follow Kepler's law for large distances and decrease as the square root of a given distance $R$:

\begin{equation}
    v=\sqrt{\frac{G \ M(R)}{R}}
    \label{eq.rotation}
\end{equation}

\begin{figure}[h!]
\centering
\includegraphics[width=0.7\textwidth]{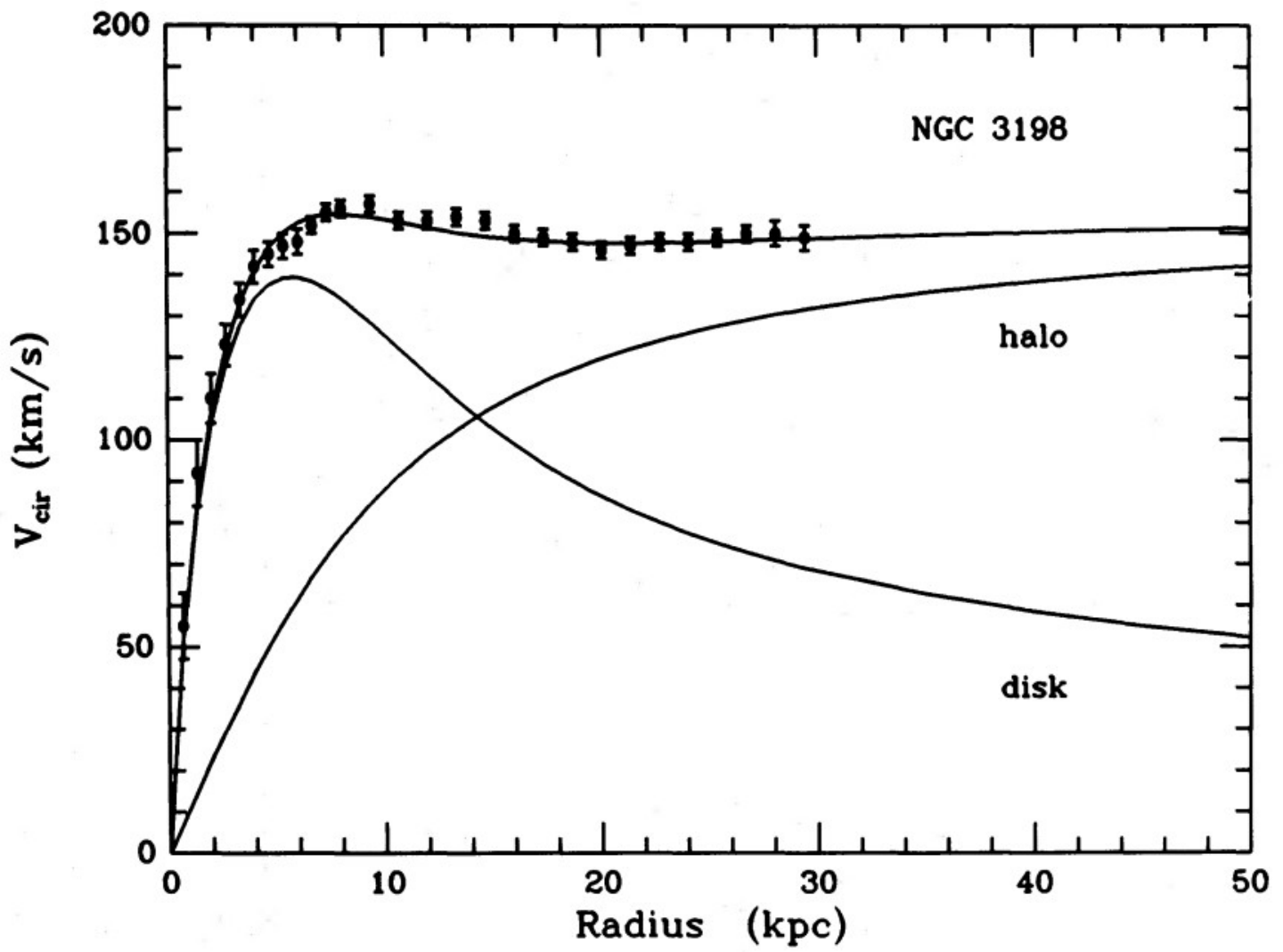}
\caption{Measured rotation velocity (data points with error bars) of the galaxy as a function of radius, i.e., the distance from the center. Overlayed is a fitting curve that assumes the existence of a baryonic mass disk and a dark matter halo. Figure extracted from \cite{van1985distribution}.}
\label{fig.rotation1}
\end{figure}

The same phenomenon was reported in \cite{van1985distribution} two years later for NGC 3198. In Figure \ref{fig.rotation1}, it can observed that the velocity begins to increase linearly as mass increases cubically. Then, instead of decreasing as in Equation \eqref{eq.rotation}, it levels off and remains nearly constant up to at least 30 kpc, with a slight dip around 20 kpc. The authors estimated the contribution of visible matter as a "disk" and that of dark matter as a "halo" (see Figure \ref{fig.rotation1}), as it appeared to be distributed around the center of the galaxy rather than at its center.

\subsection{Bullet Cluster}
\label{sec:bullet}

The Bullet Cluster is a group of galaxies (and their surrounding gas) that is actually the merger of a main cluster and a smaller one, separated by a distance of approximately 720 kpc \cite{sanders2010dark,clowe2006direct, bulletref}.
These two subclusters collided in the past at high speed (approximately 4000 km/s), so the galaxies that composed them, being much smaller compared to the overall size of the clusters, continued on their trajectory without colliding.
However, the gas surrounding each of the clusters did collide, causing it to decelerate, and after the collision was completed, it ended up at a distance shorter than the galactic centers of each cluster.

\begin{figure}[h!]
\centering
\includegraphics[width=0.9\textwidth]{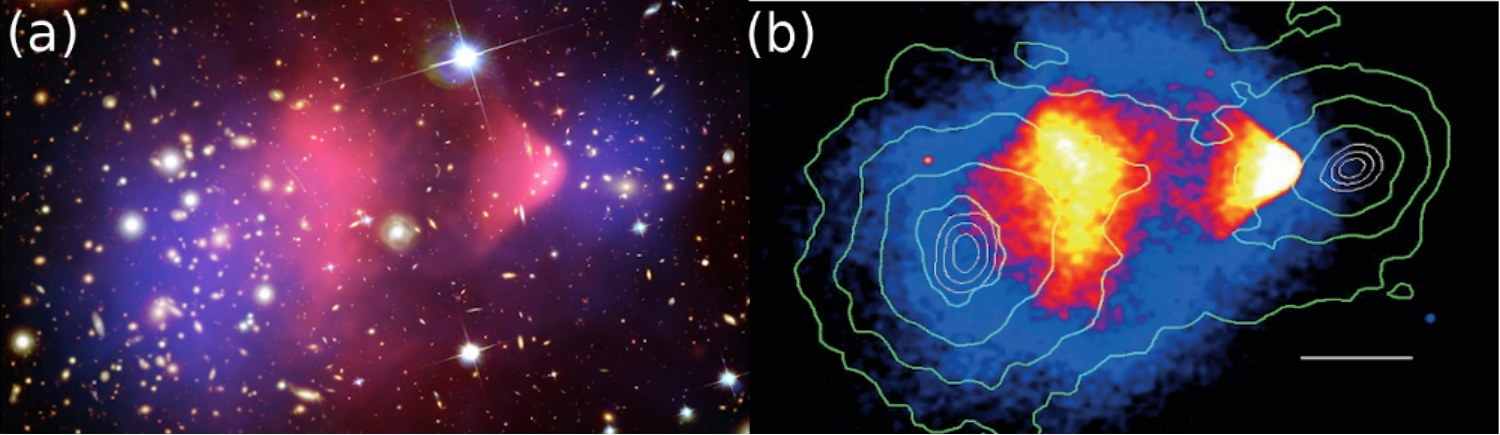}
\caption{The Bullet Cluster. (a) Optical image of the Bullet Cluster overlaid with the morphology of X-ray-emitting gas in red and the distortion generated by the mass accumulation in blue. (b) Confidence contours of mass concentrations measured by gravitational lensing in green and the X-ray emission intensity overlaid. The images are not to scale. Figures extracted from \cite{clowe2006direct, bulletref}.}
\label{fig.bullet}
\end{figure}

The Figure \ref{fig.bullet}(a) shows the optical image of the Bullet Cluster, overlaying the morphology of X-ray-emitting gas in red and the distortion generated by mass in the image due to gravitational effects, measured through gravitational lensing, in blue.
In other words, the red represents the position of the gas, where most of the baryonic mass is located, and the blue represents where the total mass is concentrated.
This is even clearer in Figure \ref{fig.bullet}(b), where confidence contours of the gravitational centers' distribution are shown in green, and X-ray emission is overlaid in red.
It is concluded that there is an invisible component of mass that does not interact electromagnetically and therefore moves virtually without colliding.

\subsection{Cosmic Microwave Background Radiation}
\label{sec:cmb}

Penzias and Wilson were the first to detect the Cosmic Microwave Background radiation (CMB) in 1965, although at that time they did not know exactly what that signal meant; they reported it as an "excess temperature" that was "3.5 K higher than expected" \cite{1965ApJ...142..419P}.
We now know that the CMB is an irreducible background that originated billions of years ago when the universe was about 380,000 years old during the recombination stage, which led to the decoupling of photons from baryonic matter.
It is also known that before recombination, the universe's temperature was above 3500 K, so protons and electrons could not join to form hydrogen and were, along with photons, in what it is called a photon-baryon fluid.
During that time, photons interacted with charged particles (electrons and protons) at such a frequency that their mean free path was so short that they could not propagate through the universe.
As recombination began, hydrogen formed, and the number of free charged particles dropped abruptly, dramatically increasing the mean free path of photons.
Approximately 13.4 billion years later, they reached Earth and, more specifically, in 1992, some of these photons were measured by the FIRAS experiment on the COBE satellite and formed what is said to be the best-fit blackbody radiation spectrum ever measured with only 9 minutes of exposure \cite{mather1990preliminary}, surpassed only by the same collaboration years later \cite{fixsen1996cosmic}.
The blackbody temperature of the CMB was 2.726(10)K with a 95$\%$ confidence level.

\begin{figure}[h!]
\centering
\includegraphics[width=0.7\textwidth]{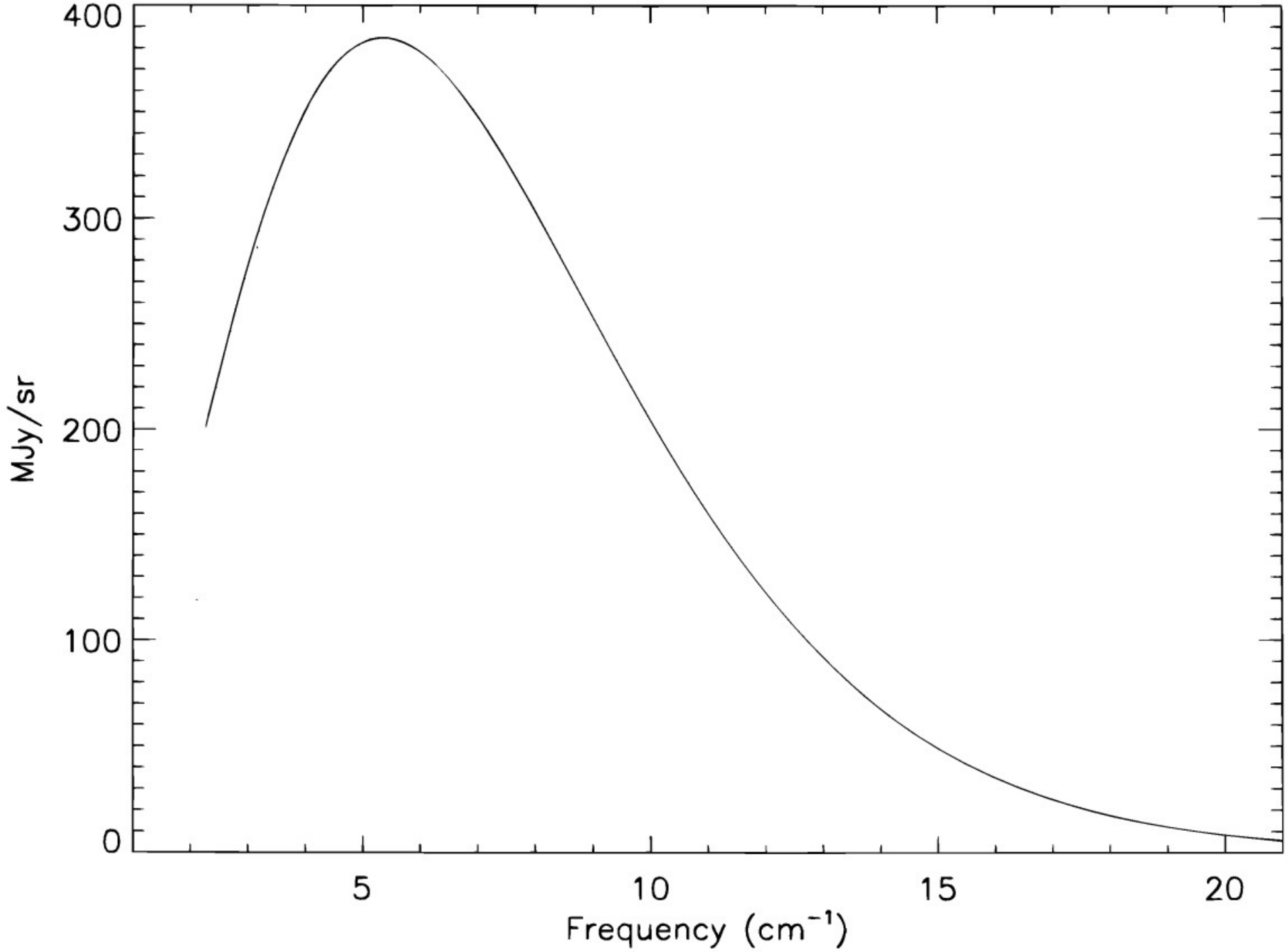}
\caption{Radiation spectrum and blackbody radiation fit for the CMB. Error bars are within the curve. Figure extracted from \cite{fixsen1996cosmic}.}
\label{fig.blackbody}
\end{figure}

\begin{figure}[h!]
\centering
\includegraphics[width=0.9\textwidth]{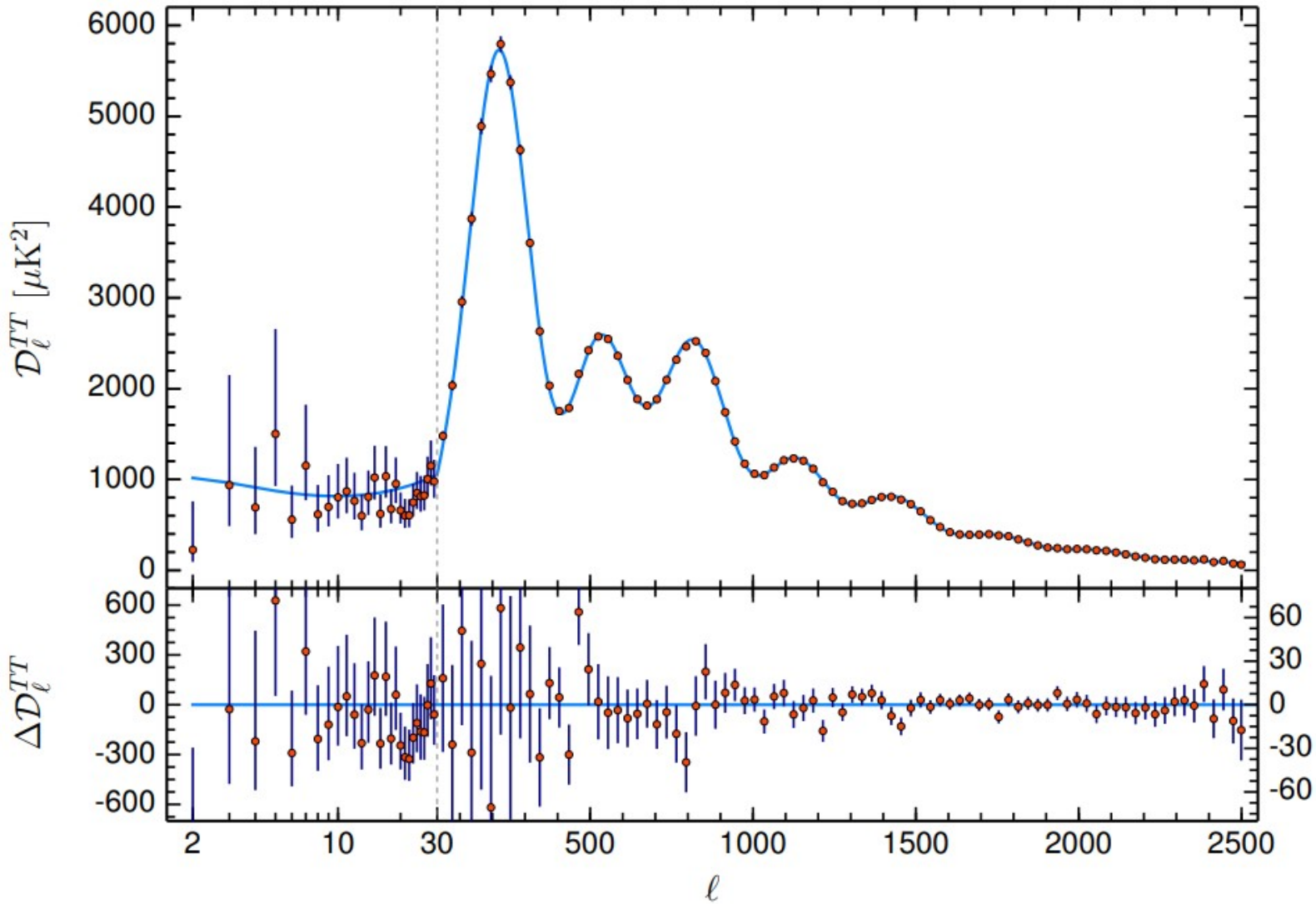}
\caption{CMB power spectrum. In the upper part, the $\ell$ axis shows the spherical harmonic modes, and the vertical axis is proportional to the intensity of each mode. In the lower part, the residuals of the fitted curve are shown. The shape of the curve fit is additional evidence for the existence of dark matter and its abundance. The figure was extracted from \cite{aghanim2020planck}.}
\label{fig.planck}
\end{figure}

However, the WMAP collaboration \cite{spergel2003first, hinshaw2013nine} and later the PLANCK collaboration \cite{aghanim2020planck} measured positional fluctuations in temperature on the order of $10^{-5}$ K. Between 1990 and 2018, tremendous advances were made both experimentally and theoretically, so that today we have excellent precision in the power spectrum of the CMB, as shown in Figure \ref{fig.planck}. This power spectrum is analogous to a Fourier transform but in spherical coordinates. Thus, the horizontal axis, $\ell$, shows the spherical harmonic modes, while the vertical axis is proportional to the intensity of that particular mode.

The Figure \ref{fig.planck} can be divided into three phases: the first, below $\ell=30$ (large-scale modes), where the intensity slightly increases from a plateau; the second, where the peaks are located; and the third, where damping occurs. The first phase is dominated by the Sachs-Wolfe effect and is strongly related to the abundance of dark energy in the universe, and the third is a consequence of the increase in the mean free path of photons, inhibiting intensity for higher modes, a damping method called Silk damping.
But how are the peaks seen in the second phase generated?
Before recombination (and during the decoupling process that lasted approximately 35,000 years), the cosmic fluid composed of baryons and photons experienced local density fluctuations: the fluid became denser where more mass was randomly located, but at the same time, that excessive density generated outward electromagnetic pressure (since baryons did not fully recombine) counteracting compression, and so on, creating oscillations. The acoustic peaks of these oscillations are the peaks found in the power spectrum. The amount of baryonic matter and photons can be inferred from the spectrum because it was generated by their oscillation, so it was found that much more mass was needed to explain what is seen in it. This is not only further evidence for the existence of dark matter but also provides cosmological parameters for the overall abundance of dark matter in the universe.

\section{General Properties of Dark Matter}
\label{sec:dmproperties}

The evidence mentioned in the preceding section provides proof of the existence of dark matter, but the following question remains unanswered: What is dark matter, and how is it constituted? Although many models predict different suitable candidates for dark matter, the vast majority agree on certain properties:

\begin{itemize}
    \item \textbf{(Quasi)-Neutrality.}
    Large-scale evidence of dark matter (such as the Bullet Cluster or the CMB) predicts a component of the universe's mass that does not electromagnetically interact with ordinary matter and is, therefore, "dark." However, dark matter could be partially charged or milli-charged \cite{goldberg1986new}, and the intensity of this charge may be approximately limited to below $10^{-7}$ in units of the electron charge for a dark matter candidate around 1 GeV \cite{particle2020review}. The restriction on this charge comes from the fact that a charged component can alter fluctuations in the baryon-photon fluid during and before recombination, resulting in deviations from the CMB power spectrum.
    
    \item \textbf{Non-interacting within the Dark Sector.}
    Because the entire abundance of dark matter may not be constituted by a single particle but by a variety of particles that make up what is called a "dark sector," DM-DM self-interactions can occur on small scales. However, such interactions are strongly constrained by observations of merging clusters (Bullet Cluster \cite{Randall_2008}) and cosmological predictions (CMB power spectrum), very similar to what was explained in the previous point.
    
    \item \textbf{Mass between $10^{-22}$ and $10^{66}$ eV.} 
    \begin{itemize}
        \item \textbf{Lower Limit:}
        For fermions, lower limits on dark matter candidates are set by the Pauli exclusion principle. Depending on the galactic system under observation, an upper limit on the velocity of system components can be inferred from data, and a minimum mass can be estimated \cite{tremaine1979dynamical}. Using data from a dwarf galaxy, a lower limit of 70 eV is established in \cite{randall2017cores}. For bosons, dark matter can be as light as $10^{-22}$ eV, entering the mass range where it behaves like a coherent field \cite{https://doi.org/10.48550/arxiv.1904.07915}. Thus, the wavelike nature of the dark matter candidate will dictate its spatial reach through the De Broglie wavelength, limiting it above the mentioned mass.
        
        \item \textbf{Upper Limit:}
        Assuming that the abundance of dark matter comes from a single candidate, its mass is established to be less than 5 solar masses or $10^{66}$ eV. Larger masses may result in disruptions to the structures of dark matter halos \cite{particle2020review}.
    \end{itemize}
    
    \item \textbf{Cold.} 
    According to cosmological observations, dark matter is cold, meaning its velocity is non-relativistic \cite{particle2020review}. This assumption is compatible with observations on cosmological scales. On smaller scales, certain models and observations predict "warm" dark matter in such a way that it prevents the buildup of high-density structures in the centers of galactic clusters \cite{tulin2018dark}.
 
    \item \textbf{Stable.} The lifetime of dark matter must be at least greater than the age of the universe \cite{audren2014strongest}.

\end{itemize}

\begin{figure}[h!]
\centering
\includegraphics[width=0.8\textwidth]{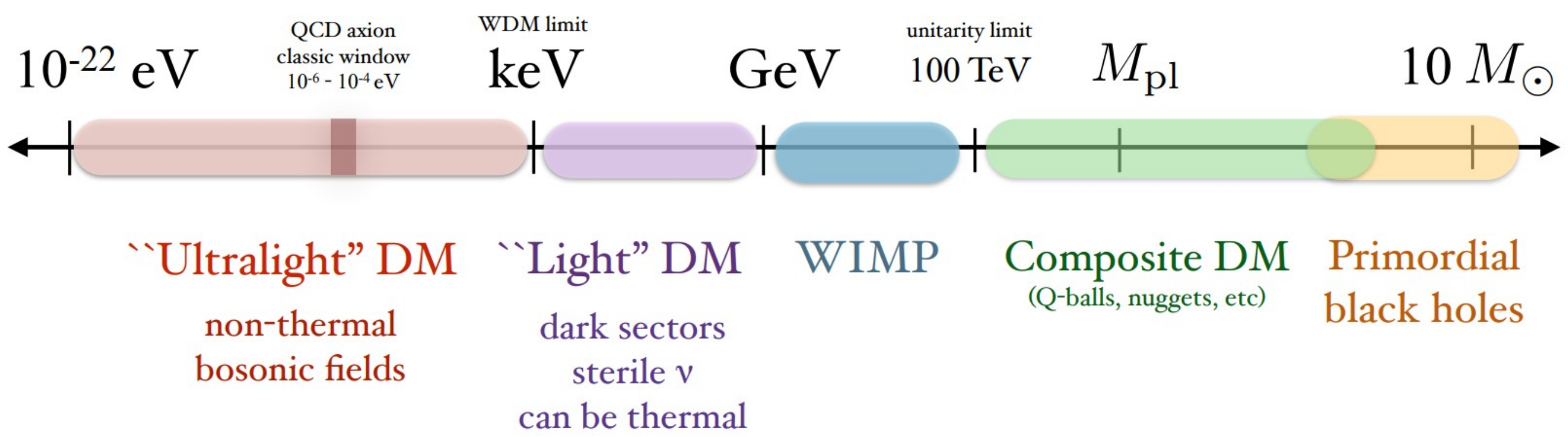}
\caption{Dark matter candidates for different mass ranges. Figure extracted from \cite{https://doi.org/10.48550/arxiv.1904.07915}.}
\label{fig.mass}
\end{figure}

\section{Dark Matter Production Mechanisms}
\label{sec:productionmechanisms}

There is consensus that the abundance of dark matter we see today originated in the early universe. Many mechanisms have been proposed to explain it, and here, some of them will be reviewed as they are relevant to this thesis.

\subsection{Freeze-Out}
\label{sec:freezeout}

\begin{figure}[h!]
\centering
\includegraphics[width=0.65\textwidth]{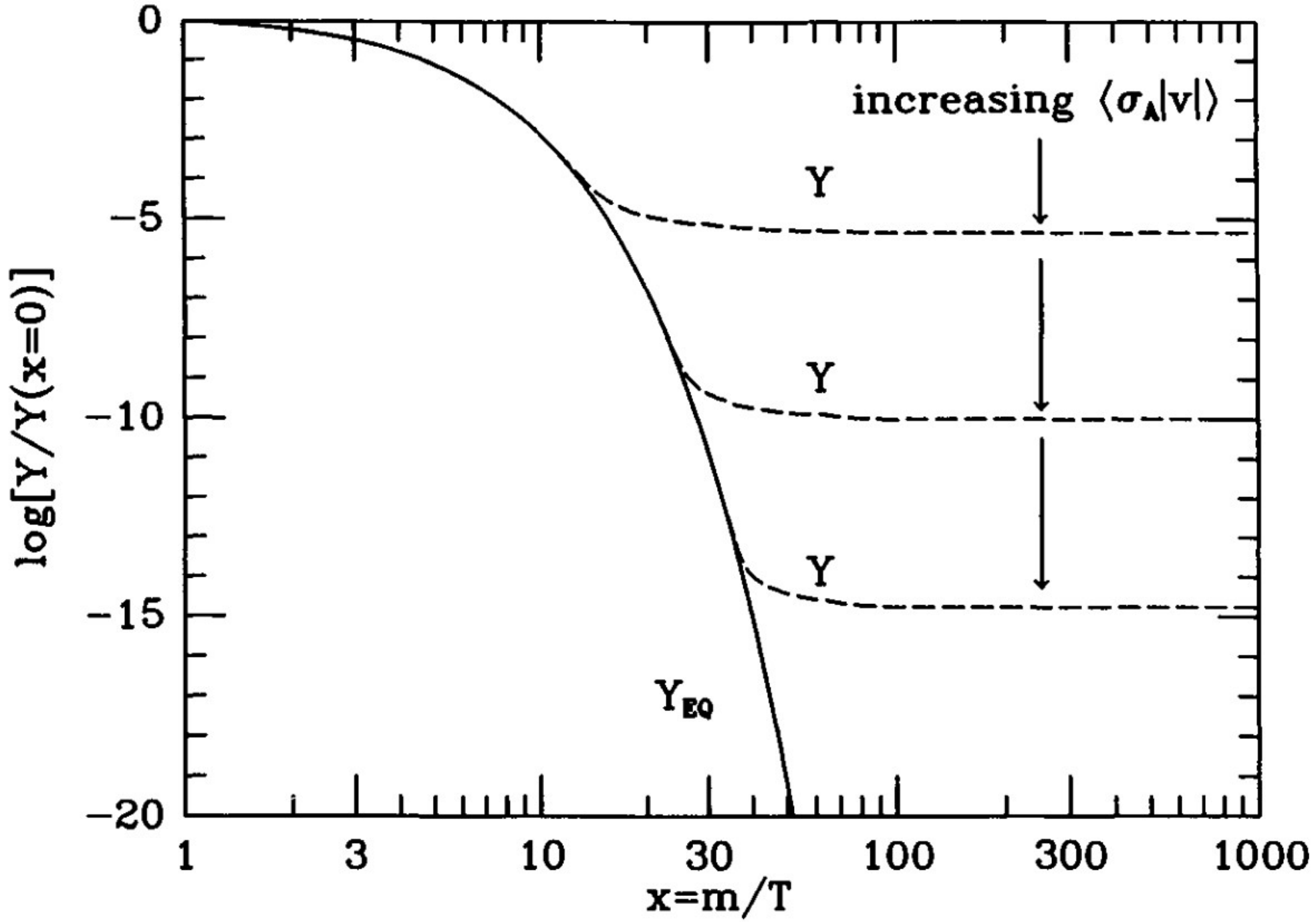}
\caption{Logarithm of the dark matter abundance versus the ratio between a single dark matter candidate's mass $m$ and the universe's temperature $T$. The abundance Y is defined as the ratio of dark matter density $n$ to entropy density $s$. Figure extracted from \cite{1990eaun.book.....K}.}
\label{fig.wimp1}
\end{figure}

In the context of "freeze-out," it is stated that dark matter was in equilibrium with the thermal bath that makes up the early universe and decoupled from it before recombination began. As it was in thermal equilibrium and there existed an interaction portal between baryonic matter and the dark sector, the pair annihilation processes of DM+DM $\longleftrightarrow$ SM+SM, where SM represents a particle of ordinary matter, were common. According to \cite{1990eaun.book.....K}, the number density $n$ of dark matter, under assumptions of homogeneity and isotropy, can be expressed as:

\begin{equation}
\frac{dn}{dt}-3Hn=-< \sigma v > \ (n^{2}-n_{eq}^2)
\label{eq.wimp1}
\end{equation}

\noindent where $H$ is the Hubble constant, $<\sigma v>$ is the mean cross-section of the process times velocity, and $n_{eq}$ is the equilibrium density. Dark matter decoupling occurs when the self-annihilation rate $\Gamma$ becomes comparable to the Hubble constant $H$, which represents the universe's expansion. When this happens, the interaction becomes very infrequent, and the abundance of dark matter "freezes out." For a non-relativistic dark matter candidate, the density $n$ in equilibrium can be expressed as:

\begin{equation}
n \sim \bigg( \frac{m_{DM}}{T} \bigg)^{3/2} \ e^{-\frac{m_{DM}}{T}}
\label{eq.wimp2}
\end{equation}

\noindent where $m_{DM}$ is the mass of dark matter, and $T$ is the temperature of the universe at a given time.
Figure \ref{fig.wimp1} from \cite{1990eaun.book.....K} shows how the abundance decreases as the fraction between mass and temperature, expressed as x, increases (for a fixed mass, this means that the temperature decreases), and then it remains quasi-static for low temperatures.
Dashed lines show three cases indicating how the abundance decreases for increasing $<\sigma v>$ values, meaning that, for a fixed velocity, a higher effective cross-section of the self-annihilation process implies a lower resulting dark matter abundance.
It is at this point where the "WIMP miracle" resides because values on the weak scale of $<\sigma v>$ and mass for a WIMP candidate (Weakly Interacting Massive Particle) coincide with the values needed to acquire the current abundance of dark matter present since the Big Bang.
It was not only the previous experience that particle physicists had in searching for particles on the weak scale (the discovery of W and Z bosons) but also the fact that WIMPs were capable of solving existing problems of the Standard Model, such as the hierarchy problem (the enormous scale difference between weak and gravitational forces), and that they arise as a natural extension of the Standard Model in theories like supersymmetry \cite{jungman1996supersymmetric}, which made WIMPs the most attractive dark matter candidate at the time of their formulation and in the subsequent decades.

\subsection{Freeze-In}
\label{sec:freezein}

The term "freeze-in" suggests that the abundance of dark matter did not originate from a prior overabundance of dark matter but rather started accumulating from very small quantities. Since this scenario considers that interactions between the dark sector and the visible sector are weak, it is said that dark matter is not in thermal equilibrium with baryonic matter, so the portal connecting them becomes available at lower temperatures, reaching its maximum at $T \sim m_{DM}$ and fading away at even lower temperatures as DM production is suppressed \cite{hall2010freeze}. The freeze-in process is of special relevance as it might be responsible for the creation of light dark matter candidates, which are one of the primary search targets of SENSEI.

\subsection{Asymmetric Dark Matter}
\label{sec:asymmetricdm}

This scenario suggests that, similarly to baryonic matter, there exists a dark matter candidate $\chi$ with its own antiparticle $\Bar{\chi}$ and an asymmetry between them that may be related to the baryonic asymmetry. If it is related and if

\begin{equation}
n_{\chi}-n_{\bar{\chi}} \sim n_{b}-n_{\bar{b}}
\label{eq.asymm}
\end{equation}

\noindent where $n$ is the density of a particular species ($\chi$ for dark matter and $b$ for baryons), then a direct relationship follows between their abundances, and knowing that the ratio of their densities $\frac{\rho_{DM}}{\rho_{B}} \sim 5$, one can estimate $m_{DM} \sim 5$ GeV \cite{zurek2014asymmetric}. Therefore, asymmetric dark matter can become a suitable candidate in the GeV range or even below for specific models (see \cite{lin2012symmetric} and its references).

\section{Candidates for Dark Matter}
\label{sec:candidates}

In this section, we will review some candidates for dark matter, with a particular focus on those that are relevant to this thesis: light dark matter (LDM) and dark photons.

\subsection{WIMPs}
\label{sec:wimpcandidate}

As mentioned earlier in Section \ref{sec:wimpcandidate}, WIMPs arise from the freeze-out mechanism because they produce the correct amount of dark matter abundance, interact weakly on the weak scale, and have a mass in the GeV-TeV range. Additionally, the introduction of WIMPs is theoretically motivated to alleviate tensions arising from the hierarchy problem between the weak force and gravitational force.

Since it became technologically possible, dozens of experiments have been constructed and operated to search for this particular candidate. In an experiment searching for WIMPs, the DM signal is typically produced through nuclear recoils induced by the DM candidate on the nucleus of the material used as the target (sometimes the same detector). This results in either light (scintillation), charge (ionization), phonons (vibrations/heat), or a combination of these three, as discussed in Section \ref{sec:directdetection}.

\begin{figure}[h!]
\centering
\includegraphics[width=0.85\textwidth]{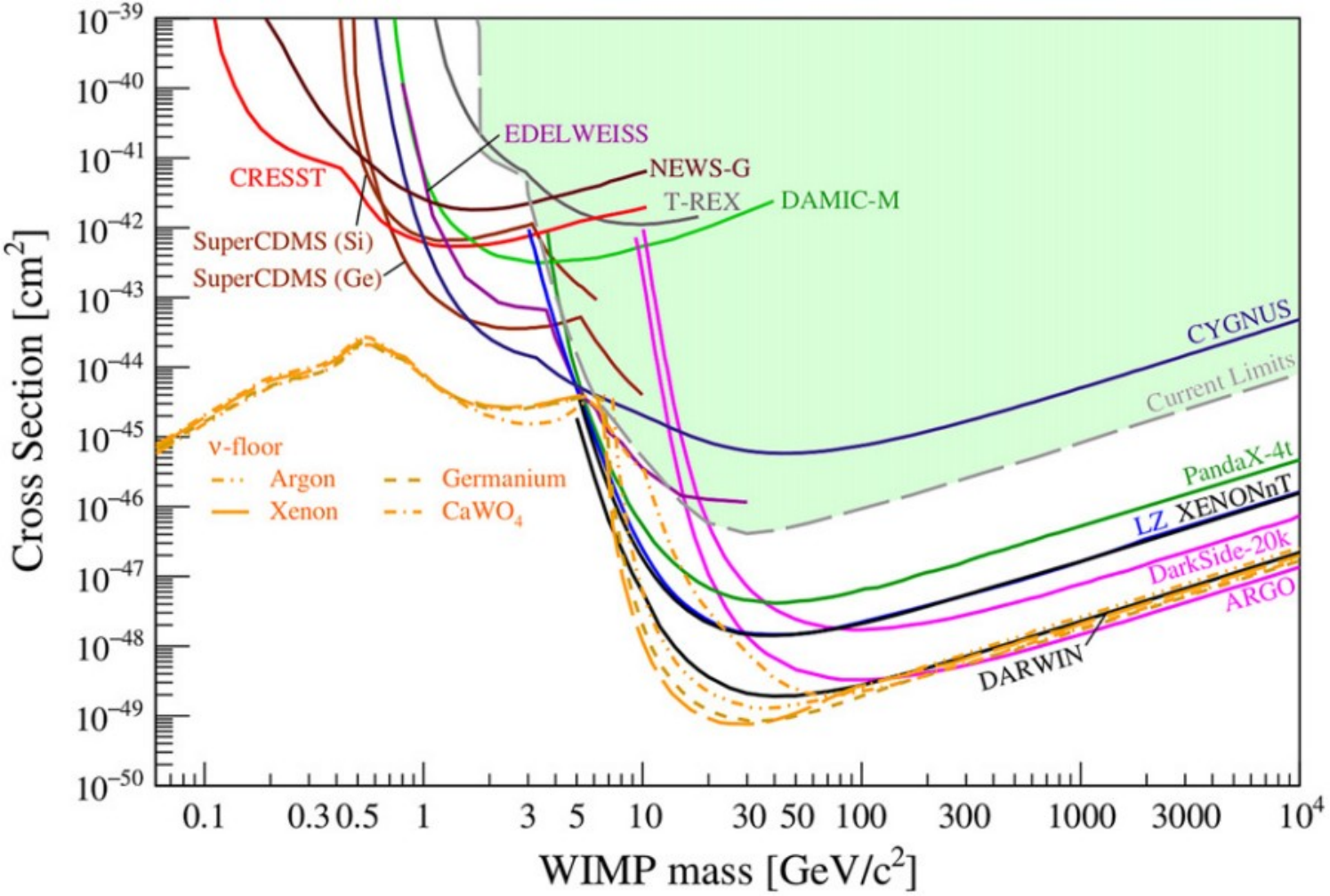}
\caption{Projections of 90$\%$ confidence sensitivity for spin-independent DM-nucleus scattering, assuming the dark matter candidate is a WIMP. Various experiments are presented, and exclusion limits as of April 2022 are shaded in green. In orange, the neutrino background for different materials is shown, limiting the search for sufficiently low cross-sections. Figure extracted from \cite{billard2022direct}.}
\label{fig.wimp2}
\end{figure}

To date, no unequivocal discovery of such a candidate has been made, and the search continues in the mass range of GeV-TeV. Figure \ref{fig.wimp2} shows exclusion limits for a WIMP-like candidate projected by different experiments, along with the current established limits. These limits represent regions in the parameter space where the presence of a dark matter candidate has not been found. The formalism used to establish these limits, taking SENSEI as an example, will be detailed in Section \ref{sec:limits}.

\subsection{Light Dark Matter (LDM)}
\label{sec:ldmcandidate}

Following Figure \ref{fig.mass}, Light Dark Matter (LDM) can exist in the range of keV-GeV, covering approximately 6 orders of magnitude. The theoretical framework that motivates the existence of LDM \cite{battaglieri2017us} allows for its production through various mechanisms, some of which have been described previously in Section \ref{sec:productionmechanisms}. LDM also motivates the existence of new mediators below the weak scale and within the dark sector, usually referred to as the hidden sector, to achieve the correct abundance \cite{https://doi.org/10.48550/arxiv.1904.07915}.

Although there are many mediators that can create a portal between both sectors, two cases are worth highlighting: the vector portal (where the dark photon resides) and the scalar portal (for a Higgs-like mediator). A candidate LDM can interact with baryonic matter (a detector) through these two portals and generate a measurable signal.

Because LDM is much lighter than WIMPs, the signal generated by the interaction of LDM with a nucleus is very small. This motivates the search for new interaction mechanisms to detect these candidates, with electronic recoils (the interaction of a dark matter candidate with the electrons of the detector) being one of the most popular and robust options.

It is important to note that while there are various LDM models that can be tested (see \cite{essig2012direct} and references therein), the approach used in the SENSEI experiment, as described in Subsection \ref{sec:electronrecoils} for dark matter scattering with electrons in silicon, allows for establishing limits on dark matter masses and cross-sections independently of the model.

\subsection{Dark Photons}
\label{sec:darkphotonsandidate}

A massive dark photon candidate, as introduced in Section \ref{sec:ldmcandidate}, can not only act as a mediator between the dark sector and the visible sector but also achieve the correct abundance of dark matter present since the Big Bang under certain assumptions, such as a very low coupling and a mass below twice the electron mass \cite{nelson2011dark}. Such a candidate can be absorbed by an electron in a given detector in a manner similar to how a photon from the Standard Model is absorbed: Through the photoelectric effect \cite{bloch2017searching, hochberg2017}. The probability of this process is mediated by the parameter $\epsilon$, which describes the strength of the mixing between dark photons and particles in the visible sector.

\subsection{Axion and Axion-Like Particles}
\label{sec:axioncandidate}

Below the keV scale, the most relevant candidates for dark matter are axions, which are theoretically motivated from the field of high-energy physics as a solution to the CP problem \cite{dine1981simple}. In particular, axions in the mass range required to account for the correct dark matter abundance are referred to as axion-like particles (ALPs). Due to their (potentially) very low mass, these candidates are expected to exhibit wave-like behavior.

\section{Dark Matter Detection Techniques}
\label{sec:dmtechniques}

\begin{figure}[h!]
\centering
\includegraphics[width=0.7\textwidth]{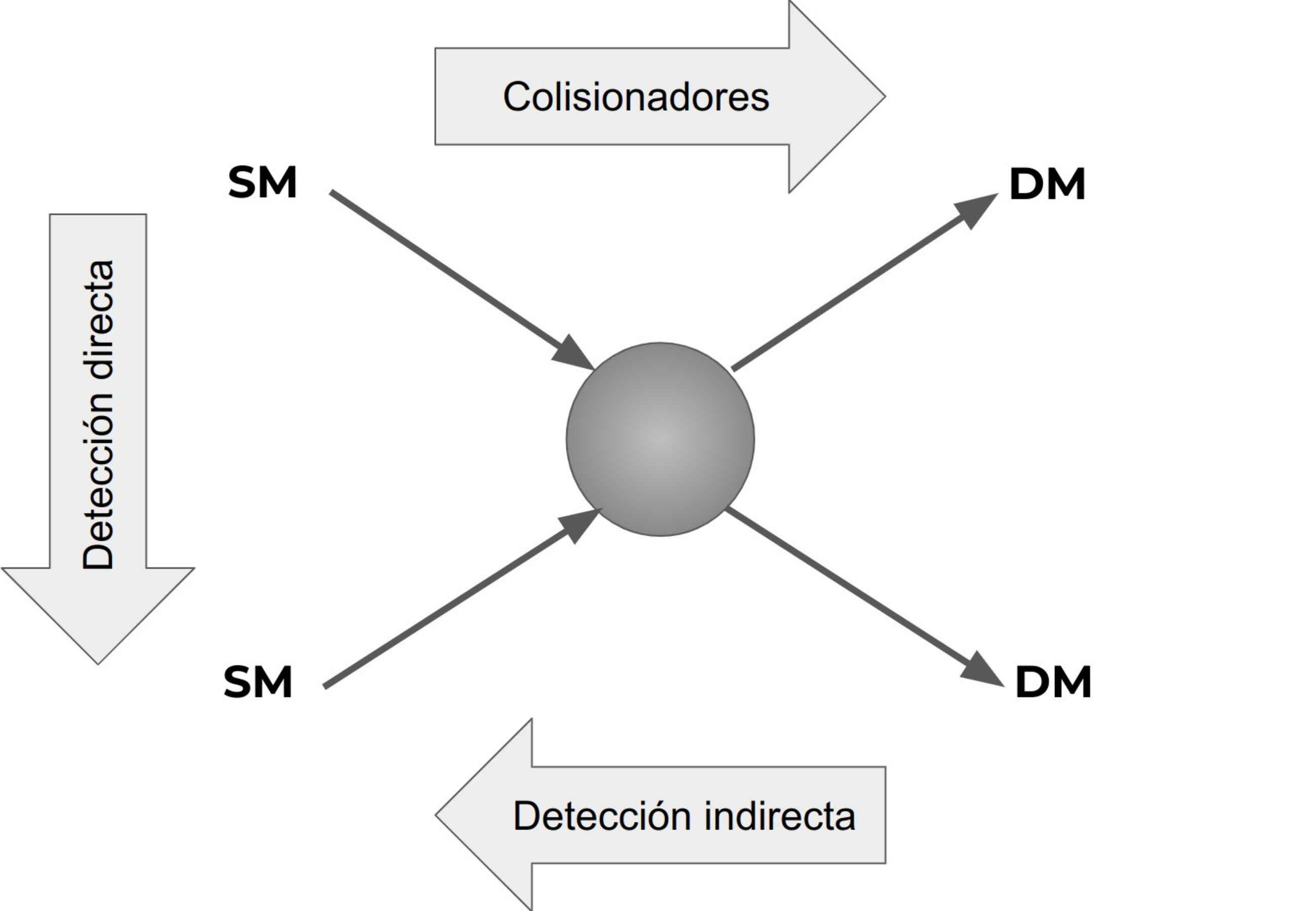}
\caption{Illustration in the form of a Feynman diagram showing the interaction channels between dark matter and Standard Model particles within each group of particles. The arrows indicate the direction of the interaction, and the name of each channel is provided.}
\label{fig.DMSM}
\end{figure}

In this section, three methods for detecting dark matter are discussed: detection at colliders, indirect detection, and direct detection. In this thesis, we will focus on the latter as it is the method used by SENSEI for dark matter searches.

\subsection{Detection at Colliders}

For dark matter searches at colliders, experiments look for missing energy signatures that result when Standard Model (SM) particles annihilate each other, creating, among other SM particles, dark matter. Since dark matter particles are by definition long-lived, stable, and have a very low probability of interaction, they escape the detector's reach, resulting in missing energy when measuring the interaction products. Similarly, this type of experiment can also test dark matter models in which, in the energy regime in which they operate, dark matter particles regenerate SM particles that are detectable.

\subsection{Indirect Detection}

Dark matter particles can decay or annihilate among themselves, creating a measurable signal of SM particles. This is the signal that indirect detection experiments are searching for and can be seen in the right-to-left direction in Figure \ref{fig.DMSM}. Since a significant portion of dark matter is found in galaxy clusters, telescope-based experiments are most relevant in this area. A signal of particular interest is the 3.5 keV X-ray signal observed in multiple galaxy clusters, the origin of which is still unknown \cite{bulbul2014detection}.

\subsection{Direct Detection}
\label{sec:directdetection}

\begin{figure}[h!]
\centering
\includegraphics[width=0.9\textwidth]{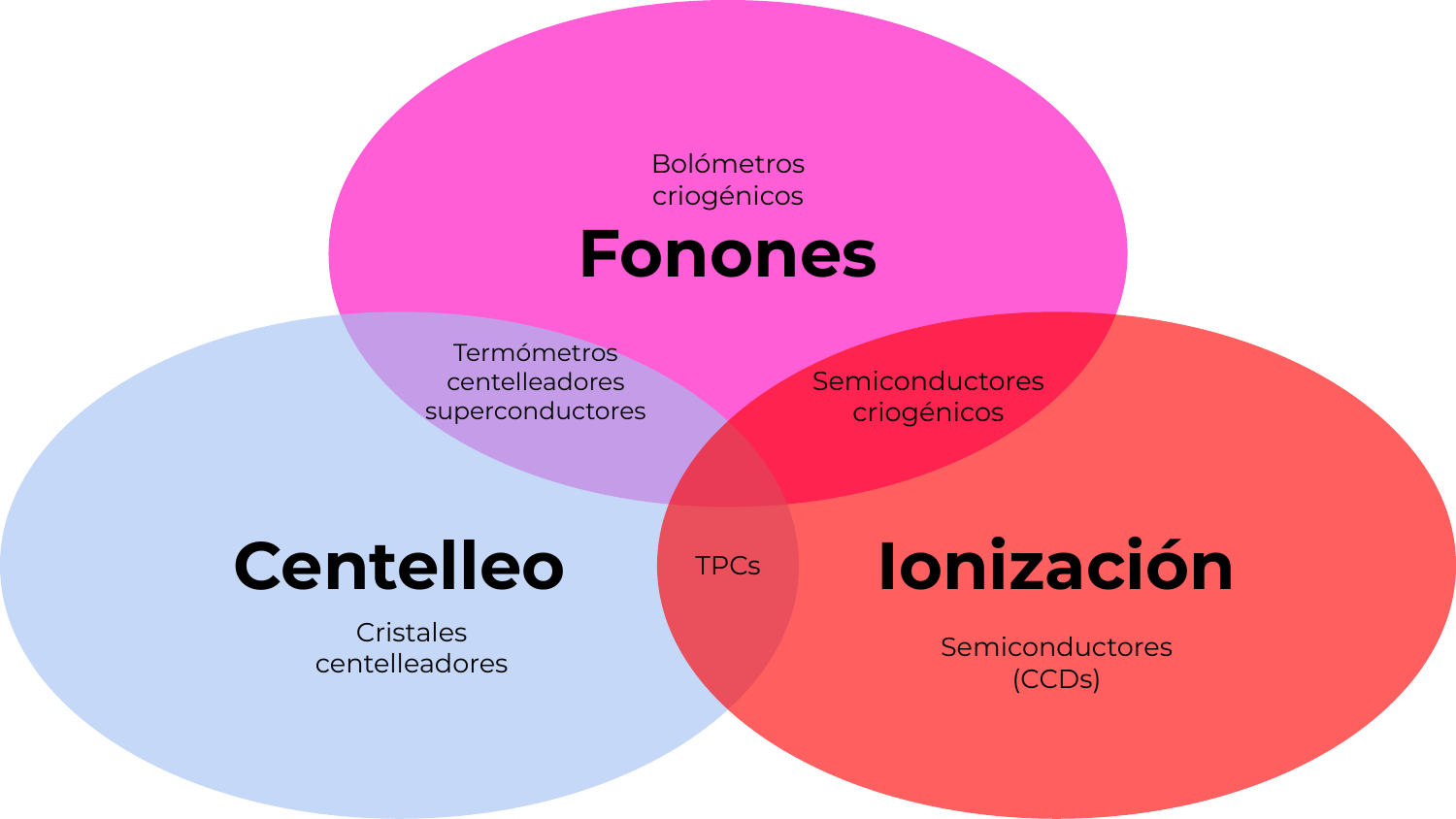}
\caption{Venn diagram illustrating the different signals produced in a detector when a particle interacts with it and some of the detectors used to capture these signals.}
\label{fig.signalgeneration}
\end{figure}

Following the direction from top to bottom in Figure \ref{fig.DMSM}, direct detection aims to detect a measurable signal generated when a dark matter candidate interacts directly with a SM particle. As mentioned earlier, this interaction can leave three types of signals in the detector: charge (ionization), phonons (vibrations/heat), or light (scintillation).
Figure \ref{fig.signalgeneration} presents a Venn diagram illustrating these three signals and some of the detector types used by dark matter search experiments to capture these signals.
In particular, for charge-coupled devices (CCDs), dark matter would ionize charges in the detector, which would then be collected by the same detector. However, not all the energy deposited in the detector is directly converted into charges. After a nuclear recoil, a significant portion of the energy will be converted into phonons, which inevitably dissipate into the detector material as heat.
A quenching factor can be calculated to account for the relationship between the amount of energy collected in the form of charge and the amount of deposited energy \cite{lindhard1963integral,chavarria2016measurement}. However, this relationship is unknown below 700 eV, and it may even be negligible at very low energies. In contrast, when energy is transferred to an electron, the efficiency with which it is used to ionize other electrons in the crystal lattice is much higher \cite{FANO,rodrigues2023unraveling}. This process is well-modeled, even at very low energies \cite{ramanathan}.

Some experiments were designed to measure more than one type of signal. This is the case for SuperCDMS (which measures both ionized charges and phonons) and XENON1T (which measures scintillation and ionized charges).

Direct detection searches operate under the assumption that dark matter particles can interact (beyond gravitational interaction) with particles in the visible sector, which could, in principle, be false. Similar assumptions are made for indirect detection and collider methods.
Although the direct detection method aims to be independent of the dark matter model (and it is, in the sense that its final exclusions do not depend on the model being tested), it does depend on certain parameters, such as the dark matter density, the average velocity, the escape velocity, among others. In other words, experiments need to know how dark matter flows around the Earth, based on cosmological and astronomical observations.

There are three types of DM-SM interactions to mention: DM-nucleus scattering, DM-electron scattering, and DM absorption.
The first one involves a scattering interaction where the DM candidate interacts with the nucleus of an atom in the target material, generating a signal (as introduced in Section \ref{sec:wimpcandidate}: light, charge, or phonons) detectable by a sensor near the target material.
The second one is similar to the first, but the scattering occurs between the DM candidate and the electrons in the detector material.
Finally, the third one involves the absorption of the DM candidate by electrons in the material, similar to the photoelectric effect for ordinary particles.
In the following sections, we will focus on the last two interactions, DM-electron scattering, and DM absorption since these are the ones used by the SENSEI experiment to search for dark matter.

\section{Light Dark Matter Electron Scattering}
\label{sec:electronrecoils}

As mentioned previously, WIMP searches have been very successful in excluding a large area of the parameter space of interest, particularly in the GeV-TeV range (see Section \ref{sec:wimpcandidate} and Figure \ref{fig.wimp2}). However, despite these extensive efforts, no WIMP candidate for dark matter has been discovered. This motivated the exploration of different theoretical models that could explain the known abundance of dark matter in the universe using a lighter dark matter candidate.

\subsection{Light Dark Matter Interaction Models}

The exploration of LDM models (see Section \ref{sec:ldmcandidate}) seemed promising not only due to the feasibility of their implementation but also because there were already experiments (such as XENON10 \cite{zeplin} and ZEPLIN-II \cite{xenon10}) that had the technology available to search for this type of candidate. However, the traditional DM-nucleus scattering channel became insufficient for detecting the small energy deposits that a MeV-scale DM candidate would leave in a target material after an interaction. As shown in \cite{essig2012direct}, the energy deposited by nuclear recoil is

\begin{equation}
    E_{nr}=\frac{q^{2}}{2 m_{N}} \sim 1 \ {\rm eV} \cdot \Big(\frac{m_{DM}}{100 \ {\rm MeV}} \Big)^{2} \cdot \Big(\frac{10 \ {\rm GeV}}{m_{N}} \Big)
    \label{eq.enr}
\end{equation}

where $m_{N}$ is the mass of the nucleus, and $q$ is the momentum transferred during the scattering (typically around $10^{-3} \ c$ since $q \sim m_{DM} \ v$). For a xenon detector (a typical target material for WIMP searches), the expected recoil energy for a 10 GeV DM candidate with a nucleus is approximately 2 keV.
However, if we attempt to explore the MeV mass range, the expected recoil is on the order of $\sim$ eV, making it unattainable for such experiments even when the total energy deposited by the interaction, which can be expressed as

\begin{equation}
    E_{total} \sim \frac{m_{DM} \ v^{2}}{2} \sim 50 \ {\rm eV} \cdot \Big( \frac{m_{DM}}{100 {\rm MeV}}\Big)
    \label{eq.energythreshold1}
\end{equation}

is above the minimum energy threshold in xenon (or silicon) for a 10 MeV (1 MeV) candidate, as will be shown below.
In the case of DM-electron scattering, the maximum allowed transferred energy can be expressed as \cite{essig2016direct}

\begin{equation}
    E_{e} \leqslant \frac{1}{2} \ {\rm eV} \cdot \Big( \frac{m_{\chi}}{{\rm MeV}}\Big)
    \label{eq.energythreshold2}
\end{equation}

so that for a xenon target, with a threshold energy as low as $\sim$12.4 eV, this process can create a detectable signal for DM masses as low as tens of MeV \footnote{These are, of course, simplified models, and the complete calculations (and how to perform them for different materials) are developed in \cite{essig2012direct} and \cite{essig2016direct}.}.
Furthermore, semiconductors such as germanium or silicon offer an energy threshold of 0.7 and 1.1 eV, respectively, making them very attractive detectors for DM-electron scattering searches.

To establish exclusion limits as in WIMP searches (see Figure \ref{fig.wimp2}), a reference cross-section between electrons and dark matter candidates must be used to parameterize the interaction's strength. The following expression, based on the non-relativistic interaction of a particle with a free electron, is derived in \cite{essig2012direct}:

\begin{equation}
    \bar{\sigma_{e}} \equiv \frac{\mu^{2}_{DM e}}{16 \pi m^{2}_{DM} m^{2}_{e}} \ \overline{|\mathcal{M}_{DMe}(q) |^{2}} \bigg\rvert_{q^{2} = \alpha^{2} m_{e}^{2}}
    \label{eq.crosssection1}
\end{equation}

where $\mu^{2}_{DM e}$ is the reduced mass of the interaction, and the variable $q$ is the momentum transfer vector, fixed here at $\alpha m{e}$, where $\alpha$ is $e^{2} / 4 \pi$. $\overline{|\mathcal{M}_{DMe}(q) |^{2}}$ is the absolute square of $\mathcal{M}$, the matrix element for dark matter-electron scattering, averaged over initial states and summed over final spin states, and follows

\begin{equation}
    \overline{|\mathcal{M}_{DMe}(q) |^{2}} \equiv \overline{|\mathcal{M}_{DMe}(\alpha m_{e}) |^{2}} \times |F_{DM}(q)|^{2}
    \label{eq.matrix}
\end{equation}

where $|F_{DM}(q)|$ is the form factor of the dark matter interaction giving the momentum transfer dependence of the scattering.
Two particular possibilities are considered for the SENSEI experiment: $F_{DM}=1$ (where a heavy vector mediator induces the interaction) and $F_{DM}=(\frac{\alpha m_{e}}{q})^{2}$ (an interaction mediated by a massless or ultra-light vector mediator).
In summary, the cross-section expressed in Equation \eqref{eq.crosssection1} depends on the form factor (type of exchange mediator), the matrix elements for that particular interaction, and primarily the mass of the dark matter candidate.
The last piece necessary to establish exclusion limits is the rate or, more specifically, the differential scattering rate. The detailed procedure for obtaining this expression from a semiconductor target is shown in \cite{essig2016direct} and leads to the following result:

\begin{equation}
\frac{\textit{d}R_{crystal}}{\textit{d} ln E_{e}} \ = \ \frac{\rho_{\chi}}{m_{\chi}} \ N_{cell} \ \overline{\sigma}_{e} \alpha \times \frac{m^{2}_{e}}{\mu^{2}_{DMe}} \int \textit{d} ln \ q \Big( \frac{E_{e}}{q} \eta(v_{min}(q,E_{e})) \Big) |F_{DM}(q)|^{2} |f_{crystal}(q,E_{e})|^{2}
\label{eq.drate}
\end{equation}

where $\rho_{DM}$ is the local dark matter density, $N_{cell}$ is the number of cells in the crystal used as the target, $\eta$ is a function carrying information about the dark matter velocity distribution, and $|f_{crystal}(q,E_{e})|$ is the crystal's form factor, providing information about the electronic structure of the crystal.

\subsection{Expected Event Rate}

\begin{figure}[h!]
\centering
\includegraphics[width=0.7\textwidth]{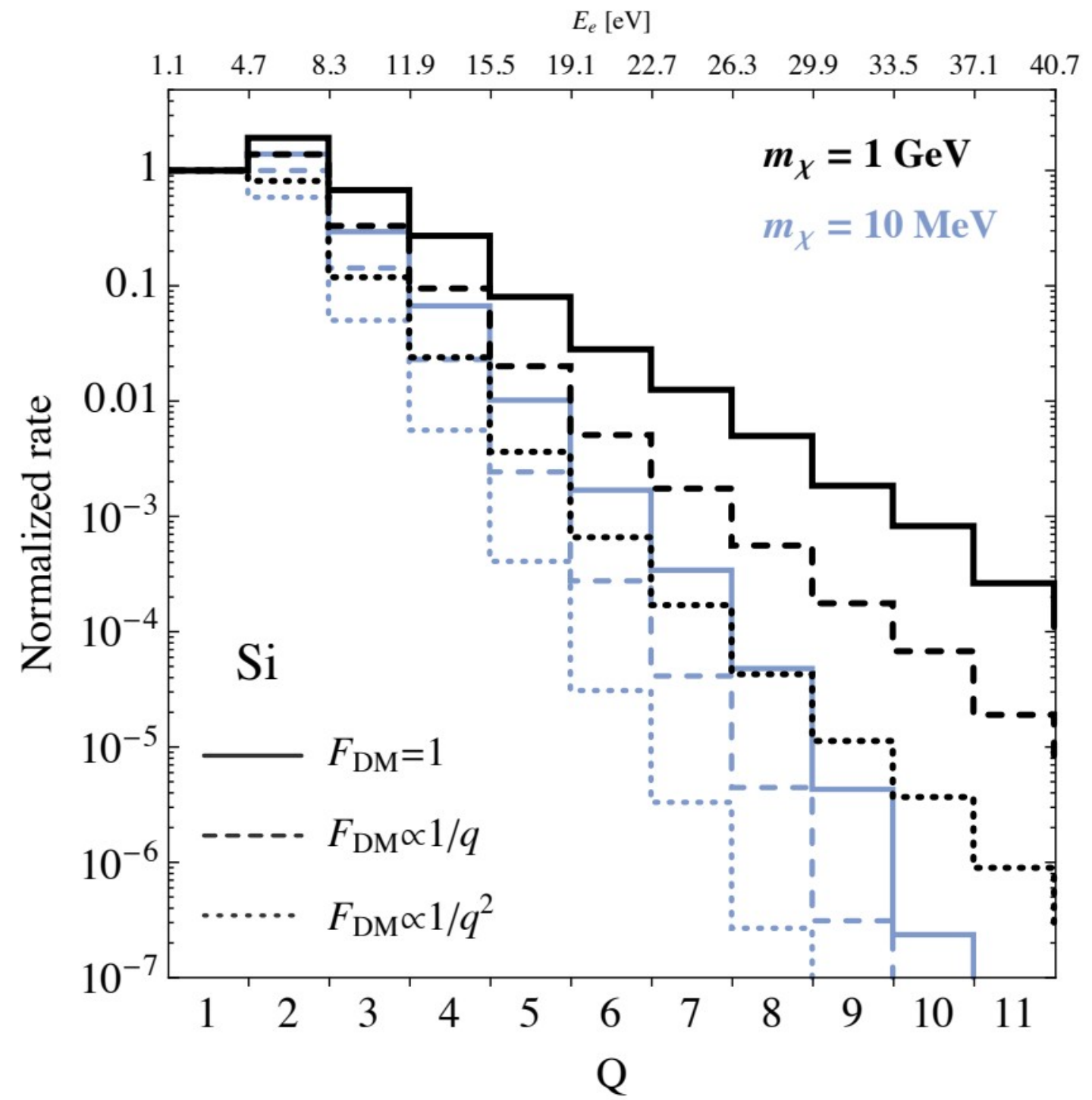}
\caption{Normalized expected event rate as a function of ionization charge in a silicon sensor for two dark matter candidate masses (1 GeV in black and 10 MeV in blue) and three different dark matter form factors. The spectrum is normalized to 1 according to the value in the first bin. Figure extracted from \cite{essig2016direct}.}
\label{fig.expectedrates}
\end{figure}

Given Equation \eqref{eq.drate}, the required values for a given material, and the local density and velocity of the dark matter halo, the expected event rate can be calculated for a specific cross-section $\bar{\sigma}_{e}$ and dark matter mass. Figure \ref{fig.expectedrates} shows the normalized expected event rate in silicon for two masses (1 GeV and 10 MeV) and three different mediators (i.e., dark matter form factors).
The x-axis can be interpreted as deposited energy (upper axis) or, equivalently, as ionization signal in electrons (lower axis). The equivalence between these two is established as follows:

\begin{equation}
Q(E_{e})=1+[(E_{e}-E_{gap})/\varepsilon]
\label{eq.Q}
\end{equation}

where $\varepsilon$ is the energy for electron-hole pair production, and $E_{gap}$ is the band-gap of silicon.
Equation \eqref{eq.Q} is a simplified model of a complex chain of secondary scattering interactions that occur during the ionization process. For more realistic models translating the measured charge $Q$ into the total deposited energy $E_{e}$, one can refer to the work published by Ramanathan et al. \cite{ramanathan}.

\section{Dark Matter Absorption}
\label{sec:dmabsorption}

Certain dark matter candidates, such as ALPs and dark photons (see Sections \ref{sec:axioncandidate} and \ref{sec:darkphotonsandidate}, respectively), can be absorbed by the detector material rather than being scattered. During this process, the entire energy of the dark matter candidate is absorbed by the detector through the absorption by an electron (specifically, a valence-band electron, for semiconductors like silicon) which is then captured, analogous to the photoelectric effect.
Considering silicon as the material and a non-relativistic flux of dark photons or ALPs, masses as low as 1.1 eV, the minimum energy required to generate a detectable signal in silicon, can be explored.

We will limit ourselves to the case of dark photons, as explored by SENSEI, but a similar approach can be followed for ALPs, developed for silicon (among other relevant targets) in \cite{bloch2017searching} and \cite{hochberg2017}.

\begin{figure}[h!]
\centering
\includegraphics[width=0.7\textwidth]{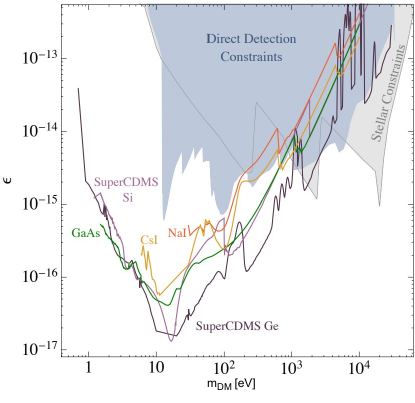}
\caption{Dark matter exclusion limits for dark photon absorption. Figure extracted from \cite{bloch2017searching}. The shaded region represents the limits at the time of the referenced work in 2017, and the colored lines represent projected limits assuming 10 kg-years of exposure and no measured signal.}
\label{fig.darkphoton}
\end{figure}

The dark matter absorption rate per atom for a given material used as the target is defined as \cite{bloch2017searching}:

\begin{equation}
R = \frac{\rho_{DM}}{m_{DM}} \epsilon^{2} \sigma_{PE}(m_{DM})
\label{eq.rate}
\end{equation}

where $\sigma_{PE}$ is the photoelectric cross-section, evaluated at the energy of the incident dark photon (in this case, non-relativistic mass), and $\epsilon$ is the coupling between the Standard Model photon and the dark photon, often called the kinetic-mixing parameter. Figure \ref{fig.darkphoton} shows sensitivity projections for various materials \cite{bloch2017searching}.
Silicon, labeled as "SuperCDMS Si," was calculated for a total exposure of 10 kg-years. It can be seen that the best performance for this material is around 10 eV, six orders of magnitude below the mass range covered for LDM scattering (see Figure \ref{fig.senseiprojectedscattering}).

\section{Dark Matter Exclusion Limits}
\label{sec:limits}

Based on the models developed earlier, the search for dark matter can be conducted by reducing the dimensionality of the problem to two parameters: the effective interaction cross-section and the mass of the candidate being sought. This exploration of the parameter space is accomplished by establishing exclusion limits on the cross-section for each possible mass of the candidate.
Figure \ref{fig.expectedrates}, as mentioned previously, depicts the number of expected events in a silicon target for two masses and three types of mediators. This profile scales linearly with the cross-section while maintaining its shape. To establish the exclusion limit for a given mass and mediator, the profile of expected events is compared to the measured profile. This comparison can be performed bin by bin or by summing the expected events for all bins of interest.
For the case of a heavy mediator ($F_{DM}=1$) and a mass of 1 GeV, for instance, one can observe that the majority of expected events fall between 4.7 and 8.3 eV, corresponding to the 2-electron channel. Additionally, it can be noted from Figure \ref{fig.expectedrates} that the number of expected events decreases exponentially as a function of the number of ionization-produced charges $Q$.
The limits on the cross-section are then established bin by bin, ranging from 1 to 4 electrons, obtaining a limit across the entire mass range for each energy bin.

\subsection{Exclusion Limit for a Given Energy Channel}

Taking the example of the 1-electron channel and the scattering of a dark matter candidate with electrons in the target material, we start with the assumption that $e_{1}$ events were measured for a detector exposure time of $t_{1}$. Additionally, since the backgrounds to the measured signal are initially unknown, it is assumed that all measured events originate from a dark matter interaction. Considering the local uniformity of the Earth's dark matter halo and the extremely low expected interaction rate, the quantity of measured events $e_{1}$ is assumed to follow a Poisson distribution.

Therefore, the upper limit of the frequentist confidence interval constructed with a 90$\%$ confidence level for $e_{1}$ is obtained as the value of $\eta_{1}$ such that:

\begin{equation}
\sum_{k>=0}^{e_{1}} P(k,\eta_{1}) = \frac{\eta_{1}^{k} e^{-\eta_{1}}}{k!} \leq 0.1
\label{eq.limit}
\end{equation}

After determining $\eta_{1}$, the cross-section for a given mass is extracted from Figure \ref{fig.expectedrates}, matching the first bin with $\eta_{1}$ events, and this cross-section is denoted as $\sigma_{1}$. The same process is repeated for the 2, 3, and 4-electron channels, yielding values for $\sigma_{2}$, $\sigma_{3}$, and $\sigma_{4}$. Finally, the lowest value of $\sigma_{i}$ is taken as the final limit for the given cross-section and mass.

This process is then repeated for each of the masses to obtain exclusion limits with a 90$\%$ confidence level across the entire range of masses of interest, as well as for all models of dark matter to be studied (heavy mediator, light mediator, and dark photon).

\begin{figure}[h!]
\centering
\includegraphics[width=0.98\textwidth]{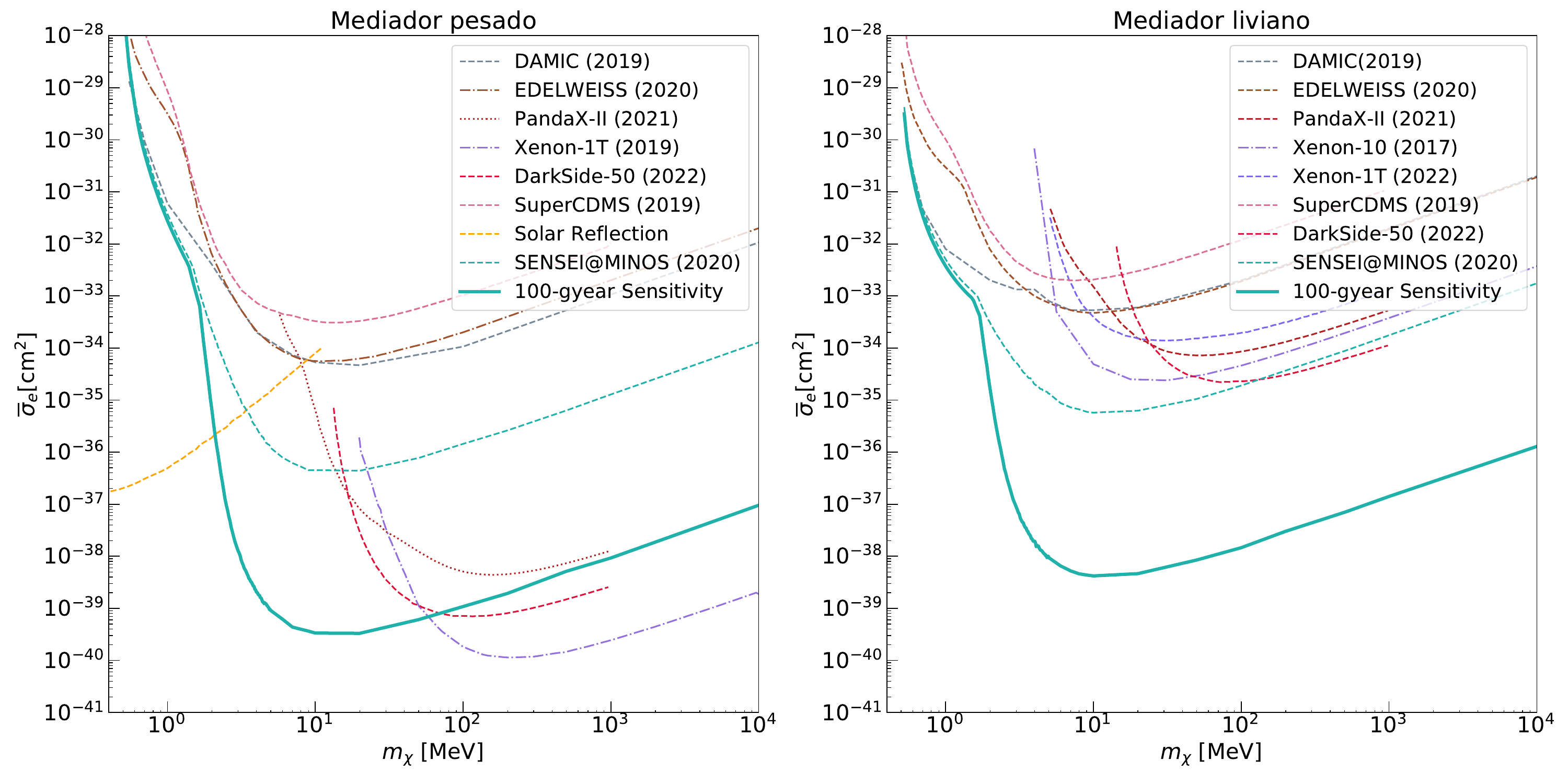}
\caption{Projected exclusion limits for dark matter scattering. On the left is the heavy mediator case ($F_{DM}=1$), and on the right is the light mediator case ($F_{DM}=(\frac{\alpha m_{e}}{q})^{2}$). The limits established to date by DAMIC \cite{DAMIC2019}, EDELWEISS \cite{EDELWEISS2020}, PandaX-II \cite{PANDA2021}, XENON-1T \cite{XENON1T2019}, DarkSide-50 \cite{agnes2023search}, SuperCDMS \cite{SUPERCDMS2018}, and solar reflection \cite{solarreflection} are shown. The projection was made using results obtained from SENSEI in 2020 \cite{SENSEI2020} and scaled for 100 grams-day, maintaining the reported efficiencies for each energy channel.}
\label{fig.senseiprojectedscattering}
\end{figure}

\begin{figure}[h!]
\centering
\includegraphics[width=0.8\textwidth]{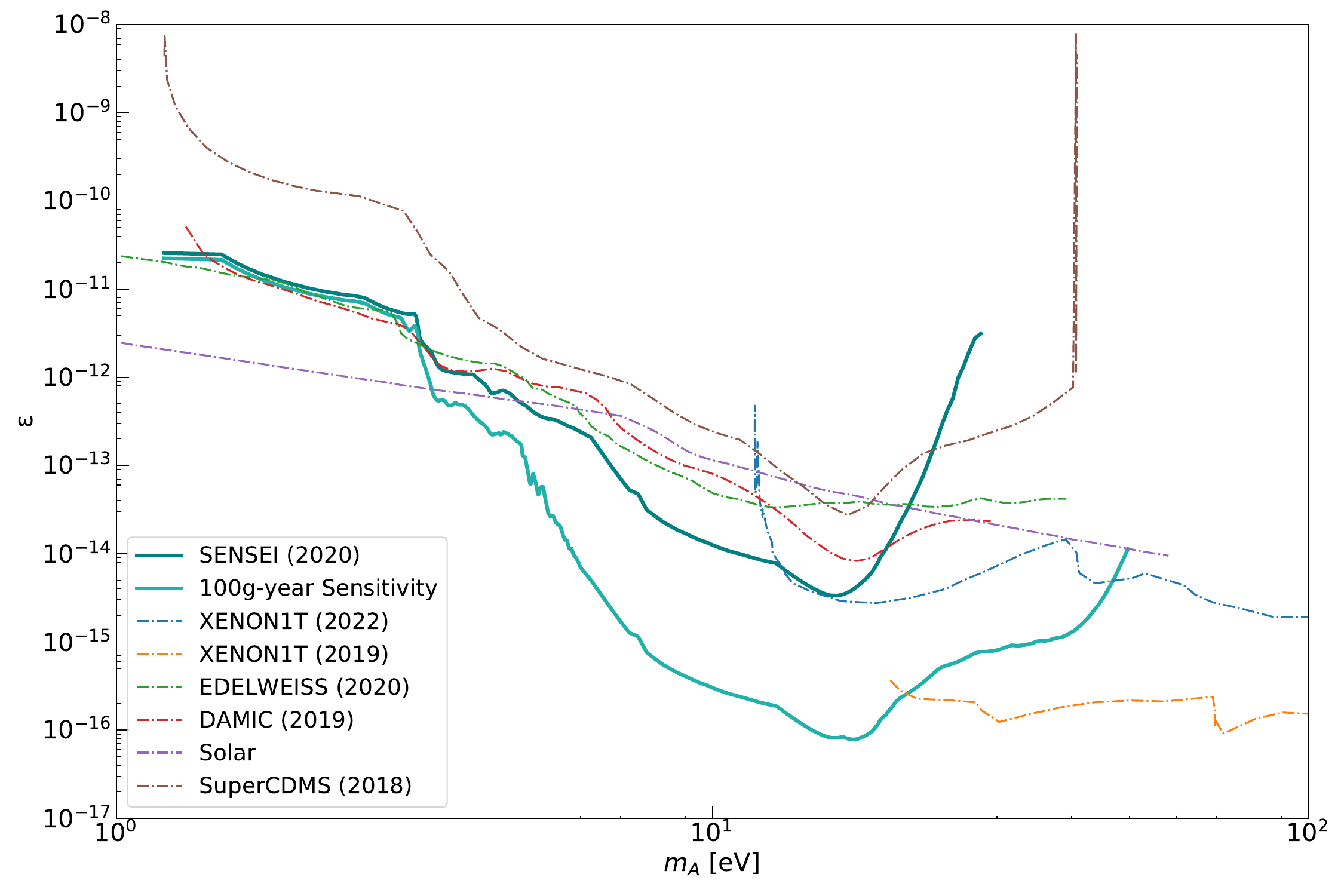}
\caption{Projected exclusion limits for dark photon absorption. The limits established by XENON1-T in 2019 \cite{XENON1T2019} and 2022 \cite{XENON1T2022}, EDELWEISS \cite{EDELWEISS2020}, DAMIC \cite{DAMIC2019}, SuperCDMS \cite{SUPERCDMS2018}, and the Sun \cite{bloch2017searching,an2013new,redondo2013solar} are shown. As in Figure \ref{fig.senseiprojectedscattering}, the projection was made using data obtained from SENSEI in 2020 \cite{SENSEI2020} and scaled for 100 grams-day of exposure, keeping the efficiencies of each energy channel fixed.}
\label{fig.senseiprojectedabsorption}
\end{figure}

Figures \ref{fig.senseiprojectedscattering} and \ref{fig.senseiprojectedabsorption} display projected dark matter exclusion limits for SENSEI, considering 0 events and utilizing an exposure of 100 grams-year (i.e., a 100-gram device exposed for one year) for dark matter scattering and dark photon absorption, respectively.

\clearpage
\newpage
\newpage
\mbox{}
\thispagestyle{empty}
\newpage

\chapter{CCDs and the Skipper technology}
\label{cap:2}

In this chapter, we will cover the fundamental concepts of CCDs (Charge-Coupled Devices) and Skipper-CCDs (SCCDs), with a specific focus on their applications in the search for dark matter. We will start with a technical description of the CCD, including its structure and operation, with an emphasis on the readout device. Following that, we will introduce Skipper technology, which enables precise measurement of the charge collected by the detector. Finally, we will discuss the impact of this technology on the search for dark matter and the objectives of the SENSEI experiment.

\section{CCDs}

Charge-Coupled Devices (CCDs) were invented in 1969 by Willard S. Boyle and George E. Smith \cite{boylenobel} as memory devices, but they turned out to be useful as light detectors, finding applications in photography and spectroscopy, especially for astronomical purposes. The Hubble Space Telescope is one of the most notable examples of their use.
More recently, CCDs have been adopted as particle detectors in the search for dark matter (see the DAMIC collaboration \cite{DAMIC2016}) and the coherent elastic scattering of neutrinos with atomic nuclei (CONNIE collaboration \cite{CONNIE2016}). In particular, DAMIC has proven to be competitive for WIMP-like dark matter candidate searches, pushing mass exclusion limits up to 1 GeV.
In the following subsections, we will introduce the different components of a CCD, its history, and its operating principles.

\subsection{The p-n Junction and MOS Capacitor}

The heart of the CCD lies in the p-n junction, an architecture designed to exploit the nature of semiconductors. These materials have the particular characteristic of being insulators that, above certain temperatures, can effectively conduct electricity, acting as conductors \cite{ashcroft28and29}. From a physics standpoint, an insulator is defined as a material whose energy bands are completely full or empty at $T = 0K$, as only partially filled energy bands contribute to conduction \cite{ashcroft12}. Thus, the band gap is defined as the energy between the top of the highest filled band (valence band) and the bottom of the lowest empty band (conduction band) and is, therefore, the energy that needs to be overcome for an electron (or hole) to move conductively within the semiconductor material. This energy to overcome the band gap can be found, in particular, in thermal excitations or electric fields applied to the material: a sufficiently hot electron (or electrically polarized) can have enough energy to cross the band gap from the valence band to the conduction band, mimicking the behavior of a metal. This is the case for semiconductors: materials that are insulators at $T = 0K$ but have a small enough band gap ($\sim$1.1 eV for silicon) that, at sufficiently high temperatures or electric fields, behave like metals.

\begin{figure}[h!]
\centering
\includegraphics[width=0.5\textwidth]{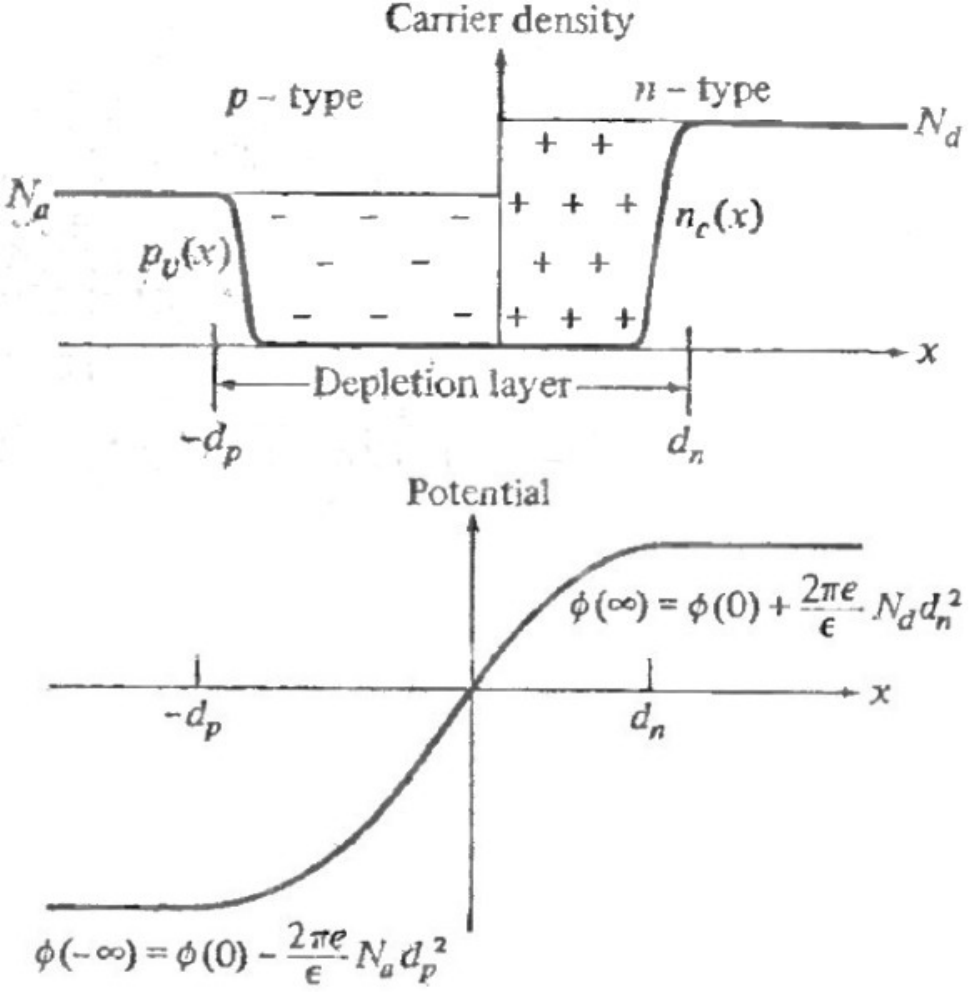}
\caption{Carrier density (top) and electric potential (bottom) as a function of position across the p-n junction. Figure extracted from Chapter 29 of \cite{ashcroft28and29}.}
\label{fig.pndiode}
\end{figure}

At the same time that an electron is promoted from the valence band to the conduction band, a hole undergoes the reverse process, ensuring that the number of holes and electrons available in the semiconductor remains constant over time. Conductivity increases as more electrons are present in the conduction band, a process that can be enhanced by adding impurities to the material.
These impurities replace a silicon atom with another atom called a donor or acceptor, depending on whether it can donate an unbound electron to the material or the opposite.
In other words, donors add electrons to the conduction bands, and acceptors add holes to the valence band, both increasing the conductivity of the material. An example of a donor (acceptor) for silicon is phosphorus (boron)\footnote{Note that adding a boron atom to a silicon crystal adds an electron to the conduction band, but since boron has one more proton than silicon, the crystal remains electrically neutral.
The reverse process applies to acceptors like phosphorus.}.
Since the dopant (donor or acceptor) is embedded in the crystal, its binding energy is very low, and, more importantly, lower than the semiconductor's band gap.
It is then very likely that charge carriers added through doping will be promoted to the conduction band in the case of donors (or the reverse process for acceptors), increasing the material's conductivity.

\begin{figure}[h!]
\centering
\includegraphics[width=0.95\textwidth]{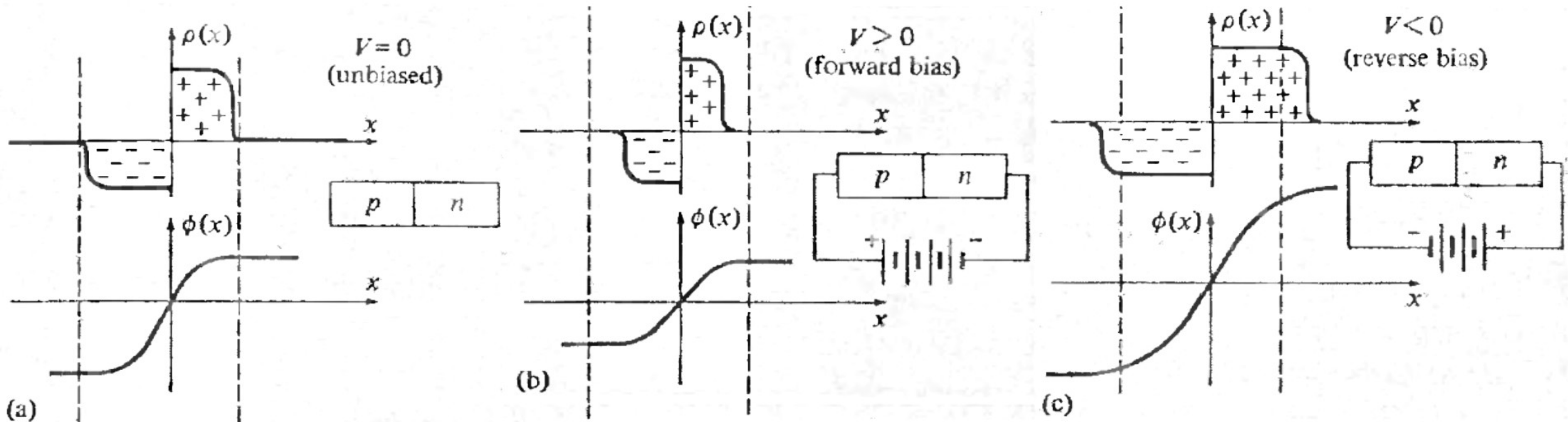}
\caption{Charge density $\rho$ and electric potential $\phi$ in the depletion layer for three cases: (a) without applying any external voltage, (b) when applying a positive bias voltage (\textit{forward bias}), and (c) when applying a negative bias voltage (\textit{reverse bias}). Figure extracted from Chapter 29 of \cite{ashcroft28and29}.}
\label{fig.forwardbias}
\end{figure}

In a CCD, the level of doping is precisely manipulated during manufacturing to create the fundamental structures that allow for the collection, transfer, and reading of collected charges.
The most important part of a CCD (and most semiconductor devices) is the p-n junction, which consists of a spatially controlled specific doping profile so that one part of the semiconductor is doped with acceptors (p-type) and an adjacent part with donors (n-type).
As a result, at the junction, and after reaching thermal equilibrium, the donors and acceptors (i.e., free electrons and holes) will combine with each other so that the p-type region will be predominantly populated by negative ions, and the n-type region by positive ions (see the upper image in Figure \ref{fig.pndiode}).
Although the entire doped volume together is electrically neutral, an electric field is generated (from n-type to p-type, see the lower image in Figure \ref{fig.pndiode}), creating a diode-type junction, an electronic structure that only allows (for very low external fields) the passage of charge particles in a specific direction.
This volume is called the depletion layer\footnote{It is often called a layer because it is assumed to be a few nanometers wide, although we will see that for fully depleted CCDs, its width can be almost a millimeter.}, and it can be extended (or reduced) by an external field, as seen in Figure \ref{fig.forwardbias}. Any charge generated within this volume will be driven by the electric field of the depletion layer, a phenomenon that will be exploited for the detection of charges generated by radiation interaction in the material.

In particular, in CCDs, it is necessary to make the depletion volume as large as possible by applying a forward bias through the junction. High voltages can have undesired effects (usually short circuits) on the electronic structure neighboring the depletion volume, so creating large and stable depletion volumes has been a challenge in the past. We will see in Section \ref{sec:ccdstructure} how this problem was resolved along with the structure used to apply forward bias to the p-n junction in CCDs.

\subsubsection{The MOS Capacitor}

\begin{figure}[h!]
\centering
\includegraphics[width=0.65\textwidth]{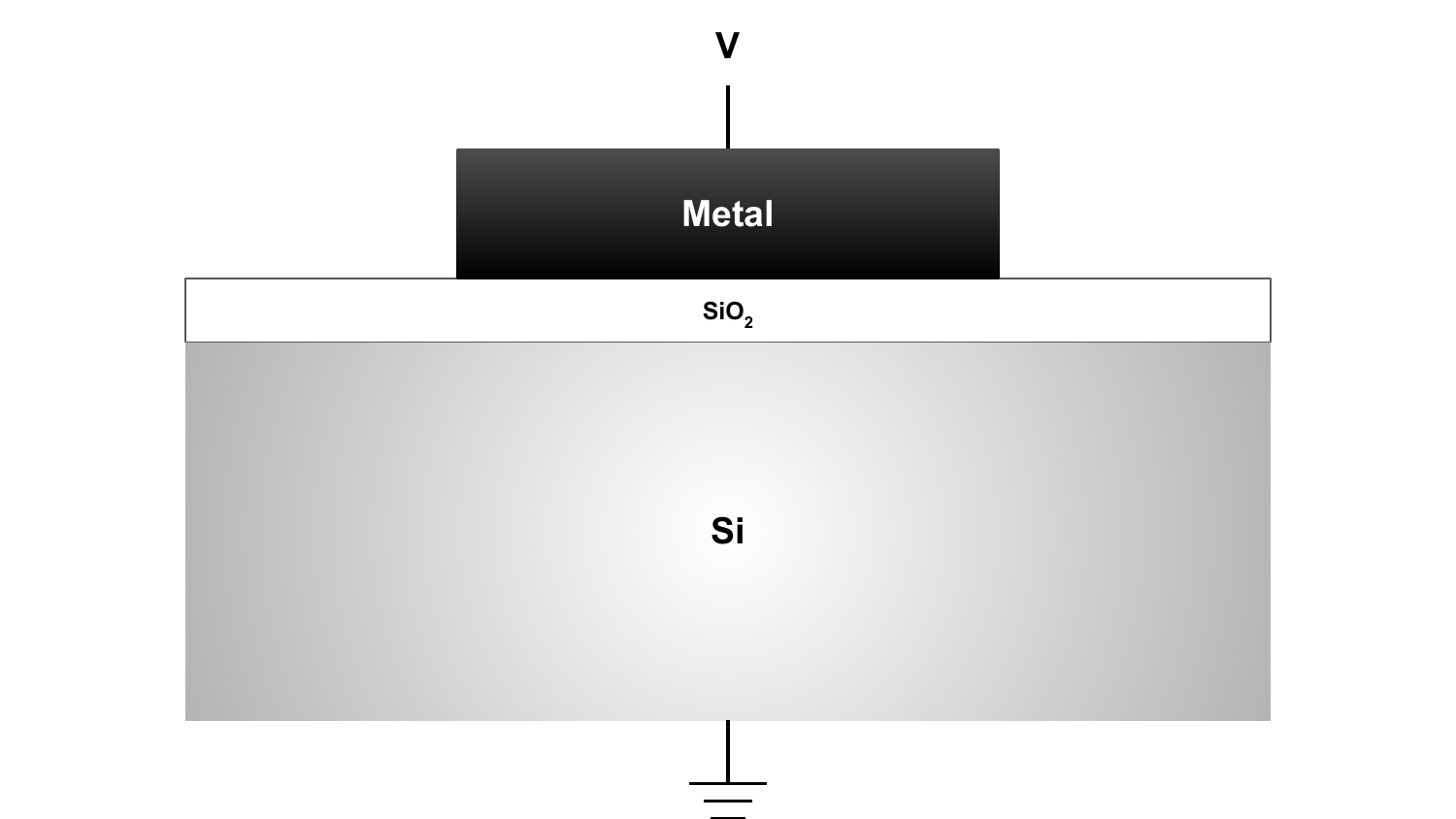}
\caption{Schematic of an MOS capacitor.}
\label{fig.moscapacitor}
\end{figure}

Another key element in CCDs is the MOS capacitor (metal-oxide-semiconductor).
Considering a silicon semiconductor (which can be of n-type or p-type), a layer of $SiO_{2}$ is grown on it, a task that can be easily achieved through thermal oxidation, i.e., by exposing the semiconductor to an oxygen-rich environment at approximately 1000 degrees Celsius for a relatively extended period, depending on the desired oxide layer thickness. A conductive material is grown on top of this layer, creating the MOS structure (see Figure \ref{fig.moscapacitor}).
Although the "M" in MOS stands for metal, for the vast majority of cases (especially considering silicon as the semiconductor component), highly doped polycrystalline silicon (or \textit{polysilicon}) is used for practical purposes, as its conductivity is sufficiently high to be used as an electrical contact.
Typically, on the opposite side of the semiconductor, another layer of polysilicon is grown, which serves as a ground contact.
As a result, a voltage applied to the contact on the oxide layer will generate an electric field through the semiconductor, which, similar to what was described for the p-n junction, will create a depletion layer beneath the $SiO_{2}$/Si interface.

In an MOS capacitor, the electrons and/or holes generated within this depletion volume will then be dragged by the electric field present at that interface. A photon (or any charged particle) whose energy is above the band-gap will generate at least one electron-hole pair through the photoelectric effect, which will be dragged (in opposite directions) by the electric field: one will be absorbed by the ground connected to the back of the capacitor, and the other will be trapped at the interface with $SiO_{2}$, depending on whether the silicon is p-type or n-type.
CCDs, as we will see below, can collect the electrons (or holes) created in the depletion volume so that they can be stored, transferred, and read in a controlled manner.

\subsection{Structure of a CCD and Operating Principles}
\label{sec:ccdstructure}

\begin{figure}[h!]
\centering
\includegraphics[width=0.7\textwidth]{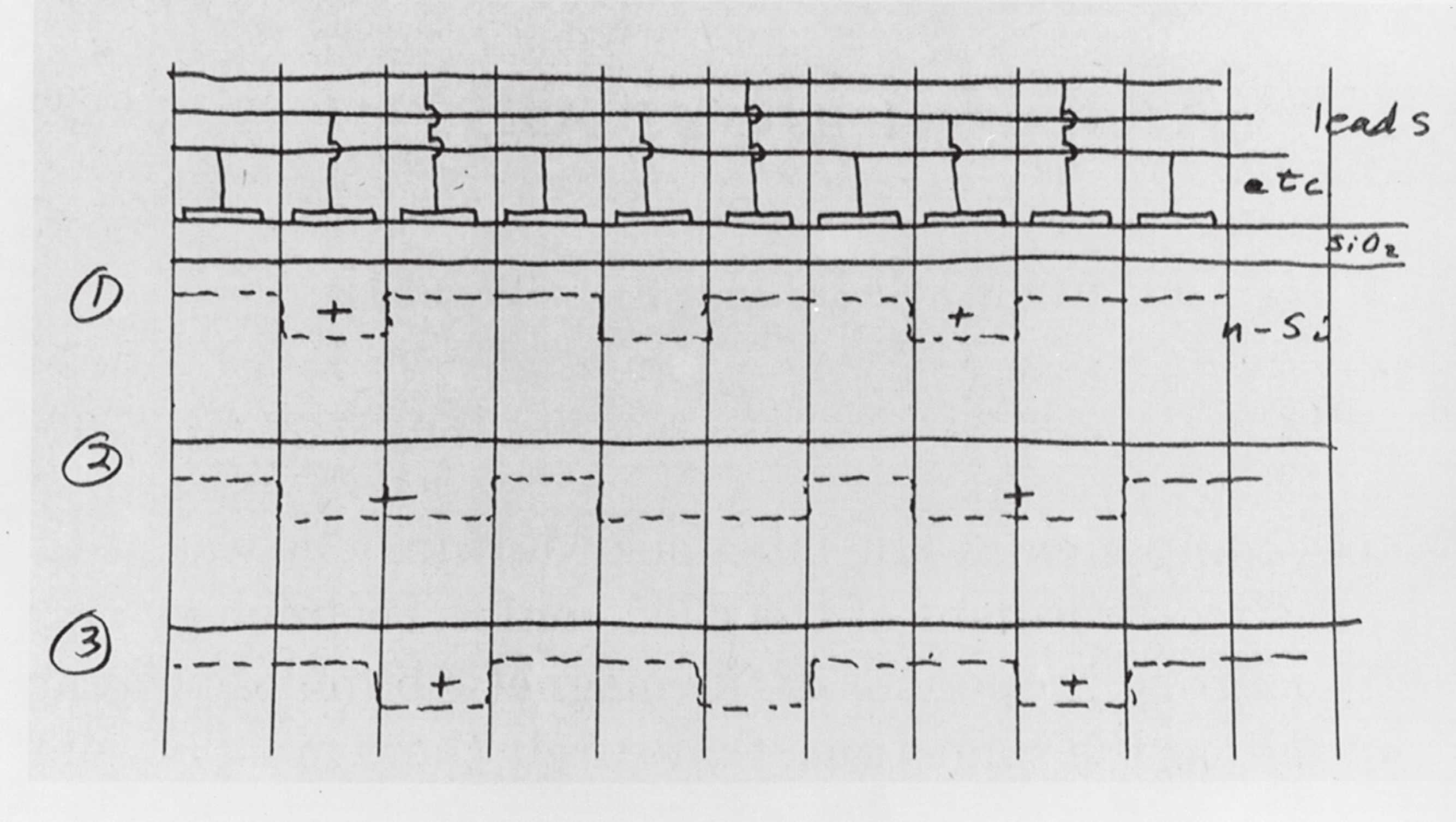}
\caption{Original schematic and temporal diagram of the CCD designed by Boyle and Smith. Figure extracted from \cite{janesick}.}
\label{fig.ccdoriginal}
\end{figure}

The original sketch of a silicon three-phase CCD and its basic structure are schematically shown in Figure \ref{fig.ccdoriginal}. This figure depicts a type-n doped Si material covered by a thin layer of SiO2. Above this layer, metal conductors are added to apply three specific electric potentials periodically throughout the device. One of these metal conductors, along with the thin layer of silicon dioxide grown over the silicon semiconductor (and another opposite metal conductor not shown in Figure \ref{fig.ccdoriginal}), constitutes the MOS capacitor, which, when operated in depletion mode, as introduced earlier, can store charges at the $SiO_{2}/Si$ interface. Suppose we want to collect holes beneath one of the MOS capacitors; then, a negative voltage must be applied to the surface (the other metal conductor can be at ground). If the semiconductor is entirely depleted, the positive (negative) charges will go to the surface (bottom) of the MOS capacitor, as close (far) as possible to the metal conductor gate. Since charges cannot travel beyond the $Si/SiO_{2}$ interface (and as long as the applied voltages remain constant), they will remain fixed at the MOS capacitor interface.
A CCD consists of a particular arrangement of MOS capacitors \emph{coupled} in such a way that charges can be transferred from one capacitor to another by applying a specific sequence of voltages to the metal conductor gates of each capacitor. In summary, positive charges will be attracted to the metal gate with the lower electric potential. Figure \ref{fig.ccdoriginal} depicts 10 of these MOS capacitors and how charges are transferred from one capacitor to another. Three potential wells (created in the second, fifth, and eighth capacitors from left to right) are drawn in the time sequence $(1)$ showing charges (represented as $+$) captured by the potentials. After applying different voltages to the metal conductors in the time sequence $(2)$, the charges move slightly to the right, and by changing the voltages again, the charges are effectively transferred to the adjacent capacitor in the time sequence $(3)$.

\begin{figure}[h!]
\centering
\includegraphics[width=0.8\textwidth]{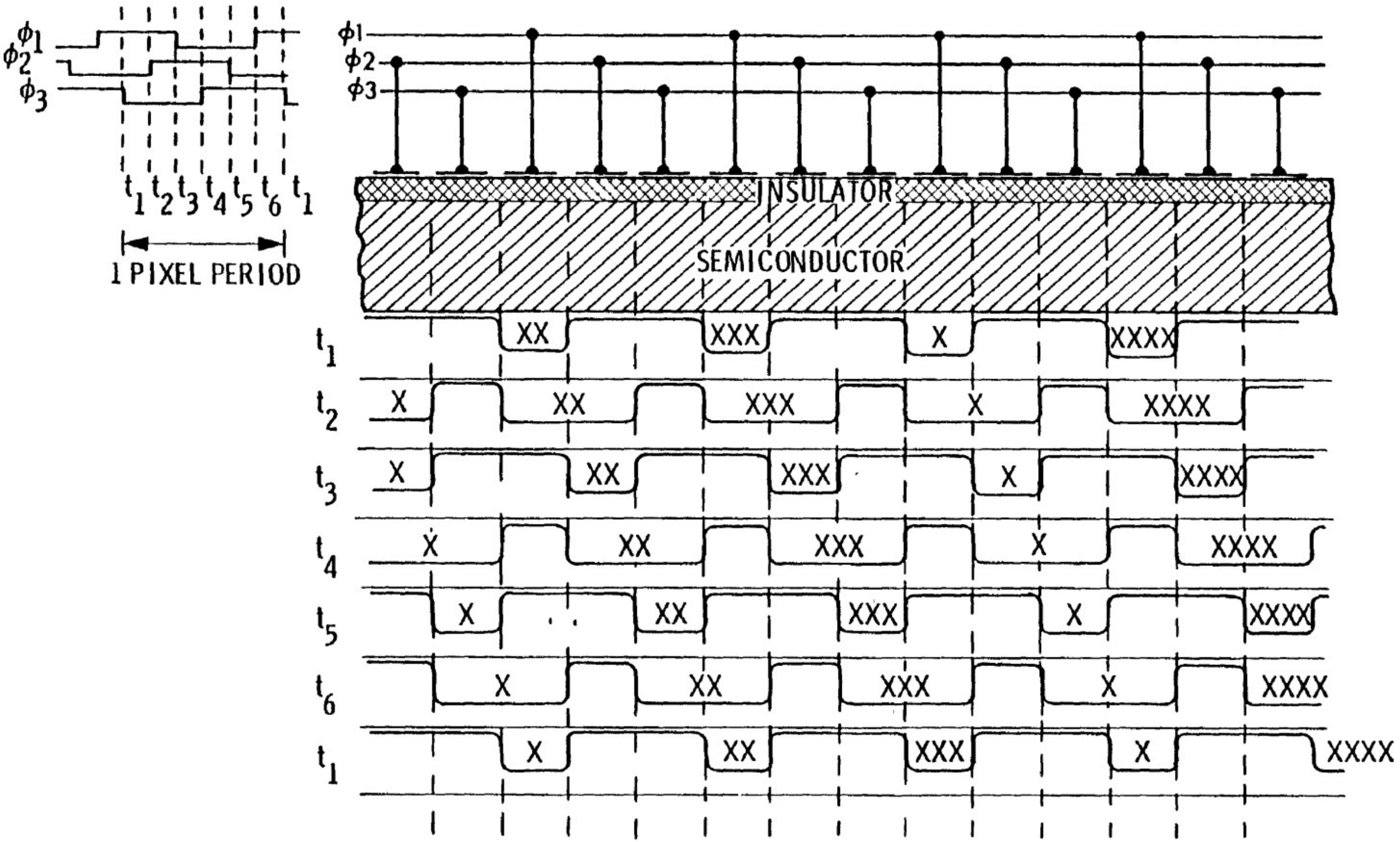}
\caption{Schematic showing the charge transfer structure of a three-phase CCD. A temporal diagram of the potential wells shows the charge being transferred from left to right as the illustrated electric potentials also change. Figure extracted from \cite{janesick}.}
\label{fig.ccdjanesick}
\end{figure}

A more detailed schematic of the CCD transfer operation principle is shown in Figure \ref{fig.ccdjanesick}, extracted from \cite{janesick}. Specifically, this diagram illustrates how three voltages ($\phi_{1}, \phi_{2}$, and $\phi_{3}$) are periodically applied (in space) to every three capacitors, defining a pixel. The complete transfer of a pixel from $t_{1}$ to $t_{6}$ is depicted, illustrating the changes in electrical potentials across each capacitor and the transferred charges (marked as $X$). It is noteworthy that the transfer occurs synchronously for each pixel, coordinated by $\phi_{1}, \phi_{2}$, and $\phi_{3}$.

\subsection{Pixel Structure and Multiple Pixels}

\begin{figure}[h!]
    \centering
    \includegraphics[width=0.5\textwidth]{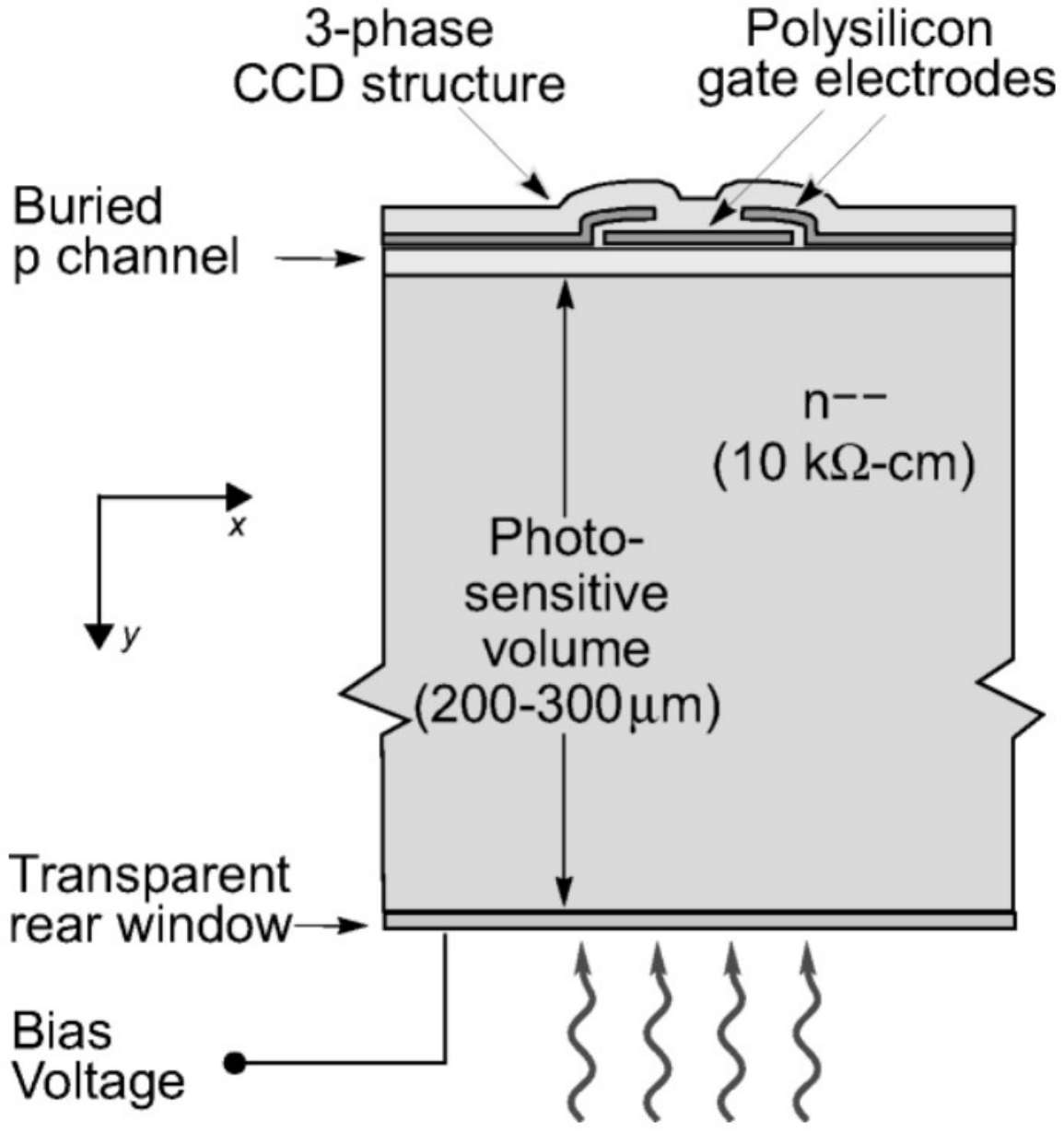}
    \caption{Diagram of the cross-section of a CCD pixel as described in \cite{holland2003fully}.}
    \label{fig.pixel}
\end{figure}

Figure \ref{fig.pixel} displays a cross-sectional view of a pixel \cite{holland2003fully}, very similar to the sensors used in the SENSEI experiment. The bulk of the pixel is made of n-type silicon \footnote{Hereafter, '--' will denote very light doping, '-' will denote light doping, '+' will denote heavy doping, and '++' will denote very heavy doping when specified.} with very low doping, and the thickness of the pixel can vary from 200 $\mu$m to 675 $\mu$m. When a positive voltage is applied to the backside (bottom in the image) of a pixel, the majority charge carriers (electrons) in the bulk are repelled from the surface, creating a depletion region beneath the $SiO_{2}/Si$. If the voltage is high enough, the entire bulk can be depleted, and the CCD can operate in a fully depleted mode. In this mode, electron-hole pairs created by incident radiation are generated throughout the bulk rather than just in isolated depletion regions, utilizing the entire device mass, which is crucial for the search for dark matter events benefiting from increased detector mass. However, several problems arise when applying high bias voltages, including CCD destruction due to a short circuit. To address this issue, it is necessary to use high-resistivity Si as the bulk (i.e., very low doping) \cite{holland2003fully, ccdinventionholland}.

\begin{figure}[h!]
    \centering
    \begin{subfigure}[c]{0.45\textwidth}
        \centering
        \caption{}
        \includegraphics[width=\textwidth]{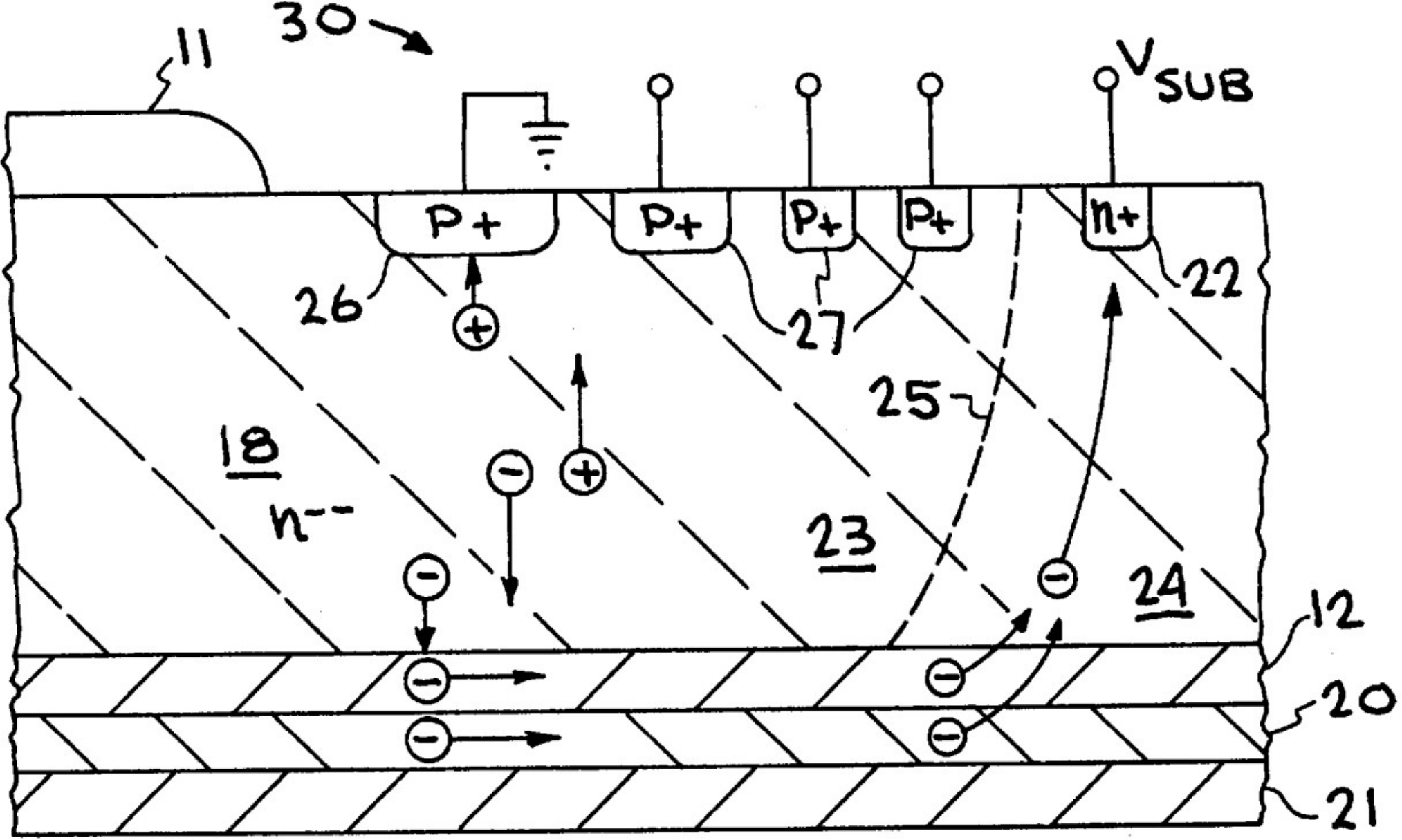}
        \label{fig.pixel2a}
    \end{subfigure}
    \hfill
    \begin{subfigure}[c]{0.45\textwidth}
        \centering
        \caption{}
        \includegraphics[width=\textwidth]{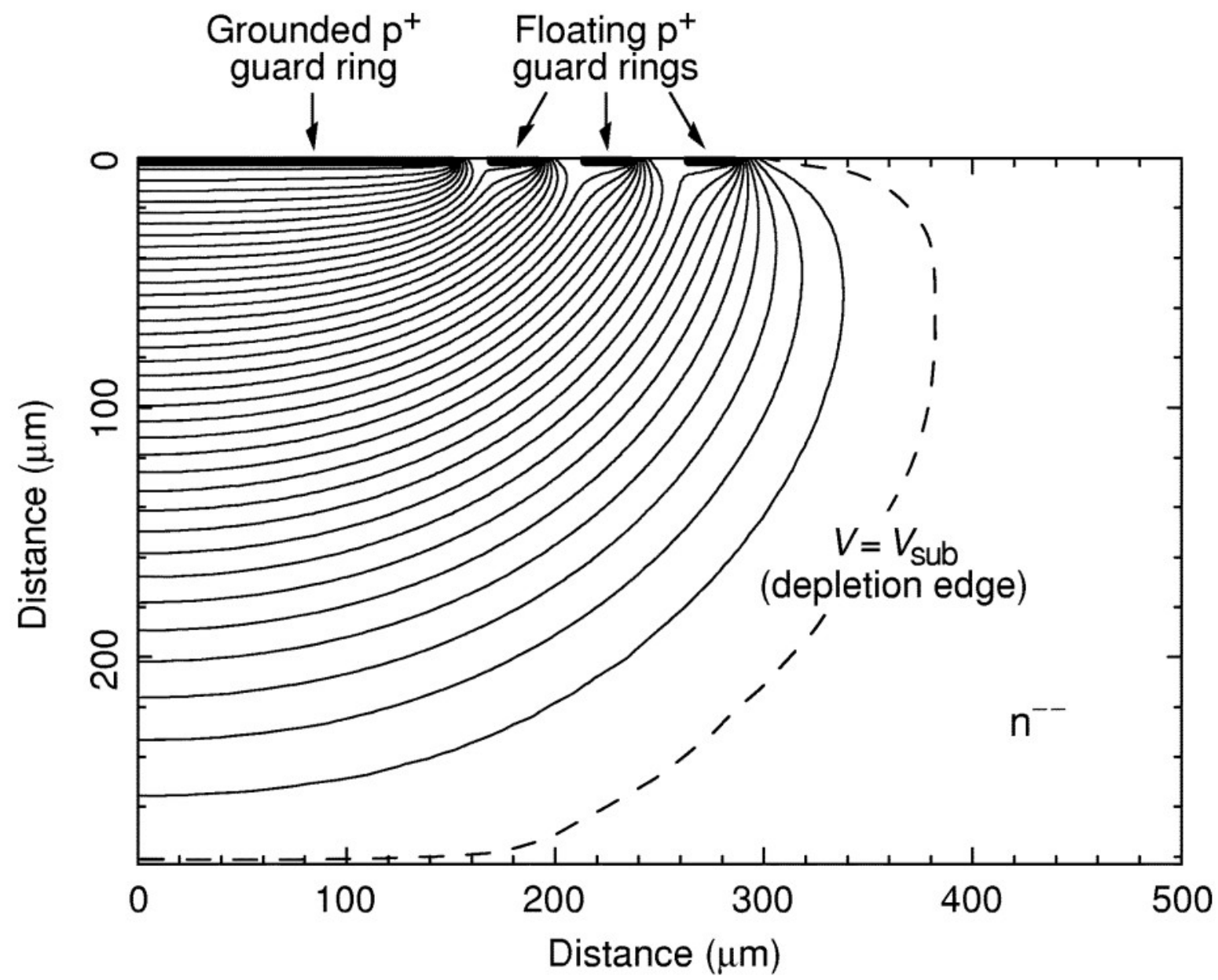}
        \label{fig.pixel2b}
    \end{subfigure}
    \caption{Figure \ref{fig.pixel2a} shows a cross-sectional diagram of an edge of a CCD as described in \cite{holland2001fully}. Figure \ref{fig.pixel2b} shows a simulation of a p-i-n diode structure, similar to what is shown in Figure \ref{fig.pixel2a}. For practicality, the bias voltage was applied on the reverse side, producing the same depletion region as if applied on the front side using an n+ contact. Equipotential lines are spaced at intervals of $1V$ \cite{holland2001fully}.}
    \label{fig.pixel2}
\end{figure}

Despite the fact that the conductive layer at the bottom of a pixel (see Figure \ref{fig.pixel}) is grown to apply the bias voltage, it is not on this conductor that the voltage is applied. Figure \ref{fig.pixel2a} shows a schematic cross-section of the edge structure of a CCD.
\textbf{11} represents a pixel as presented in Figure \ref{fig.pixel}, followed (from left to right) by four p+ implants (\textbf{26} and \textbf{27}) and an n+ implant \textbf{22}, all of them on the front surface. The voltage is applied between \textbf{22} and the p+ implant \textbf{26}, which is grounded, creating a depleted region \textbf{23} and an undepleted region \textbf{24} (separated by \textbf{25}). The p+ implants \textbf{27} control and smooth the voltage drop between these two regions.
As also illustrated in Figure \ref{fig.pixel2a}, electron-hole pairs created in the depleted region \textbf{23} (separated from the depleted region \textbf{18} beneath the pixel structure \textbf{11}) are displaced in opposite directions by the electric field to be collected by either ground \textbf{26} or the bias voltage \textbf{22}. 
In the region \textbf{18} beneath the pixel structures \textbf{11}, the holes are collected by potential wells, as described above, and the electrons are drained by the bias voltage, through the n++ implant at the bottom of the material. It should be noted that the implants \textbf{22}, \textbf{26}, and \textbf{27} are actually rings surrounding the pixel array in the CCD.

\subsubsection{The Channel Stops}

\begin{figure}[h!]
\centering
\includegraphics[width=0.7\textwidth]{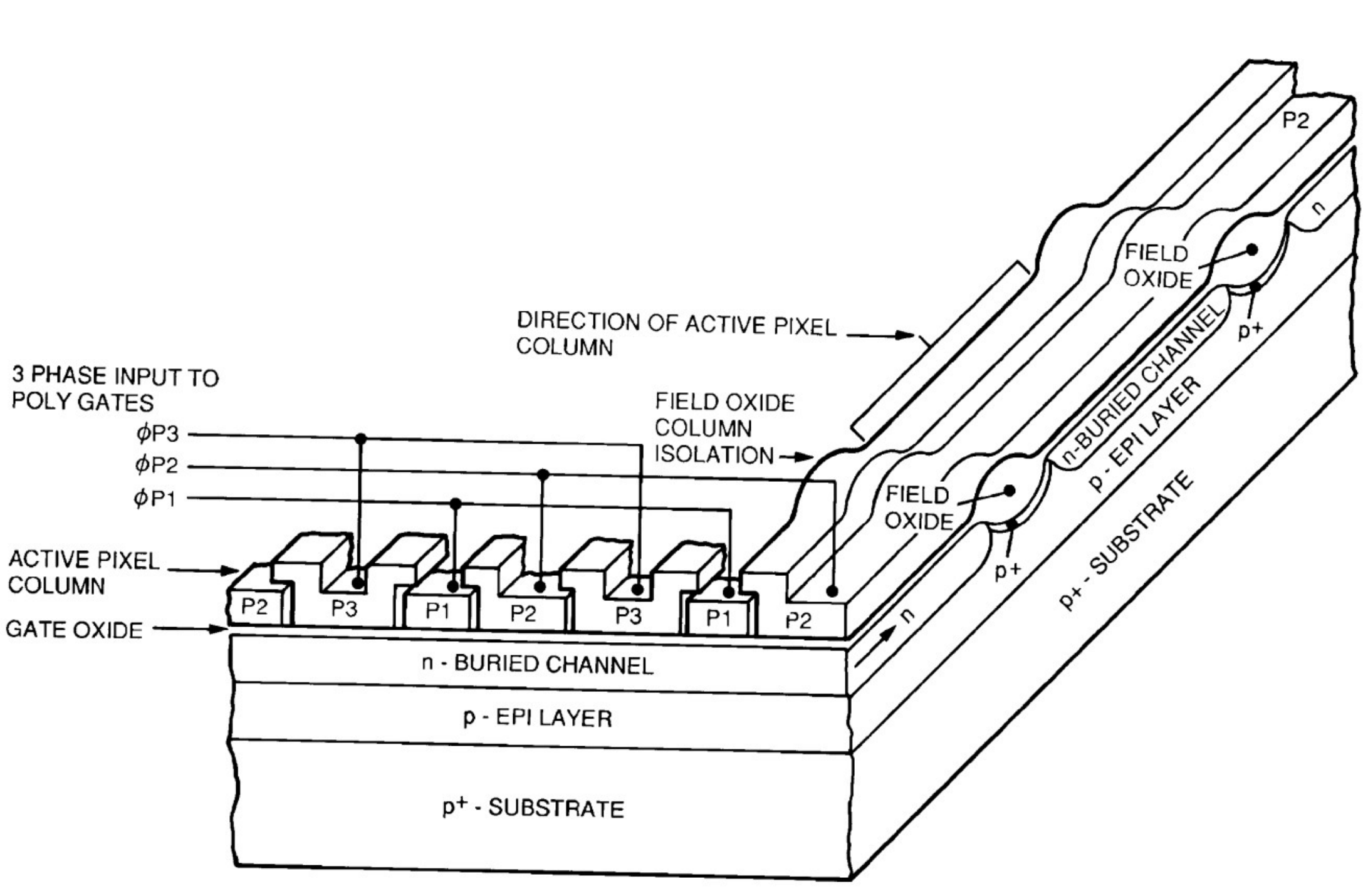}
\caption{Cross-sectional diagram showing the buried channels and the \textit{channel-stops}. It should be noted that the \textit{bulk} used for this image is p-type instead of n-type. Therefore, each \textit{p} should be replaced by an \textit{n} to match the CCD described in the text. Additionally, an \textit{EPI LAYER} is not shown in the image as presented. Adapted from \cite{janesick}.}
\label{fig.channelstop}
\end{figure}

Figures \ref{fig.ccdoriginal} and \ref{fig.ccdjanesick} show an array of $1 \times N$ pixels in a single line, or more specifically, in a column of a CCD. To create an N x N CCD, multiple lines must be arranged side by side and separated by what is known as a \textit{channel-stop}. A \textit{channel-stop} is an n-type implant that, when fully depleted by the voltage applied to the CCD, is predominantly populated by holes that repel the holes collected beneath the buried p-channels. It is important to note that the holes that populated the \textit{channel-stop} come from their donors, so they are fixed in the crystal lattice, while the holes collected beneath the buried p-channel can move as long as the electric field applied to the pixel gate allows them to do so. The structure of a \textit{channel-stop} is shown in Figure \ref{fig.channelstop}. It can be observed that they are grown periodically as strips between each channel, beneath the $SiO_{2}$ layers.

\subsubsection{The Buried Channel}

Above the bulk, silicon is doped p-type to create a buried channel, a technology adopted to reduce the trapping of charges collected at the energy levels created at the $SiO_{2}/Si$ interface. The collected holes accumulate where the potential well forms, and this occurs at the p-n interface of the two differently doped regions. This dramatically improves the Charge Transfer Efficiency (CTE), as charges are less susceptible to being trapped by the interface energy levels. It can also be stated that as CTE increases, its counterpart, Charge Transfer Inefficiency (CTI), decreases, which we will refer to multiple times in this manuscript.

The choice of a p-type buried channel and an n-type bulk is because it is preferable to collect holes as they result in lower dark current \cite{miguelthesis}.\footnote{Dark current will be discussed later, but it is defined here as the promotion of a valence electron to the conduction band, later to be collected by potential wells, due to thermal agitation.}

\subsubsection{\textit{Back-Illuminated} CCDs}

When a CCD is exposed from its reverse side, it is said to be back-illuminated. Due to this, three layers are grown beneath the bulk to enhance the Quantum Efficiency (QE), which is the ratio of absorbed energy to incident energy in the detector.
First, silicon is heavily doped to create an n++ conductive layer where the bias voltage can be applied. Second, a 60nm anti-reflective layer of indium tin oxide (ITO) is grown. Finally, a layer of $SiO_{2}$ is deposited to increase QE in the red wavelengths. 
It's also worth noting that in these CCDs, the front side is covered by a 0.5nm layer of $SiO_{2}$ on top of a 0.5nm layer of $SiN_{3}$. The gates above each pixel are made of heavily doped polycrystalline silicon or polysilicon, which acts as a conductor. The Skipper-CCDs used by SENSEI have this treatment on their underside.

\subsection{Generation, Diffusion, and Collection of Charges}
\label{sec:diffusion}

As introduced earlier in this chapter, electron-hole pairs are created in the bulk through the photoelectric effect when radiation passes through it. Since the interior is depleted of charge carriers by an external voltage (and because the bulk is of n-type), the generated electrons are transferred to the ground at the bottom of the CCD, and the holes are captured beneath the surface, specifically at the p-n junction interface between the buried p-channel and the n-type bulk.

\begin{figure}[h!]
     \centering
     \begin{subfigure}[c]{0.6\textwidth}
         \centering
         \caption{}
         \includegraphics[width=\textwidth]{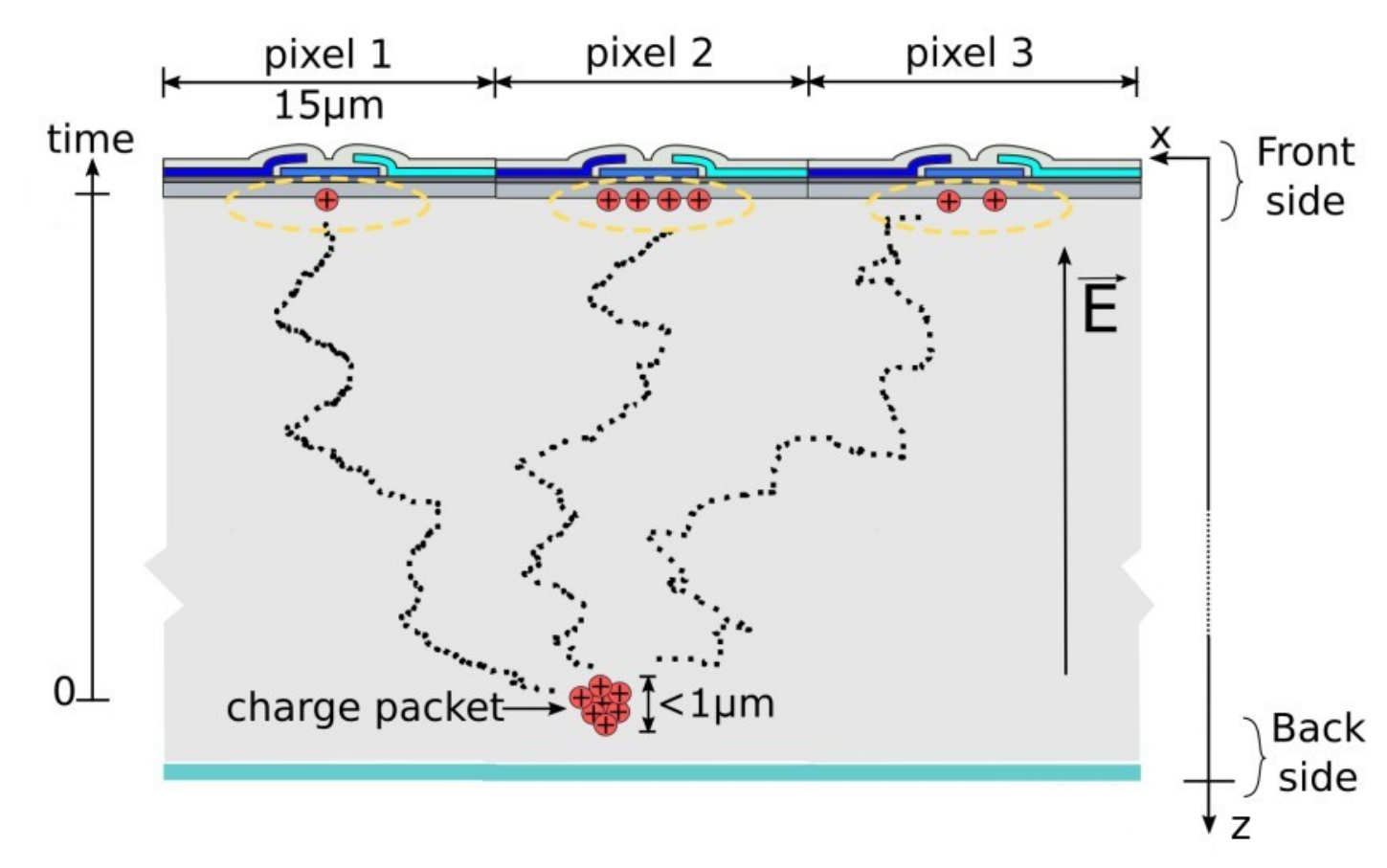}
         \label{fig.diffusion1}
     \end{subfigure}
     \hfill
     \begin{subfigure}[c]{0.35\textwidth}
         \centering
         \caption{}
         \includegraphics[width=\textwidth]{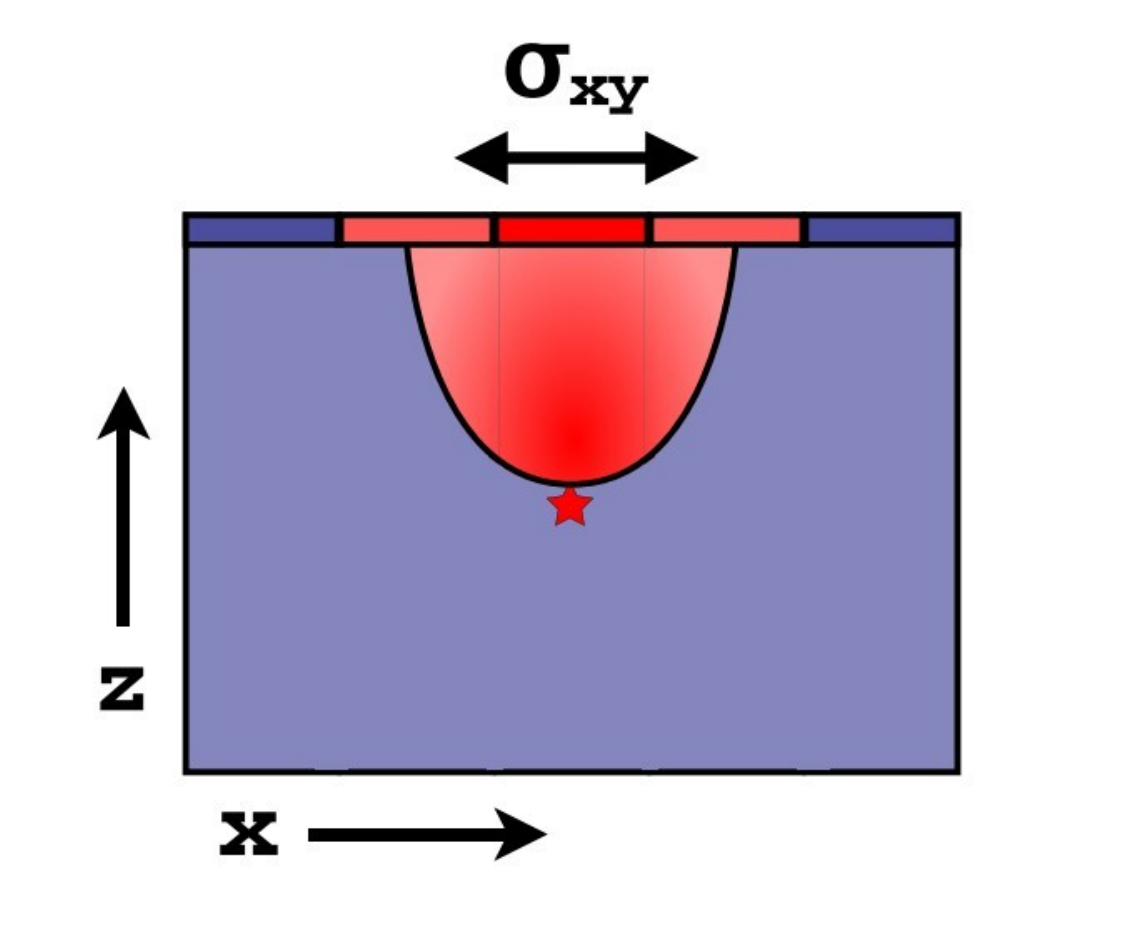}
         \label{fig.diffusion2}
     \end{subfigure}
        \caption{Figure \ref{fig.diffusion1} illustratively shows the diffusion of charges from an event of 7 holes to 3 pixels on the CCD surface. Figure \ref{fig.diffusion2} shows the distribution characterizing this diffusion, displaying the cross-sectional view of a Gaussian bell curve. Figures extracted from \cite{haro2020studies} and \cite{DAMIC2016}, respectively.}
        \label{fig.diffusion}
\end{figure}

The diffusion process begins with a charge generation event, which can be considered point-like for all practical purposes. Due to the electric field applied to the silicon bulk, the holes will migrate towards the front, specifically towards the buried channel, resulting in a spatial distribution well described by a bivariate Gaussian from the point of interaction. This is illustrated in Figures \ref{fig.diffusion1} and \ref{fig.diffusion2}.

Figure \ref{fig.diffusion1} demonstrates how seven electrons generated inside the CCD move towards its front following a Brownian motion and end up being collected beneath the potential wells of three different pixels. Figure \ref{fig.diffusion2} similarly illustrates diffusion, showing in red a cross-sectional view of a bivariate Gaussian bell curve projected onto the xz plane.

This diffusion phenomenon was extensively studied in \cite{miguelthesis} and \cite{holland2003fully}. Later, a formula relating the depth of an interaction and the variance of the charges collected on the surface in the xy plane ($\sigma_{xy}$) was derived \cite{DAMIC2016}:

\begin{equation}
    \sigma_{xy}=-A \ln |1-bz|
    \label{eq.diffusion}
\end{equation}

\noindent where A and b are parameters that depend on temperature, thickness, donor density, and the voltage present in the substrate. Given A and b, and if we know the variance of the collected charges for a specific event or trace, we can infer the depth at which the interaction occurred, and vice versa. Specifically, greater depths result in greater variances. The procedure for obtaining A and b, given a CCD and a dataset, will be explained in detail in Section \ref{sec:2020eficienciapordifusion}.

\subsection{Charge Transfer Sequence}
\label{sec:transferencia}

\begin{figure}[h!]
\centering
\includegraphics[width=0.7\textwidth]{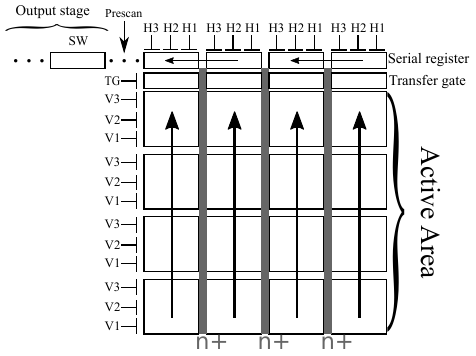}
\caption{Illustration of a $4\times4$ pixel CCD. The arrows show the direction in which the collected charges are transferred during readout. H1, H2, and H3 are the final horizontal voltages in the serial register before the Summing Well (SW). The channel stops are shaded in gray. Figure from \cite{SENSEI2022}.}
\label{fig.ccd}
\end{figure}

As seen in Figure \ref{fig.ccd}, charges generated and collected in the active area are transferred vertically, pixel by pixel, by changing the voltages in a specific sequence. The sequence used involves applying different voltages to the electrodes of the column pixels while the row pixel electrodes are held at a constant voltage. After a complete vertical transfer (two if the transfer gate is considered), the charges generated in the top row are stored beneath the pixels in the horizontal register or \textit{serial register}, which have a different voltage sequence, arranged periodically along the horizontal axis (perpendicular to the pulses in the active area). Transfer begins immediately in the horizontal register towards the output device, where each pixel is read, and the collected charge is transformed into a voltage signal through a video signal, as discussed in Section \ref{sec:dispositivodesalida}.

\subsection{The Output Device}
\label{sec:dispositivodesalida}

\begin{figure}[h!]
     \centering
     \begin{subfigure}[c]{0.45\textwidth}
         \centering
         \caption{}
         \includegraphics[width=\textwidth]{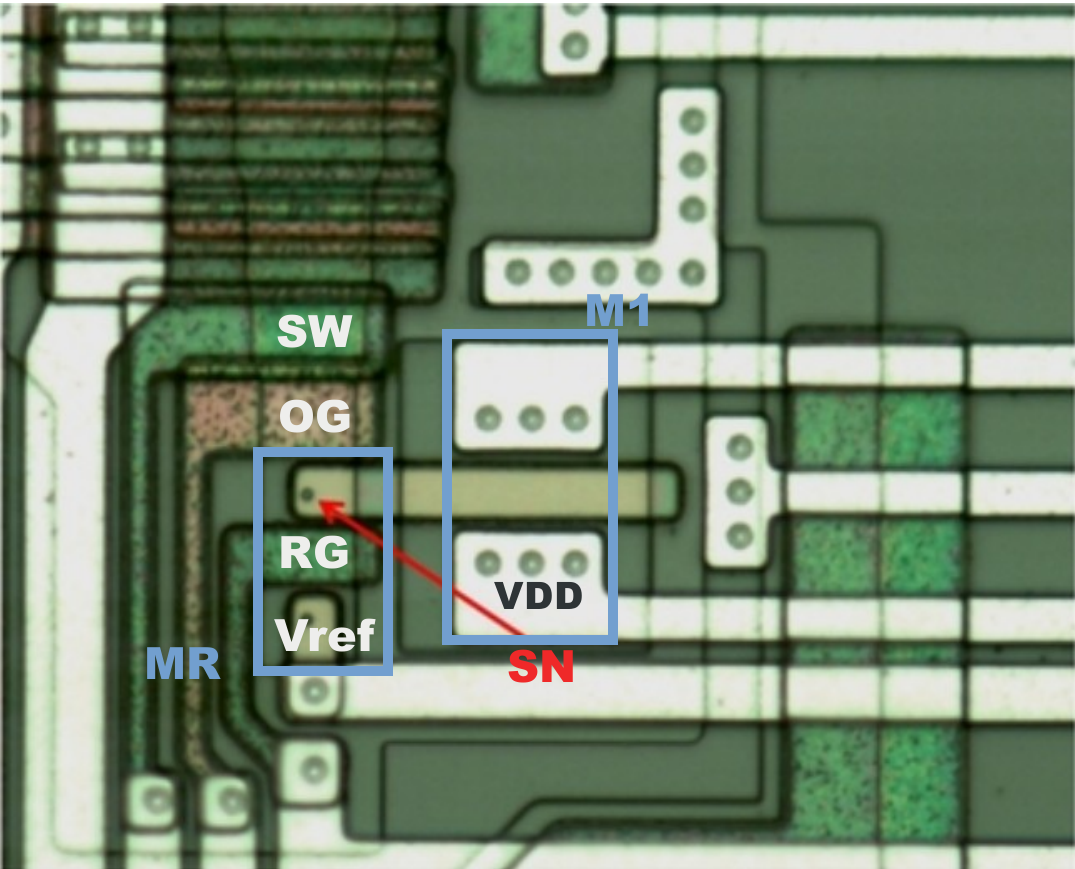}
         \label{fig.desi}
     \end{subfigure}
     \hfill
     \begin{subfigure}[c]{0.5\textwidth}
         \centering
         \caption{}
         \includegraphics[width=\textwidth]{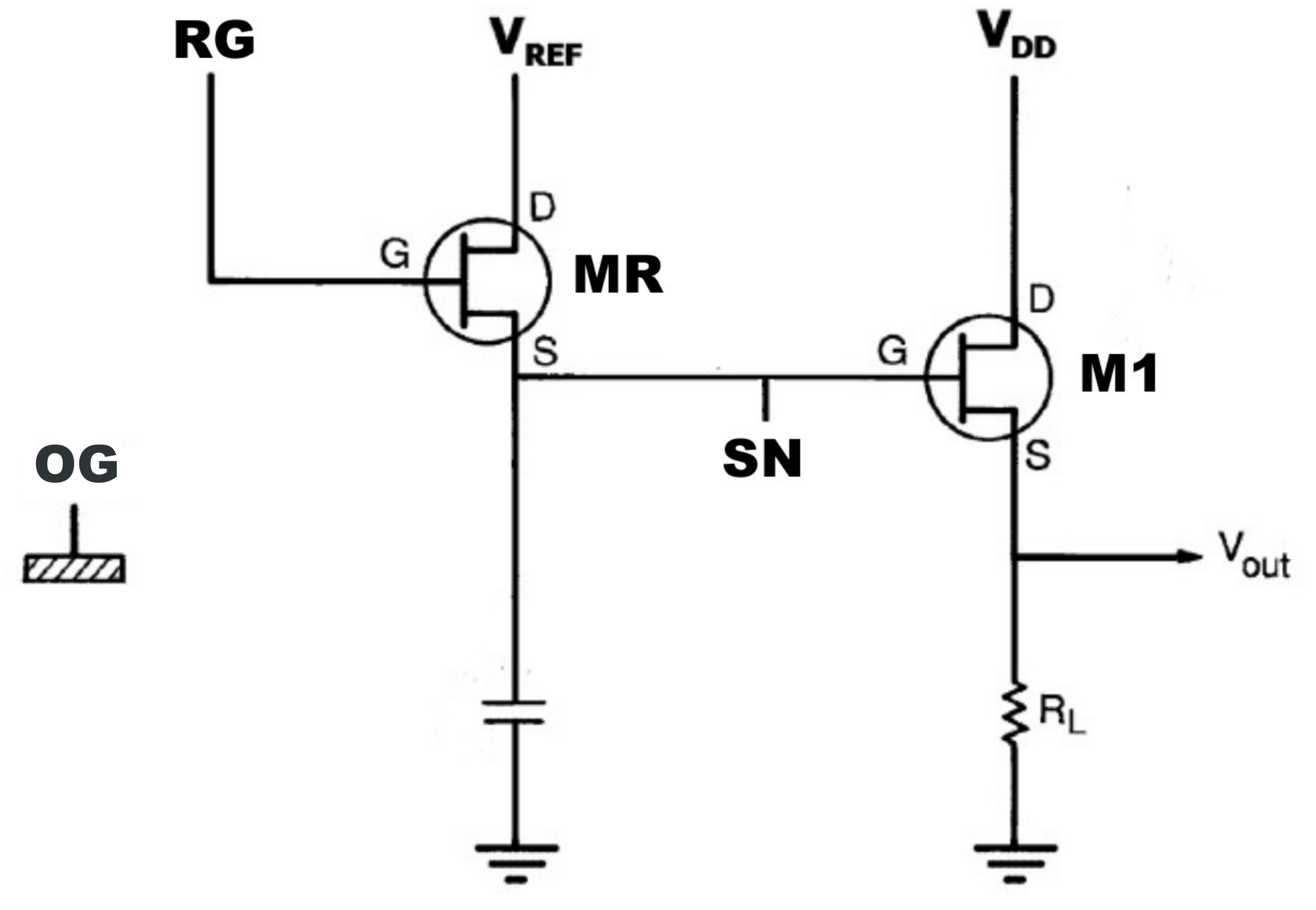}
         \label{fig.amplifier1}
     \end{subfigure}
        \caption{Figure \ref{fig.desi} shows a photograph of the output device used by the DESI experiment, very similar to the one used by the DAMIC and CONNIE collaborations. Figure \ref{fig.amplifier1} illustrates a schematic of the electronic structure used. Figures extracted from \cite{bebek2017status} and \cite{janesick}, respectively.}
        \label{fig.amplifier}
\end{figure}

The conversion of charges to voltage (i.e., from $e^-$ to $V$) is performed in the output device. Figure \ref{fig.amplifier} shows a micrograph of the output device used in the Dark Energy Spectroscopic Instrument (DESI) \cite{bebek2017status} and its electronic design, from left to right. It consists of two MOSFET transistors (M1 and MR) connected by the sense node (SN).

Once the charges are shifted to the horizontal register, they are moved, pixel by pixel, to the Summing Well (SW) by changing the voltages applied to the H1, H2, and H3 electrodes, as indicated in Figure \ref{fig.ccd}.

\begin{figure}[h!]
\centering
\includegraphics[width=0.7\textwidth]{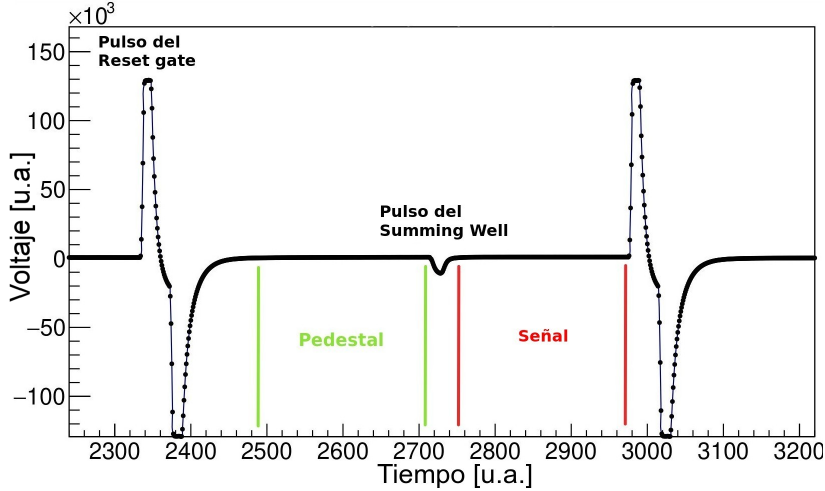}
\caption{Voltage ($V_{out}$) as a function of time, as measured at the SN. This image was obtained from a test run of SENSEI. The pedestal (green) is measured after the RG pulse, clearly visible in the image, and the signal (red) is measured after the SW pulse. This process is then repeated for the next pixel. The difference between these two is negligible in this case as the pixel was likely empty.} 
\label{fig.pedestalsignal}
\end{figure}

After the complete transfer of a pixel and while keeping the horizontal voltages fixed, charges are transferred from the Summing Well (SW) to the Output Gate (OG). As shown in Figure \ref{fig.pedestalsignal}, a pulse is sent from the Reset Gate (RG) to the Sense Node (SN) to establish a reference voltage at the SN. Since M1 is configured in source follower mode, the voltage at the SN is sensed through the source of M1, resulting in $V_{out}$. This voltage (pedestal) is measured for a period of time T before sending the charges from the OG to the SN.
After transferring the charges to the SN, the voltage measured at the SN (again, for a period of time T) will be the sum of the pedestal and the signal generated by the charges present in that pixel. Therefore, subtracting this signal from the pedestal will be proportional to the number of electrons in the SN. The charges in the SN are discarded when a new pulse is sent to the SN from the RG. This process is repeated until the entire horizontal register has been read. Then, the next row is transferred until the charge quantity in the entire CCD has been measured.

The measured signal, in volts, will depend on the total capacitance of the output device. Each of the gates will contribute to this value, so the gain of the CCD will be equal to $G_{CCD}=1/C_{eq} \ ({\rm \mu V}/e-)$ \cite{miguelthesis}. Capacitances as low as 10 fF have been achieved for devices similar to those used by SENSEI, resulting in a $G_{CCD}$ of around hundreds of ${\rm \mu V}/e-$. A higher gain will lead to lower electronic noise and increased sensitivity.

\section{Skipper-CCDs}

\subsection{Design and Operating Principles}
\label{sec:designandoperationprinciples}

Skipper-CCDs were designed in 1990 by Janesick et al. to reduce electronic noise measured by the CCD's output amplifier \cite{janesick1990new}. As mentioned earlier, in conventional CCDs, the signal (and pedestal) for each pixel is measured over a period of time T, equivalent to M measurements made by the employed data acquisition system and its intrinsic temporal resolution.
Each measurement is associated with electronic noise due to fluctuations in both the signal and pedestal, which can theoretically be reduced to zero for longer integration times T, reducing noise as the square root of the number of samples M. However, low-frequency noises (such as 1/f noise from the output device) begin to dominate for longer integration times, thwarting noise reduction efforts.
Skipper-CCDs have the capability to read the charge in a pixel multiple times and non-destructively, allowing for the reduction of low-frequency components in the measured signal.

\begin{figure}[h!]
\centering
\includegraphics[width=0.95\textwidth]{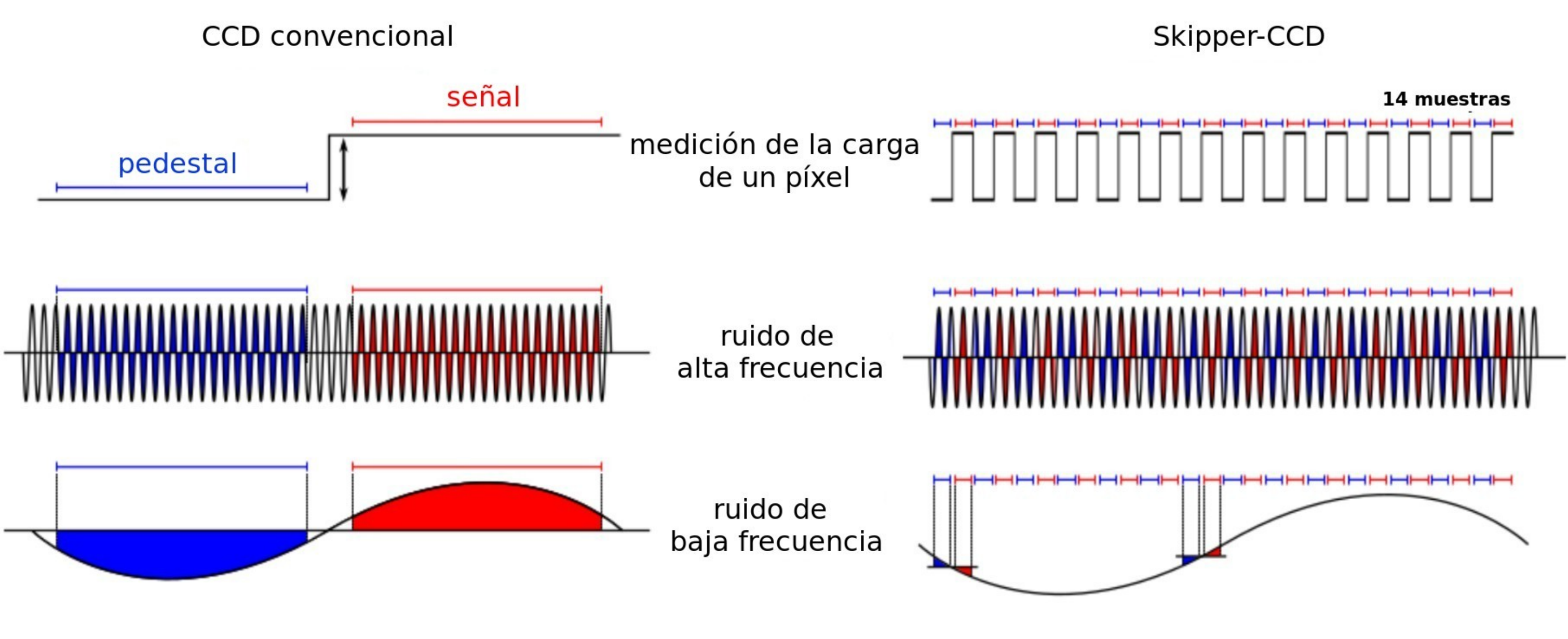}
\caption{Illustration of electronic noise contributions during the readout of a pixel for the case of a conventional CCD (left) and one with Skipper technology (right). The top part schematizes the pedestal and signal periods illustrated in Figure \ref{fig.pedestalsignal}. In the middle part, high-frequency noise ($1/f >> T$, where T is the measurement period for that pixel) is shown in the central part, and low-frequency noise ($1/f \sim T$) in the lower part. It is worth noting that the same integration time as a conventional CCD is used for the Skipper case, but divided into 14 groups, i.e., samples.}
\label{fig.skipperexample}
\end{figure}

The Figure \ref{fig.skipperexample} illustrates the limitations related to noise in conventional CCDs for low-frequency noise, showing in the left part of the Figure the pedestal (blue) and signal (red) measurements for a pixel in a conventional CCD.
In the lower part of the Figure, high-frequency noise ($1/f >> T$, where T is the measurement period for that pixel) in the central part and low-frequency noise ($1/f \sim T$) in the lower part are shown.
It can be observed that several periods of the signal produced by high-frequency noise (left-center) are measured both by the pedestal and the signal, and therefore, when both are subtracted, this noise will contribute weakly to the result. For the low-frequency scenario (bottom left), neither the pedestal nor the signal are able to measure even a single period of the noise contribution.
The illustrated case is the worst possible scenario, for a $1/f \sim T$ noise period so that the pedestal and the signal measure opposite contributions of the noise.

To the right of the Figure, the same scenario is shown for the Skipper-CCD with 14 non-destructive samples for the same pixel.
Again, the pedestal is shown in blue and the signal in red, but this time once for each sample. The pedestal/signal integration time for each sample has been reduced in such a way that the total integration time of the 14 samples in the Skipper system is identical to the total integration time of the conventional system.
The contribution of high-frequency noise remains negligible since it is repeatedly measured by both the pedestal and signal readings.
For the low-frequency scenario, the contribution of the pedestal and signal is averaged if the total integration time (equal to 14 pedestals and 14 signals) is as long as the noise period.
In the Figure, for this case (bottom right), you can see how the noise contribution measured in a pedestal-signal pair is counteracted by the measurement a few samples later.

It is clear that contributions of $1/f$-type noise can be reduced by increasing the pedestal/signal integration time and/or the number of samples N for each pixel. There is a trade-off between these two parameters. Longer pedestal/signal integration times (for a small number of samples) are inefficient in reducing $1/f$-type noise contributions, as explained earlier, making increasing the number of samples the only viable option. However, certain noise components are more efficiently reduced by increasing integration time rather than the number of samples \cite{miguelthesis}.

\begin{figure}[h!]
     \centering
     \begin{subfigure}[c]{0.4\textwidth}
         \centering
         \caption{}
         \includegraphics[width=\textwidth]{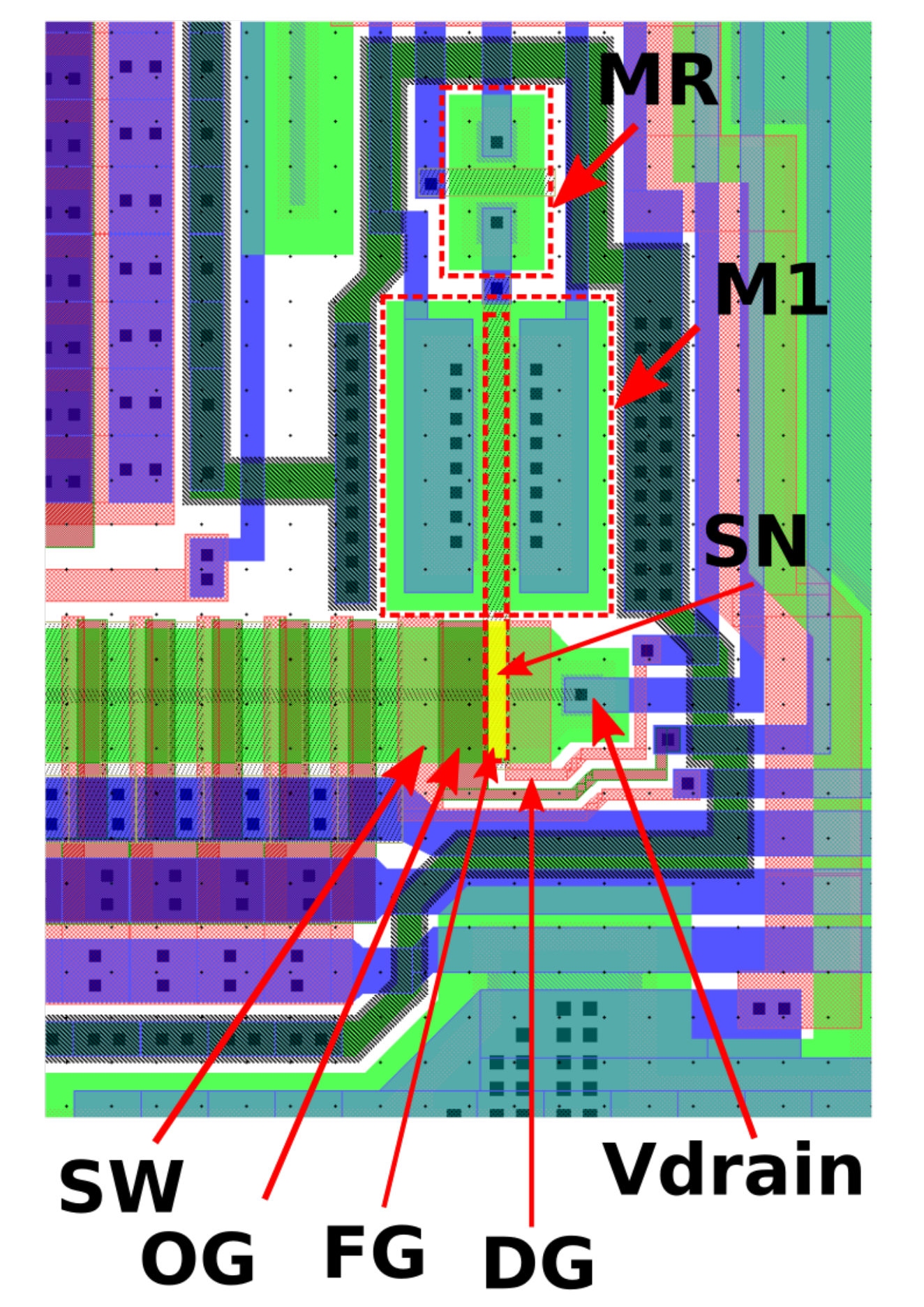}
         \label{fig.skipperoutputstage}
     \end{subfigure}
     \hfill
     \begin{subfigure}[c]{0.5\textwidth}
         \centering
         \caption{}
         \includegraphics[width=\textwidth]{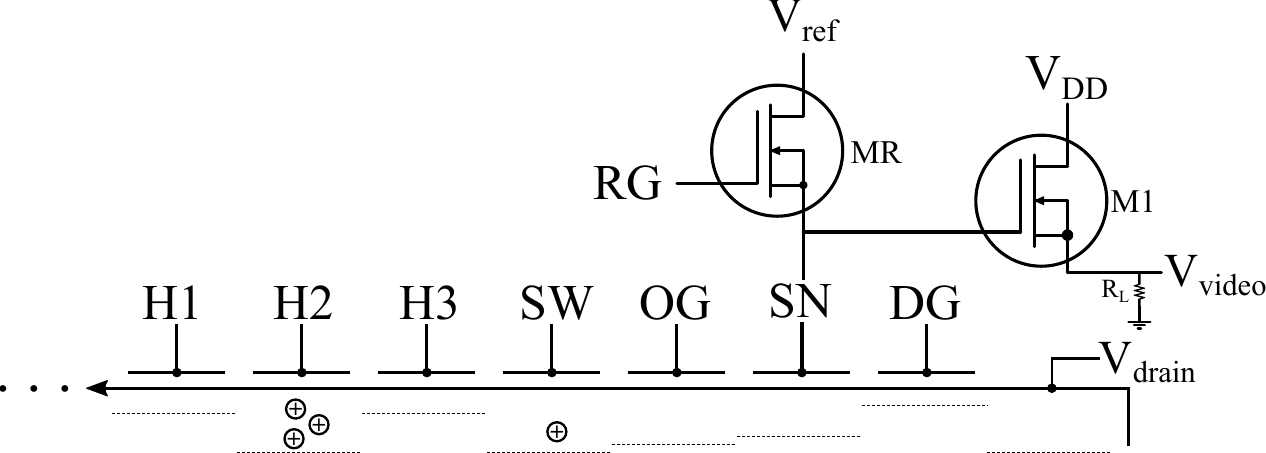}
         \label{fig.skipperoutputstagelayout}
     \end{subfigure}
        \caption{Figure \ref{fig.skipperoutputstage}, extracted from \cite{miguelthesis}, shows a schematic of the output device displaying its different components. In particular, FG represents the floating gate below SN. On the right, another schematic of the output device \cite{SENSEI2022}.}
        \label{fig.skipperamplifier}
\end{figure}

Figure \ref{fig.skipperamplifier} displays a micrograph of the Skipper's output device along with its electronic design. As mentioned earlier, the modifications enable the charge in the SN to be read multiple times in a non-destructive manner. This is achieved by using a floating gate as SN, allowing the charge to move back and forth from the SW to the SN through the OG. In a conventional CCD's output device (see Figure \ref{fig.desi}), the SN is not floating but directly connected to both transistors M1 and MR, so when the charge is transferred, it is inevitably lost.
In a Skipper-CCD, when the charge is transferred to the SN, it is capacitively coupled to both M1's gate and MR's source so that it can return to the SW by applying a voltage change in the OG (and also in the SW). After the measurement is completed, the charge is discharged by sending it to the voltage drain ($V_{drain}$) through the discharge gate (DG).

\begin{figure}[h!]
     \centering
     \begin{subfigure}[c]{0.45\textwidth}
         \centering
         \caption{}
         \includegraphics[width=\textwidth]{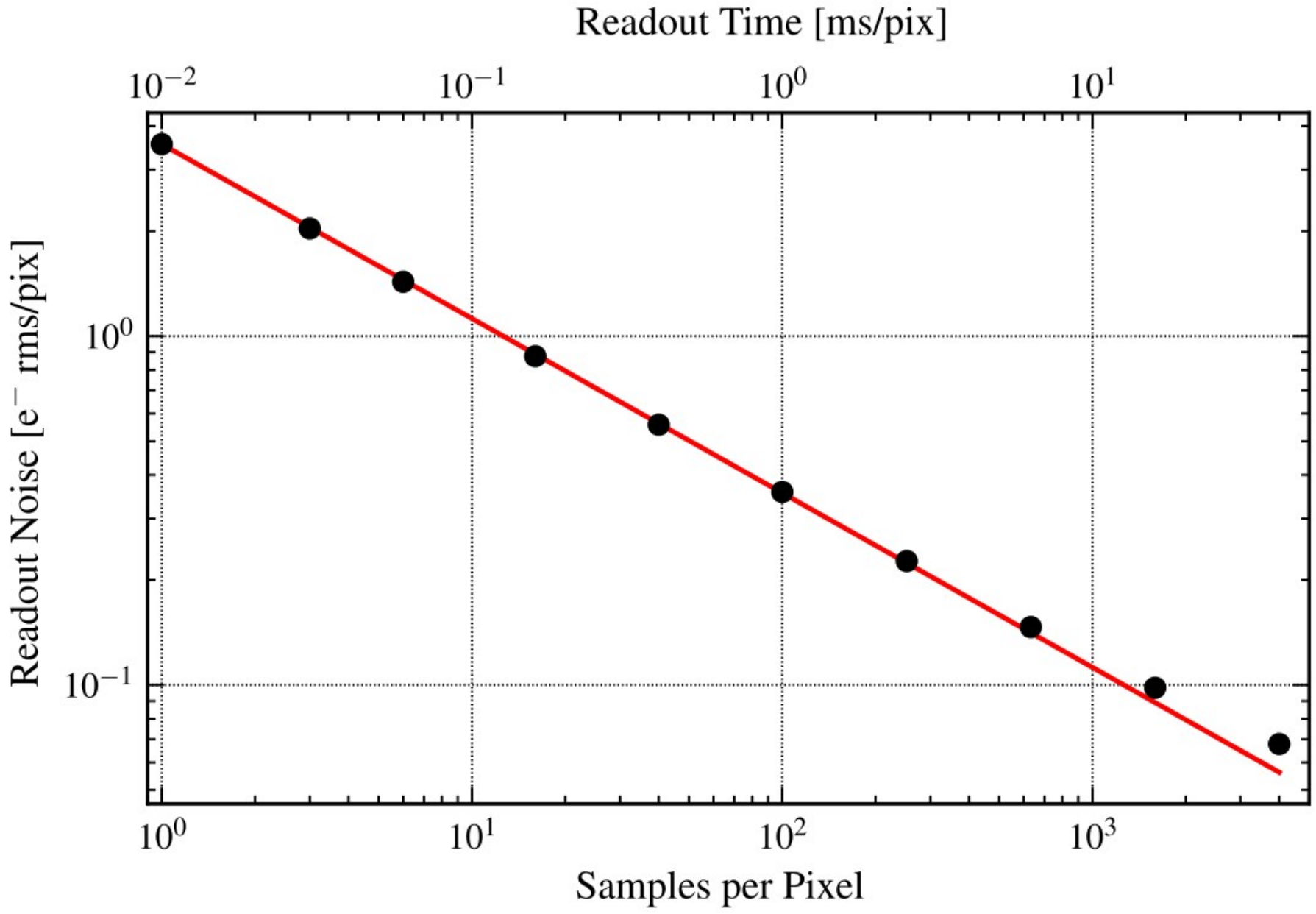}
         \label{fig.skippervsn}
     \end{subfigure}
     \hfill
     \begin{subfigure}[c]{0.5\textwidth}
         \centering
         \caption{}
         \includegraphics[width=\textwidth]{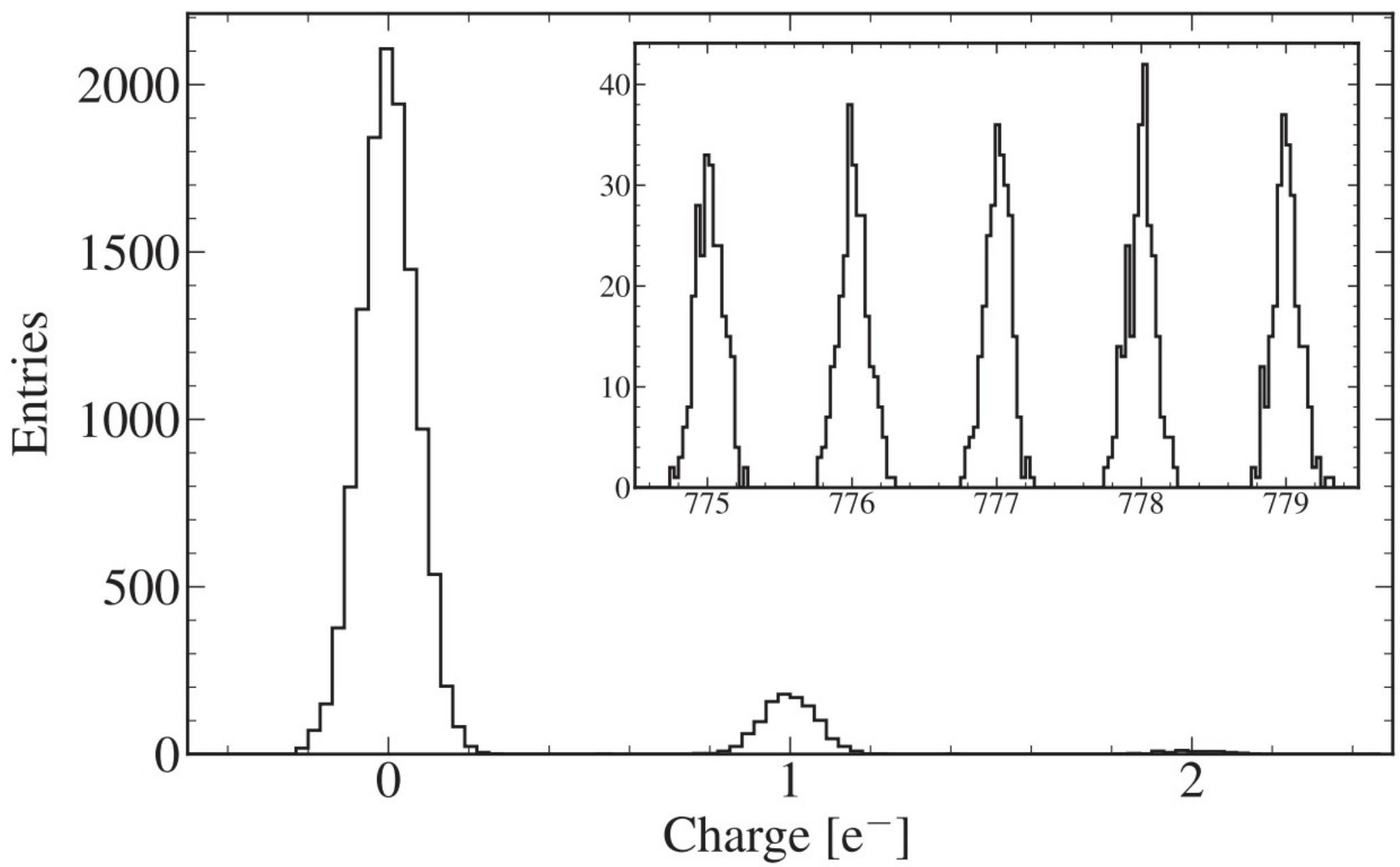}
         \label{fig.skipperpeaks}
     \end{subfigure}
        \caption{Figure \ref{fig.skippervsn} shows the readout noise as a function of the number of samples per pixel for a Skipper-CCD. In Figure \ref{fig.skipperpeaks}, the peaks of 0 and 1 electrons are clearly differentiated using 4000 samples and a noise level of 0.068 electrons. Within the same figure, peaks are distinctly separated between 775 and 779 electron charges. Figures extracted from \cite{tiffenberg2017}.}
        \label{fig.skippertiffenberg}
\end{figure}

The ability of the Skipper amplifier to read a pixel multiple times allows for reducing readout noise as the square root of the number of samples, N. For a given noise obtained with N=1 sample, $\sigma_{1}$, the noise for N samples can be estimated as

\begin{equation}
    \sigma_{N}=\frac{\sigma_{1}}{\sqrt{N}}
    \label{eq.skipper}
\end{equation}

Although the idea was originally presented in 1974 \cite{wen1974design}, Skipper-CCDs did not fully achieve their functionality until 2017 \cite{tiffenberg2017}, after multiple attempts \cite{janesick1990new,moroni2012}. In \cite{tiffenberg2017}, the authors demonstrated the use of this technology, achieving readout noise as low as 0.068 electrons for 4000 samples and up to nearly 800 electrons in each pixel. This is shown in Figure \ref{fig.skippervsn}, which illustrates the reduction in noise as the square root of the number of samples, and in Figure \ref{fig.skipperpeaks}, which demonstrates sub-electron resolution from 0 to 779 electrons.

\section{Impact of Skipper Technology on the Search for Dark Matter and Neutrinos}

In 2015, the DAMIC experiment \cite{chavarria2015damic} started using conventional CCDs for dark matter searches, looking for scattering events resulting in nuclear recoils of Weakly Interacting Massive Particle (WIMP) dark matter candidates with silicon atoms in the silicon crystal. Due to the aforementioned quenching factor and a reported readout noise of 2.5 $\e$, the experiment was insensitive to energy deposits below 0.5~keV.

In 2017, the achievement of sub-electronic noise thanks to Skipper technology marked a significant technological advancement in the use of this technology. It became possible to detect energy deposits as low as 1.1 eV, and the capability to precisely count the number of electrons in a pixel allowed for the accurate counting of dark matter events with hundreds of electrons. This breakthrough enabled the exploration of various dark matter models, such as Light Dark Matter (LDM) models (see Section \ref{sec:candidates}), where the mass of the dark matter candidate falls in the keV-MeV mass range. In such models, nuclear recoils are negligible compared to electronic recoils. Additionally, the search for dark photons via absorption was made feasible. These candidates often produce only a few ionized electrons, far below the energy threshold of conventional CCDs. The theoretical framework for electronic recoils of dark matter candidates with ordinary matter, as discussed in Section \ref{sec:electronrecoils}, is exploited by SENSEI to search for sub-GeV mass dark matter (and absorption of dark photon matter down to 1 eV, as introduced in Section \ref{sec:dmabsorption}).

Furthermore, access to sub-electronic noise also had implications for other low-occurrence event searches besides dark matter using SCCDs. CONNIE is a collaboration that uses conventional CCDs to search for neutrinos that scatter coherently with silicon nuclei \cite{freedman1974}. Sub-electronic noise is highly useful for this application as energy deposits lower than 30 ionized electrons are expected for coherent neutrino scattering \cite{CONNIE2015}. Additionally, SCCDs can have fruitful applications in the search for Earth-like exoplanets, as the photon flux from these sources can be on the order of photons per minute. Reducing electronic noise will result in lower readout times, which are crucial for space-based imaging and spectroscopy \cite{drlica2020}.

\subsection{Objectives of SENSEI}

In the framework of Skipper technology, SENSEI was born, which exploits the capability of SCCDs to search for Light Dark Matter (LDM) candidates. The SENSEI experiment had two clear objectives: (1) to conduct a 10-gram experiment in the MINOS cavern at Fermilab in Batavia, Illinois, United States, and (2) to conduct a 100-gram experiment in the deep underground facility SNOLAB in Sudbury, Canada. Both objectives were divided into smaller stages, as will be explained in the following chapters. Furthermore, although (1) was partially completed in 2020 (see Section \ref{sec:2020}) with a 2-gram detector, the excellent competitiveness of the results and the availability to install the final detector in SNOLAB starting from the same year the 2020 results were published allowed SENSEI to proceed with (2), step by step, progressively increasing the detector's mass (see Section \ref{sec:future}).

\newpage
\mbox{}
\thispagestyle{empty}
\newpage
\chapter{Experimental device}
\label{cap:3}

In this chapter, we will present the experimental devices used for the data collection conducted during this Thesis. There will be three aspects to consider in their description: the location, the vacuum chamber used, and the type of detector employed. Since both the characterization of 1-electron events and the establishment of dark matter exclusion limits were performed using virtually the same experimental setup, its description will not only serve as an introduction to the obtained results but also as an example to detail key concepts of the tools used for data acquisition and processing.

\section{Locations}

During the development of this Thesis, sensors located in two different locations were used, which we will identify as \textit{SiDet} and \textit{MINOS}.

\subsubsection{\textit{SiDet}}

\textit{SiDet} is the abbreviation for the \textit{Silicon Detector Facility} located at the Fermi National Accelerator Laboratory (FNAL) in Batavia, Illinois, USA. This facility is situated on the surface and, as its name suggests, is equipped for testing and fine-tuning silicon devices, particularly CCDs. Images taken in this facility are quickly populated by background events from cosmic rays and the Earth's atmosphere. A study of surface radiation at this location using a Skipper-CCD and conducted during the course of the Thesis can be found in \cite{moroni2022skipper}. Because this background of events covers the entire energy range of interest, this location is disadvantaged for dark matter searches, although it was used repeatedly for rapid tests without the need for underground facilities, as presented below.

\subsubsection{\textit{MINOS}}

This location gets its name from the experiment called \textit{Main Injector Neutrino Oscillation Search}, given its initials, which studied the oscillation of neutrinos produced by the Main Injector at FNAL. For the installation and commissioning of this experiment, an underground cavern at a depth of 107 meters was used, which we simply refer to as \textit{MINOS}. The 107 meters of earth drastically reduce the rate of high-energy events, particularly muons, reaching the detectors, increasing sensitivity to events with much lower interaction rates such as neutrinos or, in the case of interaction with ordinary matter, dark matter. Both the results of contributions presented in Chapter \ref{cap:4} and the exclusion limits in Section \ref{sec:2020} were obtained at this location.

\section{Vacuum Chambers}
\label{sec:camaras}

\subsection{Surface Testing}

For all surface tests (conducted entirely at \textit{SiDet}) presented in this Thesis, various commercial vacuum chambers were used interchangeably. The primary goal was to quickly achieve a high vacuum regime ($<1 \times 10^{-5}$ Torr). The chamber's volume and design shape were tailored to the requirements of the detectors being characterized.

\begin{figure}[h!]
\begin{center}
\includegraphics[width=0.6\textwidth]{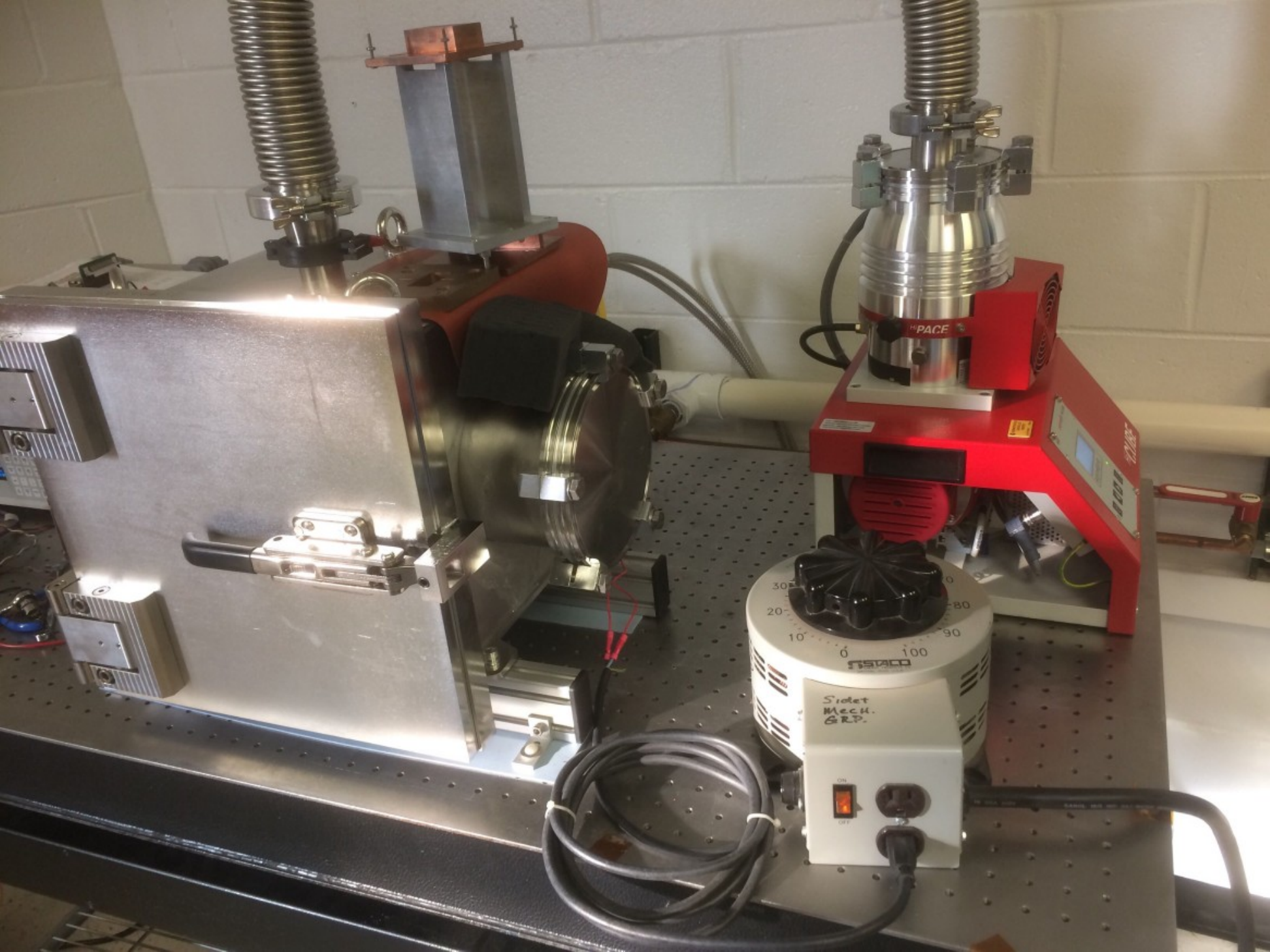}
\caption{Testing station used for the characterization of SCCDs on the surface, showing the vacuum chamber and the turbomolecular vacuum pump.}
\label{fig.supercube}
\end{center}
\end{figure}

The vacuum chambers are installed alongside a temperature and pressure controller, a cooling system, a vacuum pump, and a data acquisition system connected to a computer for data collection. These components will be described in the following section, which introduces the other vacuum chamber mentioned in this Thesis.

\subsection{\textit{MINOS Vessel}}

\begin{figure}[h!]
\begin{center}
     \begin{subfigure}[c]{0.42\textwidth}
         \centering
         \caption{}
         \includegraphics[width=\textwidth]{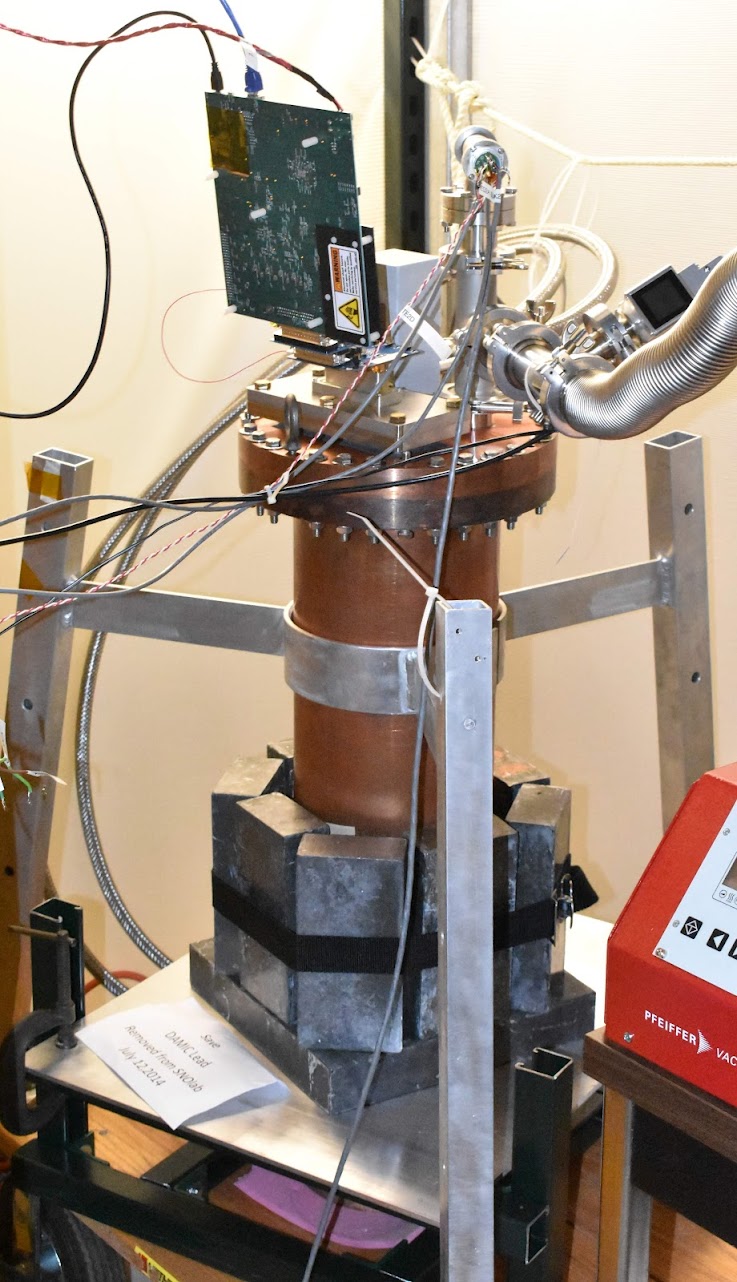}
         \label{fig.MINOS_setup1}
     \end{subfigure}
     \hfill
     \centering
     \begin{subfigure}[c]{0.46\textwidth}
         \centering
         \caption{}
         \includegraphics[width=\textwidth]{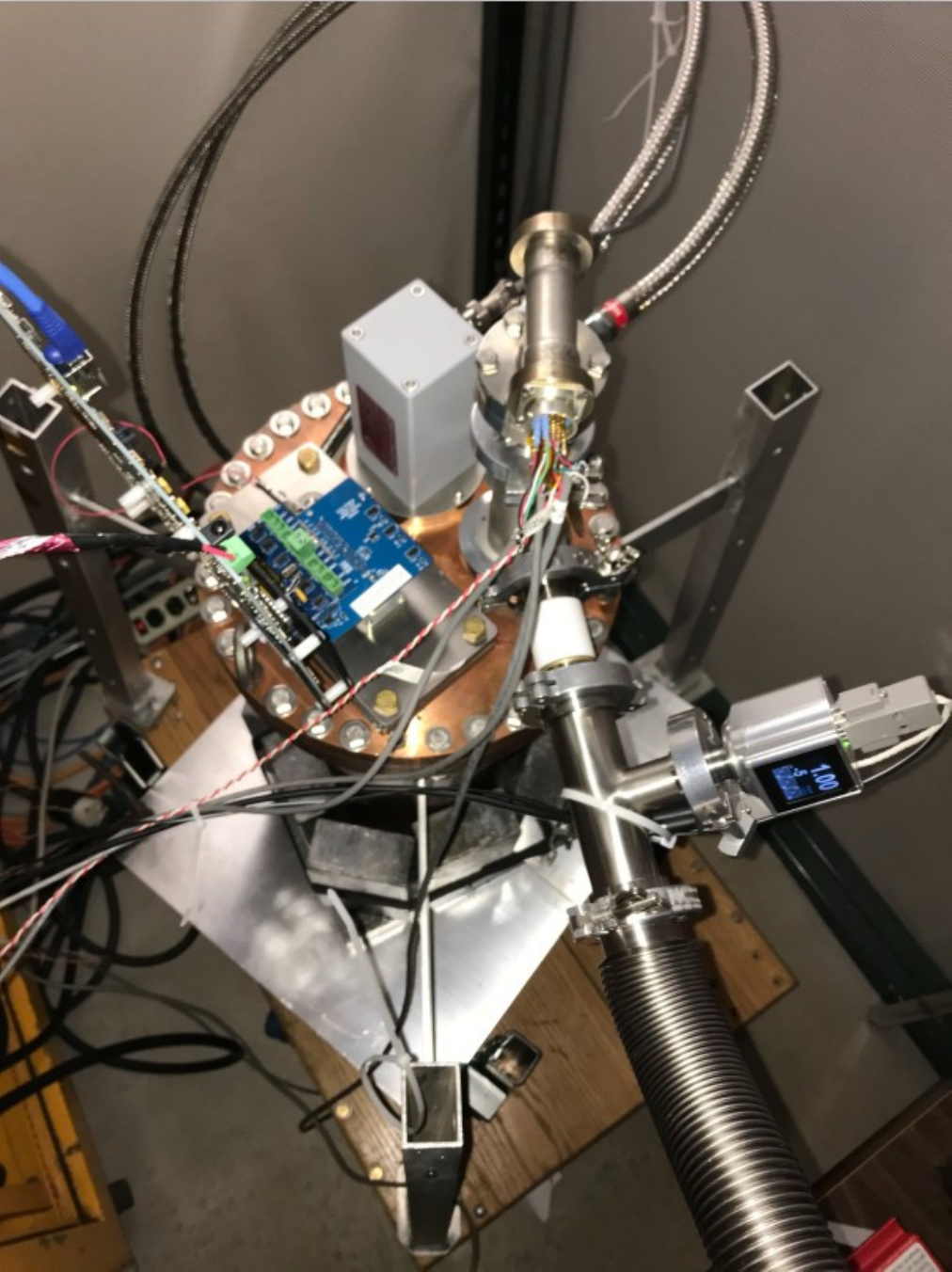}
         \label{fig.MINOS_setup_from_above}
     \end{subfigure}
\caption{Photographs of the \textit{MINOS Vessel}. In Figure \ref{fig.MINOS_setup1}, the vacuum chamber can be seen from one side, while Figure \ref{fig.MINOS_setup_from_above} shows a photograph of the top view.}
\label{fig.MINOS_setup}
\end{center}     
\end{figure}

For all results with data from the \textit{MINOS} location, the \textit{MINOS Vessel} was used. This system consists of a cylindrical vacuum chamber, as shown in the photograph in Figure \ref{fig.MINOS_setup1}. The chamber is enclosed within a tent to protect it from airborne dust. It is made of copper and opens at the top with a circular lid that is sealed shut with a series of screws around its perimeter, as seen in Figure \ref{fig.MINOS_setup_from_above}. In the latter figure, from left to right, you can see the LTA (\textit{Low-Threshold Acquisition system}) readout board (also visible in Figure \ref{fig.MINOS_setup1}), the top part of the cold finger used to cool the detector, a stainless steel structure through which the temperature sensors for the detector and cold finger are connected, and embedded in the same structure, the vacuum hose through which the chamber is evacuated, along with the corresponding pressure sensor. The turbomolecular vacuum pump can be seen in Figure \ref{fig.MINOS_setup1} on the right.

Next to the chamber is a rack of controllers responsible for providing electrical power to the entire system and controlling the temperature. Heat removal is carried out by a cryocooler installed outside the tent, allowing for a typical operating temperature of 130K for dark matter event searches. The cryocooler and vacuum pump are electrically isolated from the electrical power sources that feed the temperature controllers and, especially, the LTA readout board, in order to isolate noise sources during image readout.

The LTA board was used to obtain all the datasets presented in this manuscript. It was developed in response to the use of Skipper technology to read the CCD pixels and the consequent reduction in readout noise. With this purpose in mind, a readout system was created that could reduce all external noise contributions to the CCD and function as both a controller for its operation and a reading device. The LTA is responsible for controlling the voltages of the SCCD and its readout using sequencers or sequence routines that precisely control the state of the voltages to be delivered throughout the SCCD. The readout is performed using four 18-bit analog-to-digital converters with a sampling rate of 15 MHz, allowing parallel reading of the four video channels associated with the four output amplifiers of the SCCD. A detailed description of this board can be found in \cite{moroni2019low, cancelo2021, 8214366}. The board is connected via an Ethernet cable to the computer where the data is stored, processed, and analyzed, as described in Chapter \ref{cap:3bis}.

\begin{figure}[h!]
\begin{center}
\includegraphics[width=0.6\textwidth]{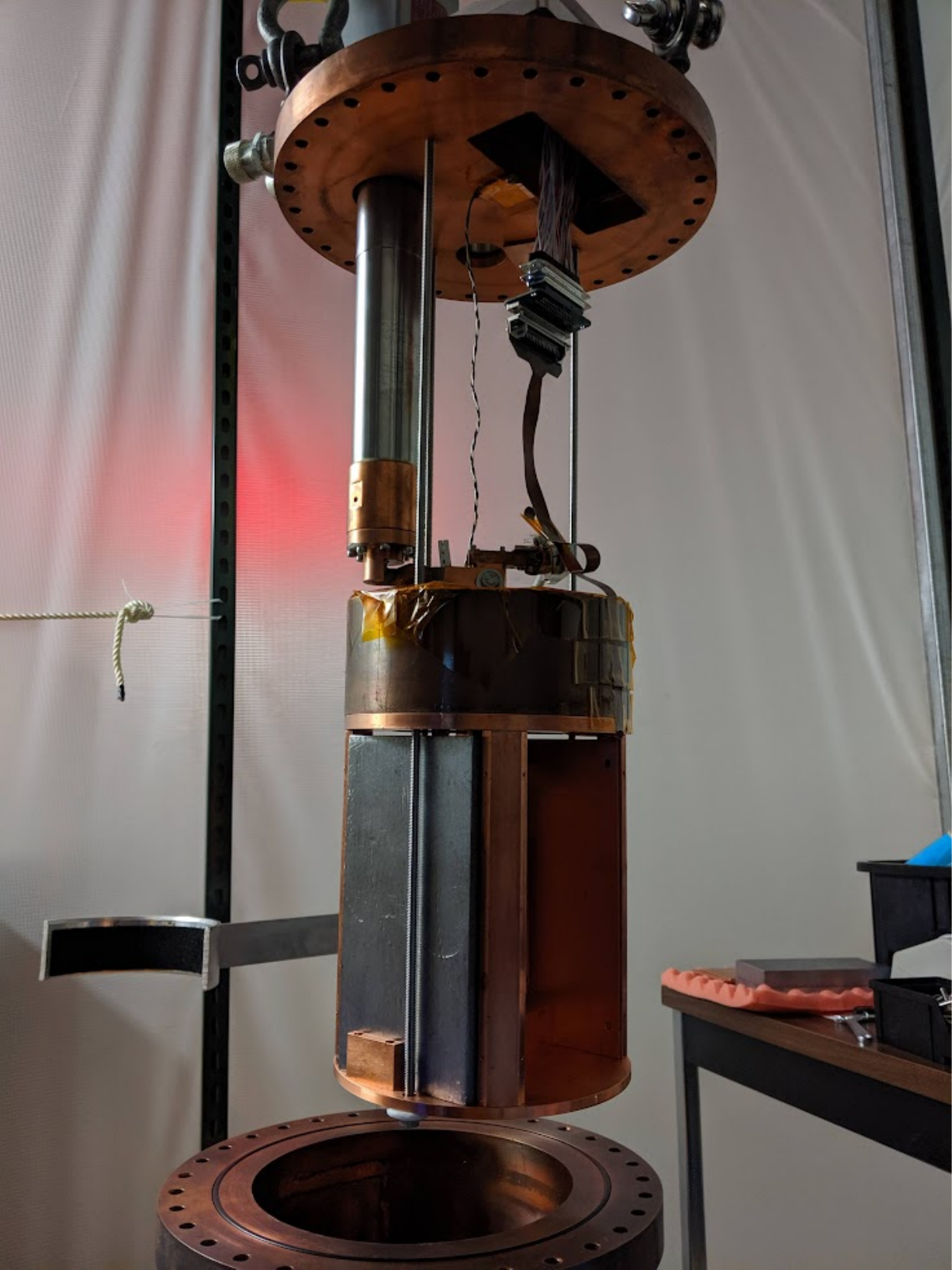}
\caption{Photograph of the open \textit{MINOS Vessel}, with the structure containing the SCCD suspended by a crane.}
\label{fig.MINOS_setup2}
\end{center}
\end{figure}

Figure \ref{fig.MINOS_setup2} shows a photograph of the open system taken during a maintenance task. To open it, due to the weight of the lid and supporting structure, a crane about two meters high is needed, which is located inside the tent. In this figure, you can see the upper lid of the vacuum chamber at the top of the image, and at the bottom, the structure where the SCCD is placed. The structure is connected to the lid by two stainless steel bars that are embedded in the lower part of it. Hanging between the structure and the top lid are the LTA cables and temperature sensor cables, which lead to the SCCD, as well as the cold finger responsible for cooling the structure where the SCCD is positioned.

\begin{figure}[h!]
\begin{center}
\includegraphics[width=0.7\textwidth]{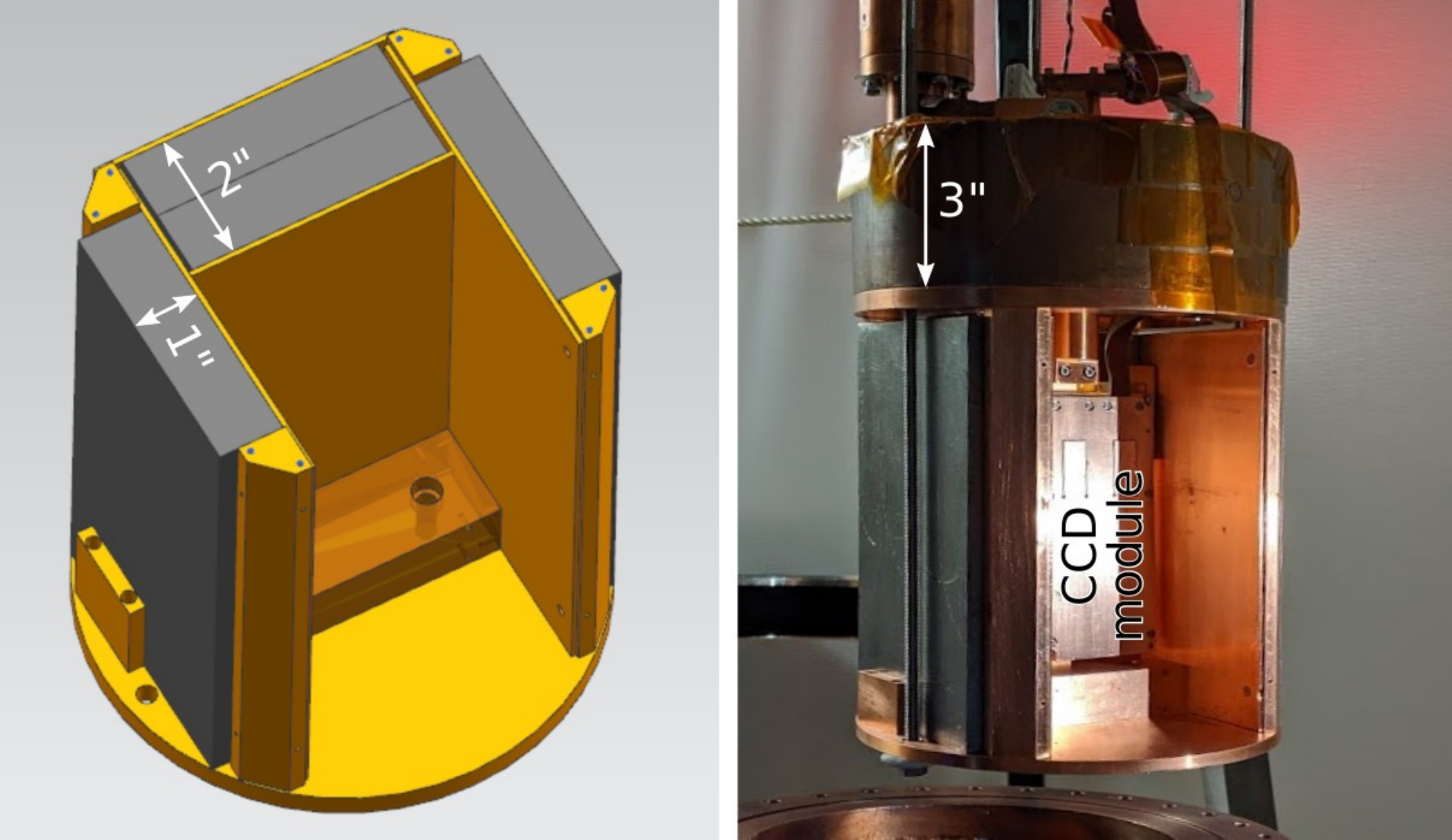}
\caption{Diagram of the structure containing the detector and its surrounding shielding (left). Photograph of the suspended structure specifying the position of the CCD (right). Figure extracted from the supplementary material of \cite{SENSEI2020}.}
\label{fig.innerShield}
\end{center}
\end{figure}

The SCCD is positioned inside a module, and this module is placed within the mentioned structure, as shown in Figure \ref{fig.innerShield}. The copper structure shields the module from infrared photons that may be produced on the walls of the vacuum chamber since it is at room temperature. It also serves as support for a 1 and 2-inch lead shield. Additionally, at the top, a 3-inch lead plate shields the upper part of the structure. The entire structure is at the same temperature as the module and the SCCD, which drastically reduces the emission of infrared photons. Figure \ref{fig.innerShield} shows the detail of the structure, both in a photograph and in a schematic, while also showing the SCCD module positioned parallel to the faces of the lead blocks. The internal shielding is completed by adding a missing 2-inch block, opposite to the one visible in the figure, which was removed to take the photograph.

Finally, with regard to the lead shielding, it's worth noting the use of additional 2-inch lead shielding outside the vacuum chamber, at the level of the detector and visible in Figure \ref{fig.MINOS_setup1}. This lead shielding was added during the course of this thesis and has special relevance in the analysis carried out on the contributions to the signals of interest developed in Chapter \ref{cap:4} and Section \ref{sec:2020}. We will refer to this shielding as the "additional shielding."

\section{Detectors Used}
\label{sec:dispositivo}

Two types of detectors were used in the development of this thesis. Both are Skipper-CCDs, but they belong to different generations, with the second being an improved version of the first. Therefore, we will focus on describing the second one in detail, but we will also mention the first prototype and its differences for completeness.

\subsection{\textit{protoSENSEI}}

The \textit{protoSENSEI} is a Skipper-CCD used until 2019. It has a structure similar to the final device but with some differences, including:

\begin{itemize}
    \item \textbf{Reduced Thickness}: It has a thickness of 200 $\mu$m compared to the 675 $\mu$m of the final device.
    \item \textbf{Fewer Pixels}: It has 225,888 pixels compared to 5,443,584 in the final device.
    \item \textbf{Lower-Quality Silicon}: The silicon used in the \textit{protoSENSEI} has lower resistivity due to doping and the presence of defects, resulting in higher intrinsic backgrounds that hinder the search for dark matter.
\end{itemize}

\noindent While this device produced competitive exclusion limits during its use, it was limited by its lower mass and the presence of background noise. The \textit{protoSENSEI} was used to obtain dark matter exclusion limits prior to the start of this thesis, and its analysis and results are reviewed as historical background in Section \ref{sec:antecedentes}.

\subsection{\textit{SENSEI-SCCD}}

The SCCD, called \textit{SENSEI-SCCD}, is made from high-quality silicon with a resistivity of 18 k$\Omega$-cm. It consist of 6144 rows and 886 columns (5,443,584 pixels of 15 $\mu$m each), resulting in an active area of 9.216 cm $\times$ 1.329 cm and a thickness of 675 $\mu$m, corresponding to an active mass of 1.926 grams.
Each pixel has a volume of 15 $\mu$m $\times$ 15 $\mu$m $\times$ 675 $\mu$m and a mass of 3.537 $\times 10^{-7}$ g.
Each SCCD has four amplifiers, one in each corner, which can read the entire CCD or each of the four quadrants (3072 rows and 443 columns) synchronously and independently, generating four sub-images, one for each quadrant, for each readout of the SCCD.

\renewcommand{\arraystretch}{1.2}
\begin{table}[h!]
\centering
\begin{tabular}{lccc}
\hline
Name                      & protoSENSEI                 & SENSEI-SCCD                                                                       &                                \\ \hline
Dimensions            & 624  $\times$  362   & 3072  $\times$  443                                                         & pixels                               \\
Pixel Size               & 15 $ \times$ 15      & 15 $ \times$ 15                                                            & $\mu$m$^{2}$                         \\
Thickness                 & 200                  & 675                                                                        & $\mu$m                               \\
Total Mass               & 0.0947  & 1.926                                                                     & grams                                    \\
Number of Amplifiers      & 4(*)           & 4(*)                                                               &                                      \\
\hline
\end{tabular}
\caption{Table summarizing the characteristics of the used detectors. (*) In both cases, for different reasons, one of the four amplifiers did not function correctly.}
\label{tab:dispositivos}
\end{table}

The architecture of the MOSFET amplifiers in the output device was chosen during the manufacturing process based on previous internal studies to maximize output gain and reduce amplifier light emission, which negatively impacts device performance, as will be discussed in Sections \ref{sec:al} and \ref{sec:2019}.
This study involved using different areas for the M1 amplifier gates in the output device, as detailed in \cite{miguelthesis}.
As a result of this, and the improved quality and reduced resistivity of the silicon used, a lower (though not entirely eliminated, as will be seen in Section \ref{sec:al}) luminescence was observed during the SCCD readout.

\begin{figure}[h!]
    \centering
    \includegraphics[width=0.7\textwidth]{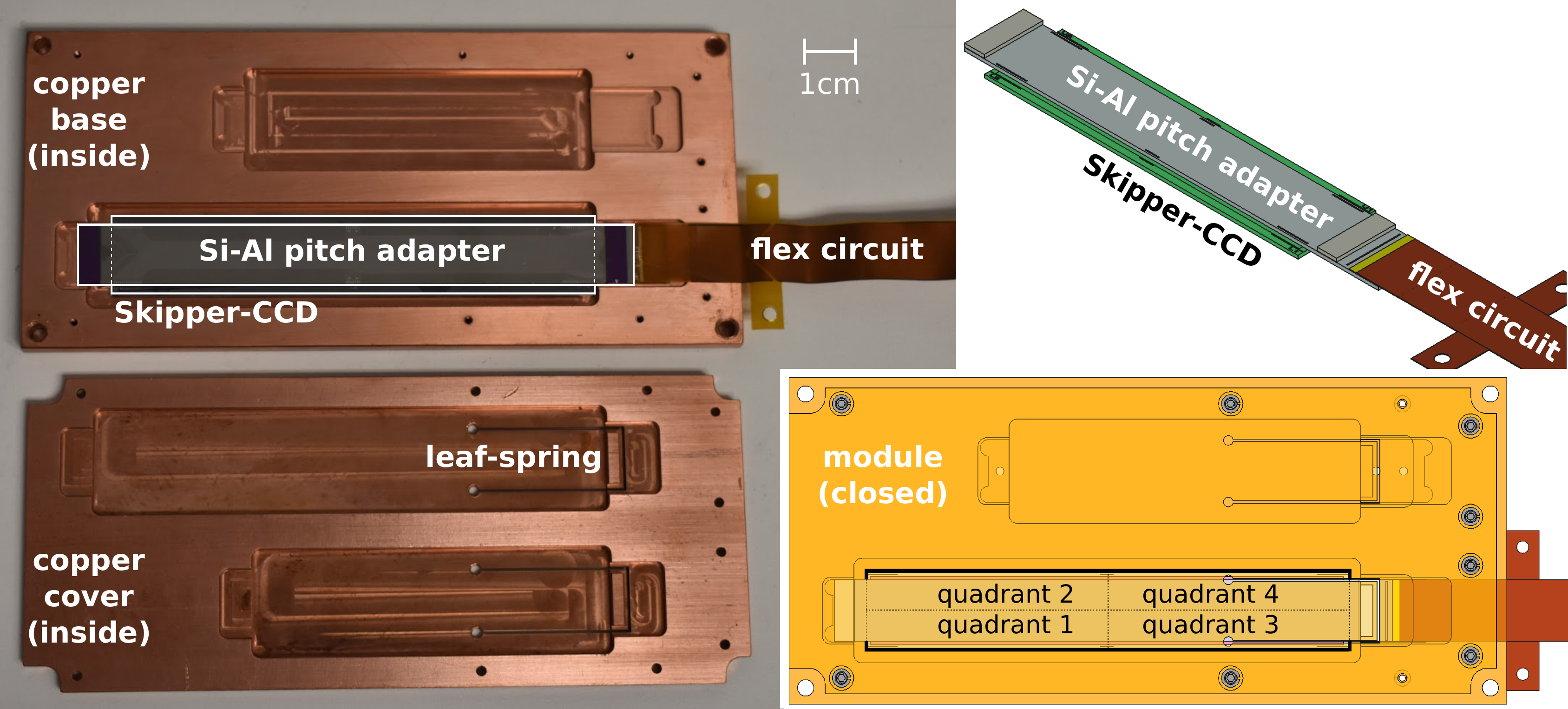}
    \caption{Diagram and photograph of the module containing the SCCD, specifying its parts. Figure extracted from \cite{SENSEI2020}.}
    \label{fig.sensei_module}
\end{figure}

The SCCD is attached to a silicon base where the pitch-adapter circuit connecting the detector to the readout electronics is located.
The SCCD's output pads are connected by gold wires through wire bonding carried out at FNAL facilities, ensuring the stability of this connection.
Subsequently, the signal is transferred via a flexible cable to the LTA.
Finally, the package is placed, as shown in Figure \ref{fig.sensei_module}, inside a copper module that provides robustness, portability, and prevents the collection of infrared photons from the surrounding environment.
A flat spring on top of the module ensures thermal contact between the SCCD and the copper module.

\newpage
\mbox{}
\thispagestyle{empty}
\newpage
\chapter{Data Acquisition and Processing}
\label{cap:3bis}

A significant part of the efforts carried out in the context of this Thesis were aimed at developing and optimizing data acquisition and processing routines to characterize the operation of the SENSEI Skipper-CCDs (\textit{SENSEI-SCCD}s) and subsequently search for dark matter. This chapter will describe these efforts and the structure of the resulting process.
While some of the tools and concepts described in this chapter predate the start of my doctoral studies, it was my task to extend the functionalities of these tools and create new concepts and solutions for the development of the SENSEI experiment. In particular, the creation of new data acquisition routines and the development of clustering and masking techniques were essential. Additionally, it's worth noting that this contribution extended beyond the SENSEI experiment, resulting in highly useful tools for other studies conducted with SCCDs. These include the characterization of Compton background \cite{comptonbotti}, the measurement of the Fano factor \cite{FANO,rodrigues2023unraveling}, and the characterization of high-energy surface backgrounds \cite{moroni2022skipper}, all of which I co-authored.

This chapter is divided into three sections encompassing the data acquisition protocol used and the data processing, with a particular focus on the event selection criteria, which is a crucial element in the search for events compatible with a dark matter signal.

\section{Data Acquisition}
\label{sec:dataacquisition}

\begin{figure}[h!]
\begin{center}
\includegraphics[width=0.7\textwidth]{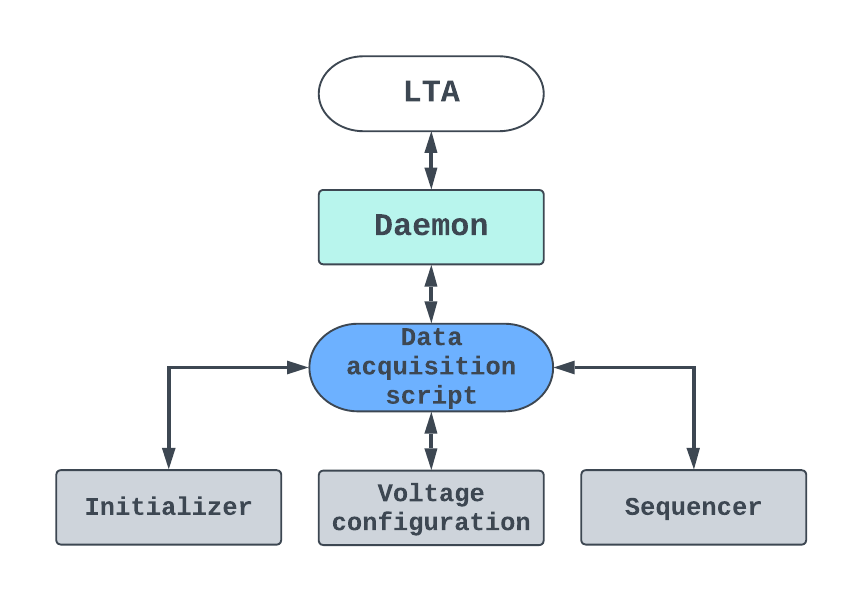}
\caption{Diagram illustrating the structure of the programs used for data acquisition by the Low Temperature Apparatus (LTA).}
\label{fig:scripts}
\end{center}
\end{figure}

Figure \ref{fig:scripts} shows the structure of the data acquisition software in conjunction with the LTA (Low Temperature Apparatus). A daemon runs in the background throughout the operation, responsible for powering on the LTA and initializing it so that the LTA, in turn, does the same for the Skipper-CCD (SCCD). The daemon runs as an executable compiled in the terminal and written in C/C++. Data acquisition begins when the Data Acquisition script, illustrated in Figure \ref{fig:scripts}, is executed. Firstly, the Initializer is called by the Data Acquisition script to create the environment for data collection (file names, write directories, etc.), read modes\footnote{It can, for example, instead of taking a complete image, which is necessary for dark matter searches, measure the output voltage as a function of time, as shown in Figure \ref{fig.pedestalsignal}.}, waiting time between voltage application, integration times for signal and pedestal, among others. The Voltage Configuration is responsible for pre-configuring the voltages that will be used at each of the SCCD gates when operating it. Some examples include the upper and lower voltages of the pixels in the active area (V1, V2, and V3) or the horizontal register (H1, H2, and H3), or the V$_{drain}$ voltage to which the charge from a pixel is transferred for final disposal after being read at the sensing node (SN).

Finally, the sequencers used for data collection are loaded. These codes are in XML format and detail step by step the reading of the SCCD after it is powered on. This process ranges from the first vertical transfer to the reading of the last pixel in the image. Additionally, there are sequencers for the cleaning phase or special data collection routines that require skipping the reading of certain rows/columns, cleaning between rows, changing the direction of vertical/horizontal charge transfer, among other reading variations. The LTA board will use these sequencers to instruct the gates of each pixel on how and when, and for how long, they should change their electrical potential.

The Data Acquisition script is responsible for encapsulating the execution of these codes and organizing the reading of the detector, both by modifying some of the parameters previously configured by the Initializer and the Voltage Configuration and by setting other parameters such as exposure time or the number of images to be read, among others.

During the readout, the data is saved in binary format and is later translated into the FITS (Flexible Image Transport System) format, where the values of the number of electrons for each sample in each pixel are stored in units of ADU (analog-to-digital unit). This unit has a relationship with the number of electrons in each pixel that depends on the output gain of the entire system (SCCD, wiring, adapters, and LTA, as a whole). FITS files also store useful metadata about the measurement (voltages, integration time, image dimensions, start and end date of the reading, among others) to be used in the future.

\begin{figure}[h!]
\begin{center}
\includegraphics[width=0.8\textwidth]{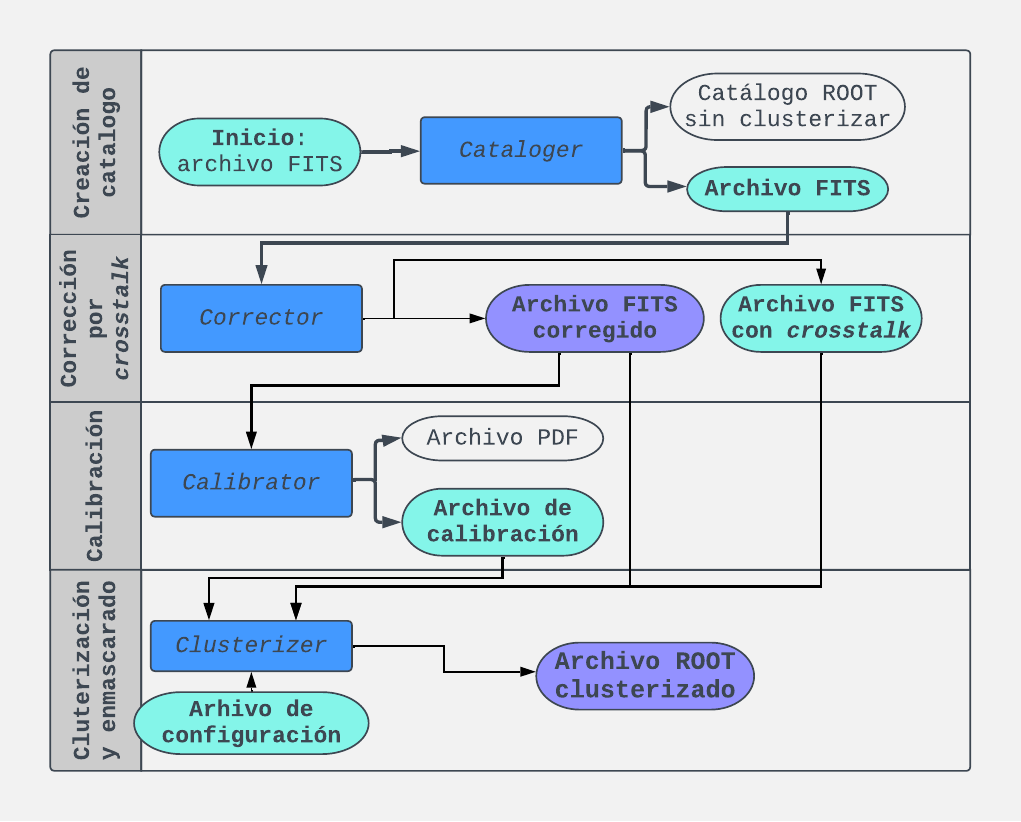}
\caption{Diagram depicting the flow of processes that an image undergoes during data processing. In blue, the parts of the process where code is executed; in light blue, elements used as inputs within these codes; in violet, products used in subsequent analysis; and in white, products not used for the same.}
\label{fig.dataprocessing}
\end{center}
\end{figure}

\section{Data Processing}

Once the FITS file containing all the samples for each pixel is obtained, data processing follows for subsequent analysis, following the process flow illustrated in Figure \ref{fig.dataprocessing}. This flow can be divided into four stages: catalog creation, crosstalk correction, calibration, and clustering and masking. In this section, each of these processes will be detailed, except for masking, which will be discussed separately in Section \ref{sec:masqueo}, due to its relevance and extent.

\subsection{Catalog Creation}
\label{sec:catalogcreation}

The catalog creation results in two files: a FITS image with the charge value for each pixel in ADUs, averaging the number of samples corresponding to each pixel, and a ROOT file/catalog with information about certain observables from the image that are useful for further analysis. The observables in this catalog include the charge per pixel, the position of that pixel, which quadrant it belongs to, and a number that identifies the processed image.

The most crucial aspect of catalog creation is undoubtedly the subtraction of the baseline for each pixel. As explained in Section \ref{sec:designandoperationprinciples}, each pixel is read multiple times, and the output signal is an analog voltage continuously read by the LTA board. Typically, such signals are susceptible to changes in the baseline over time, making it difficult to determine the actual charge value in ADUs for each pixel. Baseline subtraction allows for a consistent representation of an "empty pixel" for all CCD pixels, addressing this issue.

During catalog creation, this baseline subtraction can be performed in various ways. In particular, the simplest method involves reading a small number of "extra" pixels for each row, known as overscan pixels. These pixels do not generate charge in the active area of the SCCD, as they do not come from that region. These pixels begin to collect charge when the charges from the last pixel in the row, which does come from the active area (i.e., the one transferred vertically from it to the horizontal register), move one pixel closer to the sensing node in the horizontal register. The "lifetime" of the overscan pixels is, therefore (and in most cases), much shorter than that of the active area pixels, so the probability of these pixels collecting charge is very low (again, compared to the active area pixels). Therefore, the simplest way to subtract the baseline is on a row-by-row basis, using the average value of the overscan pixels.
For images with very low occupancy, such as those obtained in underground laboratories at cryogenic temperatures using high-purity radioactive materials, empty pixels are highly abundant. In such cases, all empty pixels in a row are typically used without the need to obtain overscan pixels for these images.

\subsection{Crosstalk Correction}
\label{sec:crosstalkcorrection}

\begin{figure}[h!]
\begin{center}
\includegraphics[width=0.7\textwidth]{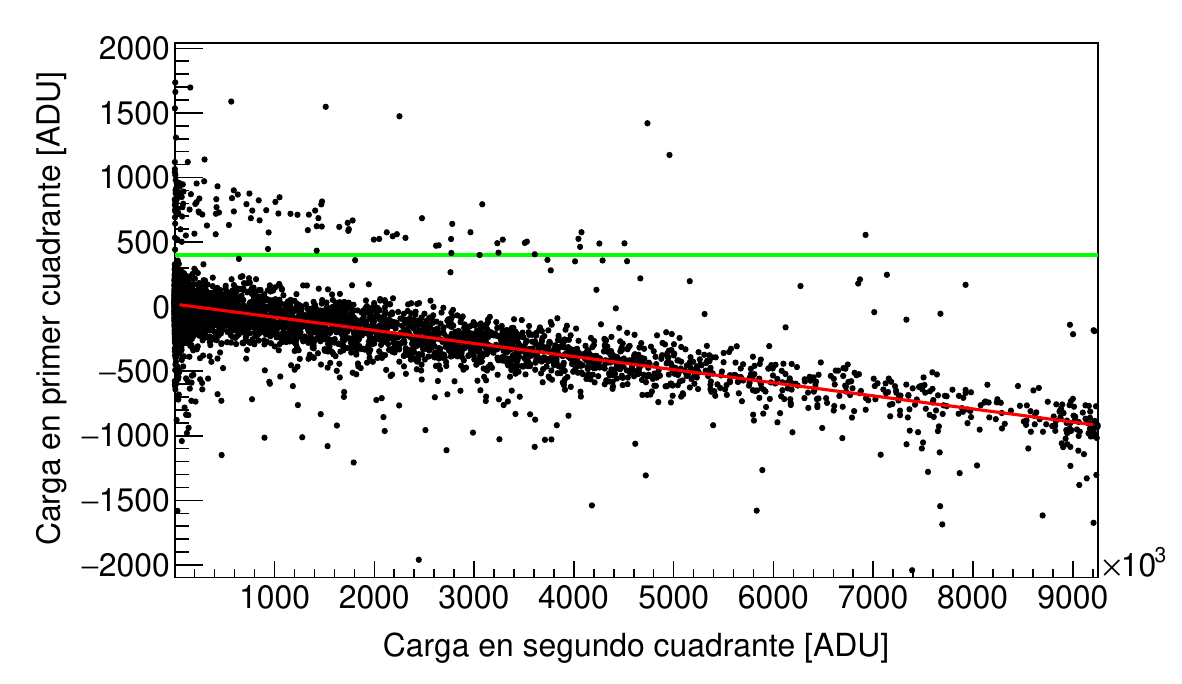}
\caption{Charge in the first quadrant relative to the charge in the second quadrant, expressed in ADUs. The value of 1 electron corresponds to approximately 800 ADUs. Due to crosstalk, the charge read in the first quadrant exhibits a slope concerning the charge collected in pixels read synchronously in the second quadrant. A linear fit to the cluster of empty pixels in the first quadrant relative to the second quadrant yields a slope of $(-1.02\pm0.02) \times 10^{-4}$. A green line is drawn at y=400 to separate empty pixels from non-empty ones in the first quadrant. Image extracted from the \textbf{SC} dataset (see Appendix \ref{cap:glossaryofdatasets}).}
\label{fig.crosstalk}
\end{center}
\end{figure}

After creating the initial catalog, the next step is crosstalk correction. This involves correcting the effect of signal coupling between one quadrant and another \cite{bernstein2017instrumental}. This effect can also occur between multiple SCCDs that share electrical wiring. Signal coupling has several consequences, but we will focus on the most significant one, illustrated in Figure \ref{fig.crosstalk}.
In this figure, you can see the charge signal in one quadrant relative to another quadrant, both read from the same LTA. Each point represents a pair of pixels read synchronously. At the origin of the coordinates, you can observe a cluster of points corresponding to pairs of pixels with a charge close to 0 ADUs, i.e., empty pixel pairs. Along the y-axis, about 800 ADUs above the cluster, there are empty pixels from the second quadrant, but with 1 electron in the first quadrant. The same occurs when moving along the x-axis, in the second quadrant: the pixels in the first quadrant remain empty while those in the second quadrant increase in their charge number, up to approximately 9,000,000 ADUs, which is roughly equivalent to 10,000 electrons in charge.
However, a non-zero slope can be observed between the charge in one quadrant and the other, originating from crosstalk. Furthermore, if a line is drawn at $y = 400$ ADU (equivalent to 0.5 $\epsilon$), approximately 20 pixels appear to come from the 1-electron peak in the first quadrant, and they would be classified as empty pixels\footnote{It is not the standard practice to use 0.5$\epsilon$ as the threshold between 0 and 1 electron pixels, although it serves as an example to explain the phenomenon}.

The effect produced by this slope, although small, is corrected for searches for events compatible with dark matter, resulting in a new FITS image, now corrected for crosstalk. This process involves calculating the slopes for each pair of quadrants and correcting them to obtain a zero slope. Additionally, to be conservative when selecting the data to use, the positions of the pixels that, according to the calculated slope, could generate enough crosstalk to subtract or add 1 electron in adjacent quadrants and/or SCCDs are saved and discarded from the dataset for analysis. This is performed in conjunction with masking, as discussed in Section \ref{sec:masqueo}.

\subsection{Image Calibration}
\label{sec:calibration}

\begin{figure}[h!]
\begin{center}
\includegraphics[width=0.7\textwidth]{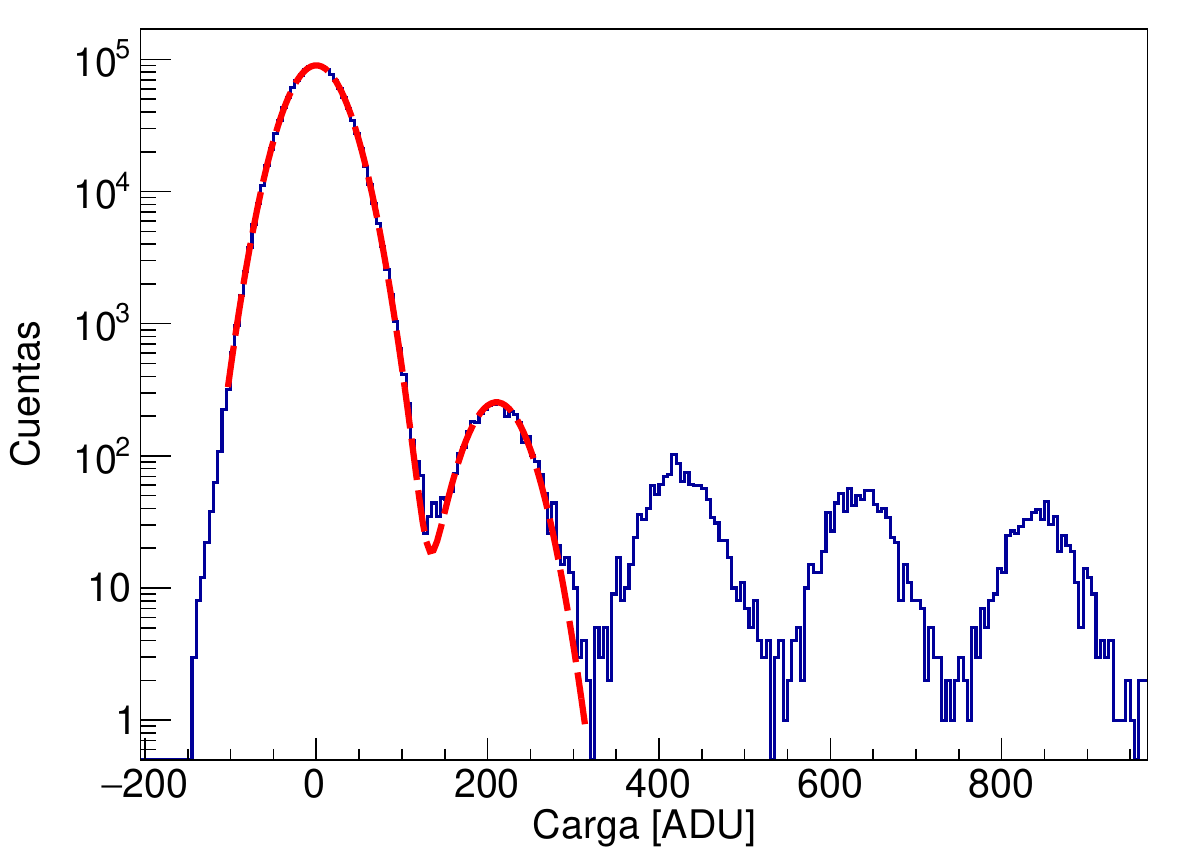}
\caption{Histogram of the number of pixels with charge between -200 and 900 ADUs. For this image (extracted from the calibration images, see Appendix \ref{cap:glossaryofdatasets}), 1 electron corresponds to approximately 220 ADUs. A fit, like that in Equation \eqref{eq.1efit}, is performed on the first two peaks of this histogram to obtain gain, electronic noise, and the rate of events of 1 electron per pixel.}
\label{fig.cal}
\end{center}
\end{figure}

Figure \ref{fig.cal} shows a screenshot of the PDF file resulting from the calibration of an image. This calibration is performed on the ROOT catalog generated by reprocessing the FITS image, which has already been corrected for crosstalk, using the catalog creator described earlier.

A histogram of the charge in ADUs for each pixel in the image is created within the given charge range. Due to electronic noise, the charge value of the pixels follows a Gaussian distribution around their actual charge value, with the dispersion equal to the electronic noise. Therefore, the area under each bell-shaped curve provides an estimate of the number of pixels in the image with a charge corresponding to the center of that curve.

In order to characterize the electronic noise of the image and the gain of that measurement, a fit of two Gaussian distributions is performed, one centered at 0 and the other at the expected gain value. Additionally, both curves are convolved with a Poisson distribution whose parameter is an estimator of the ratio of the area between the two bell-shaped curves. The form of this fit is described in the following equation:

\begin{equation}
    F(x|\mu,\sigma)=\sum_{k=0,1} Pois(k|\mu) \ \times \ Gaus(x|k,\sigma)
    \label{eq.1efit}
\end{equation}

\noindent where, $\mu$ is the rate of pixels with 1 electron divided by the total number of pixels, $\sigma$ is the electronic noise, $Pois$ is a Poisson distribution, and $Gaus$ is a normalizable Gaussian distribution. As a result, four parameters are obtained: the mean value of the peak at 0 electrons, the gain, the electronic noise ($\sigma$), and the characteristic parameter of the Poisson distribution ($\mu$).

This last parameter, anecdotal for calibration, is of vital importance when the same fit is used in the search for dark matter since it allows estimating the number of single electron events (SEEs) in an image without introducing any bias and without resorting to pixel clustering, a procedure that will be detailed later.
In other words, it is not necessary to set a threshold between the 0 and 1-electron peaks to obtain the rate and, therefore, the area under the 1-electron peak, equivalent to the number of pixels with 1 electron in the image. Thus, whenever this work refers to an estimation of the rate of 1-electron events or SEEs (single electron events) per image for a given set of images unless otherwise specified, it refers to this counting technique.

Once the gain calibration value is obtained for each image, clustering and masking are performed, both within the "Clusterizer" code (see Figure \ref{fig.dataprocessing}). To run this code, three files are needed as input: the FITS file corrected for crosstalk, a file with the pixel values to be masked for crosstalk, and a configuration file specifying parameters to be used for masking and pixel clustering, as well as other secondary configuration parameters.

\subsection{Clusterization}
\label{sec:clusterizacion}

Following Figure \ref{fig.dataprocessing}, we continue with the clustering process performed by the "Clusterizer." Clustering involves grouping non-empty pixels into the same cluster. This allows for organizing the analysis by associating multiple pixels with the same event or phenomenon, such as the interaction of an electron, a muon, or an event resulting from dark matter interaction with the detector.
Among the clustering parameters, the most important one is the threshold that specifies when a pixel is considered empty and when it is not (i.e., with 1 electron or more). As mentioned earlier in Section \ref{sec:calibration}, this value is irrelevant when calculating the number of single electron events (SEEs) in an image since this observable is calculated through a nonlinear fit of two Gaussian distributions convoluted with a Poisson distribution.
However, to identify events with 2 or more electrons, in one or more pixels, it is crucial to understand when a pixel is considered empty or not. In Figure \ref{fig.cal}, it can be seen that due to the electronic noise still present in the images, it is not clear where to set the threshold that separates the 0-electron peak from the 1-electron peak, or the 2-electron peak from the 3-electron peak, and so on.

This phenomenon and its consequences in the analysis were studied during this thesis and resulted in a tool that calculates the threshold in such a way that the number of pixels originating from the 0-electron peak is identical to that of pixels originating from the 1-electron peak, finding the cut-off value \(c\) that satisfies the following equality:

\[
\frac{1 - \int_{-\infty}^{\frac{c}{\sigma}} N}{\int_{-\infty}^{\frac{c-1}{\sigma}} N} = \mu
\]

where \(N\) is the Normal distribution, \(\sigma\) is the electronic noise, and \(\mu\) is the rate of single electron events (SEEs) per pixel. The rate effectively acts as the ratio between the area of pixels from the 1-electron peak divided by the area of pixels from the 0-electron peak. Using the calibration fit data, one can input the value of electronic noise present in the image \(\sigma\) and the rate of SEEs \(\mu\) into this equation. In practice, this procedure to obtain \(c\) is performed for a set of calibrated images simultaneously.

In general, for a rate of SEEs per pixel \(\mu\) on the order of about \(1 \times 10^{-4} \, \epix\) and electronic noise \(\sigma\) of \(0.16 \, \e\), the cut-off between 0 and 1 electrons is between 0.6 and 0.7 e. The same applies to successive energy levels.

\section{Event Selection Criteria or Masking}
\label{sec:masqueo}

In addition to the functions described in the preceding subsections, the \textit{Clusterizer} code is responsible for event selection criteria or masking, the development of which was a fundamental part of the efforts made during this thesis. The criteria are quality cuts that exclude elements from the dataset associated with deficiencies during image reading, the presence of defects in the detector material, or the effects of high-energy radiation incident on the material, considering that these three factors have the ability to contaminate the dark matter signal being sought, as will be detailed below. Additionally, the criteria can simply reduce the dataset in order to use the fraction of data compatible with the signal being studied, which we will identify as a specific selection.

The four mentioned groups (deficiencies during reading, defects in the material, signals associated with high-energy backgrounds, and specific selection) define the following subsections. Each of them organizes the quality criteria that, as an internal index, are presented in Table \ref{tab:cortes}.

\renewcommand{\arraystretch}{1.7}
\begin{longtable}{lp{8cm}}

\centering
\textbf{Name}                          & \textbf{Description}                                                                                                                                          \\ 
\midrule
\multicolumn{2}{c}{\textbf{Deficiencies during reading}} \\
\textbf{Noise}                           & Exclude pixels whose electronic noise is incompatible with the characteristic noise of the dataset.                         \\
\textbf{\textit{Crosstalk}}                        & Exclude pixels whose electronic value is affected by the signal simultaneously collected in another quadrant.                                           \\
\textbf{Bleeding}                        & Exclude pixels in the upper and right areas with 100 or more electrons to prevent counting SEEs generated by CTI.    \\
\midrule
\multicolumn{2}{c}{\textbf{Material defects}} \\
\textbf{Hot pixels}               & Exclude pixels with a SEE rate that is abnormally higher than the average. \\
\textbf{Hot columns}              & Exclude columns with a SEE rate that is abnormally higher than the average. \\
\textbf{\textit{Loose cluster}}          & Exclude areas of the dataset where an excess of SEEs is observed for the analysis of 2, 3, and 4-electron events. \\
\midrule
\multicolumn{2}{c}{\textbf{High-energy backgrounds}} \\
\textbf{High-energy halo}            & Exclude pixels around events with 100 or more electrons to prevent counting spatially correlated SEEs with high-energy events.          \\
\textbf{Edge}                           & Exclude pixels on the edges of the active area to avoid SEEs from high-energy events that interact near the edges.  \\
\textbf{Horizontal register events} & Exclude pixels in the same line or neighboring lines of events generated when impacting the horizontal register or below it.       \\ \midrule
\multicolumn{2}{c}{\textbf{Specific selection}} \\
\textbf{Single-pixel events}               & Exclude pixels whose neighbors are non-empty pixels.                \\
\textbf{Low-energy cluster}         & Exclude areas of the dataset where clusters of 2 to 100 electrons that are too close are observed.               \\
\bottomrule
\caption{Glossary of quality criteria used. }
\label{tab:cortes}\\

\end{longtable}

\subsection{Cuts for Erratic Behavior}

\subsubsection{Noise}
\label{sec:noise}

The first criterion in this group, noise, involves excluding elements from the dataset to be analyzed whose electronic noise is incompatible with the characteristic noise of a given quadrant. This criterion is sensitive to the specific system being worked on as noise sources can be very diverse. For example, for a set of SCCDs, an upper limit on acceptable noise for each quadrant can be established, excluding quadrants that exceed this noise. Additionally, if the noise is intermittent, pixels where the noise spikes and then stabilizes again can be excluded. This stability can be measured in terms of the Median Absolute Deviation (MAD), which is calculated as the median of the absolute deviation of a dataset from its median.

\subsubsection{Crosstalk}
\label{sec:crosstalk}

\begin{figure}[h!]
\begin{center}
     \begin{subfigure}[c]{0.3\textwidth}
         \centering
         \caption{First quadrant.}
         \includegraphics[width=\textwidth]{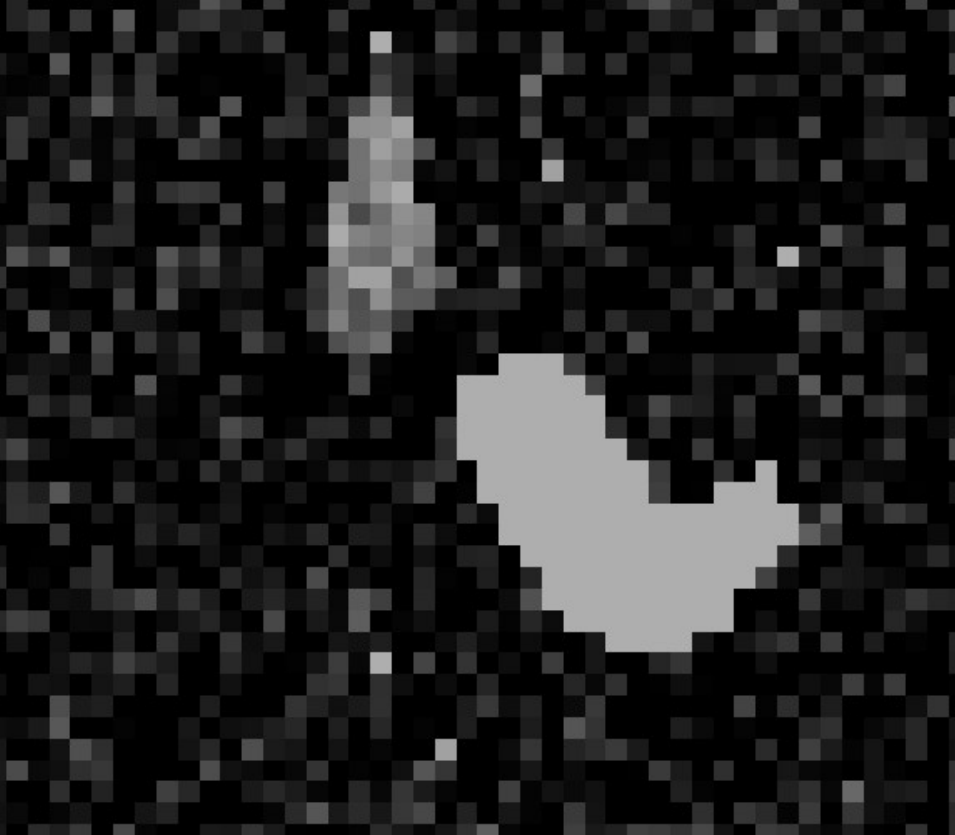}
     \end{subfigure}
     \hfill
     \centering
      \begin{subfigure}[c]{0.3\textwidth}
         \centering
         \caption{Second quadrant.}
         \includegraphics[width=\textwidth]{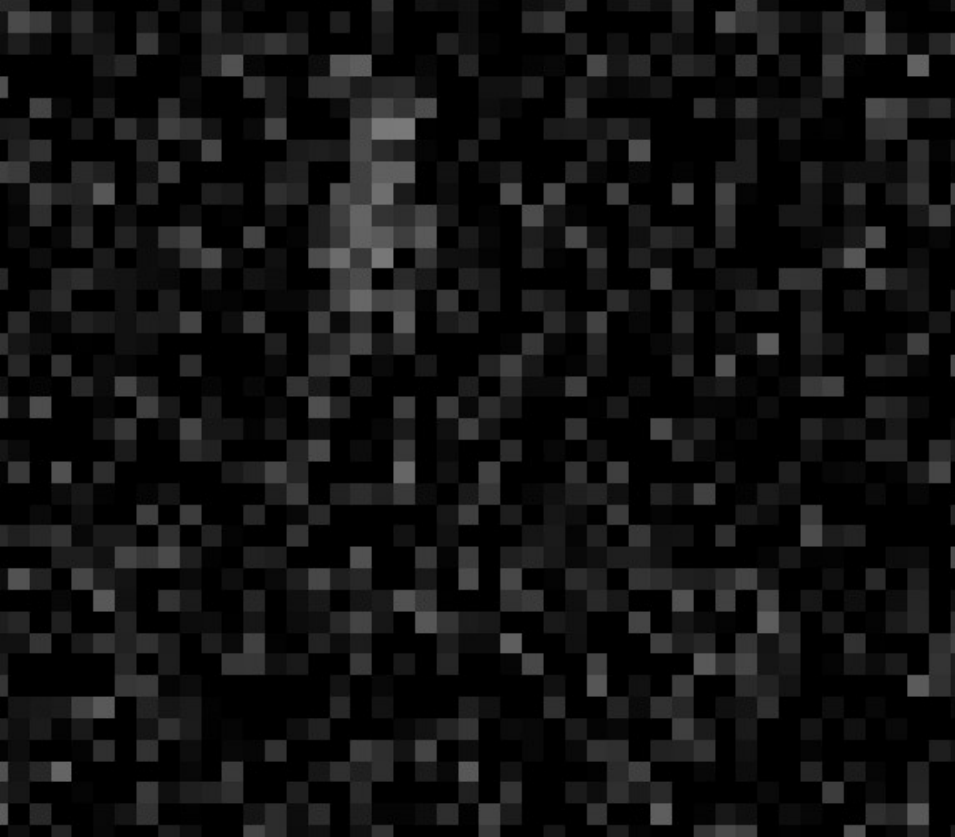}
     \end{subfigure}
     \hfill
     \centering
     \begin{subfigure}[c]{0.3\textwidth}
         \centering
         \caption{Third quadrant, where the event generating crosstalk can be observed.}
         \includegraphics[width=\textwidth]{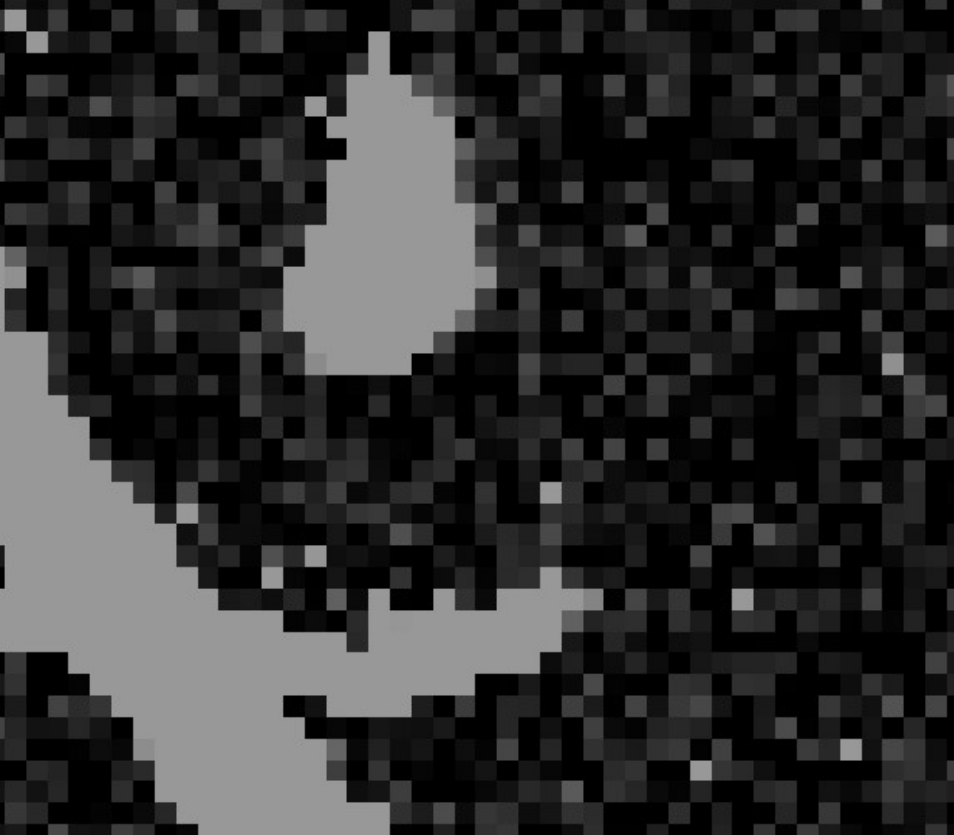}
     \end{subfigure}
\caption{The same section of three different quadrants showing the effect of crosstalk caused by the impact of an event of approximately 2.5 MeV on 100 pixels. Both in the first and second quadrants, the shadow of this event can be seen. In the first quadrant, the result is a ghost event of 25 electrons.}
\label{fig.crosstalkimages}
\end{center}     
\end{figure}

The second quality criterion in this group, crosstalk, has already been introduced in depth in Section \ref{sec:crosstalkcorrection}. In summary, pixels whose measured signal may be affected by crosstalk will be excluded from the dataset (see Figure \ref{fig.crosstalkimages}).

\subsubsection{Bleeding}
\label{sec:sangrado}

Bleeding, introduced earlier in Section \ref{sec:transferencia}, accounts for the charge transfer inefficiency (CTI). In the case of astronomical images, CTI will create shifted versions of the collected image since the charge "lags behind" during a charge transfer, either vertically or horizontally.

As a result, the shift will be in the opposite direction of the readout, and it will be more pronounced in pixels with higher electron occupancy. For dark matter searches, pixels with high occupancy, often associated with high-energy events, will leave a tail of 1 or 2 electrons separate from the high-energy event. These tails can be mistakenly classified as dark matter-compatible events generating 1 or 2 electrons through ionization.

\begin{figure}[h!]
\begin{center}
     \begin{subfigure}[c]{0.58\textwidth}
         \centering
         \caption{Charge due to CTI up to 30 pixels away from a pixel charged with 1000 electrons for three different CTI values.}
         \includegraphics[width=\textwidth]{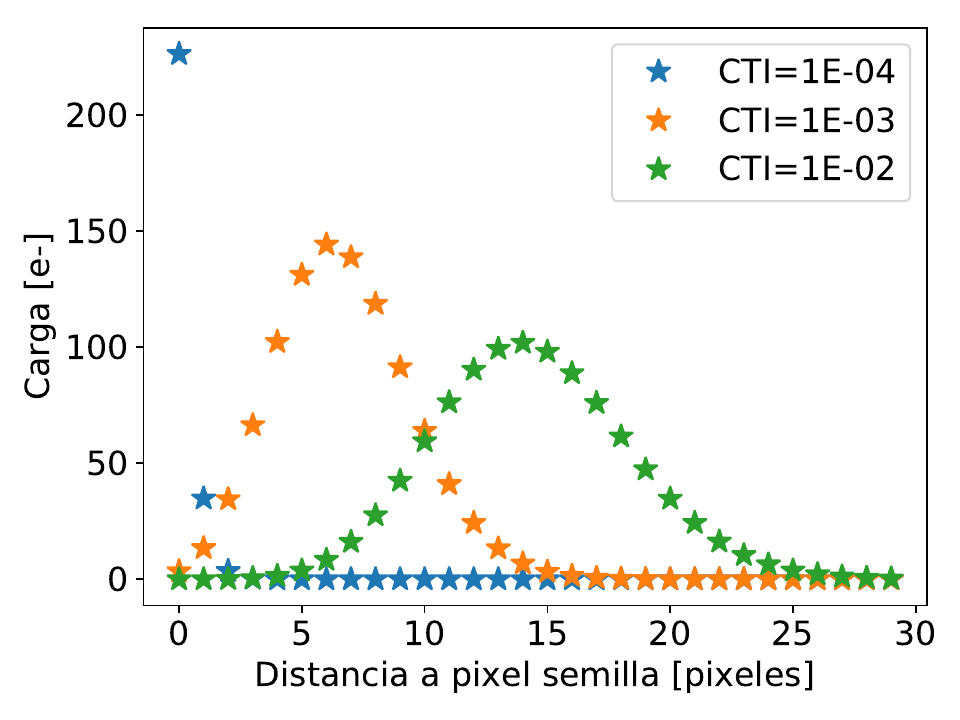}
         \label{fig.cti1.a}
     \end{subfigure}
     \hfill
     \centering
    \begin{subfigure}[c]{0.4\textwidth}
         \centering
         \caption{Illustration of an image with pronounced CTI.}
         \includegraphics[width=\textwidth]{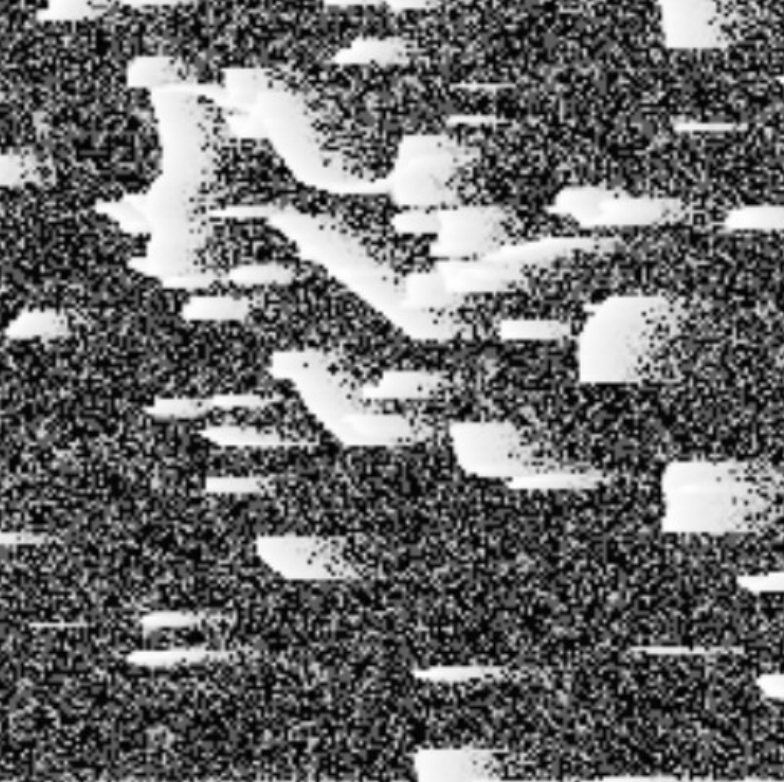}
         \label{fig.cti1.b}
    \end{subfigure}
\caption{The Figure \ref{fig.cti1.a} shows the charge due to CTI up to 30 pixels away from a pixel charged with 1000 electrons for three different CTI values. Figure \ref{fig.cti1.b} is an illustration of an image with pronounced CTI.}
\label{fig.cti1}
\end{center}     
\end{figure}

There is a model that predicts the number of electrons $n$ pixels away from a pixel due to CTI (see Chapter 5 of \cite{janesick}):

\begin{equation}
    S_{N_{P}+n} \ = \ \frac{S_{i} (N_{P} CTI)^{n}}{n!} \exp(-N_{P} CTI)
    \label{eq.cti}
\end{equation}

where $N_{P}$ is the number of transfers for the pixel, $S_{i}$ is the charge the pixel would have if CTI were virtually 0, $S_{N_{P}+n}$ is the charge contained in the $N_{P}+n$-th pixel, and CTI is the CTI factor ranging from 0 to 1. Figure \ref{fig.cti1.a} shows different values of equation \eqref{eq.cti} for different CTI values.

In Figure \ref{fig.cti1.b}, you can see an image with pronounced CTI, which creates a tail of events with a few electrons to the right of the high-occupancy events. CTI is more pronounced in the horizontal direction than in the vertical direction, in part because the horizontal register of this particular SCCD is approximately 6 times longer than the number of rows per column. However, the aspect ratio between the two dimensions is not the only factor affecting the relationship between horizontal and vertical CTI.

Firstly, the pixels in the horizontal register are different from those in the active area, with the latter being three times longer than the former. Therefore, the applied electric fields are different, and their comparison is not straightforward. 
Secondly, CTI is highly dependent on three factors concerning the applied voltage during transfer: the time charges wait beneath each gate between transfers, the amplitude between the upper and lower voltage of each gate, and the waveform used to change the voltage applied to each gate from its lower to upper state and vice versa. The first factor is related to a relaxation time that charges require to be transferred and the time charges wait between transfers. If this waiting time is too short, the transfer efficiency decreases. Regarding the second factor, a lower voltage amplitude implies less electrical attraction of the deeper potential with respect to the charges of the higher potential due to a smaller voltage difference between the potentials of each gate. Increasing this amplitude reduces CTI but results in more spurious charge generation, as discussed in Section \ref{sec:ctivssc}. Finally, the third factor considers controlling the shape, especially the speed, at which the voltage changes between the upper and lower voltage (and vice versa), and also affects the generation of spurious charge, as mentioned earlier.

These three factors are clearly interconnected, and their control is crucial for high-quality data acquisition. Optimizing these parameters is crucial and must be done separately for the active area and the horizontal register, as well as for the pixels of the output device.

Furthermore, since thermal electron diffusion decreases with temperature, CTI decreases at higher temperatures. However, there is a trade-off between CTI and dark current, as increasing the temperature will decrease the former but increase the latter. Even after optimizing these parameters and having scientific-grade SCCDs, with CTI factors that can be as low as $1-0.999999$ \cite{janesick}, a tail of 1 or 2 electron events can often be observed after high-energy pixels. Therefore, a specific zone above and to the right of pixels with 100 electrons or more is excluded, and this zone will change depending on the specifics of the dataset and the SCCDs being used.

\subsection{Cuts associated to defects in the material}
\label{sec:columnasypixelescalientes}

\subsubsection{Hot Pixels and Columns}

Hot pixels and columns are the result of defects in the silicon that generate an excess of SEEs in the recorded images. Depending on the intensity of this background signal, a defect can produce an excess of SEEs in a particular pixel (if the characteristic time for generating SEEs is compatible with the combined exposure and readout time of the SCCD) or in the entire column where the defect is present (if the characteristic time is much shorter).

The algorithm that performs this cut first identifies the pixels and then the columns that show a higher level of electronic occupancy than the rest.

\begin{figure}[h!]
\begin{center}
\includegraphics[width=0.55\textwidth]{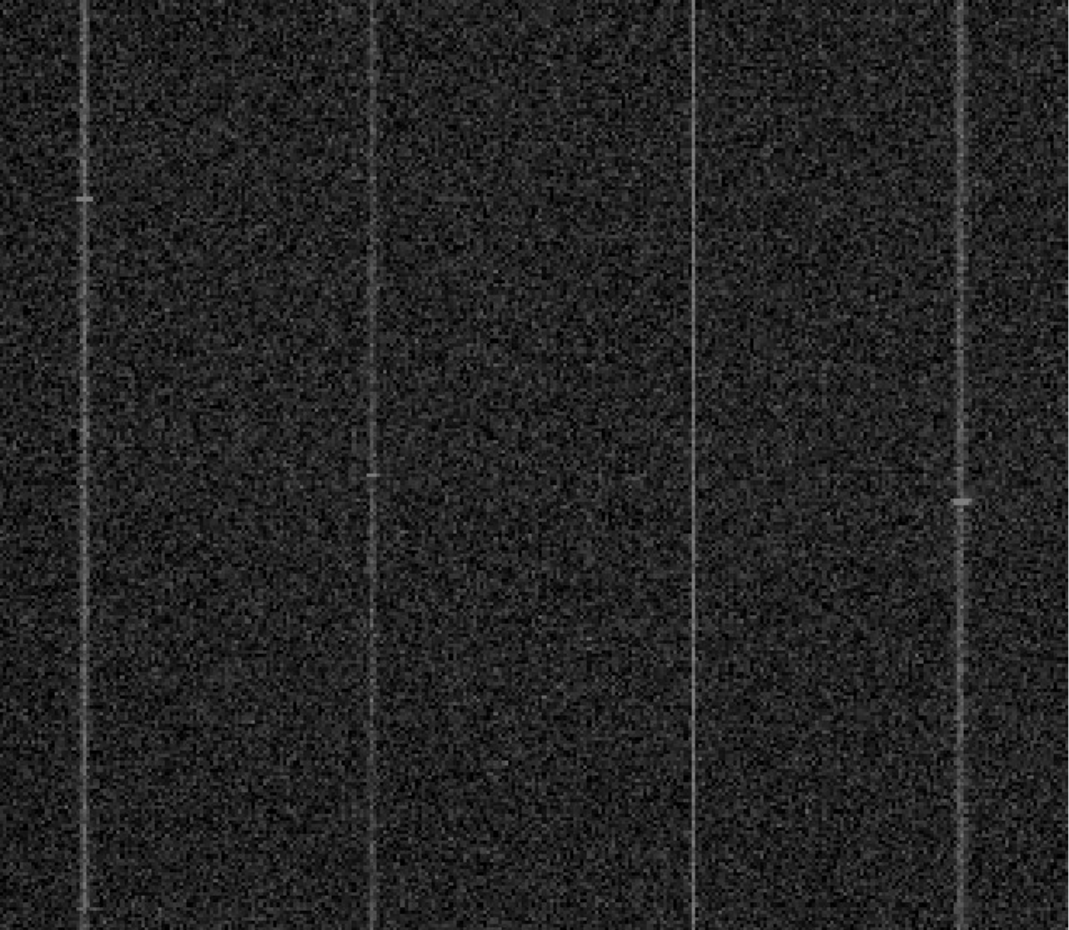}
\caption{Image taken at 210K to identify dark spikes.}
\label{fig.darkspikes}
\end{center}
\end{figure}

Additionally, SCCDs can be heated (in our case, to approximately 210K) to obtain images like the one shown in Figure \ref{fig.darkspikes}. In this image, it is filled with SEEs due to dark current, even though the image was read in just 1 minute. Four columns called "dark spikes" can also be observed in which the dark current was anomalously high, a phenomenon that is amplified at higher temperatures. These columns are also excluded from the analysis as hot columns, even if they are not selected by the algorithm mentioned earlier.

\subsubsection{Loose Cluster}

The "Loose cluster" criterion was created to exclude hot zones from the analysis of 2, 3, and 4 electron events, where there is an excess of SEEs. The existence of such excesses can be attributed to defects in the silicon of the detector or some other unknown source associated with an extrinsic or intrinsic radiation background. Additionally, it is considered that these areas are not captured by the algorithm that generates the hot pixel and column cut. The parameters used for this cut, as well as a detailed description of how it works, will be discussed in Section \ref{sec:2020criterio}.

\subsection{Cuts associated to high-energy events} 

\subsubsection{High-Energy Halo and Edge}
\label{sec:halo}

\begin{figure}[h!]
    \begin{center}
        \includegraphics[width=0.75\textwidth]{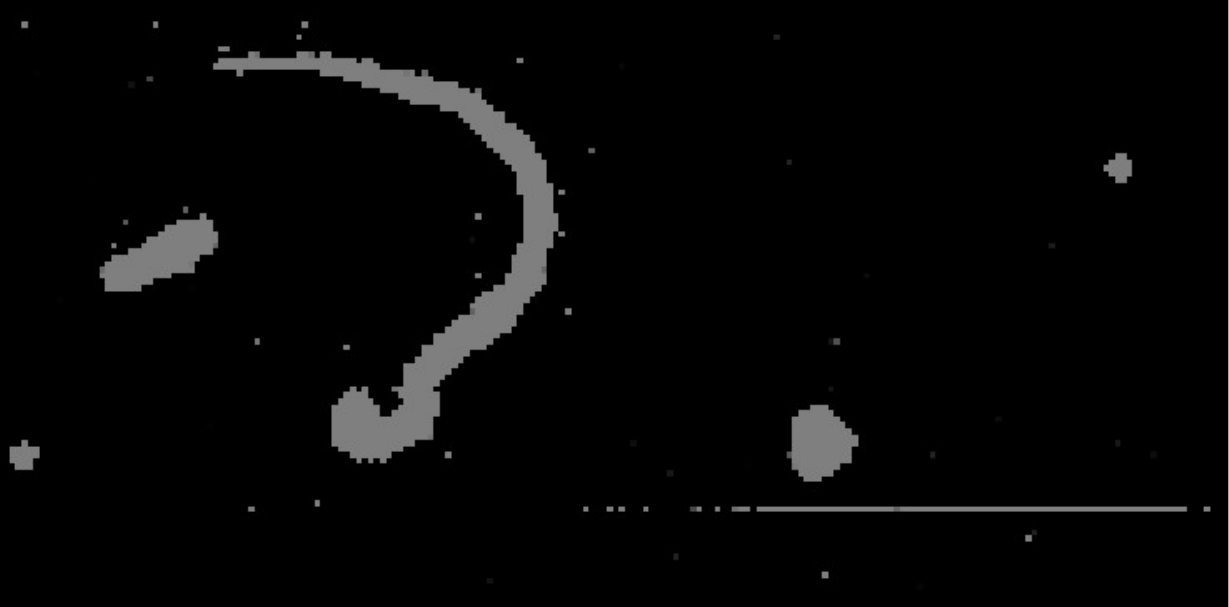}
        \caption{Image used to introduce the high-energy halo and horizontal registry events. In the center of the image, you can see a worm-shaped trace corresponding to a 64~keV electron. Around it, you can see SEEs surrounding the trace of this electron. Below the electron, to the right, there is a horizontal registry event, easily identifiable by its completely horizontal shape, resulting from interaction with the SCCD's horizontal registry during the reading phase.}
        \label{fig.sampleimage}
    \end{center}
\end{figure}

The high-energy halo corresponds to a spatial correlation between high-energy events interacting with the silicon of the SCCD and 1 or 2 electron events. This spatial correlation is not related to bleeding, as 1 or 2 electron events are also found to the left and below the high-energy events. It is also not related to highly scattered events, as the effect can be observed even up to 60 pixels away from a high-energy pixel. In Figure \ref{fig.sampleimage}, you can see the trace of an electron of approximately 64~keV with its Bragg peak and that of another charged particle to its left, of approximately 5.4~keV. While the charged particle exhibits isolated white pixels, corresponding to 1 or 2 electron events in a single pixel, only in the area above its trace (typical of bleeding), these low-energy events can be observed around the entire trace of the electron.

The authors of \cite{du2022sources} attribute this effect to Cherenkov radiation and/or radioactive recombination, both consequences of the interaction of high-energy events with silicon. Cherenkov radiation is well-known and consists of the emission of low-energy photons at a certain angle when a charged particle traverses a non-conductive material (like silicon) at a speed higher than the speed of light in that material. This radiation will exist for charged particles (mostly muons and electrons) with sufficiently high energy to move at the required speed.
On the other hand, radioactive recombination involves the de-excitation of 1 electron from the conduction band to the valence band and the associated emission of a photon with an energy equal to the silicon's band-gap. The ionization of electron-hole pairs resulting from an electromagnetic interaction with the material is a process that can trigger radioactive recombination in silicon. In particular, the presence of acceptors (donors) acts as a catalyst for this effect by providing additional charge carriers for recombination with the electrons (holes) created.
Due to the fact that the bulk of silicon has very low doping, such that recombination does not produce a noticeable photon emission, the authors of \cite{du2022sources} explain how the phosphorus present in the few microns of the backside of the material for generating electrical contact is capable of generating an excess of 1 or 2 electron events near high-energy events.

As will be seen in the data analysis in Section \ref{sec:2020criterio}, this spatial correlation decreases with distance, such that the excess of 1 or 2 electron events disappears for sufficiently large distances. Thus, a "halo" or circle of pixels around pixels with 100 or more electrons is excluded in order not to count this excess. Furthermore, and conservatively, the radius of this halo is used to exclude pixels at the edges of each quadrant since the interaction of high-energy events beyond the edge of the active area but within the SCCD can generate an excess of 1 or 2 electron events in the active area.
The correlation between high-energy and very low-energy events will be discussed in more detail in Section \ref{sec:dc}.

\subsubsection{Horizontal Register Events}
\label{sec:SRevents}

Horizontal register events consist of interactions of charged particles with the horizontal register, or its surrounding areas, during the readout phase. Excluding the moments when charge is transferred vertically, which represent a very small portion of the readout time, the gate voltage or TG (see Figure \ref{fig.ccd}) is always high so that an interaction in the bulk of the horizontal register can only diffuse horizontally along it\footnote{During the first weeks of operation of the 675$\mu$m SCCDs, the TG voltage in its upper position was slightly higher than that of the horizontal register in the upper position, and charges diffused erratically into the active area. This issue was resolved by increasing the TG voltage in its upper position.}. Because of this, horizontal register events are elongated, as can be seen in Figure \ref{fig.sampleimage}. Additionally, they often exhibit isolated pixels in the same row, which are attributed to the diffusion of electrons from the same interaction. Consequently, when such an event is detected, the entire row is excluded from the dataset.

\subsection{Specific-selection cuts} 

\subsubsection{Single-Pixel Events}
\label{sec:monopixelybajaenergia}

The single-pixel event selection criterion is intimately related to the concept of charge diffusion introduced in Section \ref{sec:diffusion}. In that section, it is explained how charges generated in the SCCD bulk are transported by the electric potential towards the surface, following a bivariate Gaussian distribution. The dispersion of this distribution will depend on the energy of the incident particle and the depth at which the interaction occurred, among other factors.

Events produced by dark matter interactions that generate 1 electron of charge will always diffuse into 1 pixel and will be counted unless masked by some event selection criteria. However, for higher energies or events that produce some electrons but are far from the surface, the probability that an event deposits all its charge in a single pixel begins to decrease. This effect worsens as the energy increases. This concept will be discussed in detail in Section \ref{sec:2020eficienciapordifusion}.

The mentioned quality criterion (single-pixel event) consists of selecting, as part of the dataset to be analyzed, pixels, empty or non-empty, that have another non-empty pixel as a neighbor. Consequently, the surviving pixels under this mask will mostly be isolated empty pixels and isolated non-empty pixels (mostly containing 1 electron). This criterion is useful for analyzing single-electron events or events with higher electron multiplicities but located in a single pixel.

\subsubsection{Low-Energy Cluster}
\label{sec:lowenergycluster}

The low-energy cluster cut excludes areas in the dataset where clusters between 2 (5) and 100 electrons are observed for the 3 and 4 (1 and 2) electron channels that are too close together and may result from an unknown background and/or are incompatible with the dark matter signal under analysis. Similar to the high-energy halo, this cut generates a halo around certain events but in a lower energy range. These groupings are considered incompatible with a dark matter signal because the probability of such groupings occurring is very low for the channels of interest.

As an example, if you are looking for 4-electron events and find a 2-electron event a few pixels away, the 4-electron event will be discarded\footnote{It is worth noting that the 2-electron event is not discarded since it is not relevant to the search for 4-electron events.} because it is highly likely that its origin is related to joint diffusion with the 2-electron event. Thus, it is presumed that the 4-electron event is the result of an interaction that generated 6 electrons and may or may not be compatible with dark matter, but our analysis does not investigate it (since we are looking for events between 1 and 4 electrons).

The cut is motivated by the presence of low-energy groupings in the dataset analyzed in Chapter \ref{cap:5}, where the cut and the potential biases it may introduce will be discussed in depth.

\newpage
\mbox{}
\thispagestyle{empty}
\newpage

\chapter{Origin and characterization of SEEs}
\label{cap:4}

In the following chapter, we will discuss the work carried out within the framework of this thesis to characterize single-electron events (SEEs) since they constitute the main source of background in the search for light dark matter using SCCDs. The objective of this study is to understand and model the sources of SEEs in order to develop strategies to mitigate their effects. To achieve this, we developed a semi-empirical model that encompasses the main contributions of SEEs in the measured images, characterizing them based on their spatial and temporal dependencies. Additionally, we created a measurement protocol that allowed us to determine these contributions.

\section{Motivation}
\label{sec:contributions}

As introduced earlier, SENSEI achieves its highest sensitivity in its dark matter exclusion limits between 2 and 4 electrons (see Figures \ref{fig.senseiprojectedscattering} and \ref{fig.senseiprojectedabsorption} and discussions in the text). There are two backgrounds that are of primary relevance for the analysis of these channels: the Compton background and the SEE background.

The Compton background originates from the scattering of relatively high-energy photons with electrons in the surrounding material of the CCD or within the CCD itself. Incident photons lose energy depending on the angle of interaction and their initial energy. In the case of interaction with the CCD, they result in a continuous energy spectrum, spanning from the initial energy of the incident photon up to 1.1 eV for a silicon detector. A study of this background for CCDs can be found in \cite{PhysRevD.96.042002}, and a study conducted by my co-author using SCCDs during the course of this thesis can be found in \cite{comptonbotti}.

The second contribution to the background, on which special emphasis will be placed in this chapter, is produced by SEEs. As mentioned earlier, SEEs are events with only one electron, presumed to originate from a phenomenon or effect that generates exactly one electron. By definition, these events are isolated single-electron events. However, if a sufficient number of SEEs are present in the same image, two or more SEEs can occur in the same pixel and "pile up." As explained in the Data Processing Section (Section \ref{sec:clusterizacion}), the image clustering algorithm would recognize this stacking of two single-electron background events as a two-electron event. The same can happen at higher energy levels.
This background will be referred to as "pile-up," and it will be the primary background contribution for the channels of interest to be studied, particularly for the two-electron channel. This is because the probability of stacking $k$ single-electron events in the same pixel ($P(k)$), given a process of SEE generation, is well described by a Poisson distribution:

\begin{equation}
\label{eq.poisson1e}
P(k)=\frac{\mu^{k} e^{-\mu}}{k!}
\end{equation}

Where the parameter $\mu$, introduced in Section \ref{sec:calibration}, represents the rate of pixels with one electron divided by the total number of pixels, a crucial parameter for calculating the number of SEEs in the acquired images. Additionally, because the clustering algorithm groups neighboring events into the same cluster, two or more SEEs generated with a rate of $\mu$ can be classified into a multipixel cluster (i.e., with two or more pixels) of 2 or more electrons.
 
\begin{figure}[h!]
    \centering
    \includegraphics[width=0.75\textwidth]{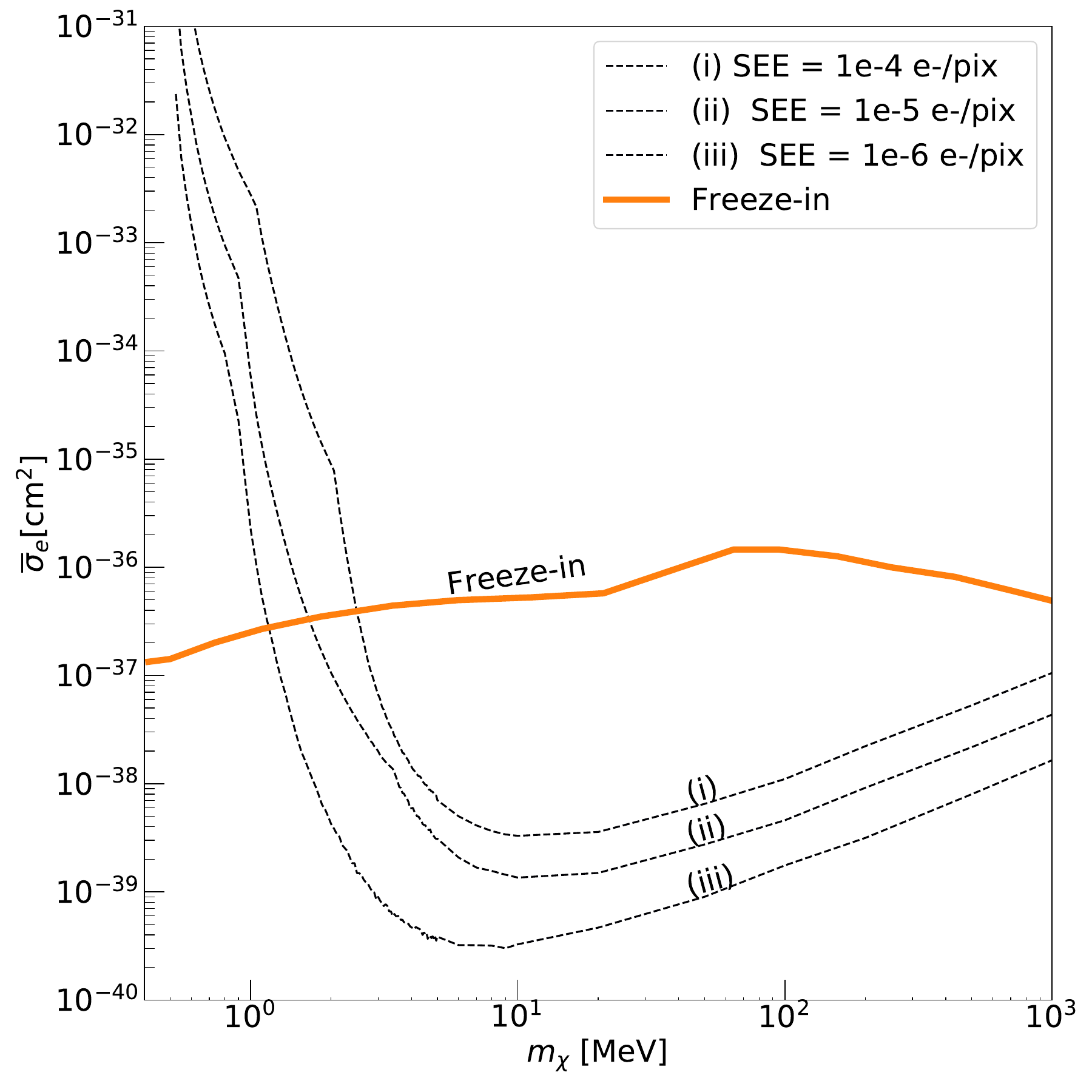}
    \caption{Projected exclusion limits for SENSEI with 100 grams-year of exposure and three different SEE rates per pixel: (i) $1 \times 10^{-4} \epix$, (ii) $1 \times 10^{-5} \epix$, and (iii) $1 \times 10^{-6} \epix$. It is assumed that the experiment has 100 grams of mass, performs 365 measurements with one day of exposure each.}
    \label{fig.dcsensitivity}
\end{figure}

It is clear that the parameter governing the behavior of this background is $\mu$, the SEE rate per pixel. It will be shown in this chapter and the next that the lowest value obtained for this rate is $\sim 1 \times 10^{-4} \epix$, and it was achieved using SENSEI-SCCDs. Despite being the lowest value recorded to date, it is about two orders of magnitude higher than projected by theoretical estimates\footnote{This estimate will be introduced shortly in Section \ref{sec:semiempiricodc}.}. Figure \ref{fig.dcsensitivity} shows the impact of the SEE production rate in SCCDs on the exclusion limits estimated for SENSEI.

There are different contributions that generate SEEs in the observed images. In the following sections, we will focus on describing these contributions, their origins, and, whenever possible, their mitigation.

\section{Data Acquisition Protocol and SEEs Contribution Model}

\label{sec:protocolodemedicion}

In order to study the contributions of SEEs in the \textit{SENSEI-SCCDs}, a semi-empirical model was designed that aims to discriminate which events were created during the device's exposure and which during reading. The data acquisition protocol used in the SENSEI experiment can be structured into the following three stages: \textit{Cleaning}, \textit{Exposure}, and \textit{Readout}.

\subsubsection{Cleaning}
\label{sec:limpieza}

The cleaning process has two objectives: firstly, to clear the image of events that have deposited energy in the SCCD prior to a new measurement. This process is called \textit{emptying}. Emptying is done by reading the active area of the SCCD without saving the signal obtained in the video channels and measuring with a single sample in the Skipper output device. The process can be repeated multiple times to ensure that there is no residual memory of energy deposits in the active area of the detector prior to data acquisition.

Secondly, and performed before emptying, a procedure called \textit{erase and purge} is carried out. This procedure is used to reduce one of the most relevant contributions to SEEs, which will be covered in Section \ref{sec:dc}: the surface dark current. Since it is necessary to understand what surface dark current is, how it relates to SEEs, and its mechanism of production in advance, it should be noted for now that \textit{erase and purge} is performed prior to emptying and is part of the image cleaning process. The procedure itself and the reason for its operation will be discussed in Section \ref{sec:eraseandpurge}.

\subsubsection{Exposure}
\label{sec:exposicion}

During this process, all voltages in the active area are fixed so that charges generated by any charged particle or radiation passing through the SCCD are captured beneath the pixels of the active area. Additionally, the drain voltage of the M1 amplifier of the output device is set to 0V to reduce photon emission during exposure, a phenomenon that will be discussed in detail in Section \ref{sec:al}.

\subsubsection{Readout}
\label{sec:lectura}

As introduced in Section \ref{sec:transferencia}, after collecting charges during Exposure, the transfer process follows. This process is performed row by row vertically towards the horizontal register, and between each vertical transfer, pixel by pixel along the register towards the sense node, where the signal is converted into voltage, as explained in Section \ref{sec:dispositivodesalida}.

\subsection{Semi-Empirical Model}

SEEs are generated by different mechanisms during the three stages mentioned in the previous section. In particular, contributions of SEEs that depend on time can be generated either during exposure or during readout. For exposure time $t_{\rm EXP}$ and readout time $t_{\rm RO}$, both contributions will be expressed in terms of their rates $\mu_{\rm{EXP}}$ and $\mu_{\rm{RO}}$, respectively, in units of ($\epix$). Additionally, an independent production rate ($\mu_{0}$) is considered.

Following these definitions, the total contributions to SEE can be modeled using the following equation:

\begin{equation}
\begin{split}
\label{eq.model}
\mu(t_{\rm{EXP}},t_{\rm{RO}}) & =\mu_{\rm{EXP}} (t_{\rm{EXP}})+\mu_{\rm{RO}}(t_{\rm{RO}})+\mu_{0} \\
                             & = \lambda_{\rm{EXP}} t_{\rm{EXP}}+ \lambda_{\rm{RO}} t_{\rm{RO}} +  \mu_{0}\,,
\end{split}
\end{equation}

\noindent where, in the second line, it is assumed that both \(\mu_{\text{EXP}}\) and \(\mu_{\text{RO}}\) scale linearly with time. 
The parameters $\mu_{\rm EXP}$ and $\mu_{\rm RO}$ are the SEE production rates during exposure and readout, respectively, in units of events per pixel (\epix)), while the times ($t_{\rm EXP}$ and $t_{\rm RO}$) are expressed in days.

There are three known contributions that generate SEEs: dark current, amplifier light, and spurious charge. 
In the following subsections, each of these contributions will be discussed in detail, contrasting the proposed model with the obtained results. A detailed study will be conducted to separate these contributions and find ways to mitigate them, whenever possible. Contributions will be classified based on the following characteristics: spatial distribution (localized or uniform) and temporal dependence.

\section{Dark Current}
\label{sec:dc}

Dark current, as defined in the literature, consists of the generation of electron-hole pairs due to thermal agitation \cite{janesick}. 
In Chapter \ref{cap:2}, it was emphasized that an incident charged particle with energy greater than the silicon bandgap, around 1.12 eV, is capable of producing at least one electron-hole pair through ionization.
Similarly, if the device is at a temperature where thermal agitation in the material (whose average energy scales as $kT$, where $k$ is Boltzmann's constant evaluated at approximately $8.62 \times 10^{-5}$ eV~K$^{-1}$) is energetically comparable to the silicon bandgap, one electron in the valence band can be thermally excited to the conduction band and then captured by the electric potential present in the material. 
This process is mediated by traps or defects in the material that introduce intermediate energy levels in the material's bandgap. 

Thus, the process of promoting one electron from dark current occurs in stages: first, the electron is thermally excited to an intermediate energy state within the bandgap, and then it is promoted to the conduction band and captured by the electric potential. 
At the same time, a hole is generated in the valence band, maintaining the carrier density, and it is captured by the material's ground~\footnote{In the case of an n-type bulk, as is the case with the SCCDs used by SENSEI, the hole is collected, and the electron is captured.}.

\subsection{Bulk and Surface Dark Current}

There are multiple contributions to dark current, depending on which region of the CCD the charges are generated. In particular, two dominate for the SENSEI-CCD: surface dark current and bulk dark current.

Bulk dark current is mediated by traps and defects that are generated throughout the volume of the CCD. In SENSEI, the silicon used has a very low level of crystal defects, which drastically reduces the level of bulk traps. On the other hand, at the surface, there is a high number of defects that occur at the interface between silicon and $SiO_{2}$ because silicon has a specific crystal lattice while $SiO_{2}$ is amorphous. Therefore, for high-quality silicon crystals like those used by SENSEI, the dominant dark current is at the surface due to the high presence of traps at the interface.

The equation that governs the thermal behavior of dark current (both surface and bulk) as a function of temperature is given by the following equation, taken from Section 7.1.1.9 of \cite{janesick}:

\begin{equation}
    DC \ = \ 2.5 \times 10^{-15} \ P_{S} \ D_{FM}  \ T^{1.5} \ e^{-\frac{E_{g}}{2kT}}
    \label{eq.dc}
\end{equation}

where $P_{S}$ is the pixel area and $D_{FM}$ is the value of dark current at room temperature, i.e., 300K. Equation \eqref{eq.dc} can then be used to estimate dark current values at a given temperature, knowing values at higher or lower temperatures. However, measuring the different dark current components separately is a significant challenge, as the one with the higher value tends to dominate. In the case of SENSEI's SCCDs, the surface current is orders of magnitude higher than the bulk current, as will be seen below.

\subsection{The \textit{Erase and Purge} Protocol and Surface Dark Current}
\label{sec:eraseandpurge}

There exists a method to reduce surface dark current to its lowest possible value \cite{burke1991,widenhorn2002temperature,holland2003fully}. This method involves turning off the applied voltage to the CCD, temporarily stopping the complete depletion of its volume, and applying a positive voltage of 9V to the CCD's surface gates to fill the intermediate energy levels in the band-gap with electrons from the channel stops adjacent to each pixel. Then, the appropriate voltage is reapplied for the correct reading of the detector.
With the intermediate energy levels filled, they no longer contribute to surface dark current since electrons from the valence band cannot use the intermediate levels to be promoted to the conduction band. According to estimates made using the model developed in \cite{burke1991}, at 135K, less than 0.01$\%$ of the surface dark current is recovered in the first 24 hours after the erase and purge, making bulk dark current dominant. In Figure 14 of reference \cite{holland2003fully}, a two-order-of-magnitude reduction in dark current can be observed after performing an erase and purge. After filling the intermediate energy levels, the CCD is depleted again for its subsequent operation.

\begin{figure}[h!]
    \centering
    \includegraphics[width=0.7\textwidth]{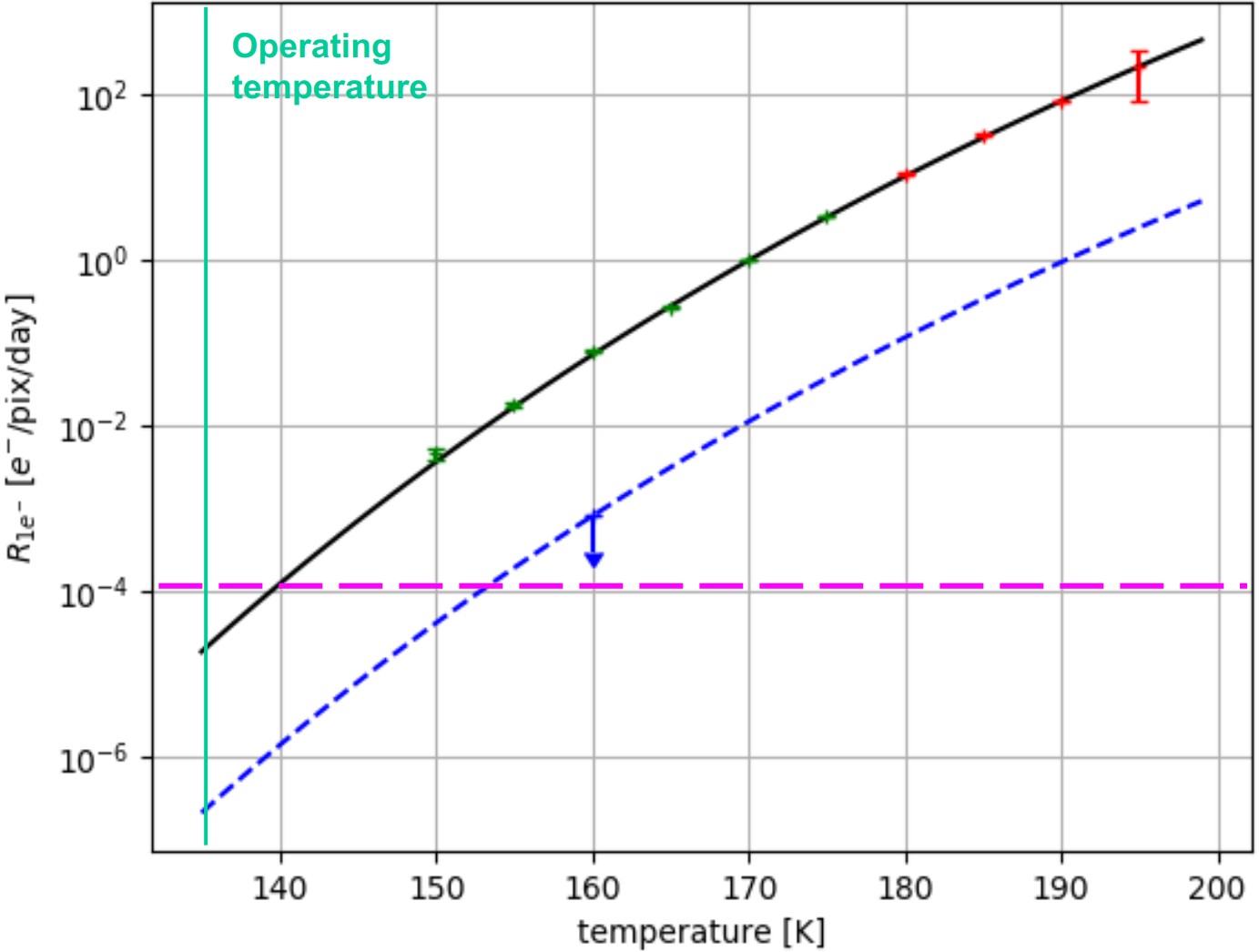}
    \caption{Dependence of the rate of events produced by dark current on temperature. Measurements from 150 to 175 K in blue (green) are from an SCCD installed in SNOLAB, taken with (without) erase and purge. Measurements from 180 to 195 K (red) are from an SCCD of the same production batch as the one previously mentioned, taken on the surface and, once again, without reducing surface dark current. The black curve is a fit to the green and red points using the theoretical model expressed in Equation \eqref{eq.dc} \cite{janesick}. Figure extracted from \cite{SENSEI2020}.}
    \label{fig.shodc}
\end{figure}

Figure \ref{fig.shodc} illustrates the effect of the erase and purge method on the SENSEI-CCD installed in MINOS. The red and green points correspond to dark current values measured without using the erase and purge method, and thus, these values predominantly represent surface dark current. Additionally, the theoretical model proposed in Equation \eqref{eq.dc}, illustrated with a black curve in the same figure, perfectly describes the measured dark current values. When using the erase and purge method at a temperature of 160K, the point marked with an arrow pointing towards the negative domain of the vertical axis is extracted, indicating that the measured value is an upper bound estimate of the dark current at that temperature.

This result not only demonstrates the effect of the erase and purge method for SENSEI-SCCDs but also highlights the discrepancy between the theoretical model and experimental results. In \cite{SENSEI2020}, as discussed in Section \ref{sec:2020conteo1e}, a dark current at 135K of $(1.59 \pm 0.16) \times 10^{-4}$ $\epixdia$ is reported, illustrated by a pink dashed line in Figure \ref{fig.shodc}. However, following the blue curve, the dark current was expected to be at least 2 orders of magnitude lower, below $\sim 1 \times 10^{-6}$ $\epixdia$. There is no conclusive analysis in the literature that explains the difference between the measured and theoretical dark current for such low values.

\subsection{Semi-Empirical Dark Current Model}
\label{sec:semiempiricodc}

Assuming a uniform distribution of defects in all directions, both localized defects at the silicon-SiO$_{2}$ interface and bulk defects, the generation of SEE due to dark current for a given pixel can be described by a Poisson distribution with a characteristic parameter $\lambda_{DC}$, in units of $\epixdia$. Furthermore, given the assumption of a uniform distribution of defects, $\lambda_{DC}$ is assumed to be uniform in space and constant over time for fixed control parameters (temperature, operating voltages, etc.).

It is worth noting that this assumption of spatial uniformity is a sensitive one because $\lambda_{DC}$, without applying any quality cuts to the dataset to be used, may not be constant in space. A clear example of this non-uniformity can be seen in the high-energy halo event selection criterion developed in Section \ref{sec:masqueo}, which indicates a spatial correlation between SEE and high-energy events. The uniform SEE criterion would not be possible without the use of this cut as well as others mentioned in that section (hot pixels and columns, crosstalk, and single-pixel events are some other clear examples).

Returning to the semi-empirical model, given a Poissonian description of SEE generation due to dark current and the mentioned assumptions, the average number of SEE expected due to dark current will scale linearly with time, and its event rate, $\mu_{DC}$, expressed in $\epix$, will also:

\begin{equation}
    \mu_{\rm DC} \ = \ \lambda_{\rm DC} \ t_{\rm EXP}
    \label{eq.dc2}
\end{equation}

Dark current contributes during both the exposure and readout phases of data acquisition. The number of events originating from dark current scales linearly with time during the exposure phase, and thus, $\lambda_{\rm DC}$, the SEE rate due to dark current, will be one of the components of $\lambda_{\rm EXP}$. During the readout phase, the exposure of pixels is not uniform, as the last pixel read has an additional exposure time of $t_{\rm RO}$ compared to the pixel read first. Consequently, the average contribution of dark current during the readout phase is given by $\lambda_{\rm DC}/2$.

\subsubsection{Intrinsic and Extrinsic Dark Current}

The issue introduced in this section (\ref{sec:eraseandpurge}) in which the dark current at 135K is approximately two orders of magnitude higher than expected according to theoretical predictions needs to be considered. There are two relevant points to mention regarding this matter: the first is that the temperature dependence of dark current is a well-studied phenomenon in the CCD community, and the validation of the theoretical model has been a cornerstone in characterizing these devices and advancing their development. The second, complementary to the first, is that there are no comprehensive studies, outside of those presented in this thesis and in \cite{SENSEI2022}, on dark current for such low values. Dark current has been extensively studied for values greater than $1 \epixdia$, well above what has been reported by our collaboration.

For now, and even though there are currently no reasons to doubt the validity of the temperature-dependent dark current model expressed in equation \eqref{eq.dc}, there is evidence, as illustrated in Figure \ref{fig.shodc}, of a discrepancy between theoretical values and those measured by our SCCDs. Thus, dark current is divided into two categories: \textit{intrinsic} and \textit{extrinsic}, depending on whether the charges generated by it are produced by the SCCD independently or by an interaction of the SCCD with the environment, respectively. Likewise, dark current (intrinsic or extrinsic) is defined as the generation of SEEs that are collected by the SCCD in the absence of light, with no relation to the movement of charges pixel by pixel, during the exposure and readout phases, and whose charge accumulation is linear with time and spatial distribution, uniform.
In this way, intrinsic dark current encompasses what was previously defined as dark current (i.e., the generation of electron-hole pairs due to thermal agitation in a semiconductor material), and its temperature dependence is encapsulated in equation \eqref{eq.dc}. Extrinsic dark current, on the other hand, arises as an explanation for the discrepancy between measured dark current values and the theoretical model, without breaking the paradigm of pair creation due to thermal agitation. Additionally, as pointed out earlier in Section \ref{sec:halo}, which covers the quality criterion that excludes areas near high-energy events from the dataset to be used, the model developed in \cite{du2022sources} appears to provide a plausible explanation for this discrepancy.

For practical purposes, we will continue in the following subsections to study the total dark current, without distinguishing between intrinsic or extrinsic contributions, in order to compare this contribution against spurious charge and light produced in the amplifier.

\section{Amplifier Light}
\label{sec:al}

As the number of SEEs per pixel decreases with the reduction of temperature and background radiation, as well as with the improved detector performance, additional signals of a few electrons appear that are not included in the definition of dark current. The second contribution to consider consists of light emitted from the output device during the operation of amplifier M1 (see Figure \ref{fig.skipperoutputstage}).

The origin of luminescence includes various mechanisms discussed in \cite{toriumi1987} and more recently reviewed in \cite{venter2013reach}. Among them are (1) transitions within the conduction band, (2) phonon-assisted recombination, and (3) bremsstrahlung of hot charge carriers. While the second and third mechanisms were ruled out as main mechanisms in \cite{PhysRevB.52.10993} and \cite{bude1992,199363,499199}, respectively, the first one was ruled out in \cite{obeidat1997model}, which at the same time praised recombination as the main production mechanism.

The aim of this thesis is not to shed light on the mechanism by which photons are emitted but to study their dependence on control parameters and, likewise, reduce their impact on the rate of collected SEEs produced by this mechanism, considering that all cited sources agree that hot charge carriers are the mediators of this radiation. The studies presented here and published in \cite{SENSEI2022} could contribute to its characterization and the planning of new amplifiers that mitigate this effect.

The corresponding production rate is expressed as $\lambda_{AL}$ (in units of events per pixel per day). Since this contribution is localized near the amplifier, $\lambda_{AL}$ depends not only on the distance to the output device but also on the specific zone of the CCD being studied. For simplicity, $\lambda_{AL}$ is averaged over the entire CCD, and its spatial dependence is not studied.

Since ${\rm V_{DD}}$ is set to 0~V during exposure, the SEEs collected due to amplifier light only occur during the readout and cleaning phases. It should be noted that if ${\rm V_{DD}}$ is kept on during the exposure phase, an additional non-negligible contribution of amplifier light must be considered. This contribution may be different, as the voltage on the floating gate (which is also the gate of transistor M1; see Figure \ref{fig.skipperoutputstagelayout}) is constant during exposure but changes rapidly during readout. Additionally, as the active area is read, and the pixels are transported to the sense node, the spatial dependence of amplifier light during the readout and exposure phases is different.

The origin of luminescence includes various mechanisms discussed in \cite{toriumi1987} and more recently reviewed in \cite{venter2013reach}. Among them are (1) transitions within the conduction band, (2) phonon-assisted recombination, and (3) bremsstrahlung of hot charge carriers. While the second and third mechanisms were ruled out as main mechanisms in \cite{PhysRevB.52.10993} and \cite{bude1992,199363,499199}, respectively, the first one was ruled out in \cite{obeidat1997model}, which at the same time praised recombination as the main production mechanism.

The aim of this thesis is not to shed light on the mechanism by which photons are emitted but to study their dependence on control parameters and, likewise, reduce their impact on the rate of collected SEEs produced by this mechanism, considering that all cited sources agree that hot charge carriers are the mediators of this radiation. The studies presented here and published in \cite{SENSEI2022} could contribute to its characterization and the planning of new amplifiers that mitigate this effect.

The corresponding production rate is expressed as $\lambda_{AL}$ (in units of events per pixel per day). Since this contribution is localized near the amplifier, $\lambda_{AL}$ depends not only on the distance to the output device but also on the specific zone of the CCD being studied. For simplicity, $\lambda_{AL}$ is averaged over the entire CCD, and its spatial dependence is not studied.

Since ${\rm V_{DD}}$ is set to 0~V during exposure, the SEEs collected due to amplifier light only occur during the readout and cleaning phases. It should be noted that if ${\rm V_{DD}}$ is kept on during the exposure phase, an additional non-negligible contribution of amplifier light must be considered. This contribution may be different, as the voltage on the floating gate (which is also the gate of transistor M1; see Figure \ref{fig.skipperoutputstagelayout}) is constant during exposure but changes rapidly during readout. Additionally, as the active area is read, and the pixels are transported to the sense node, the spatial dependence of amplifier light during the readout and exposure phases is different.

For now, $\lambda_{\rm AL}$ will be a contribution that scales linearly with time and is located near the output device.

\section{Spurious Charge}
\label{sec:sc}

Spurious charge is generated when transferring charges from one pixel to its neighboring pixel or, more specifically, from one gate to its neighboring gate.
In Chapter 7 of \cite{janesick}, spurious charge is defined as the charge collected due to the generation of electron-hole pairs resulting from the collision of charge carriers that accumulate on the detector's surface when the gates leave the inversion mode.
Assuming that the gates collect holes, as is the case in our setup, when transferring the voltage, the gate where these holes are located (V1) applies a higher voltage than the gate (V2) where it is desired to transfer the holes (assuming, of course, that the other adjacent gate V0 is applying the same voltage as V1). The holes are attracted to a lower voltage and leave gate V1. As discussed in the section detailing the "blooming" cut (see Section \ref{sec:sangrado}), the difference between these two voltages, among other factors, can affect charge transfer, resulting in CTI (Charge Transfer Inefficiency).

Usually, as described in \cite{janesick}, the gate voltage is raised in such a way that the silicon immediately beneath it enters accumulation mode, accumulating holes from the channel-stops. This allows for greater hole repulsion when collected and, consequently, lower CTI when transferring them to the adjacent gate.
However, when changing the gate voltage, in the subsequent transfer, the holes are repelled from the gate and return to the channel-stops, colliding in their trajectory with the crystal containing them with sufficient energy to create electron-hole pairs, whose holes are collected by the same gate as it enters depletion mode again.

Additionally, the methods described in \cite{janesick} to reduce spurious charge have worked satisfactorily in our SCCDs, making it an effect of the same order of magnitude as the other contributions, as will be seen in Section \ref{sec:resultados2022}.
These methods were introduced in Section \ref{sec:sangrado} and consist of reducing the voltage difference used to transfer charges both horizontally and vertically and changing the waveform used for this voltage change. The first is intuitive since a greater voltage difference will result in a higher electric field and greater kinetic energy for the holes that return to the channel-stops after a readout. On the other hand, the second is related to the abrupt voltage change after completing the transfer because charges from the channel-stops tend to become energetically trapped in the traps under the gates, and an abrupt voltage change does not allow enough time for the charges to be released, forcing them, and resulting in much higher kinetic energy than necessary.

Thus, the generation of spurious charge does not depend on the exposure or readout time but on the number of times the voltage in a pixel changes to perform the transfer to another adjacent pixel. It should be noted that strictly speaking, this transfer occurs both during cleaning (when the CCD is cleared of residual charges from the "erase and purge" method) and during readout (when the charges are transferred to the sensing node). Since cleaning involves reading the CCD multiple times before starting the exposure, the number of times a pixel in an image is transferred will be the same for all pixels.

It is considered that spurious charge is the only component independent of time and therefore,

\begin{equation}
    \mu_{\rm 0} \ = \ \mu_{\rm SC}
    \label{eq.sc1}
\end{equation}

where $\mu_{SC}$ is the rate of SEEs per pixel from spurious charge.

\subsection{CTI and Spurious Charge in the Horizontal Register}
\label{sec:ctivssc}

\begin{figure}[h!]
    \centering
    \includegraphics[width=0.8\textwidth]{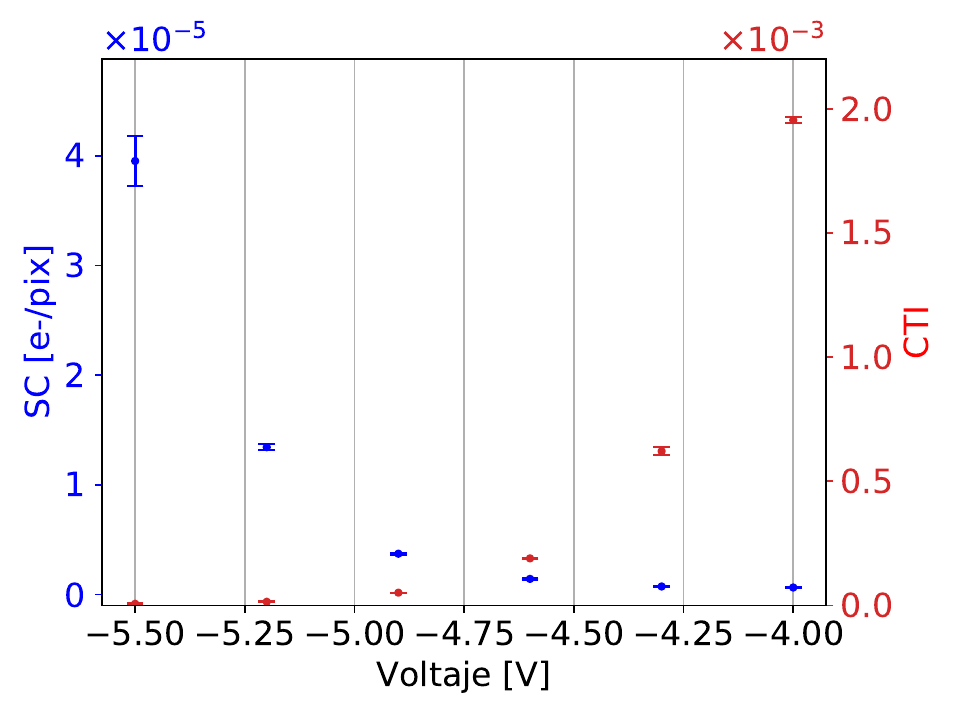}
    \caption{CTI (blue) and spurious charge (red) as a function of the applied voltage when the horizontal register gate is at its lowest voltage. For more details on how these values were obtained, refer to Appendix \ref{appendix:cti}.}
    \label{fig.CTIvsSC}
\end{figure}

Figure \ref{fig.CTIvsSC} quantitatively shows the relationship between the generation of spurious charge and CTI as a function of the applied voltage during charge transfer. The study presented is part of the quality testing routine that SENSEI-CCDs undergo before being used for scientific measurements. The details of conducting this study are specified in Appendix \ref{appendix:cti}. It is worth mentioning that the purpose of showing this study is not to highlight the obtained CTI estimation value or analyze its degree of accuracy in detail but to show its trend with different voltages used. Furthermore, what is shown is not the total CTI but primarily the horizontal CTI, which occurs when moving the charge in the horizontal register. A comprehensive study of CTI in SCCDs is pending for the future.

Despite this, the values obtained for the CTE are similar to those reported in the literature \cite{drlica2020}, ranging between 0.9999 and 0.99999. Regarding the estimation of spurious charge, something similar happens: reliable and repeatable estimates have been obtained, as will be detailed later. Still, the purpose of Figure \ref{fig.CTIvsSC} is to show the trend with voltage. Likewise, the estimated values of spurious charge are only for charge generated in the horizontal register.

The change in voltage was controlled by varying the voltage used in the lower state, i.e., when it is desired for the holes to transfer to the gate with that voltage. The upper voltage, on the other hand, was fixed at -2 Volts, as was usual for the operating parameters at that time. The crucial part of the study is to observe that there exists, for this SCCD and particular quadrant, a "valley" between CTI and the spurious charge generated in the horizontal register. This trade-off relationship is crucial when choosing which configuration to set for the devices to be used.

While a higher CTI will result in a longer tail of SEEs behind high-energy events, if this region is properly masked, and at the risk of reducing the total exposure due to over-masking, spurious charge can be significantly reduced, thereby reducing the total SEE rate per pixel. On the other hand, a lower CTI will result in less bleeding and masking but a higher rate of SEEs due to spurious charge, generating a background not only for the 1-electron channel but also for higher-energy multiplicity channels.

\section{Determination of Contributions}
\label{sec:resultados2022}

Given the description of the contributions mentioned above, Equation \eqref{eq.model} can be rewritten as

\begin{eqnarray}
\label{eq.model2}
\mu_{(t_{\rm EXP},t_{\rm RO})}&=& \lambda_{\rm DC} t_{\rm EXP} \nonumber \\& +& \biggl(\frac{\lambda_{\rm DC}}{2} +\lambda_{\rm AL}\biggr) t_{\rm RO} \\ 
&+& \mu_{\rm SC}\,. \nonumber
\end{eqnarray}

\noindent where the first line comprises contributions during exposure, the second during the readout phase, and the third contributions independent of time. Table \ref{table1.2} summarizes the characteristics of each contribution and its spatial and temporal dependence.

\renewcommand{\arraystretch}{1.7}
\begin{table*}[t!]
\centering
\caption{Charge contributions and their properties, following Equation \eqref{eq.model2}. The units for $\lambda_{\rm DC}$ and $\lambda_{\rm AL}$ are $\epixdia$, while $\mu_{\rm SC}$ is in $\epix$.}

\label{table1.2}
\begin{tabular}{|c|c|c|c|c|c|} 
\hline
\multicolumn{2}{|c|}{\multirow{3}{*}{\begin{tabular}[c]{@{}c@{}}\textbf{Contribution}\\\textbf{($e^{-}/{\rm pix}$)}\end{tabular}}}                        & \multicolumn{3}{c|}{\textbf{Temporal dependence}}                                                                                                          & \multirow{3}{*}{\begin{tabular}[c]{@{}c@{}}\textbf{Spatial}\\\textbf{distribution}\end{tabular}}  \\ 
\cline{3-5}
\multicolumn{2}{|c|}{}                                                              & \multicolumn{2}{c|}{\textbf{\textbf{Linear}}}                                       & \multirow{2}{*}{\textbf{\textbf{\textbf{\textbf{Independent}}}}} &                                                                                                   \\ 
\cline{3-4}
\multicolumn{2}{|c|}{}                                                              & \ Exposure \                  & \ Readout \                             &                                                                  &                                                                                                   \\ 
\hline
\multirow{2}{*}{\begin{tabular}[c]{@{}c@{}}Dark\\ Current\end{tabular}} & Intrinsic & \multirow{2}{*}{$\lambda_{\rm DC} \ t_{\rm EXP}$}  & \multirow{2}{*}{$\frac{\lambda_{\rm DC}}{2} \ t_{\rm RO}$} & \multirow{2}{*}{-}                                               & Uniform                                                                                           \\ 
\cline{2-2}\cline{6-6}
                                                                        & Extrinsic &                                     &                                               &                                                                  & Uniform                                                                                           \\ 
\hline
\multicolumn{2}{|c|}{Amplifier Light}                                       & -                 & $\lambda_{\rm AL} \ t_{\rm RO}$                     & -                                                                & Localized                                                                                         \\ 
\hline
\multicolumn{2}{|c|}{Spurious Charge}                                               & -                                   & -                                             & $\mu_{\rm SC}$                                                   & Uniform                                                                                           \\ 
\hline
\multicolumn{2}{l}{}                                                                & \multicolumn{1}{l}{}                & \multicolumn{1}{l}{}                          & \multicolumn{1}{l}{}                                             & \multicolumn{1}{l}{}                                                                             
\end{tabular}
\end{table*}

\subsection{Protocols for Determining SEE Contributions}
\label{sec:determination}

Considering the model expressed in Equation \eqref{eq.model2}, two methods are specified for measuring the three contributions:

\begin{enumerate}

\item[I.] \textbf{Determination of $\lambda_{\rm DC}$.} Obtain a set of images with different exposure times and a fixed readout time. For each image (with a specific exposure time), measure the SEE rate per pixel using the method described in Section \ref{sec:calibration}. Next, plot this rate as a function of exposure time and perform a linear fit. The slope of this fit corresponds to $\lambda_{\rm DC}$ as shown in Equation \eqref{eq2.model3}, and the y-intercept will be the SEE rate per pixel from the readout stage ($\mu_{\rm RO}$ for a fixed readout time $t_{\rm RO}$) plus the spurious charge ($\mu_{\rm SC}$):

\begin{equation}
\label{eq2.model3}
\mu(t_{\rm EXP})=  \lambda_{\rm DC} \ t_{\rm EXP} + ( \mu_{\rm RO}+  \mu_{\rm SC})\, .
\end{equation}

\item[II.] \textbf{Determination of $\lambda_{\rm AL}$ and $\mu_{\rm SC}$.} Using the measured value of $\lambda_{\rm DC}$ obtained from the previous procedure, $\lambda_{\rm AL}$ and $\mu_{\rm SC}$ can be measured by taking multiple images with different readout times and zero exposure time.
To avoid changing the geometry of the active area (and therefore the value of $\lambda_{\rm AL}$ due to the spatial non-uniformity of this contribution, see Section \ref{sec:al}), $t_{\rm RO}$ is varied by changing the number of samples taken per pixel.
For each image (with a specific readout time), the number of SEEs per pixel is measured.
Next, the SEE rate is plotted as a function of $t_{\rm RO}$ and a linear fit is performed.
The slope of this linear function is $\frac{\lambda_{\rm DC}}{2} +\lambda_{\rm AL}$ as shown in Equation \eqref{eq2.model4}, while the y-intercept is $\mu_{\rm SC}$:

\begin{equation}
\label{eq2.model4}
\mu(t_{\rm RO})=  \biggl(\frac{\lambda_{\rm DC}}{2} +\lambda_{\rm AL} \biggr) t_{\rm RO}  +  \mu_{\rm SC}\, .
\end{equation}
\end{enumerate}

These two proposed methods will be used to determine the values of the three contributions and serve as a model for developing new methods specific to different types of images and configurations. 
In particular, method \textbf{II} requires zero exposure time and variable readout time, which can be achieved in various ways. 
A very common example is to use a single image (or a group of images with identical settings) and exploit the fact that each row has a unique mean lifetime, equal to the exposure time plus the readout time of that row and all preceding rows. 
Thus, the same linear fit can be performed on this group of images.

\subsection{Studies on the Contributions}

In the following sections, we will detail the studies conducted on the contributions described above using the model expressed in Equation \eqref{eq.model2}. The measurements were carried out using a \textit{SENSEI-SCCD}, which is part of the final batch that SENSEI will use for its dark matter exclusion results at SNOLAB. These devices were also used for the study that will be discussed in Chapter \ref{cap:5} and led to the publication of dark matter exclusion limits in 2020 \cite{SENSEI2020}. The device was placed in \textit{MINOS}, and the installation process used the \textit{MINOS Vessel}, similar to the previous publication mentioned earlier. For more details about the experimental device, please refer to Chapter \ref{cap:3}.
We will utilize datasets A, B, C, and D as described in Appendix \ref{cap:glossaryofdatasets}. It is worth noting that certain datasets involve additional shielding, and different values of ${\rm V_{\rm DD}}$ were applied to the output device. This variation is due to the limited time available for data characterization of the devices in facilities like \textit{MINOS}, where most of the detector's time is allocated to scientific data collection.

Furthermore, the data selection criteria used for all mentioned datasets are specified in Table \ref{tab:cortes2022}. Detailed justifications for the number of pixels excluded due to bleeding, high-energy halo, and edge criteria will be discussed in Section \ref{sec:2020criterio}, where the same device and configuration were employed.

\renewcommand{\arraystretch}{1.7}
\begin{longtable}{lp{8cm}}
\centering
\textbf{Name}                           & \textbf{Description}     \\ 
\midrule
\textbf{Single-pixel Events}           & Exclude pixels whose neighbor is a non-empty pixel.    \\
\textbf{Bleeding}                      & Exclude 100 pixels to the right and above pixels with 100 or more electron charge.    \\
\textbf{High-Energy Halo}              & Exclude 40 pixels around any pixel with 100 or more electrons.          \\
\textbf{Edge}                          & Exclude 40 pixels around the edge of the active area.  \\
\bottomrule
\caption{Glossary of quality criteria used in datasets A, B, C, and D.}
\label{tab:cortes2022}
\end{longtable}

\subsection{Determination of Dark Current}

\begin{figure}
    \centering
  \includegraphics[width=.8\linewidth]{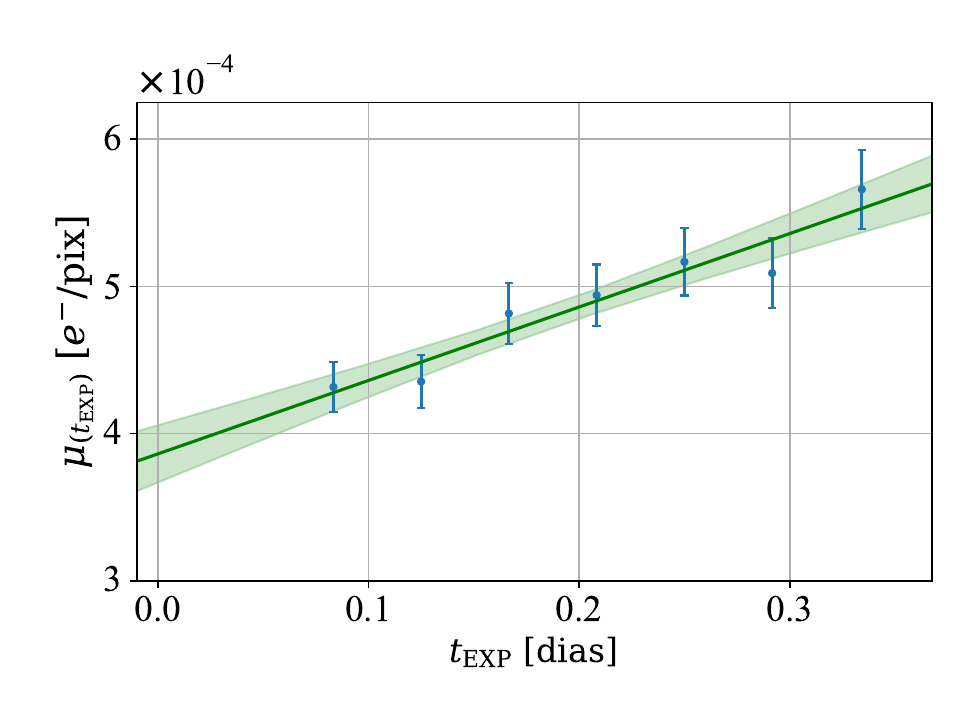}
  \caption{Estimation of $\lambda_{\rm DC}$. The figure shows the SEE rate per pixel as a function of exposure time $t_{\rm EXP}$ for dataset A. A linear regression to the measured points, along with their respective uncertainty bands, is illustrated in green. The slope is estimated to be $(5.89 \pm 0.77) \times 10^{-4} \epixdia$, while the intercept is $(3.69 \pm 0.13) \times 10^{-4} \epix$.}
  \label{fig.DCfit}
\end{figure}

Using dataset A, the number of SEE events per pixel is extracted for each image, and a linear regression is performed using Equation \eqref{eq2.model3}, taking the exposure time of each image $t_{\rm EXP}$ as the independent variable. As shown in Figure \ref{fig.DCfit}, a dark current rate of $\lambda_{\rm DC}= (5.89 \pm 0.77) \times 10^{-4} \epixdia$ is obtained. It is worth noting that this estimate of $\lambda_{\rm DC}$ differs from the one reported in \cite{SENSEI2020}, which will be discussed in detail in Chapter \ref{cap:5}, due to various reasons, including the absence of additional shielding and the use of event selection criteria with a narrower scope to increase the available statistics for the study presented here.

This estimation identifies a significant value for the term $(\mu_{\rm RO}+\mu_{\rm SC})$ independent of time (see the last term of Equation \eqref{eq2.model3}). This term dominates the number of SEE events for exposures of less than 15 hours. To trace its origin and unravel the contributions of $\mu_{\rm RO}$ and $\mu_{\rm SC}$, we first estimate the contributions of spurious charge and amplifier light in the following section.

\subsection{Determination of Spurious Charge and $\lambda_{\rm AL}$}

\begin{figure}
  \centering
  \includegraphics[width=.8\linewidth]{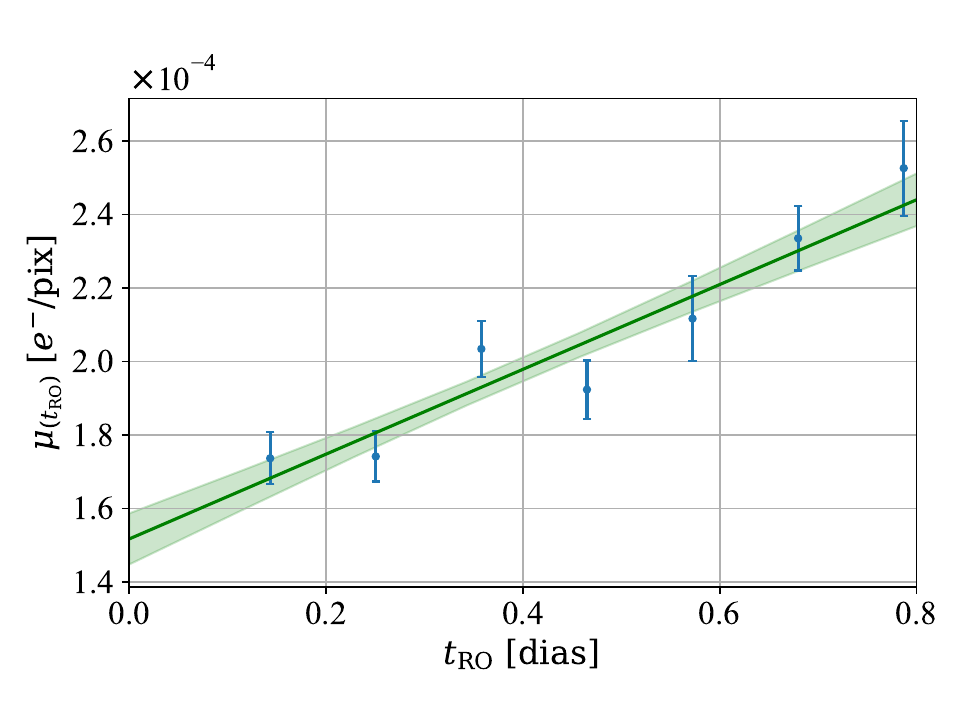}
  \caption{Estimation of $\lambda_{\rm AL}$ and $\mu_{\rm SC}$. SEE rate per pixel as a function of readout time $t_{\rm RO}$ (blue points) for images from dataset D. In green, a linear regression with its corresponding confidence interval is shown. The extracted slope is $(1.15 \ \pm \ 0.16) \times 10^{-4} \epixdia$, while the intercept is $(1.52 \pm 0.10) \times 10^{-4} \epix $.}
  \label{fig.SCfit}
\end{figure}

In order to estimate $\mu_{\rm SC}$ and $\lambda_{\rm AL}$, the number of SEE events per pixel was extracted for each readout time $t_{\rm RO}$, for datasets C and D, and a linear regression was performed as detailed in Section \ref{sec:determination}. Figure \ref{fig.SCfit} illustrates this result for dataset D. Following Equation \eqref{eq2.model4}, the slope of the regression is the sum of $\lambda_{\rm DC}/2$ and $\lambda_{\rm AL}$, and the intercept is the spurious charge rate ($\mu_{\rm SC}$).

The value of spurious charge, $\mu_{\rm SC}$, is obtained as the intercept of the linear fit performed for each dataset. Both estimates overlap, resulting in $(1.59\pm0.12) \times 10^{-4} \epix$ for dataset C and $(1.52\pm0.10) \times 10^{-4} \epix$ for dataset D. This confirms the hypothesis that neither the V$_{\rm DD}$ voltage applied to the output device nor the presence of additional shielding have an effect on spurious charge, as expected.

Regarding $\lambda_{\rm AL}$, for dataset C ($V_{\rm DD} =-22$ V), and using the value of $\lambda_{\rm DC}$ obtained from dataset A, $\lambda_{\rm AL}$ is estimated to be $(19.9\pm1.3) \times 10^{-4} \epixdia$.

On the other hand, for dataset D ($V_{\rm DD} =-21$ V), the dark current reported in \cite{SENSEI2020}, estimated at $(1.59 \pm 0.16) \times 10^{-4} \epixdia$, is used as a reference value for $\lambda_{\rm DC}$ since both experimental configurations are identical. Thus, $\lambda_{\rm AL}$ for $V_{\rm DD} =-22$ V is estimated to be $(0.36\pm0.18) \times 10^{-4} \epixdia$, a value approximately 2 orders of magnitude lower than that of dataset C. While part of this decrease can be attributed to the presence of additional shielding for dataset D, it is primarily attributed to the change in $V_{\rm DD}$.

Next, a dedicated study to understand SEEs produced by the amplifier's light emission is detailed. This characterization effort is analyzed along with how it leads to the optimization of operating parameters.

\subsection{Operation parameters optimization}
\label{sec:optimizaciondeparametrosdeoperacion}

To characterize the contribution of amplifier light, we study the impact of varying the voltages of the output device on the number of SEEs produced. We focus on the gate voltage V$_{\rm DD}$, which controls the drain voltage of output transistor M1 (see Figure \ref{fig.skipperoutputstagelayout}), as previous work \cite{toriumi1987,lanzoni1991} has shown that an increase in the current between the drain and source leads to an increase in light emission.

\begin{figure}[t!]
    \centering
  \includegraphics[width=.7\linewidth]{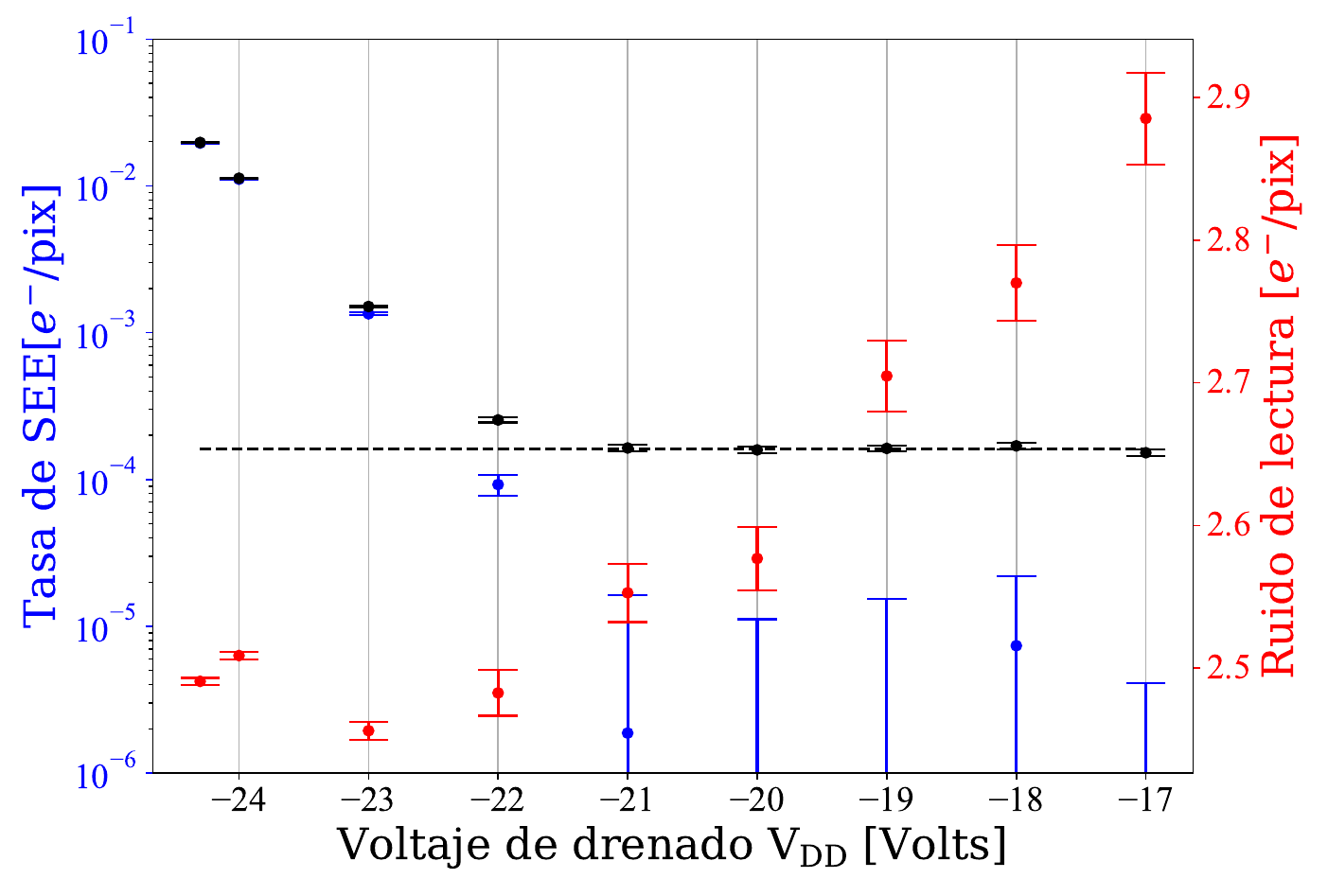}
    \caption{SEE rate per pixel (left axis) and single-sample readout noise (red, right axis) as a function of the drain voltage ${\rm V_{\rm DD}}$ of transistor M1. In black, the SEE rate per pixel collected for each voltage ($\mu_{\rm RO}$) is shown, while in blue, the contribution of amplifier light ($\mu_{\rm AL}$) estimated using Equation \eqref{eq2.model4}. The dashed black line shows the estimate of $\mu_{\rm SC}$ extracted from the fit in Figure \ref{fig.SCfit}. Images from dataset B.}
  \label{fig.vdd_pra_v3}
\end{figure}

To measure the contribution of the amplifier light, zero-exposure images were taken for nine different ${\rm V_{\rm DD}}$ values (dataset B) to verify if this change influences the number of collected SEEs. Taking zero-exposure images makes it possible to measure $\mu_{\rm (t_{RO})}$ as a function of ${\rm V_{\rm DD}}$.
Following Equation \eqref{eq2.model4}, and using the estimated values of $\lambda_{\rm DC}$ and $\mu_{\rm SC}$ (extracted from Table \ref{tablacontribuciones} for the corresponding ${\rm V_{\rm DD}}$ values, both with and without external shielding), $\mu_{\rm (t_{RO})}$ can be extracted as a function of ${\rm V_{\rm DD}}$.

In Figure \ref{fig.vdd_pra_v3}, we show, for each ${\rm V_{\rm DD}}$ voltage in dataset B, the rate of collected SEEs per pixel ($\mu_{(t_{\rm RO})}$) in black and the contribution of amplifier light ($\mu_{\rm AL}$) extracted for each of these voltages in blue, estimated as $\lambda_{\rm AL} \cdot t_{\rm RO}$. It can be observed that the contribution of amplifier light to the total SEE rate decreases drastically between $-24$~V and $-21$~V, while both signals reach a plateau above $-21$~V. This plateau, corresponding to $\mu_{\rm SC}$ (dashed black line) for the variable $\mu_{(t_{\rm RO})}$, is due to a decrease in light emission in M1 as it transitions from its saturation region (below $-21$~V) to its linear operating region.
At the same time, a reduction in the value of ${\rm V_{\rm DD}}$ increases the electronic readout noise, reducing the signal-to-noise ratio. This technique allows for optimizing the operating conditions depending on the ${\rm V_{\rm DD}}$ application.

\renewcommand{\arraystretch}{1.7}
\begin{table}[h!]
\centering
\caption{Summary of SEE contributions and results for ${\rm V_{DD}}$ $=-21$~V and $-22$~V. $\lambda_{\rm AL}$ and $\mu_{\rm SC}$ in the first two rows are extracted from Figure \ref{fig.SCfit}. The value of $\lambda_{\rm DC}$ in the first row is extracted from \cite{SENSEI2020}, while in the second row, it's from Figure \ref{fig.DCfit}. In the third row, the value of $\lambda_{\rm AL}$ is extracted from Figure \ref{fig.vdd_pra_v3}. Both $\lambda_{\rm AL}$ and $\lambda_{\rm DC}$ are in units of $10^{-4} \epixdia$, while $\mu_{\rm SC}$ is in $10^{-4} \epix$.}
\label{tablacontribuciones}
\begin{tabular}{ccccc}
\hline
${\rm V_{DD}}$ & Additional Shielding &  $\lambda_{\rm DC}$ & $\lambda_{\rm AL}$  & $\mu_{\rm SC}$      \\ \hline
$-21$ & Yes &  $(1.59\pm0.16)$  & $(0.36\pm0.18)$  & $(1.52\pm0.10)$ \\
$-22$ & No &  $(5.89\pm0.77)$  & $(19.9\pm1.3)$ & $(1.59\pm0.12)$ \\
$-22$ & Yes &  $-$  & $(23.9\pm3.9)$ & $-$ \\ \hline
\end{tabular}
\end{table}

The first two rows of Table \ref{tablacontribuciones} summarize the results obtained by combining Equation \eqref{eq2.model4} and the values obtained from linear regression for $V_{\rm DD} =-21$~V and $V_{\rm DD} =-22$~V. 
As shown, the parameter optimization method based on the developed semi-empirical model allowed for the optimization of the voltage $V_{\rm DD}$ at $-21$~V, reducing the number of events from the amplifier light by two orders of magnitude. 
To verify that this reduction is not a consequence of the presence of additional shielding, $\lambda_{\rm AL}$ is measured at $-22$~V with additional shielding, extracting the value of this contribution from Figure \ref{fig.vdd_pra_v3}. 
In Table~\ref{tablacontribuciones}, it can be seen that both estimates of $\lambda_{\rm AL}$ are compatible. As for the spurious charge contribution, the values obtained for both datasets C and D (the first two rows of Table~\ref{tablacontribuciones}) are compatible and do not depend on ${\rm V_{DD}}$ or the additional shielding, as expected.

\section{Conclusions}
\label{sec:contribucionesconclusiones}

In this chapter, we presented the different contributions to the SEE signal obtained in a Skipper-CCD, providing a detailed definition of each contribution and its properties, as well as an introduction to its historical context and current state.
To characterize these contributions, we introduced a semi-empirical model that separates and estimates their values and an image acquisition protocol for their determination.
As a result of this characterization, we managed to reduce the contribution of amplifier light to values close to zero, significantly reducing the contribution of a highly relevant SEE background. Additionally, we estimated the lowest ever measured value for spurious charge with an uncertainty of around $1\%$, resulting in $(1.52\pm0.10) \epix$.
Accurate estimation of spurious charge is crucial for subtracting this background from signals compatible with dark matter, as detailed in Section \ref{sec:2020conteo1e}. Currently, efforts are being focused at SENSEI on developing methods and techniques to further reduce spurious charge by modeling the voltage transfer profile changes.

Finally, the developed protocol allowed us to measure dark currents on the order of $1 \times 10^{-4} \epixdia$ (approximately 1 electron per pixel every 27 years), the lowest dark current value ever reported and about 5 orders of magnitude lower than what was reported in the literature just a few years ago. Obtaining a low dark current value is crucial for setting dark matter limits, as will be discussed in the next chapter, not only by examining the 1-electron event channel but also for higher energy multiplicity events that will be practically free from backgrounds originating from this contribution.

\newpage
\mbox{}
\thispagestyle{empty}
\newpage
\chapter{Search for Dark Matter with Skipper-CCDs}
\label{cap:5}

In this chapter, we will discuss the results of the dark matter search published by the SENSEI collaboration to date, covering the works published in 2018 \cite{SENSEI2018}, 2019 \cite{SENSEI2019}, and 2020 \cite{SENSEI2020}.
The first two works used the prototype device called \textit{protoSENSEI}, which was smaller and had lower-quality silicon compared to the final device used in 2020, referred to as \textit{SENSEI-SCCD}.

Since the first two collaboration works were published before the completion of this thesis, they are presented as background information, as they will serve as an introduction to the results published within the framework of this thesis.

\section{Background}
\label{sec:antecedentes}

\subsection{\textit{protoSENSEI-SCCD} on the Surface} 
\label{sec:2018}

Shortly after the Skipper technology demonstrated the capability to achieve sub-electron readout noise in 2017 \cite{tiffenberg2017}, the first experiment to explore dark matter models using this new technology was conducted.
For this purpose, a prototype SCCD, \textit{protoSENSEI}, was used, manufactured in parallel with detectors for astronomical purposes.
Table \ref{tab:2018vs2019a} presents the specifications of the experiment, highlighting that the SCCD used was thinner ($200 \mu$m instead of $675 \mu$m) and smaller ($\sim 0.9$ Mpix instead of $5.5/60$ Mpix) compared to the \textit{SENSEI-SCCDs} designed and manufactured later for installation in the final stage of SENSEI (see Table \ref{tab:dispositivos}). 
The \textit{protoSENSEI} devices had a detector mass approximately 20 times smaller and were made from lower-quality silicon. \newline

\renewcommand{\arraystretch}{1.2}
\begin{table}[h!]
\centering
\begin{tabular}{lccc}
\hline
Year                      & 2018                 & 2019                                                                       & Units                                \\ \hline
Location                  & SiDet      & \textit{MINOS} &                                      \\
Dimensions            & 624  $\times$  362   & 624  $\times$  362                                                         & pixels                               \\
Pixel Size               & 15 $ \times$ 15      & 15 $ \times$ 15                                                            & $\mu$m$^{2}$                         \\
Thickness                 & 200                  & 200                                                                        & $\mu$m                               \\
Total Mass                & 0.0947          & 0.0947                                                                     & grams                                    \\
Temperature           & 130                  & 130                                                                        & K                                    \\
Number of Amplifiers      & 4 (3 used)           & 4 (2/3 used)                                                               &                                      \\
Readout Time (1 sample)   & 24.44              & 24.44                                                                    & $\mu$s                               \\
Readout Noise (1 sample)  & 4                    & 4                                                                          & $e^{-} \ {\rm rms} \ / \ {\rm pix} $ \\
Skipper Sample Count & 800                  & 800                                                                        & \multicolumn{1}{l}{}                 \\
Readout Noise (n samples) & $\frac{4}{\sqrt{n}}$ & $\frac{4}{\sqrt{n}}$                                                       & $e^{-} \ {\rm rms} \ / \ {\rm pix}$  \\ \hline
\end{tabular}
\caption{Main characteristics of the Skipper-CCDs used in \cite{SENSEI2018} and \cite{SENSEI2019}. In 2018, all three amplifiers were used to read the detector, while in 2019, as detailed later, 2 or 3 amplifiers were used depending on the measurement protocol.}
\label{tab:2018vs2019a}
\end{table}

Additionally, the exposure is very low compared to the final exposure goal set by the collaboration (100 grams-year versus 0.019 grams-day). This is mainly due to two reasons.
Firstly, as this was an experiment conducted with a prototype, the goal was to demonstrate the feasibility of the experiment and its scientific reach even with exposure orders of magnitude lower than its main competitors. For example, the DarkSide-50 experiment in 2018 published dark matter limits using 6780.0 kilogram-days and obtained limits for light dark matter scattering through a heavy mediator that were 4 orders of magnitude more stringent (around 10 MeV) than those presented here, despite using 8 orders of magnitude more exposure \cite{agnes2018constraints}.
Secondly, due to the relatively high dark current ($\sim 1 \epixdia$), the images were dominated by SEE backgrounds. This aspect was resolved for the \textit{SENSEI-CCDs} as the dark current was reduced by approximately 4 orders of magnitude, as will be discussed in Section \ref{sec:2020}.

On the other hand, conducting measurements on the surface allowed for the establishment of exclusion limits for dark matter candidates that interact strongly with ordinary matter, a phenomenon that cannot be investigated in underground experiments. The exclusion limits were established after considering the attenuation that occurs in the Earth's atmosphere and immediately above the detector \footnote{And I quote: "... an elevation of $\sim 220$ meters above sea level and a roof composed of approximately $7.6$ cm of concrete, $2$ mm of aluminum, and $1$ cm of wood." Translation from \cite{SENSEI2018}.}.

\subsubsection{Results and Exclusion Limits}

\begin{figure}[h!]
    \centering
    \begin{subfigure}[c]{0.32\textwidth}
        \centering
        \includegraphics[width=\textwidth]{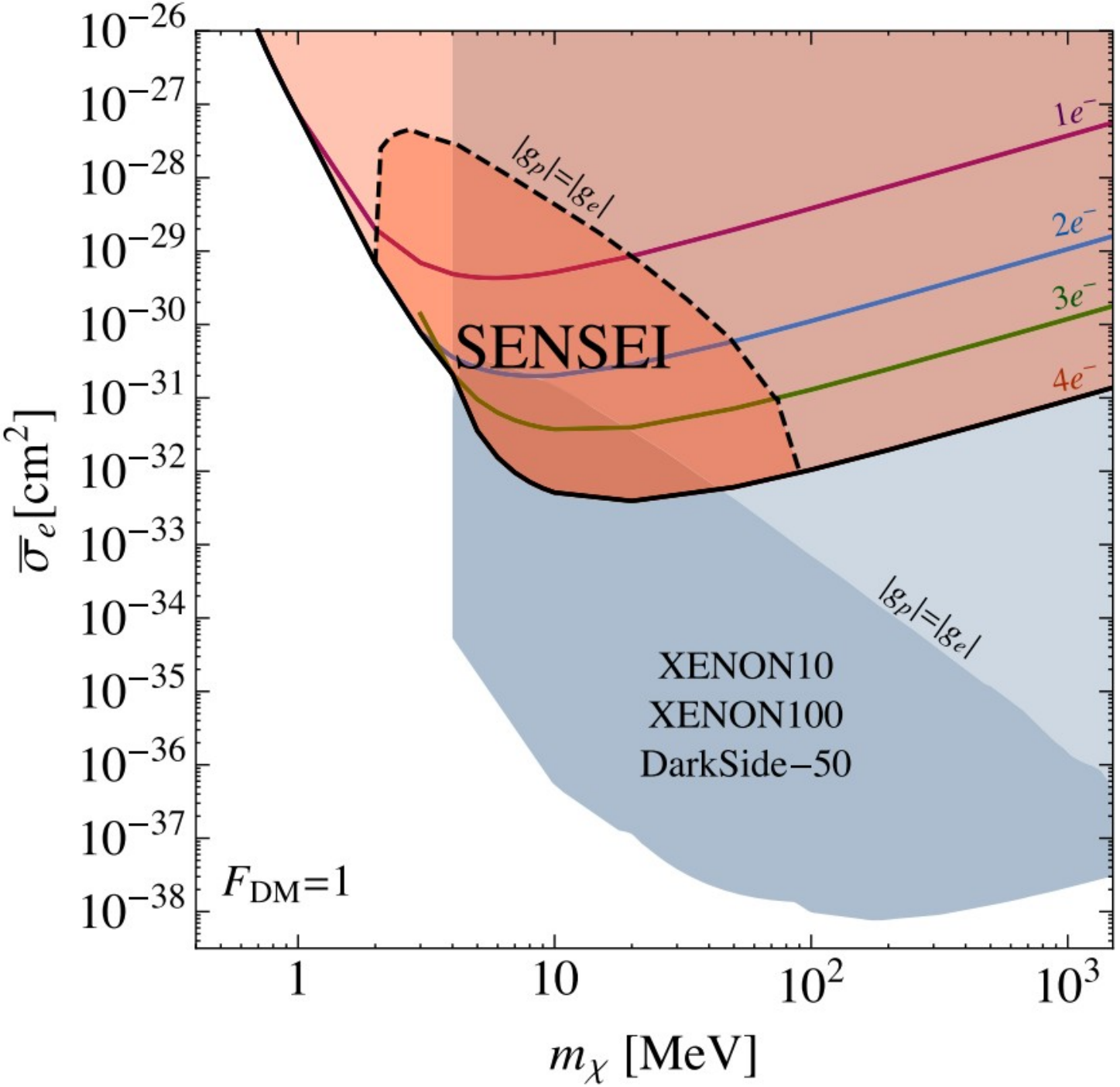}
        \caption{}
        \label{fig.SENSEI2018heavymediator}
    \end{subfigure}
    \hfill
    \centering
    \begin{subfigure}[c]{0.32\textwidth}
        \centering
        \includegraphics[width=\textwidth]{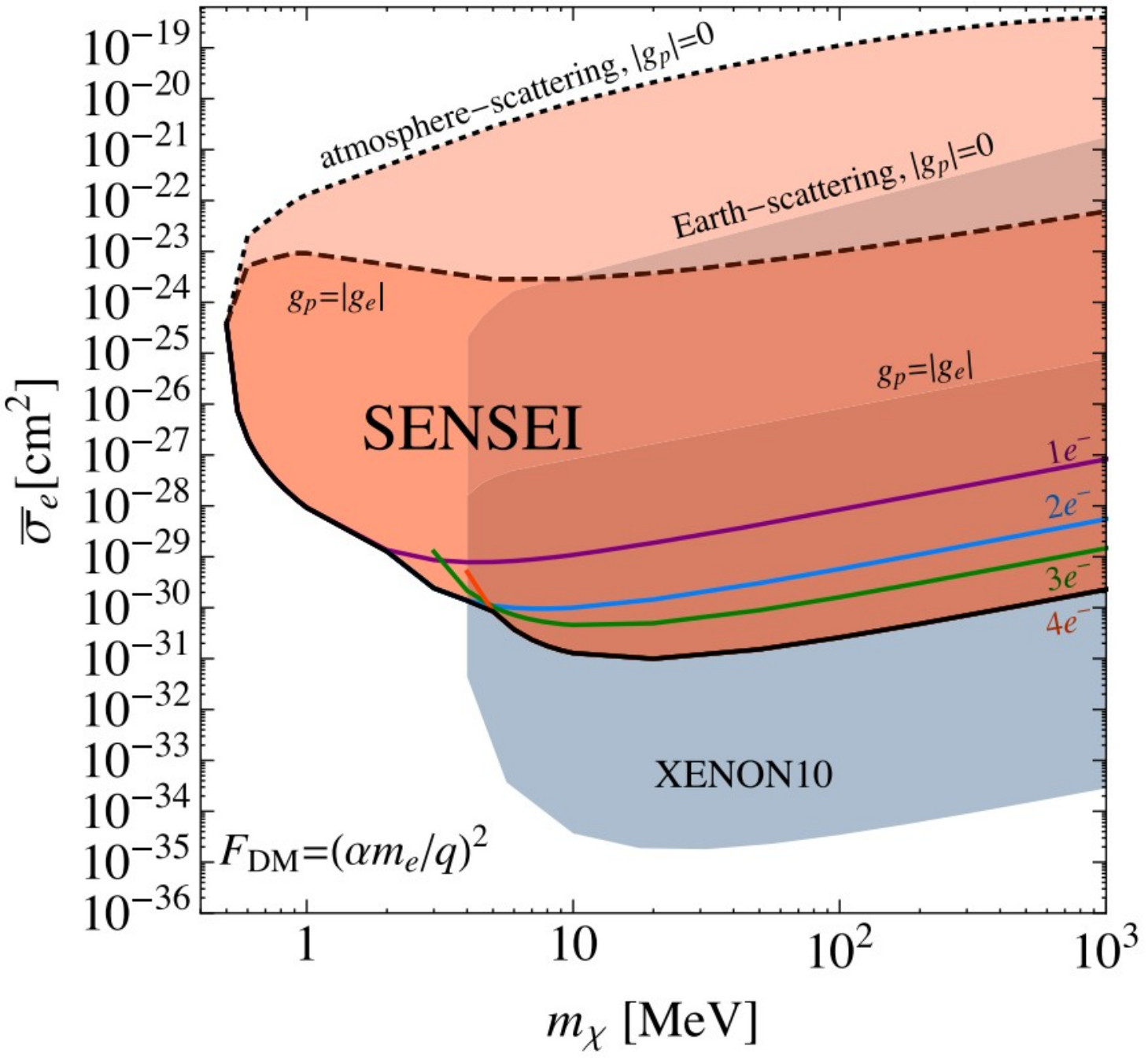}
        \caption{}
        \label{fig.SENSEI2018lightmediator}
    \end{subfigure}
    \hfill
    \begin{subfigure}[c]{0.32\textwidth}
        \centering
        \includegraphics[width=\textwidth]{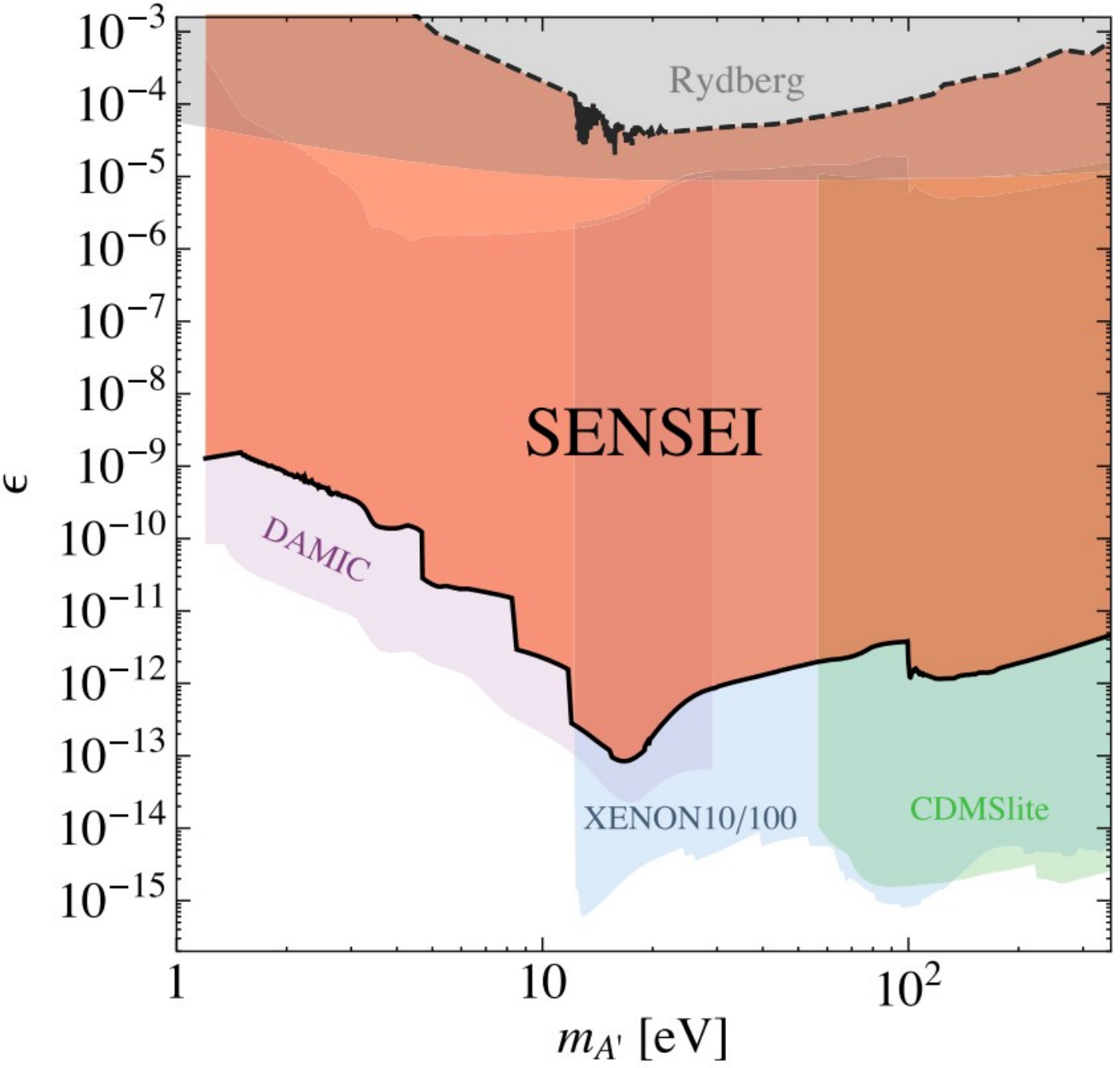}
        \caption{}
        \label{fig.SENSEI2018absorption}
    \end{subfigure}
    \caption{Figures \ref{fig.SENSEI2018heavymediator} and \ref{fig.SENSEI2018lightmediator} display exclusion limits on the effective cross-sections of DM-electron scattering as a function of dark matter mass ($m_{\chi}$). The purple, blue, green, and red lines represent limits for 1, 2, 3, or 4 electrons channels, respectively, with the black line denoting the minimum among these. The blue shaded regions represent current constraints from XENON10, XENON100, and DarkSide-50 \cite{XENON10100,agnes2018constraints}. Figure \ref{fig.SENSEI2018absorption} presents limits on dark matter absorption. The x-axis represents the dark photon mass ($m_{A'}$), and the y-axis represents the kinetic mixing constant ($\epsilon$). The shaded regions illustrate limits established by DAMIC \cite{DAMIC2017}, XENON10/100, and CDMSlite \cite{bloch2017searching}. Figures extracted from \cite{SENSEI2018}.}
    \label{fig.SENSEI2018}
\end{figure}

Figure \ref{fig.SENSEI2018} presents exclusion limits imposed at a 90$\%$ confidence level on the effective cross-sections of DM-electron scattering, $\sigma_{e}$, as a function of dark matter mass, $m_{\chi}$ (for Figures \ref{fig.SENSEI2018heavymediator} and \ref{fig.SENSEI2018lightmediator}), and on the parameter $\epsilon$ of kinetic mixing for dark matter absorption as a function of the dark photon mass, $m_{A'}$ (for Figure \ref{fig.SENSEI2018absorption}).

For Figures \ref{fig.SENSEI2018heavymediator} and \ref{fig.SENSEI2018lightmediator}, the attenuation of dark matter candidates by the Earth's atmosphere is calculated in two ways: one considering a mediator that only couples to electrons (labeled as $|g_{p}|=0$), and the other (labeled as $|g_{p}|=|g_{e}|$) assuming that the dark matter candidate interacts with ordinary matter through either a heavy mediator (Figure \ref{fig.SENSEI2018heavymediator}) or a light mediator (Figure \ref{fig.SENSEI2018lightmediator}).

Parallely, for the case of dark matter absorption (Figure \ref{fig.SENSEI2018absorption}), a maximum coupling $\epsilon_{max}$ is estimated for each experiment, taking into account the attenuation of the dark photon in the Earth's atmosphere (or in the crust when necessary). The maximum coupling for SENSEI is represented by a dashed line that bridges the gap between direct detection experiments and limits obtained from measurements of a Rydberg constant.

\renewcommand{\arraystretch}{1.4}
\begin{table}[h!]
\centering
\begin{tabular}{clccccc}
\cline{3-7}
\multicolumn{1}{l}{}  &          & \multicolumn{5}{c}{$N_{e}$} \\ \hline
\multicolumn{1}{l}{}  &          & 1   & 2   & 3   & 4   & 5   \\ \hline
2018 & Events   & 140,302   & 4,676   & 131   & 1   & 0   \\
                      & Exposure [g days] & 0.0127   & 0.0078   & 0.0061   & 0.0051   & 0.0046   \\ \hline
2019 & Events   & 2,353   & 1   & 0   & 0   & 0   \\ 
                      & Exposure [g days] & 0.069   & 0.043   & 0.118   & 0.073   & 0.064   \\ \hline
\end{tabular}
\caption{Number of events and effective exposure used for the dark matter exclusion limits published in \cite{SENSEI2018} (2018) and \cite{SENSEI2019} (2019). \label{tab:2018vs2019b}}
\end{table}

\subsection{\textit{protoSENSEI-SCCD} at 104 Meters Underground}
\label{sec:2019}

In 2019, new dark matter exclusion limits were published. The same sensor used in 2018 was installed at a depth of 104 meters underground in the \textit{MINOS} cavern at the Fermi National Accelerator Laboratory (FNAL) (see Table \ref{tab:2018vs2019a}). This allowed for a drastic reduction in high-energy background, especially from cosmic and atmospheric muons. The number of traces was reduced, and the exposure increased by approximately an order of magnitude. Additionally, a new vacuum vessel, as described in Chapter \ref{cap:3}, was used.

Table \ref{tab:2018vs2019b} shows the increase in exposure for each energy channel between 2018 and 2019, as well as the reduction in the number of events compatible with dark matter, from 1 to 5 electrons. This reduction can be attributed to the 104 meters of rock between the detector and the atmosphere.

\subsubsection{Readout Modes and Data Acquisition: Excess SEE from Amplifier Light}

While in 2018, images were read from each of the amplifiers (three of which were operational), for the 2019 analysis, two different image reading techniques were used.

The first one, \textit{continuous readout}, involved reading all four quadrants of the SCCD in parallel and independently until a certain number of rows was reached (with the size of the file being the only limiting factor). After excluding the first 624 rows of the first image of the data acquisition, the exposure for each pixel is determined by the SCCD's readout time, which is approximately 4400 seconds. A total of 0.27 grams-days of data were taken in this way over a 3.8-day data acquisition period.

\begin{figure}[h!]
     \centering
     \begin{subfigure}[c]{0.47\textwidth}
         \centering
         \caption{}
         \includegraphics[width=\textwidth]{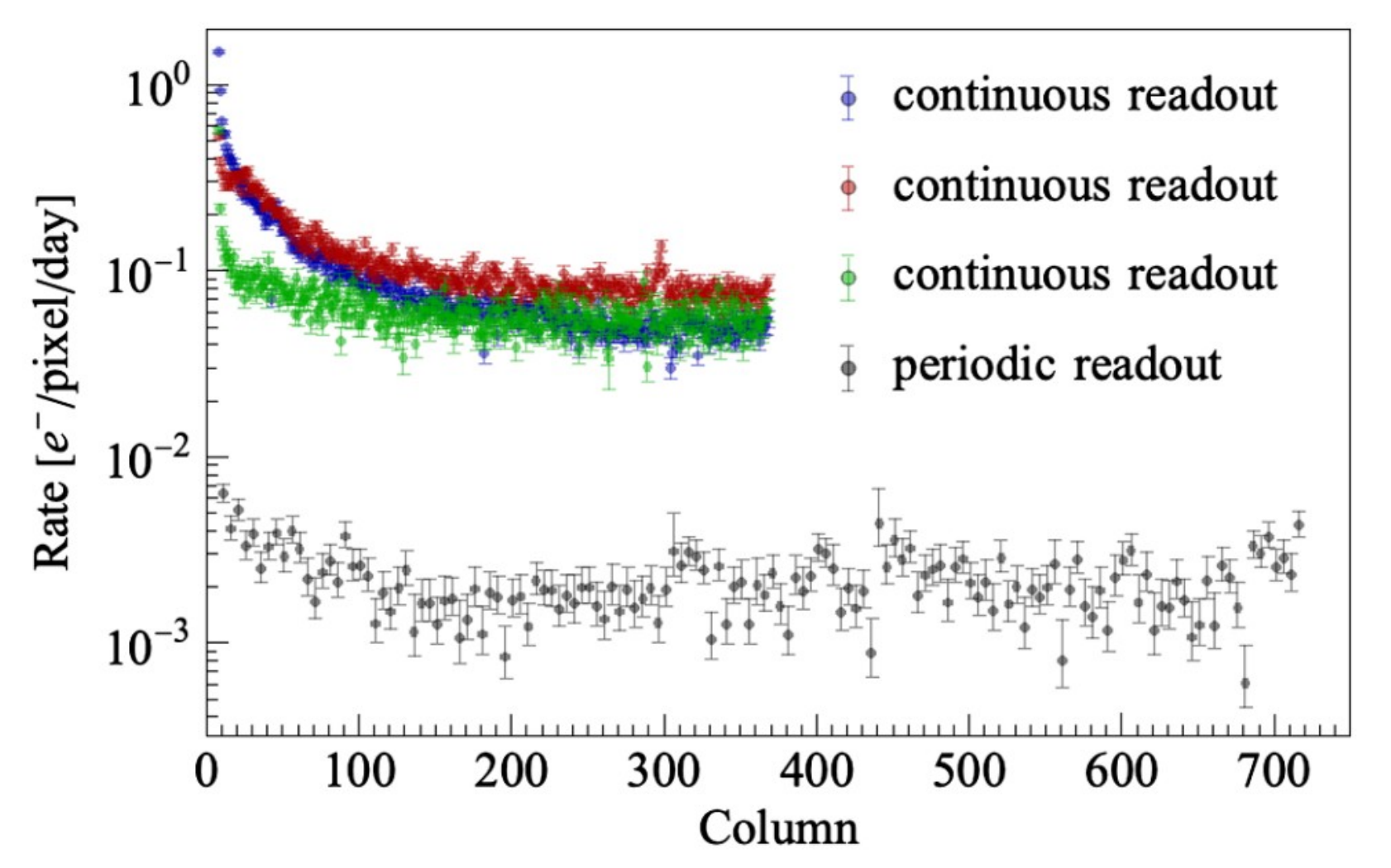}
         \label{fig.amplifier2019}
     \end{subfigure}
     \hfill
    \centering
     \begin{subfigure}[c]{0.47\textwidth}
         \centering
         \caption{}
         \includegraphics[width=\textwidth]{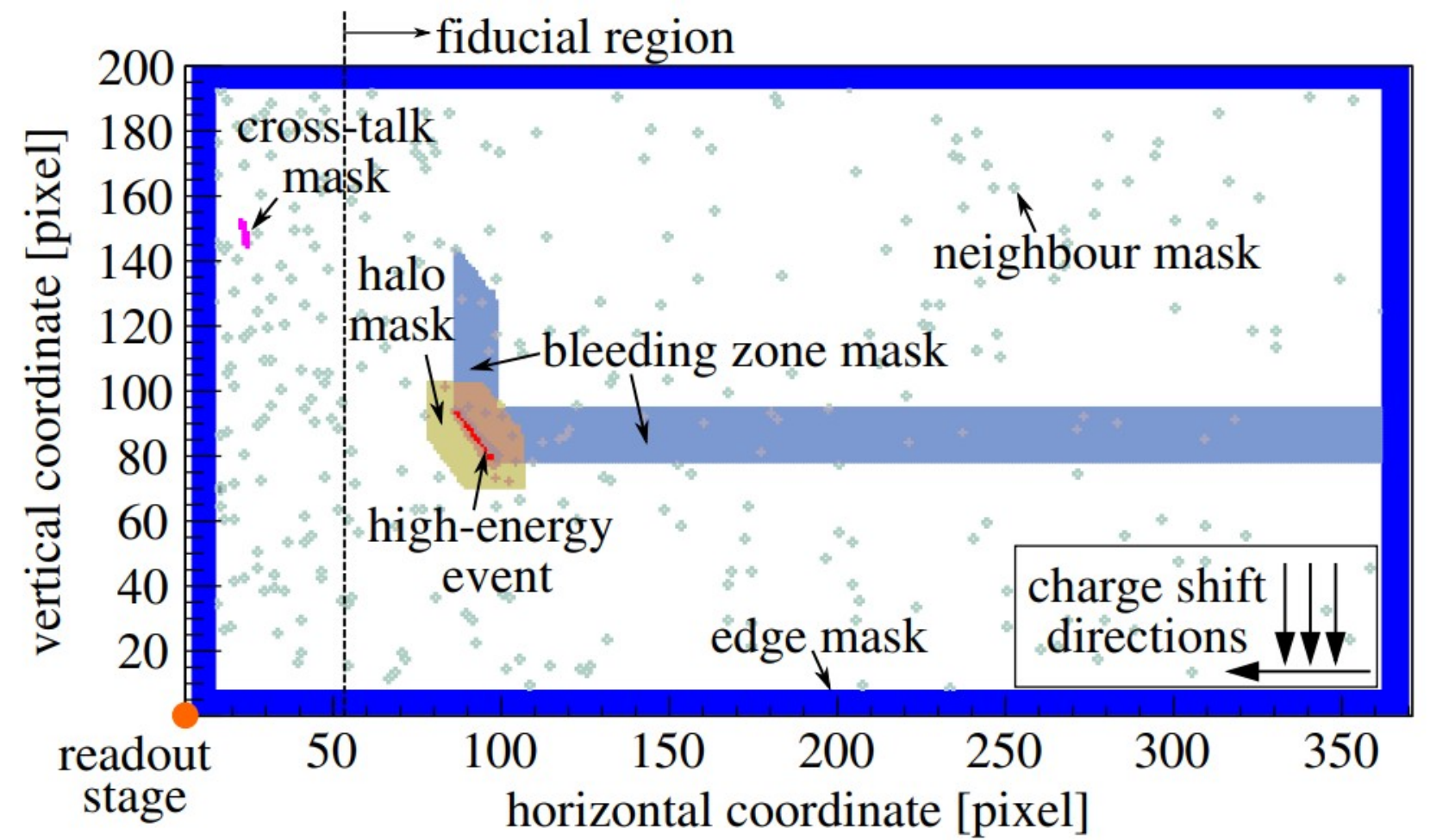}
         \label{fig.SENSEI2019image}
     \end{subfigure}
        \caption{Excess of 1 and 2-electron events due to amplifier light and the solution used for their rejection. Figure \ref{fig.amplifier2019} shows the SEE rate in $\epixdia$ as a function of the column number for the three amplifiers (purple, red, and green) in continuous readout mode. In gray, the same result is shown for the best amplifier in periodic readout mode. Figure \ref{fig.SENSEI2019image} illustrates the collected SEEs in an image and the applied image cuts. It highlights the selection of a fiducial region that is excluded due to the excess of SEEs near the readout amplifier. Figures extracted from \cite{SENSEI2019}.}
        \label{SENSEI2019}
\end{figure}

As can be seen in Figure \ref{fig.amplifier2019}, an excess of 1 and 2-electron events was found in the first columns of the images. This excess is related to the light emitted in the output amplifier, a phenomenon that was extensively studied in Section \ref{sec:al}. As expected, the impact of this source of light decreases for larger distances from the amplifier, so excluding pixels up to a certain column number effectively removes this excess (see also Figure \ref{fig.SENSEI2019image}). However, it should be noted that the amplifier light (AL) also creates an excess independent of the 1 and 2-electron event space, as each pixel will spend a fixed amount of time in the output stage where the contribution of amplifier light is stronger. This is the main reason why continuous readout mode produces almost two orders of magnitude more SEEs than periodic readout mode (see Figure \ref{fig.amplifier2019}), which will be introduced shortly.

Due to this overall excess, the columns closest to the amplifiers of each quadrant were excluded in continuous readout mode for the DM search in the energy range from 3 to 100 electrons. Additionally, searches for events between 1 and 2 electrons of energy were completely excluded for this readout mode. The maximum number of columns in the excluded region was chosen such that the probability of finding more than 0.5 events of 3 electrons for any column in the non-excluded region was low enough.

The other data-taking method, periodic readout, turned out to be an alternative to reduce the impact of the background light in 1 and 2-electron events. This method involved exposing the CCD for 120,000 seconds before reading, not from all four amplifiers, but from two of them, addressing the issue of having only three amplifiers in operation. Additionally, during the 120,000 seconds of exposure, the amplifier is turned off and, therefore, no light is emitted. Now, most of the exposure for each pixel will come from the 120,000-second exposure when the amplifier is off, and it will only be turned on during the readout for approximately 8,800 seconds. Figure \ref{fig.amplifier2019} shows the drastic reduction in the 1-electron event rate in units of $\epixdia$. Final event selection criteria were used to blindly select images with the lowest rates of 1, 2, and 3 electrons, using a data-driven analysis. This resulted in a final exposure, without additional cuts, of 0.069 gram-days.

\subsubsection{Results and Exclusion Limits}

\begin{figure}[h!]
    \centering
    \includegraphics[width=0.97\textwidth]{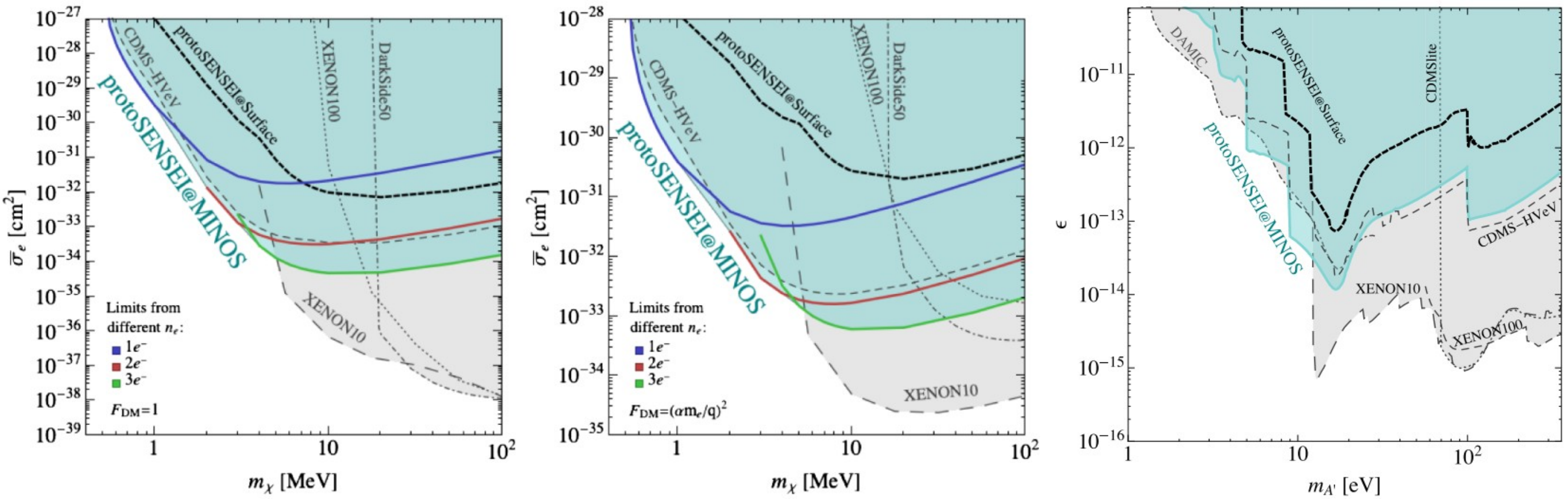}
    \caption{90\% confidence level exclusion limits using data obtained from \textit{protoSENSEI} at a depth of 104 meters in MINOS. Limits are presented for the effective cross-section considering DM-electron scattering for heavy mediator ($F_{DM}=1$, left) and light mediator ($F_{DM}=(\frac{\alpha m_{e}}{q})^{2}$, center) as well as for the \textit{kinetic mixing} parameter $\epsilon$ (right). For the scattering, the blue and red lines represent the limits using the 1 and 2-electron channels, respectively, from the periodic-readout dataset, while the green line is the limit for the 3-electron channel combining both readout methods. Limits published by XENON 10 and 100 \cite{XENON10100}, DarkSide-50 \cite{agnes2018constraints}, CDMS-HVeV \cite{SUPERCDMS2018}, DAMIC \cite{DAMIC2017}, and CDMSlite \cite{bloch2017searching} are also shown. Figure extracted from \cite{SENSEI2019}.}
    \label{fig.SENSEI2019limits}
\end{figure}

The results of the counts and the exposure obtained are shown in Table \ref{tab:2018vs2019b}. In it, the approximate increase of an order of magnitude in the obtained exposure is highlighted, and more importantly, the decrease in the number of 1, 2, and 3-electron events. This reduction is reflected in Figure \ref{fig.SENSEI2019limits}, where a decrease of several orders of magnitude is observed compared to the 2018 results.

Installing the device at a depth of 104 meters, together with a deeper understanding of the operation of SCCDs, led to an improvement of several orders of magnitude in the limits imposed on theoretical models of dark matter for both dark matter absorption and DM-electron scattering, as seen in Figure \ref{fig.SENSEI2019limits}. In particular, SENSEI has led the world in the search for light dark matter with masses below 5 MeV for both electron-mediated scattering with light and heavy mediators. A one-order-of-magnitude improvement in dark matter absorption constraints for dark photons was also achieved.

\section{\textit{SENSEI-SCCD} at 104 Meters Underground}
\label{sec:2020}

In mid-2019, a \textit{SENSEI-SCCD} was installed in \textit{MINOS}, in the same location and with the same vacuum chamber used in 2019. The configuration used is described in Section \ref{sec:dispositivo}, while the specifications of the \textit{SENSEI-SCCD} used can be found in Table \ref{tab:dispositivos}. Compared to \textit{protoSENSEI}, the most significant improvement was the increase in mass, from 0.0947 to 1.926 grams per sensor. Additionally, dark current was reduced by several orders of magnitude, mainly through the use of very high-quality silicon (which reduces defects in both the bulk and the surface) and very high resistivity (approximately 18 k$\Omega$-cm), as well as the development of event selection criteria described and developed in this thesis. It is also worth noting the addition of a lead plate (the so-called "additional shielding," see Section \ref{sec:camaras}) in the area outside the vacuum chamber, which achieved a reduction in dark current, as will be seen in Section \ref{sec:2020dc}, and the use of the LTA readout board.

This section is divided into six subsections. The first will discuss the data acquisition design, including details of the configuration used for image acquisition and the operational parameters of the SENSEI-SCCD. The second subsection will discuss the event selection criteria as introduced in Section \ref{sec:masqueo} but focusing on the steps and procedures used to set the parameters of these criteria using a calibration dataset. After establishing the event selection criteria, the third subsection will present results regarding contributions from spurious charge and dark current for the analyzed dataset. The fourth and fifth subsections will detail the procedure used for event counting and exposure calculation, necessary for setting the dark matter exclusion limits. Finally, the sixth subsection will conclude the discussion by presenting the final results.

\subsection{Data Acquisition}
\label{sec:2020tomadedatos}

The design and execution of the data acquisition, in which I played the main role and was responsible, consisted of a total of 29 images. Out of these, 7 (\textit{commissioning images}) were analyzed separately to establish the data selection criteria that were subsequently applied to the other 22 images (\textit{science images}) (see Appendix \ref{cap:glossaryofdatasets}). The dark matter exclusion limits obtained from the science images will be presented near the end of this section. The 29 images were divided into their respective four quadrants, each synchronously read by its corresponding output devices. Quadrants 1 and 2 operated almost indistinguishably from each other since their installation, with a readout noise of 2.5 $\e$ per sample and a SEE rate per pixel (after applying the event selection criteria, which will be detailed later) of approximately $1\times 10^{-4} \epix$. However, the third quadrant exhibited an excess of SEEs in the initial columns, which was attributed to the absorption of infrared photons coming from the vacuum chamber through a slit, part of the flat spring that composed the copper module containing the SCCD (see Figure \ref{fig.sensei_module} and the discussion in the text). As detailed later, different event selection criteria were applied to this quadrant to mitigate the issue without losing its entire data set. Unfortunately, the fourth quadrant was unusable due to a short circuit that occurred during the SCCD packaging and the gold wire bonding process to the pads.

Regarding the control parameters of the images, we favored the use of long exposure intervals, minimizing the time during which the active area pixels are read. This decision was partially made to reduce the impact of amplifier light on the number of SEEs collected during reading because this contribution is null during exposure since the amplifier is turned off. On the other hand, the implementation of the cleaning routine between images, introduced in Section \ref{sec:protocolodemedicion}, and crucial in reducing the SEE rate per pixel collected per image, made it impossible to use continuous-readout mode. Due to the increased thickness of the SCCD used compared to 2019, the exposure time was reduced from 120k to 72k seconds since the number of high-energy events collected increased proportionally with this thickness. It is worth noting that high-energy events not only occupy the pixels in which the charges are ionized in the sensor volume (and the pixels in which this ionized charge diffuses) but also their surroundings due to the spatial correlation between SEEs and high-energy events described in Section \ref{sec:dc}, which can be up to several tens of pixels away.

\subsubsection{Voltage Selection}

The voltages used to operate the SENSEI-SCCD were optimized and selected before the data acquisition. Thanks to previous experiences with SCCDs in the collaboration, there was a predefined voltage configuration to be used, especially in the active area of the CCD and the horizontal register. However, multiple changes were made to this voltage configuration. For example, the drain voltage $V_{drain}$, used to "discard" charges after the complete readout of a pixel, was reduced from -21 to -24 Volts because the charges did not undergo a complete transfer from the sensing node to the pixel where this discarding is performed (see Figure \ref{fig.amplifier}). This type of CTI created elongated events in the horizontal axis \textit{y}, and more critically, a higher number of SEEs per pixel. Another example, mentioned earlier in Section \ref{sec:SRevents}, is the increase in the upper TG voltage to prevent the diffusion of charges from the horizontal register to the active area of the SCCD when high-energy events impact that register during reading.

Regarding the output amplifier, the voltage optimization in this device was carried out by Dr. Ing. Miguel Sofo Haro, a member of the team responsible for its design, while the selection of the voltage ${\rm V_{DD}}$ for the amplifier was made following the details in Section \ref{sec:optimizaciondeparametrosdeoperacion}. The transfer voltages for the active area and horizontal register were chosen based on the analysis described in Section \ref{sec:ctivssc}. After achieving the lowest levels of dark current and spurious charge ever recorded, future work suggests conducting an exhaustive study that details the relationship between the voltages used and the rate of SEE per pixel.

\subsubsection{Integration Time and Readout Noise Selection}

The integration times used to measure the number of electrons collected per pixel, as well as the number of samples per pixel, were selected following the description in Section \ref{sec:designandoperationprinciples}, finding the minimum readout noise trade-off in these two control parameters. In this way, 300 samples of each pixel were taken, with each sample taking 42.83 $\mu$s, resulting in a total readout time for the SENSEI-SCCD of 5 hours and 12 minutes per image. The dimensions of the measured images slightly exceeded those of the active area: 3100$\times$470 versus 3072$\times$443 pixels (rows $\times$ columns), respectively. These "extra" spaces were used for overscan checks.
On the other hand, the baseline was extracted using all the empty pixels in each row (see the discussion in Section \ref{sec:catalogcreation}).

The readout noise, for this number of samples and the selected integration time, is approximately 0.15 $\e$, so the probability of misclassifying an event of \textit{n} electrons as one of $n\pm$ 1 is approximately 3.3~$\sigma$ or $\sim$ 0.1$\%$.

The data acquisition took a little over a month and was halted due to the quarantine measures imposed by the COVID-19 pandemic. Once the first 7 images, which were used as "test images," were obtained, they were processed as detailed in Chapter \ref{cap:3} and analyzed to establish the event selection criteria that will be detailed in Section \ref{sec:2020criterio}. On the other hand, the "science images" were left unprocessed (blind) until the event selection criteria were finalized.

\subsection{Event Selection Criteria}
\label{sec:2020criterio}

Table \ref{tab:cortes2020}, at the end of this section, summarizes the event selection criteria in a manner similar to what was detailed in Table \ref{tab:cortes} in Chapter \ref{cap:3bis}. The theoretical foundations of all the cuts were already described in detail in Section \ref{sec:masqueo}. Additionally, we will detail the analysis performed to establish the criteria using the 7 test images, which were then used for the final analysis in the 22 science images obtained.

Similar to what was published in 2019 (protoSENSEI at 104 meters depth, Section \ref{sec:2019}), we took advantage of the ability to use different event selection criteria for the four electronic channels to be analyzed, ranging from 1 to 4 $\e$. In the following sections, we will provide details of these criteria, indicating, whenever necessary, the differences in each criterion for each electronic channel. In particular, we will divide the electronic channels into two subgroups: single-electron and multi-electron (1 $\e$ or more, respectively). This separation is primarily based on the fact that we expect the 1-electron channel to have an event rate between two and three orders of magnitude higher than the channels in the multi-electron group, as indicated by the analysis of test images and later confirmed when analyzing the science images. The low occurrence of events for the multi-electron channels is an indication of low background content for those channels, allowing us to use more lenient selection cuts and increase their exposure.

Another point to consider is the solution taken to exclude the excess of SEEs in the first columns of quadrant 3, mentioned in Section \ref{sec:2020tomadedatos}. Since this background compromised the sensitivity to dark matter interactions for the 1 and 2 $\e$ channels, the data extracted from this quadrant was completely excluded from these analyses. For the remaining channels in the multi-electron group, i.e., 3 and 4 $\e$, we chose to include quadrant 3 in the analysis but excluded the first 93 columns of that quadrant, where the majority of the background events were located.

In the following Section, details on the used event selection criteria will be given. 

\subsubsection{Noise and Crosstalk}

\begin{figure}[h!]
    \centering
    \includegraphics[width=0.7\textwidth]{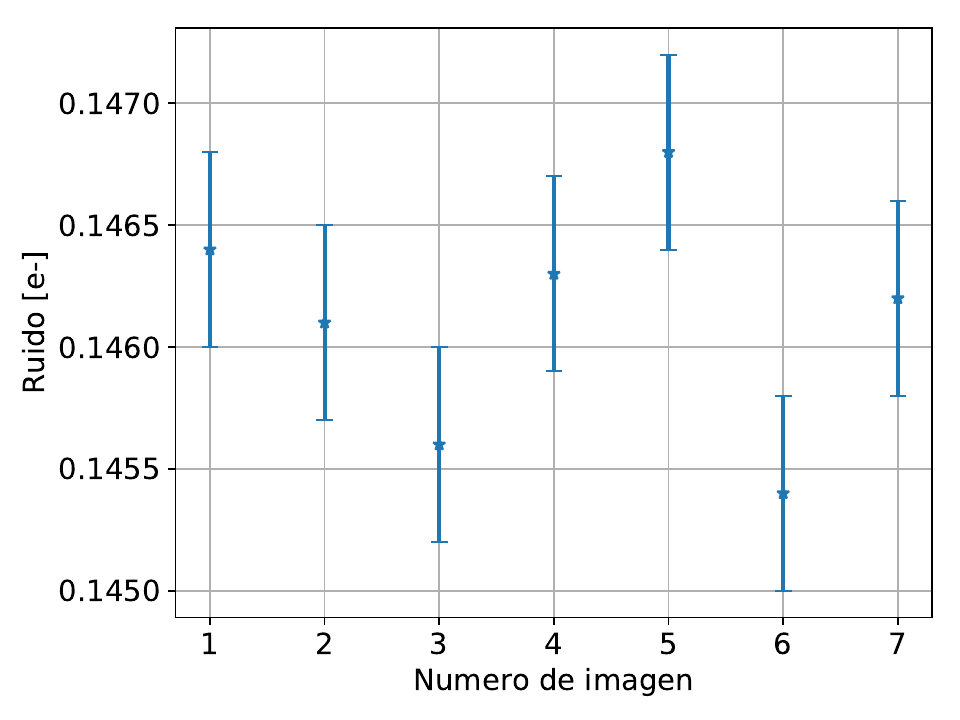}
    \caption{Readout electronic noise for the first quadrant in the 7 test images obtained, arranged chronologically. No quality cut was applied to obtain these values, using all pixels in the image. The noise value for each image is determined by the standard deviation of the 0-electron peak, which results from electronic noise.}
    \label{fig.noise}
\end{figure}

Due to advancements in the operation of SCCDs and their manufacturing, the readout noise is not only acceptably low (approximately 0.16 $\e$) but also highly stable. Figure \ref{fig.noise} illustrates the readout noise for the 7 test images in the first quadrant. Therefore, since no anomalous behavior was found in the readout noise, the selection criterion used, which dictates discarding images with noise exceeding 30$\%$ of the standard value, did not reject any image.

The value chosen for the crosstalk criterion takes into account that 700 electrons in one quadrant, at most, generate a signal of 0.1 electrons in neighboring quadrants. Therefore, this is a conservative measure with a negligible effect on the excluded area of the images.

\subsubsection{High-Energy Halo and Edge Criterion}

The high-energy halo criterion, previously discussed in Section \ref{sec:halo}, is responsible for excluding pixels near high-energy events due to a spatial correlation that exists between these events and 1 or 2 electron events (mostly SEEs). It is worth noting that a possible mechanism for the production of SEEs is Cherenkov radiation and/or radioactive recombination resulting from the interaction of high-energy events with the detector.

\begin{figure}[h!]
    \centering
    \includegraphics[width=0.9\textwidth]{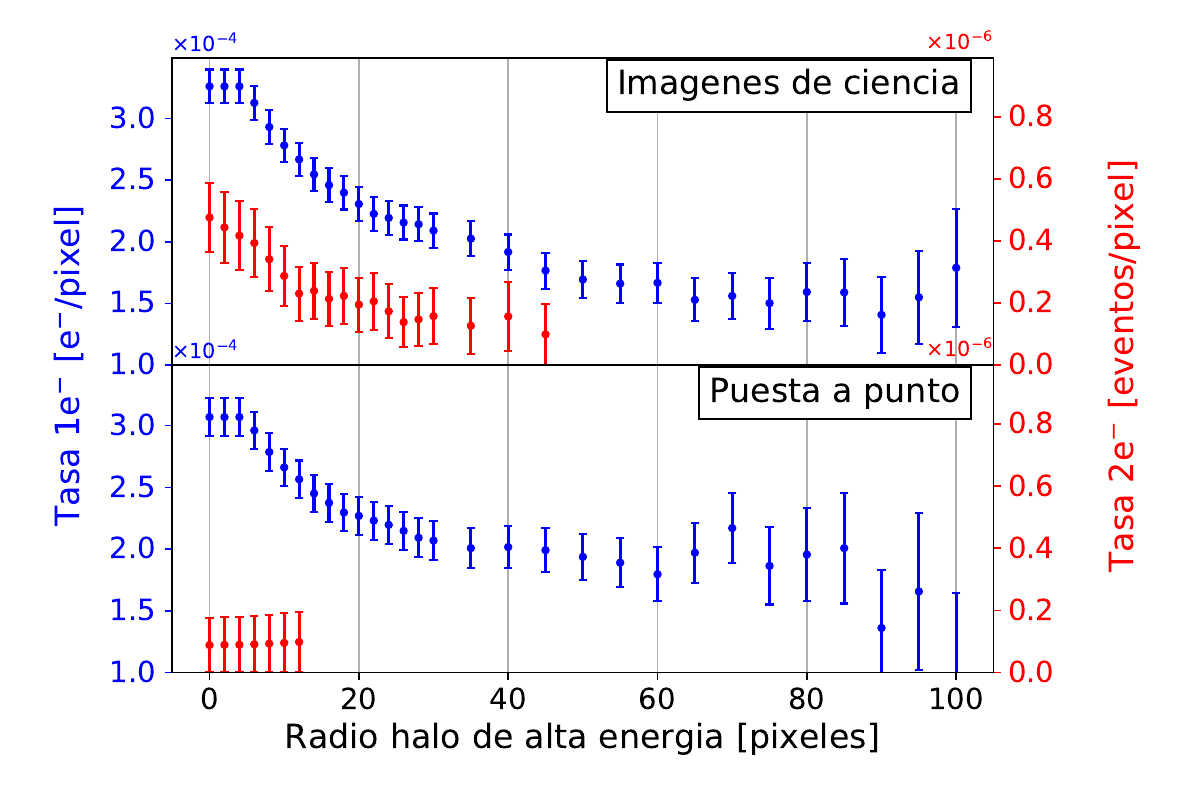}
    \caption{Event rate per pixel for single-pixel events of 1 $\e$ (blue) and 2 electrons (red) for science images (\textit{DM search data}) and commissioning images (\textit{commissioning data}), respectively, as a function of the radius of the high-energy halo used. The rate is given for the region outside the halo created by the mask mentioned in the text, i.e., the unmasked region. Figure extracted from \cite{SENSEI2022}.}
    \label{fig.Halomerge}
\end{figure}

Figure \ref{fig.Halomerge} illustrates this spatial correlation. It is divided into two subfigures, showing the results obtained for science images and commissioning images separately. The x-axis represents the radius of the halos generated around all pixels with 100 or more electrons, while the y-axis represents the rate of 1-electron events outside this halo, shown in blue (2 electrons in a single pixel in red). It can be seen that the SEE rate monotonically decreases up to 55-60 pixels for both datasets and then remains constant within the available statistical margins. For 2-electron single-pixel events, no events are recorded for halos larger than $\sim$ 15 pixels in radius for the commissioning images, so a conservative halo of 20 pixels in radius was selected. It can also be observed that, after selecting all event selection criteria and analyzing the  channel for science images, the rate for events in that channel monotonically decreases to approximately 15 pixels in radius, confirming the initial hypothesis.

Flattening of the event rates for 1 and 2 electrons for sufficiently large halos is a necessary condition for a SEE regime with a Poisson-like statistical distribution. Fulfilling this condition is crucial when studying potential dark matter signals, as energy deposits from dark matter are expected to also follow a Poisson distribution.

\begin{figure}[h!]
    \centering
    \includegraphics[width=0.7\textwidth]{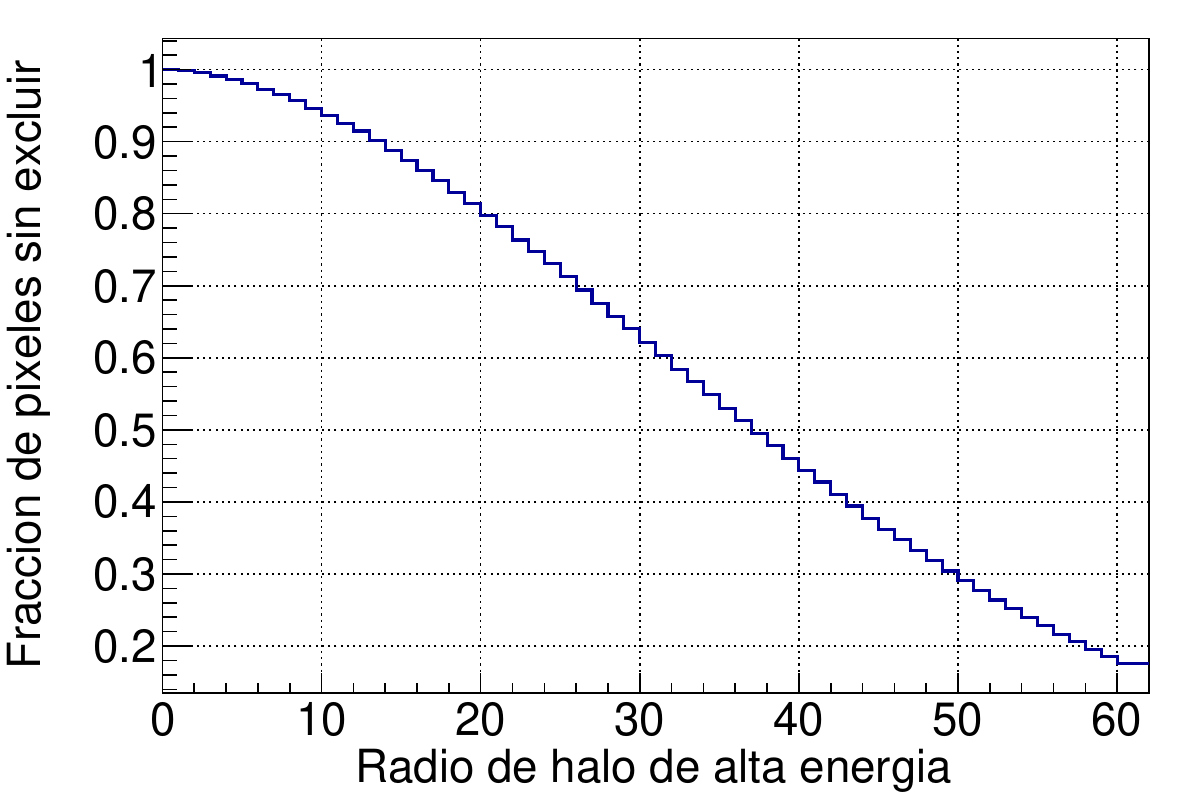}
    \caption{Fraction of pixels not excluded by the high-energy halo as a function of the radius of the halo for quadrants 1 and 2 of the science images, without applying any data quality cuts to the dataset.}
    \label{fig.Haloexp}
\end{figure}

Another noteworthy aspect of this cut is the cost it incurs in terms of exposure when creating halos of up to 60 pixels around all pixels with 100 or more electrons. Figure \ref{fig.Haloexp} illustrates this cost as the fraction of pixels not excluded as a function of the halo radius. After a halo of 60 pixels, approximately 90$\%$ of the dataset is excluded from the analysis. This percentage depends on the number of high-energy events present and may be higher for shorter exposure times and lower sources of high-energy background signals.

The edge cut is conservatively set at 60 (20) pixels for the analysis of 1 (2 or more) electrons to exclude effects of potential high-energy interactions beyond the active area of the SCCD.

\subsubsection{Bleeding}

\begin{figure}[h!]
     \centering
     \begin{subfigure}[c]{0.77\textwidth}
         \centering
         \caption{Bleeding for the 1-electron channel.}
         \includegraphics[width=\textwidth]{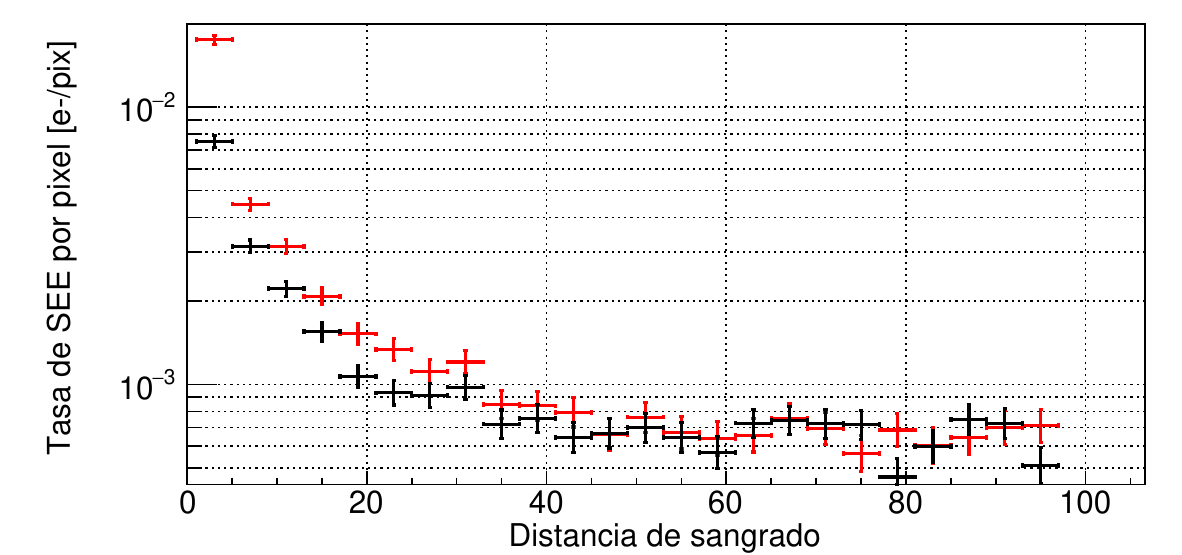}
         \label{fig.bleedXY_1e}
     \end{subfigure}
     \hfill
     \begin{subfigure}[c]{0.77\textwidth}
         \centering
         \caption{Bleeding for the 2-electron channel.}
         \includegraphics[width=\textwidth]{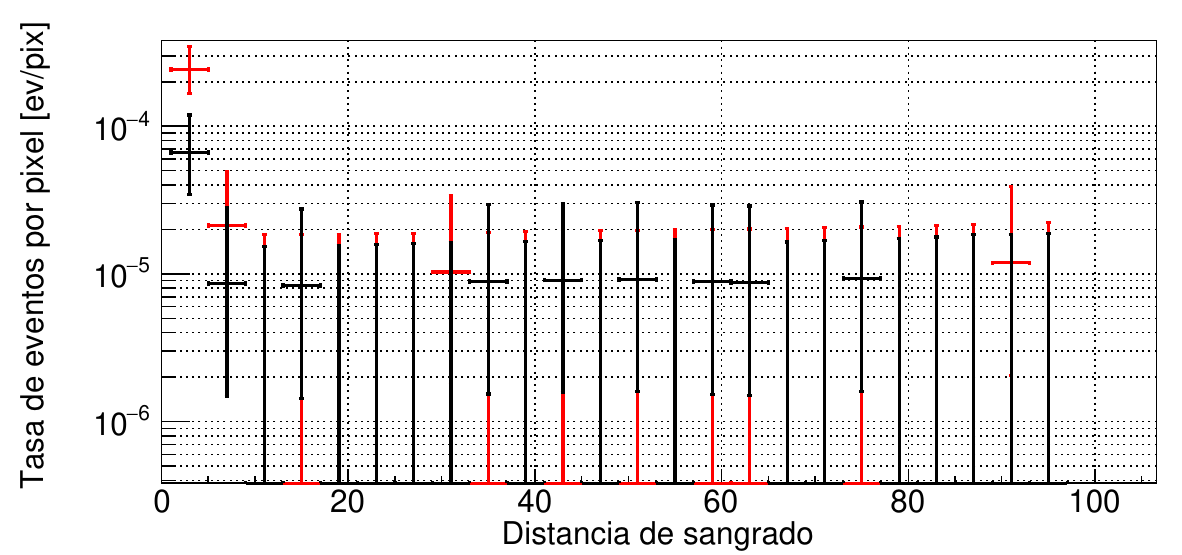}
         \label{fig.bleedXY_2e}
     \end{subfigure}
        \caption{Event rate per pixel measured within the selected bleed area for the 1-electron channel (\ref{fig.bleedXY_1e}) and 2 electrons in one pixel (\ref{fig.bleedXY_2e}). The \textit{x}-axis shows the number of pixels selected to the right (black points, bleeding in "x") or above (red points, bleeding in "y") all events with 100 or more electrons. The dataset used is the commissioning data.}
        \label{fig.bleedXY}
\end{figure}

Bleeding is the result of a non-zero CTI in the SCCD used (see the discussion in Section \ref{sec:sangrado}). This CTI generates a tail of 1 or 2-electron events to the right and above high-energy events. Therefore, it is necessary to exclude these areas from the analysis. Figure \ref{fig.bleedXY} shows the event rate per pixel measured within the selected bleeding area for both the 1-electron and 2-electron channels. Additionally, it separates pixels to the right of high-energy events (black points) and above them (red points). For example, the first black (or red) point in either of the two subfigures shows the measured rate by selecting pixels located at a distance of 1 to 5 pixels from any event with 100 electrons or more (excluding pixels neighboring a non-empty pixel). The next point is between 5 and 9, and so on.

For the 1-electron channel (Figure \ref{fig.bleedXY_1e}), both rates flatten out at around $\sim$ 55-60 pixels of distance. Additionally, they take similar values, as expected, of approximately $7 \times 10^{-4} \epix$ per image. This result, in accordance with what was measured for the high-energy halo, is expected because the pixels selected to calculate the rate will be significantly affected by this contribution. In order to discriminate between both contributions, the bleeding distance can be taken as the distance at which the red and black points start to agree, around 20-30 pixels apart, although this agreement seems to occur around 50 pixels. Conservatively, a bleeding distance of 100 pixels was chosen for the 1-electron channel. For the 2-electron channel, which loses statistics beyond 20 pixels, a bleeding distance of 50 pixels was chosen.

Additionally, twice as many pixels are excluded for the columns whose bleeding, measured in SEEs per pixel within that column, is more than 3.71 MAD (Median Absolute Deviation) from the average of all columns in that quadrant, considering the entire dataset.

\subsubsection{Horizontal Register Events}

Horizontal register events occur when a high-energy event impacts the horizontal register during readout, creating elongated events along the x-axis of the obtained images (see discussion in Section \ref{sec:SRevents}). Therefore, since they are not excluded, some of these events can be mistaken for low-electron events compatible with dark matter. Exploiting their nearly one-dimensional nature due to their diffusion along the horizontal register, a protocol was created for their exclusion.

This protocol essentially involves searching for events of 3 or more electrons in the same row with no events in the adjacent rows above and below.

Table \ref{tab:srevents} shows the events with 2 to 4 electrons, inclusive, that diffused only vertically (first column) or horizontally (second column) in the science images. In the first row, no event selection criteria are applied, while the second row shows the number of events counted after applying the horizontal register event exclusion. It is noteworthy that there is agreement between the two types of signals (purely horizontal and purely vertical events), as well as a greater reduction in horizontal events compared to vertical events, as expected.

\begin{table}[]
\centering
\begin{tabular}{@{}ccc@{}}
Is Exclusion Applied? & Vertical Events & Horizontal Events \\ \midrule
No                    & 325            & 469              \\
Yes                   & 221            & 229              \\ \bottomrule
\end{tabular}
\caption{Number of purely horizontal or vertical events measured, with and without applying the horizontal register event exclusion criterion. The event counts are obtained from the science images to increase statistics.}
\label{tab:srevents}
\end{table}

\subsubsection{Hot Columns and Pixels}

Due to the presence of defects in the silicon of the detector, certain pixels and/or columns may exhibit an anomalously excessive number of 1 and 2 electron events (see discussion in Section \ref{sec:columnasypixelescalientes}). Although this effect decreases at low temperatures, and columns known as "dark spikes" were previously selected for exclusion using another dataset at higher temperatures (see Figure \ref{fig.darkspikes} and discussion in the text), such defects can affect the signal under analysis.

\begin{figure}[h!]
     \centering
     \begin{subfigure}[c]{0.47\textwidth}
         \centering
         \caption{Two-dimensional histogram of the position of 1 electron events in the first quadrant of the 22 science images.}
         \includegraphics[width=\textwidth]{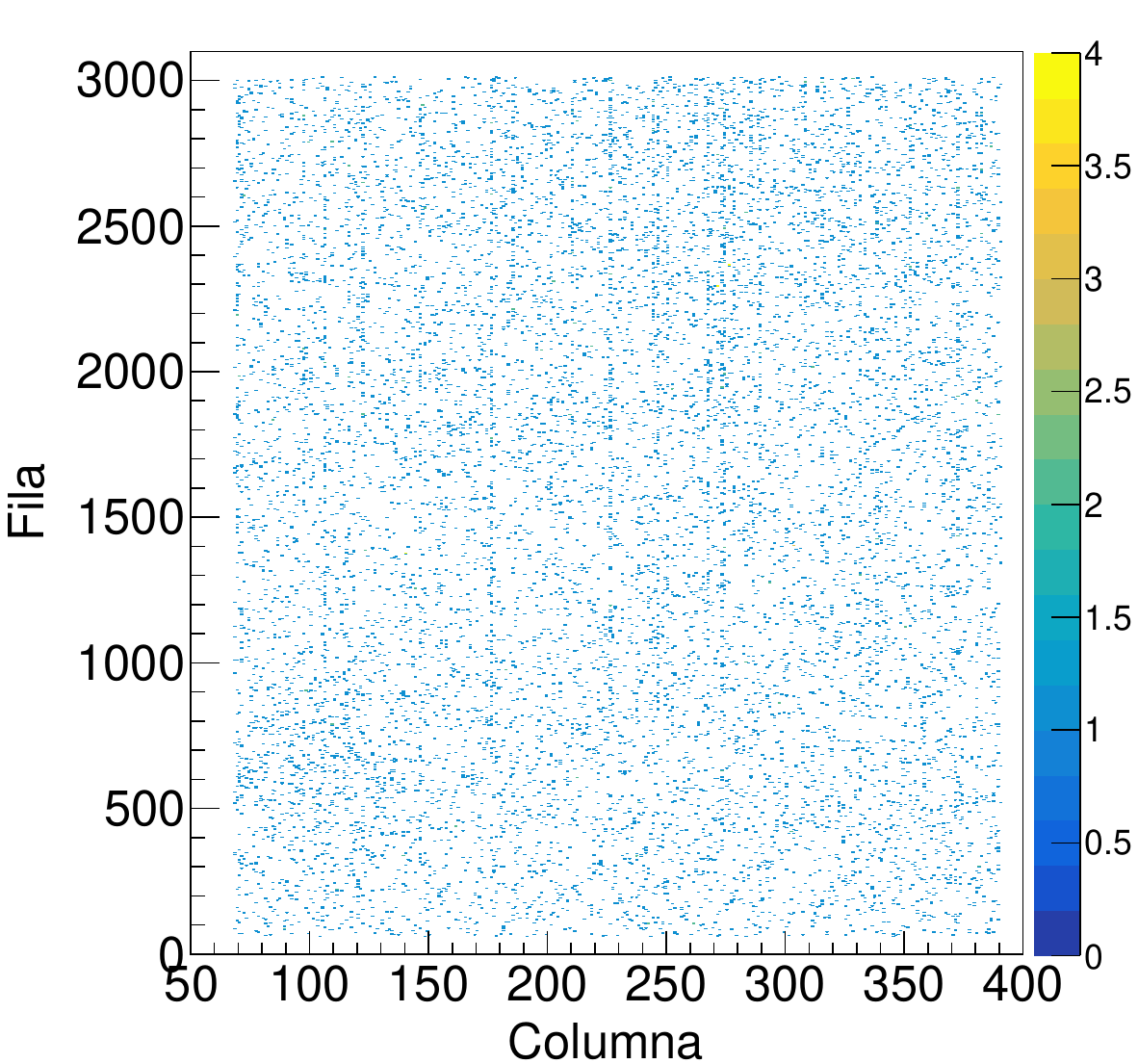}
         \label{fig.hotpix}
     \end{subfigure}
     \hfill
    \centering
     \begin{subfigure}[c]{0.47\textwidth}
         \centering
         \caption{Projection along the x-axis of \ref{fig.hotpix}.}
         \includegraphics[width=\textwidth]{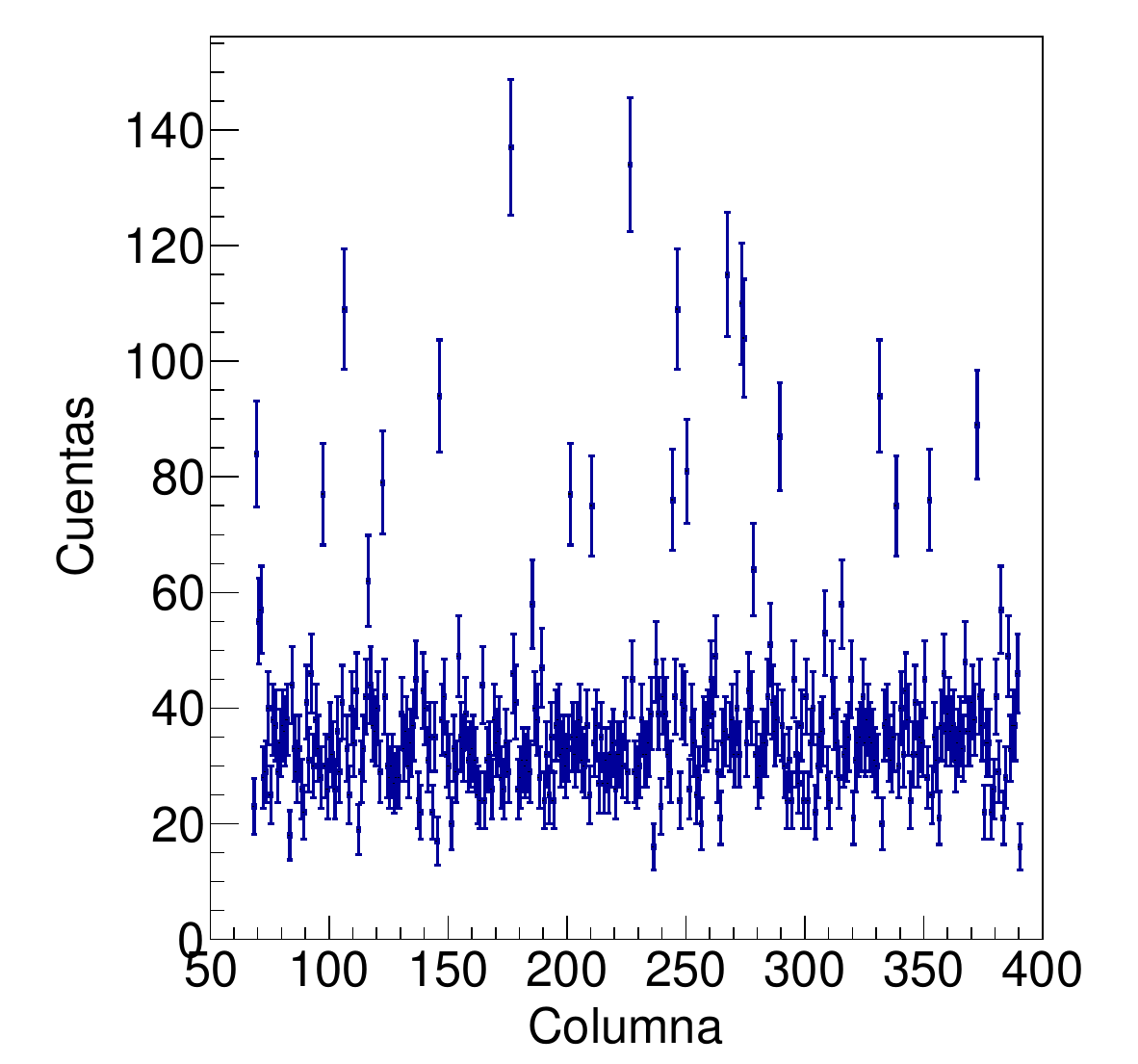}
         \label{fig.hotcols}
     \end{subfigure}
        \caption{Hot pixels and columns in the first quadrant of the 22 science images. The x-axis of both images represents the column number, while the y-axis in \ref{fig.hotpix} represents the row number, and in \ref{fig.hotcols}, the number of 1 electron events for each column.}
        \label{fig.hotpixandcols}
\end{figure}

This can be seen in Figure \ref{fig.hotpixandcols}, which illustrates the 1-electron events found in the 22 science images (after excluding pixels due to crosstalk, edge \footnote{A 20-pixel edge was used, compatible with two or more electron analyses.}, and horizontal register events mentioned earlier) in two different ways. In Figure \ref{fig.hotpix}, the image is a two-dimensional histogram showing the position in columns and rows of these events, accompanied by a color scale on the right side of the image. Figure \ref{fig.hotcols} is the projection of the previous figure along the x-axis, showing the number of 1-electron events per column.

From Figure \ref{fig.hotpixandcols}, two conclusions can be drawn. The first, which is more evident in Figure \ref{fig.hotcols}, is that there are certain columns with an excess of events. The second is that, although not visible to the naked eye, there are pixels with 1-electron events up to 4 times in just 22 images (see the color scale on the right side of Figure \ref{fig.hotpix}), indicating the need for a hot pixel cut.

\subsubsection{Hot Pixels}

The criterion consisted of classifying any pixel in the dataset of science images as a hot pixel if it exhibits at least one electron event three times or two times events totaling 3 or more electrons. However, it is possible to classify a pixel as hot when it is, in fact, a normal pixel. Assuming the null hypothesis that the pixel is normal, the probability of erroneously rejecting it (Type 1 error) is given by:

\begin{equation}
    I = \sum_{i>=3} Poisson(i|\mu^{total}_{1e})
    \label{eq.type1}
\end{equation}

Since the SEE rate with these selection criteria is approximately $5.6 \times 10^{-4} \, \text{e/pixel}$ in the commissioning images, it can be estimated for any pixel considering 17 images, $\mu_{1e}=5.6 \times 10^{-4} \, \text{e/pixel} \times 17$. Thus:

\begin{equation}
    I = \sum_{i>=3} Poisson(i|0.0109) = 2.13\times 10^{-7}
    \label{eq.type1bis}
\end{equation}

Considering that the first quadrant for the 22 images (and the selection criteria used) has a total of approximately $3.21\times 10^{7}$ pixels, i.e., $1.13\times 10^{6}$ pixels per image and quadrant, the estimated number of pixels classified erroneously (i.e., normal pixels classified as hot) is 0.16. A similar value is found for quadrants 2 and 3, where hot pixels were sought. When running the algorithm on the science images, it found 10 hot pixels, meaning that each pixel classified as hot has a probability close to 5$\%$ of being classified as hot when it is not. Similarly, quadrants two and three are in a similar position (15 and 18 hot pixels, respectively).

\subsubsection{Hot Columns}

Following the illustration in Figure \ref{fig.hotcols}, a criterion is set to exclude columns from the dataset that show an excess of events due to defects in the detector. Similar to what was developed for hot pixels, this criterion will be related to the SEE rate in each column, removing columns whose rate is double (or more) the average rate in all pixels.

The same masks as those provided for the classification of hot pixels are used, additionally adding the bleeding mask. Thus, the number of unmasked pixels in the 22 images is approximately $1.96\times 10^{7}$, about 26 thousand pixels per column.

When estimating the SEE rate at $4.7 \times 10^{-4} \, \text{e/pixel}$, it is expected that there are approximately 24.3 events for each column. Therefore, the probability of classifying a column as hot when it is not can be estimated as:

\begin{equation}
    I = \sum_{i>=24.3} Poisson(2 \times i|24.3) = 0.003
    \label{eq.type1_columna}
\end{equation}

Habiendo 403 columnas disponibles (20 fueron descartadas por el corte de borde), la cantidad de columnas que serán catalogadas erróneamente es aproximadamente 1 y, habiendo encontrado 10 columnas calientes para cada uno de los cuadrantes, se puede estimar que el 10$\%$ de las columnas clasificadas como calientes no lo son.

\subsubsection{Single-pixel Events}

This criterion was discussed in depth in Section \ref{sec:monopixelybajaenergia}.
To understand if this cut generates bias in the analysis of single-electron events, 20 empty images were simulated for a quadrant of the same dimensions as the CCD described in this Section, and single-electron events were injected with a rate of $1.0 \times 10^{-3} \, \text{e/pixel}$/image. Then, the SEE rate was calculated for both cases: before and after applying the cut, and it was found that without the cut, the rate is $(1.0037\pm0.0058) \times 10^{-3} \, \text{e/pixel}$, and with the cut, it is $(1.0046\pm0.0058) \times 10^{-3} \, \text{e/pixel}$. This concludes that the cut does not introduce any bias in the estimation of the SEE rate, as expected.

Regarding the multielectron analysis, this cut is useful for the analysis of two-electron events presented in this Section since single-pixel two-electron events will be used for exclusion limit calculations (see Section \ref{sec:2020conteomulti}).

\subsubsection{Low-Energy Cluster}

Due to the existence of clusters of 2 to 100 electrons whose proximity was incompatible with the dark matter signal under investigation, the use of this cut was established. In this subsection, the parameters used for each of the energy channels to be analyzed will be specified, potential sources of bias introduced by the cut, and how they were treated.

\begin{figure}[h!]
    \centering
    \includegraphics[width=0.7\textwidth]{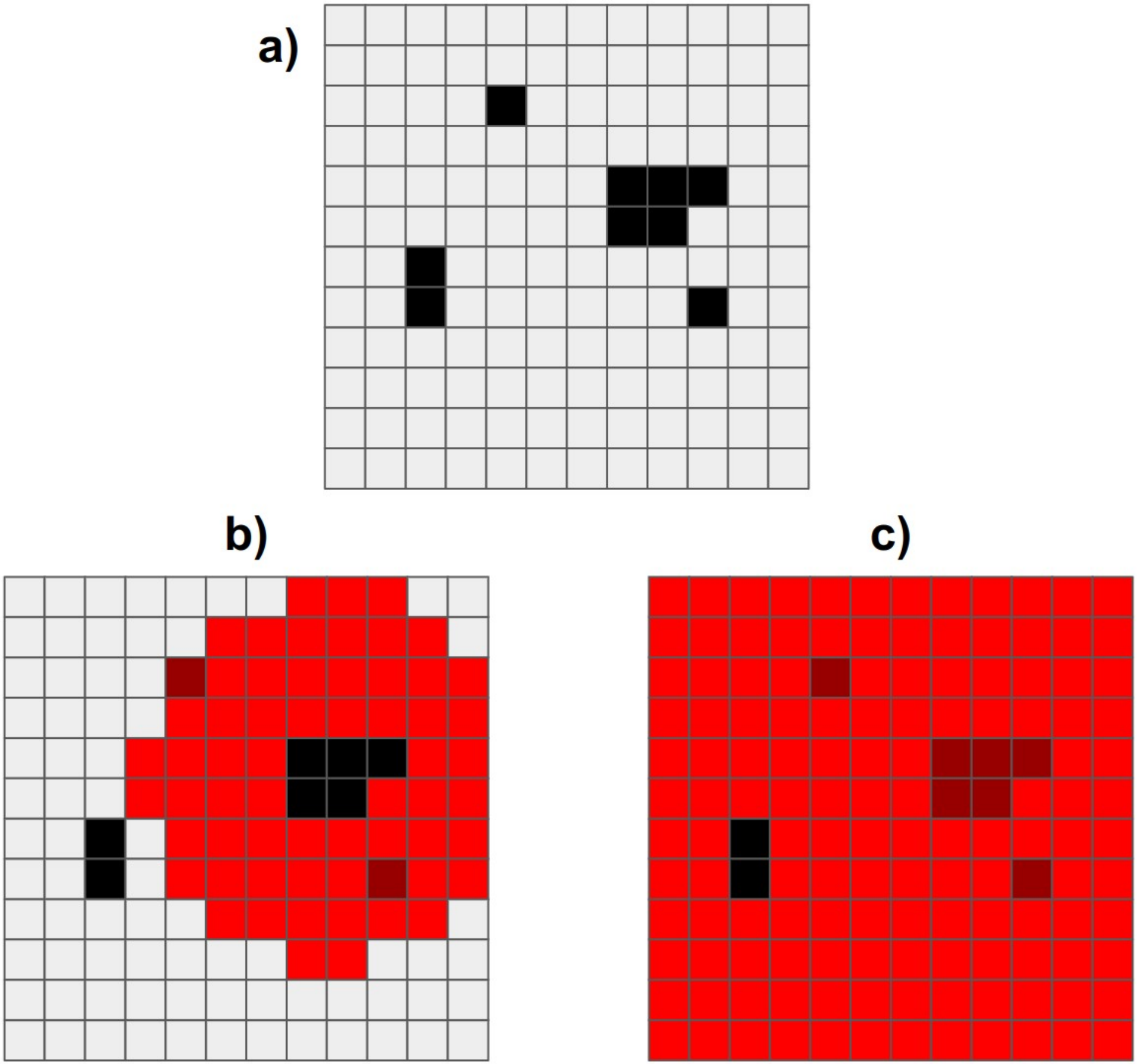}
    \caption{Illustrative diagram of the Low-Energy Cluster. (a) shows 4 events (two single-electron events, one two-electron event, and one five-electron event) in a 12$\times$12 pixel image. (b) shows the cut for the 1 and 2 electron channels (i.e., a radius of 4 pixels), while (c) shows it for the 3 and 4 electron channels (a radius of 20 pixels). In red (burgundy), the empty (non-empty) pixels that are excluded from the analysis.}
    \label{fig.lowecluster}
\end{figure}

For the analysis of 1 and 2 electrons, 4 pixels around all the pixels that make up a cluster of at least 5 electrons are discarded. This cut overlaps with the high-energy halo cut when any of the pixels in these clusters has at least 100 electrons. For the 1-electron event channel, it is not expected that this cut will introduce any bias, as there are only 4 events between 4 and 100 electrons in the commissioning data images and approximately 2000 single-electron events after applying a series of cuts\footnote{The applied cuts include all those mentioned so far, except for the single-pixel event: bleeding, high-energy halo, crosstalk, edge, and hot pixels and columns.}. In fact, the SEE rate measured in the first quadrant for the commissioning data images before and after applying the Low-Energy Cluster cut changes from $(3.792\pm0.085) \times 10^{-4} \epix$ to $(3.795\pm0.085) \times 10^{-4} \epix$, which allows us to conclude that this cut does not introduce any bias for the 1-electron channel.

For the 2-electron channel, a similar rationale applies as in the 1-electron channel but with a much lower occupancy. In fact, for the 7 commissioning data images, 0 events with 2 electrons in a single pixel were recorded, making it impossible to perform an analytical check as done for the 1-electron channel. However, since we expect a uniform interaction of dark matter with the detector, there is no reason to believe that excluding areas around events with 5 or more electrons would affect the signal we want to measure in the 2-electron channel. If there were a spatial correlation between events with 5 or more electrons and events with 2 electrons (for example, a dark matter event could deposit 7 electrons and diffuse into one with 2 and another with 5, within 4 pixels of distance), this criterion could introduce a bias for channels with 7 or more electrons~\footnote{It is worth noting that our analysis is not sensitive to such a signal, as according to our clustering algorithm, we would see one event with 5 electrons and another with 2. Additionally, our analysis is restricted to energy channels of 4 electrons or lower.}, but not for the 2-electron channel (nor, by the way, for the 5-electron channel). However, it is plausible to think that any background event with 7 or more electrons may diffuse into one with 5 or more (or multiple clusters with lower multiplicity) and another with 2. The goal of the cut is then to discard the areas around such events, as they would affect our dark matter signal measurement, specifically for the 2-electron channel, without the risk of introducing a bias to the signal of interest in that channel. It is worth noting that the same reasoning applies to the 1-electron channel, thinking of events with 6 or more electrons as seeds.

The radius for both (3 and 4 electrons) was set to 4 pixels to avoid losing exposure, although it could be larger. Additionally, it's important to note that the cut itself does not eliminate the seed cluster that generates the halo, but rather the pixels around it.

For the analysis of 3 and 4 electrons, 20 pixels are discarded around all the pixels that make up a cluster of at least 2 electrons. In the same way as for the 2-electron channel, there is no reason to think that a candidate depositing 3 (4) electrons in our detector is spatially correlated with one of 2 or more electrons. Again, this criterion could introduce bias if used for channels with higher electron multiplicities (in this case, 5 (6) electrons or more, which are not part of our analysis either).

The radius was set to 20 pixels to match the high-energy halo, and the seed clusters were not discarded.

\subsubsection{\textit{Loose cluster}}

The Loose cluster criterion will exclude areas in the images where two 1-electron events are within 20 pixels of each other. The excluded area will be defined by a circle with a radius of 20 pixels around these 1-electron events. Additionally, the 1-electron events that generate these areas (seed events) can only be used as seeds if they have not been previously excluded by the masks for bleeding, crosstalk, noise, horizontal register events, Low-Energy Cluster, and hot columns and pixels.

It's worth noting that for the 2-electron channel, the selection criterion for 1-electron seed events also includes 1-electron events that are part of clusters with higher electron multiplicities. This includes clusters of 2 or more electrons that have at least two 1-electron events. Therefore, this criterion overlaps with the Low-Energy Cluster cut for clusters of 5 or more electrons that have at least two 1-electron events~\footnote{In fact, it exceeds it, as it creates halos of 20 pixels around the clusters compared to the 4-pixel halos created by the Low-Energy Cluster cut}.

The reasoning to understand that this cut does not introduce bias to the signal under investigation is similar to what was detailed for the low-energy cluster in the previous section. There is no reason to believe that an event with 2 (3 or 4) electrons is spatially correlated with an event with 1 electron except by chance. It is identical for the event rate of 2 (3 or 4) electrons to discard the region created by two events with 1 electron close together as any other randomly located region in the detector.

This was verified by simulating 1000 images of a quadrant, with dimensions identical to those of a quadrant of the SENSEI-SCCD, with a SEE rate of $4 \times 10^{-4} \epix$. Due to the clustering of events with 1 electron in a single pixel, there are a significant number of single-pixel 2-electron events in these images, but they do not appreciably affect the total number of 1-electron events (for every 5000 1-electron events, we have one single-pixel 2-electron event, according to Equation \eqref{eq.poisson1e}). The result can be compared with and without a mask. Under the assumption that these 2-electron events are uniformly distributed, which is expected since they were generated from randomly distributed 1-electron events, it can be checked whether the mask used introduces bias to the multi-electron channels.

The rate of single-pixel 2-electron events changes from $(7.52\pm0.69) \times 10^{-9} \epix$ to $(7.54\pm0.74) \times 10^{-9} \epix$ when the mask is applied, indicating that the mask does not introduce any bias to a Poisson-type 2-electron signal.

It is crucial to emphasize that this cut is not applied to the 1-electron channel as it would clearly introduce a bias by excluding regions where, by chance, two 1-electron events happened to be neighboring. In particular, for the simulated images, the SEE rate per pixel changes from $(3.995\pm0.011) \times 10^{-4} \epix$ to $(2.795\pm0.012) \times 10^{-4} \epix$ after applying the mask.

For the analysis of 3 and 4 electrons, the same reasoning as for the 2-electron channel is followed, but only using single-pixel 1-electron events as seeds.

\renewcommand{\arraystretch}{1.7}
\begin{longtable}{lp{8cm}}
\centering
\textbf{Name}                          & \textbf{Description}                                                                                                                                          \\ 
\midrule
\textbf{Noise}                           & Exclude images in which the readout noise is 30$\%$ higher than expected, as inferred from the overscan of each image.                          \\
\textbf{\textit{Crosstalk}}              & Exclude pixels with a charge greater than 700 $\e$ (equivalent to 0.1 electrons produced by \textit{crosstalk}). Additionally, we exclude pixels in the rest of the quadrants that have been read synchronously with that pixel.                                           \\ 
\textbf{Single-Pixel Events}               & Exclude pixels whose neighbors are non-empty for the analysis of 1 and 2 electrons.                \\
\textbf{Bleeding}                        & Exclude 100 (50) pixels to the right and above pixels with 100 or more electron charges for the analysis of 1 (2 or more) electron(s). Additionally, we exclude twice the number of pixels for columns whose bleeding, measured in SEE per pixel within that column, is more than 3.71 MAD from the average of all columns in that quadrant, summing over the entire dataset.           \\
\textbf{High-Energy Halo}            & Exclude 60 (20) pixels around any pixel with 100 or more electrons for the analysis of the 1-electron (2 or more electrons) channel.          \\
\textbf{Edge}                           & Exclude 60 (20) pixels around the edge of the active area for the analysis of the 1-electron (2 or more electrons) channel.  \\
\textbf{Horizontal Register Events} & Eclude rows that have groups of five consecutive pixels with at least four non-empty pixels, and their adjacent pixels (in the rows above and below) are empty.       \\ 
\textbf{Low-Energy Cluster}         & Exclude 4 (20 pixels) around any single or multi-pixel event with at least 5 (2) electrons for the analysis of 1 and 2 (3 and 4) electrons channels. The event itself is not removed.            \\ 
\textbf{Hot Pixels}               & Exclude pixels that have presented (summed over the entire dataset) a single-pixel event of 3 electrons and/or two single-pixel events totaling at least 3 electrons \footnote{The set of pixels classified as hot remains after applying the cuts for \textit{crosstalk}, horizontal register events, low-energy cluster, and a 20-pixel edge mask. This is because pixels from these contributions can be confused with hot pixels.}. \\
\textbf{Hot Columns}              & Exclude columns whose SEE rate per pixel (averaged over the entire dataset) is double the average rate in their respective quadrant and/or any column that has at least two single-pixel events of 2 electrons \footnote{The criteria used are the same as for hot pixels, adding bleeding.}. \\
\textbf{\textit{Loose Cluster}}              & Exclude 20 pixels around a 1-electron event only if another 1-electron event is found within that region \footnote{Events with 1 electron that are not affected by the following masks are considered seeds: bleeding, \textit{crosstalk}, noise, horizontal register events, low-energy cluster, and hot columns and pixels.}. This criterion is only applied for the analysis of 2 or more electrons. For the analysis of 3 and 4 electrons, only single-pixel 1-electron events are used as seeds. \\
\bottomrule
\caption{Glossary of quality criteria to be used in science images.}
\label{tab:cortes2020}\\

\end{longtable}

\subsection{Spurious Charge and Dark Current}
\label{sec:2020dcandfriends}

In this section, we discuss the contribution of spurious charge and dark current for the \textit{SENSEI-SCCD}, as well as the counting of 1-electron background events used when setting the limits for dark matter exclusion. For more information on the various contributions, please refer to Chapter \ref{cap:4}.

\subsubsection{Spurious Charge}

Spurious charge is the result of the generation of Single Event Effects (SEEs) due to changes in voltage when transferring charges from one pixel to an adjacent pixel during readout. Since this contribution does not scale with exposure time, as expected for a signal compatible with dark matter, it acts as a background to the signal of interest under study. Efforts to mitigate this contribution were detailed in Section \ref{sec:ctivssc}, achieving a reduction of this signal to an unprecedented value on the order of approximately $2 \times 10^{-4} \epix$. However, this value is approximately 50\% of the SEE signal measured in the calibration and science images analyzed in this chapter.

\begin{figure}[h!]
    \centering
    \includegraphics[width=0.7\textwidth]{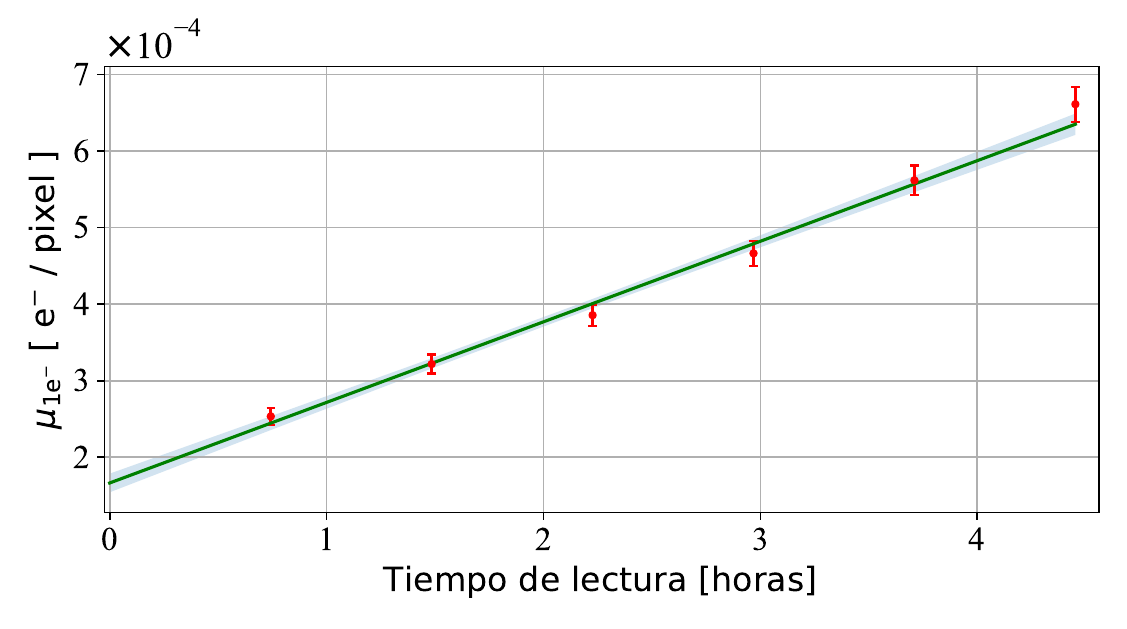}
    \caption{Rate of SEEs per pixel observed as a function of readout time (red dots). The green line represents the fitted curve, and the shaded region corresponds to the 1 $\sigma$ confidence interval.}
    \label{fig.SC_MINOS2020}
\end{figure}

To estimate the contribution of spurious charge to the measured signal, a series of images were taken following the procedure detailed for the determination of $\lambda_{\rm AL}$ and $\mu_{\rm SC}$ in Section \ref{sec:resultados2022}. Figure \ref{fig.SC_MINOS2020} shows the result of this procedure, for which a spurious charge contribution of $\mu_{\rm SC}=(1.66 \pm 0.12)\times 10^{-4} \epix$ was estimated.

\subsubsection{Dark Current}
\label{sec:2020dc}

We have defined dark current as any contribution to Single Event Effects (SEEs) that scales linearly with time and has a spatially uniform charge distribution on the detector surface (see Section \ref{sec:dc}). Therefore, it could be the result of different mechanisms of SEE production, in addition to the typically studied thermal agitation, as long as they exhibit these properties of spatial uniformity and temporal linearity. Furthermore, in the publication from 2020 \cite{SENSEI2020}, we reported a dark current significantly higher than the theoretically motivated one, and no explanation has been found for this discrepancy, despite it being the lowest ever reported in the history of silicon semiconductor detectors.

The objective of this thesis and the publications stemming from it was not to find an interpretation for this discrepancy but rather to mitigate it using the event selection criteria discussed in Section \ref{sec:2020criterio}. Additionally, an additional lead layer was added outside the vacuum chamber and around the detector (see Chapter \ref{cap:3}). This not only reduced the background of high-energy events, allowing for an increase in the exposure of the measured images but also caused a three-fold reduction in the measured dark current.

\begin{figure}[h!]
    \centering
    \begin{subfigure}[c]{0.75\textwidth}
        \centering
        \caption{}
        \includegraphics[width=\textwidth]{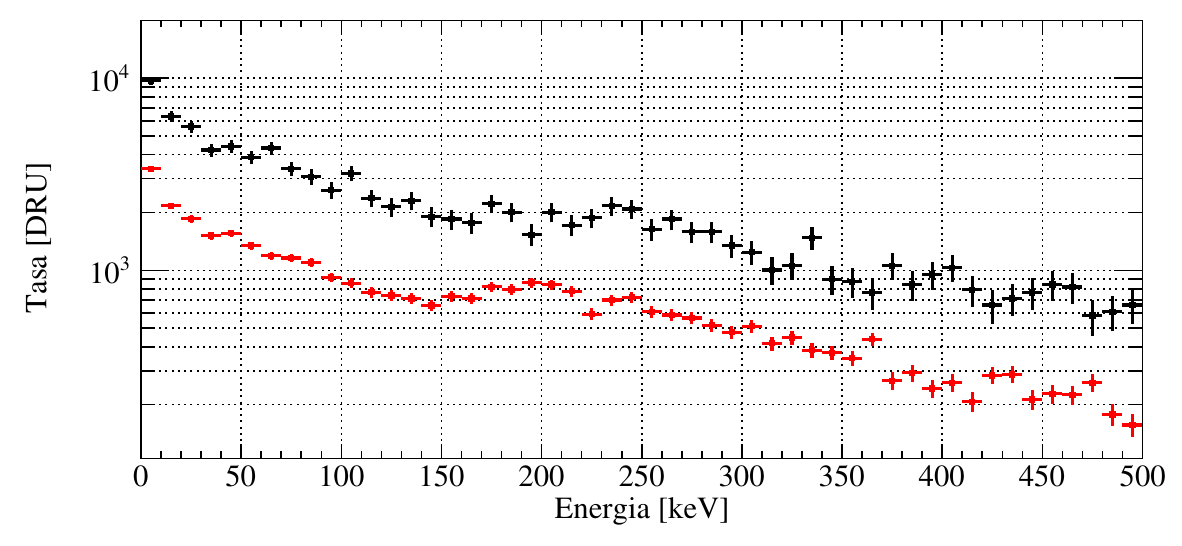}
        \label{fig.HEspectra}
    \end{subfigure}
    \hfill
    \centering
    \begin{subfigure}[c]{0.75\textwidth}
        \centering
        \caption{}
        \includegraphics[width=\textwidth]{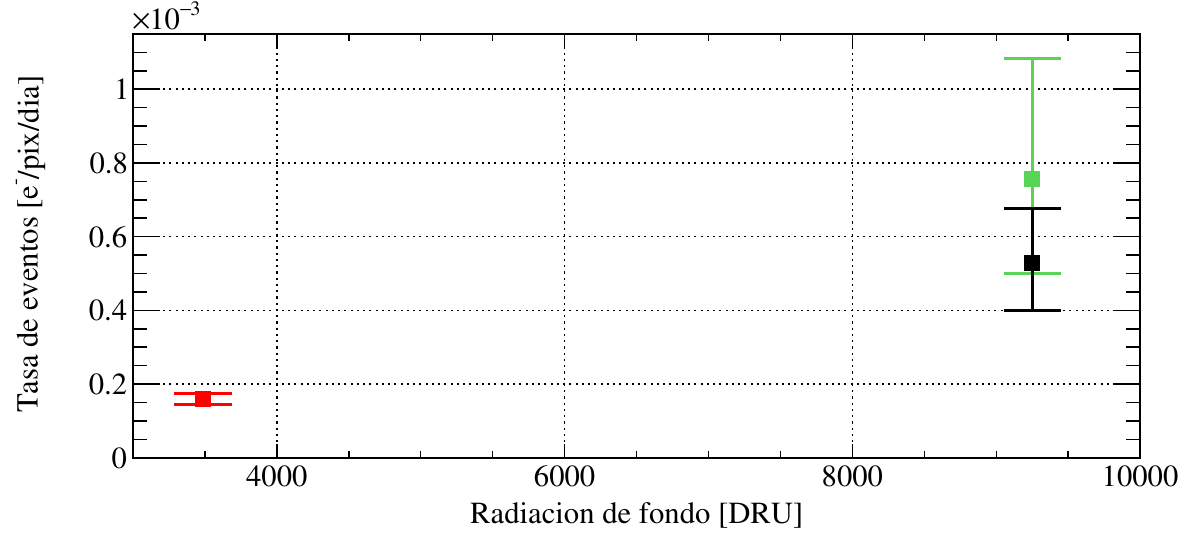}
        \label{fig.Background_in}
    \end{subfigure}
    \caption{Reduction in high-energy background and its effect on dark current. Figure \ref{fig.HEspectra} shows the high-energy spectrum from 10~keV to 1~MeV measured with (red points) and without (black points) the additional lead shielding outside the vacuum chamber. A reduction in counts is observed for all energies. Additionally, Figure \ref{fig.Background_in} shows the decrease in SEE rate measured in $\epixdia$ with (red) and without the additional lead shielding (black and green, depending on the dataset used, as explained in the text).}
    \label{fig.1eratevsbackground}
\end{figure}

The effect of adding the additional lead layer on the measured images is shown in Figure \ref{fig.1eratevsbackground}. On one hand, Figure \ref{fig.HEspectra} displays the reduction in the count of high-energy events with and without the additional lead shielding. The reduction factor varies with energy, so the reference point is taken as the first bin in the plot, which encompasses events with energies between 500~eV and 10~keV. The event rate in this bin, measured in DRU (1~DRU is 1~event/kg/day/keV), decreases from 9700 to 3370 DRU.

On the other hand, Figure \ref{fig.Background_in} illustrates the decrease in the measured dark current with (red points) and without (green and black points) additional lead shielding. The reason there are two different points in the figure published in 2020 \cite{SENSEI2020} for the configuration without additional lead is that the green point represents the estimation of the dark current obtained with an image taken with the amplifier turned off during the exposure, while the black point is the combination of the previous result and three other images taken with the amplifier turned on during the exposure (while keeping the same conditions in other parameters as the science and calibration data). It should also be noted that the red point comes from directly extracting the dark current from the science images, as explained below.

To obtain the value of the dark current measured in the science images (particularly in quadrants 1 and 2, which were used for the counting of 1-electron events), the SEE rate in these images was calculated. After applying the event selection criteria mentioned and performing the fitting introduced in Section \ref{sec:calibration}, the spurious charge's SEE rate was subtracted from this rate. The resulting rate was then divided by the exposure time of the science images, which is 20 hours. This estimation leads to a dark current $\lambda_{\rm DC}=(1.59 \pm 0.16)\times 10^{-4} \epixdia$, the lowest ever recorded in silicon semiconductors to date, indicated by the red point in Figure \ref{fig.Background_in}. On the other hand, the black point represents a dark current of $(5.3^{+1.5}_{-1.3})\times 10^{-4} \epixdia$, slightly more than three times higher.

Currently, studies are underway to determine the relationship between the high-energy event rate and the SEE rate measured in SCCDs.

\subsection{Event Counting}
\label{sec:2020conteo}

Once the event selection criteria were established, and the SEE rate due to spurious charge $\mu_{\rm SC}$ and the total SEE event rate in the science images were obtained, event counting was carried out for each of the channels used in the dark matter exclusion limits.

\subsubsection{Single-Electron Events}
\label{sec:2020conteo1e}

\begin{figure}[h!]
    \centering
    \includegraphics[width=0.75\textwidth]{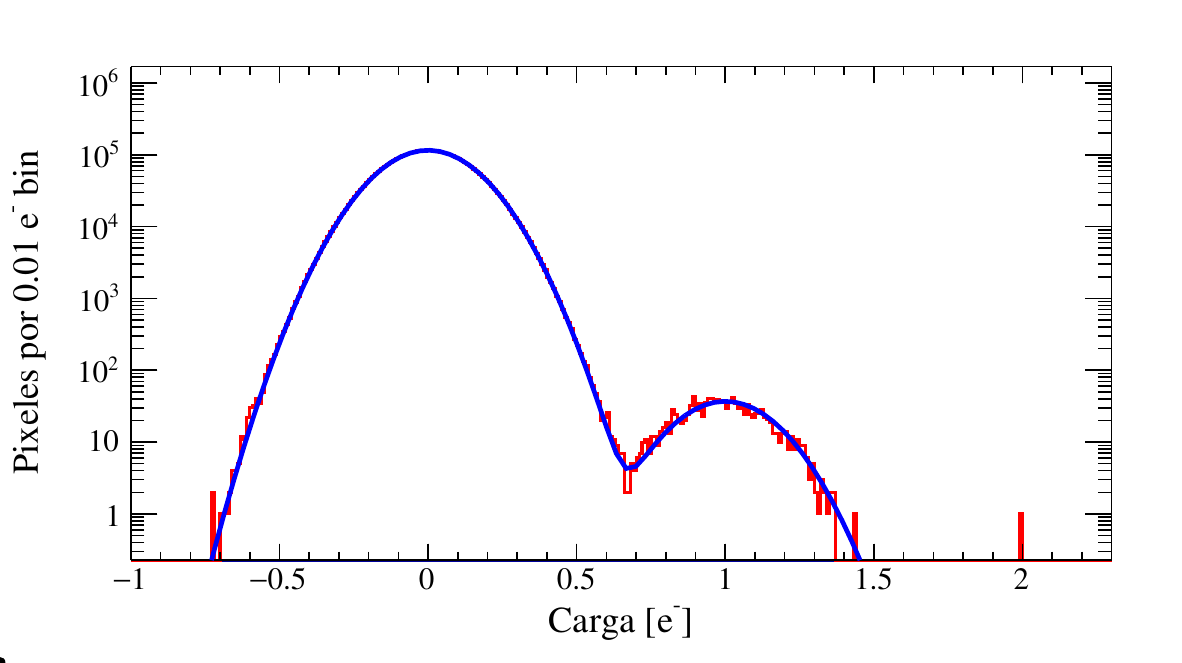}
    \caption{Charge histogram for the unmasked pixels after applying event selection criteria for the 1-electron channel, shown in red. The blue curve represents the fit of two Gaussian distributions convoluted by a Poisson distribution to the peaks corresponding to 0 and 1 electrons.}
    \label{fig.spectrum_mod}
\end{figure}

To measure the quantity of 1-electron events, the unmasked pixels were extracted, and a fit of two Gaussian distributions convoluted by a Poisson distribution, as mentioned in Section \ref{sec:calibration}, was performed. This fit is shown in Figure \ref{fig.spectrum_mod}, resulting in a SEE rate of $(3.363\pm0.095)\times 10^{-4} \epix$. There are a total of 3,900,318 pixels in the unmasked region, with 1,311.7 1-electron events. It is important to note that this value still contains an uncorrected background contribution, which we have separately characterized: spurious charge.

The upper limit was calculated as the difference between the total number of events (1311.7) and the events originating from the estimated spurious charge (649), but subtracting 2$\sigma$ from this latter component, where $\sigma$ is the statistical error associated with the spurious charge estimation (with an absolute error of the rate being $0.122\times 10^{-4} \epix$, $\sigma$ is estimated at 47.5 events). This procedure yields an upper limit estimate of 749 events compatible with dark matter-induced SEEs.

It is worth noting that this technique was chosen to be conservative and include the statistical error of these estimations, which is around 10$\%$. Section \ref{sec:2020exposición} will provide details on obtaining the exposure for this channel and those with higher electron multiplicities to determine the final SEE rate obtained.

\subsubsection{Multi-Electron Events}
\label{sec:2020conteomulti}

Counting multi-electron events is done differently than SEE events, primarily due to their rare occurrence and consequently low statistics. The clustering process plays a crucial role in this, as it groups adjacent pixels into events of a given energy. This process is detailed in Section \ref{sec:clusterizacion}, where it is specified that the clustering thresholds were set at 0.63, 1.63, and 2.5, with increments of 1 electron for higher multiplicities, for the images analyzed in this thesis.

Counting is straightforward, involving applying the established event selection criteria and counting how many events with 2, 3, and 4 electrons are present in the images. Table \ref{tab:SENSEI2020results} summarizes the events for these channels, considering all continuous configurations for 3 and 4 electrons, and only single-pixel events for 2-electron events.

It was decided to discard the multi-pixel configurations from the 2-electron analysis because the probability of finding such events originating from the accidental coincidence of SEEs was high, and there was no formalism available for handling this background.

\subsection{Exposure Calculation}
\label{sec:2020exposición}

This section discusses the calculation of exposure for each channel, i.e., how long our detector was exposed to dark matter candidates that deposit 1, 2, 3, or 4 electrons after an interaction. To perform this calculation, it is necessary to consider the concept of detection efficiency and the types of efficiency that exist.

We will refer to efficiency as the efficiency that our detector has for detecting signals deposited in each of the channels of interest. As an example, the 1-electron channel, after applying all event selection criteria, has an exposure of 1.38 gram-days when the total exposure time of the detector is known to be 19.93 gram-days. To calculate this latter value, the following quantities are needed: the number of images (22), the number of pixels per image (for the 1-electron case, two quadrants are 2721792 pixels), the mass of each pixel ($3.53716875 \times 10^{-7}$ grams), and the average exposure time per image (22.58 hours or 0.94075 days). However, this exposure is reduced to 1.38 gram-days when applying the aforementioned quality criteria and counting, pixel by pixel, the unique exposure of each pixel in the fiducial region. We will call this efficiency \textit{cut-based efficiency} since, as mentioned, it affects the total exposure for a specific channel after applying event selection criteria or cuts. For the 1-electron channel, this efficiency is very low, 0.069, although this was expected due to the aggressiveness of the cuts used, particularly the high-energy halo.

\subsubsection{Diffusion Efficiency}
\label{sec:2020eficienciapordifusion}

Diffusion efficiency is the efficiency that arises because the electrons generated after a dark matter interaction may or may not diffuse to neighboring pixels. In the case that a given interaction generates more than one cluster, our reconstruction algorithm will not recognize it as a single event but rather as two or more clusters of lower energy, resulting in some loss of efficiency. The concept of charge diffusion was discussed in Section \ref{sec:diffusion}, and in Section \ref{sec:monopixelybajaenergia}, when introducing the single-pixel events cut.

\begin{figure}[h!]
    \centering
    \includegraphics[width=0.7\textwidth]{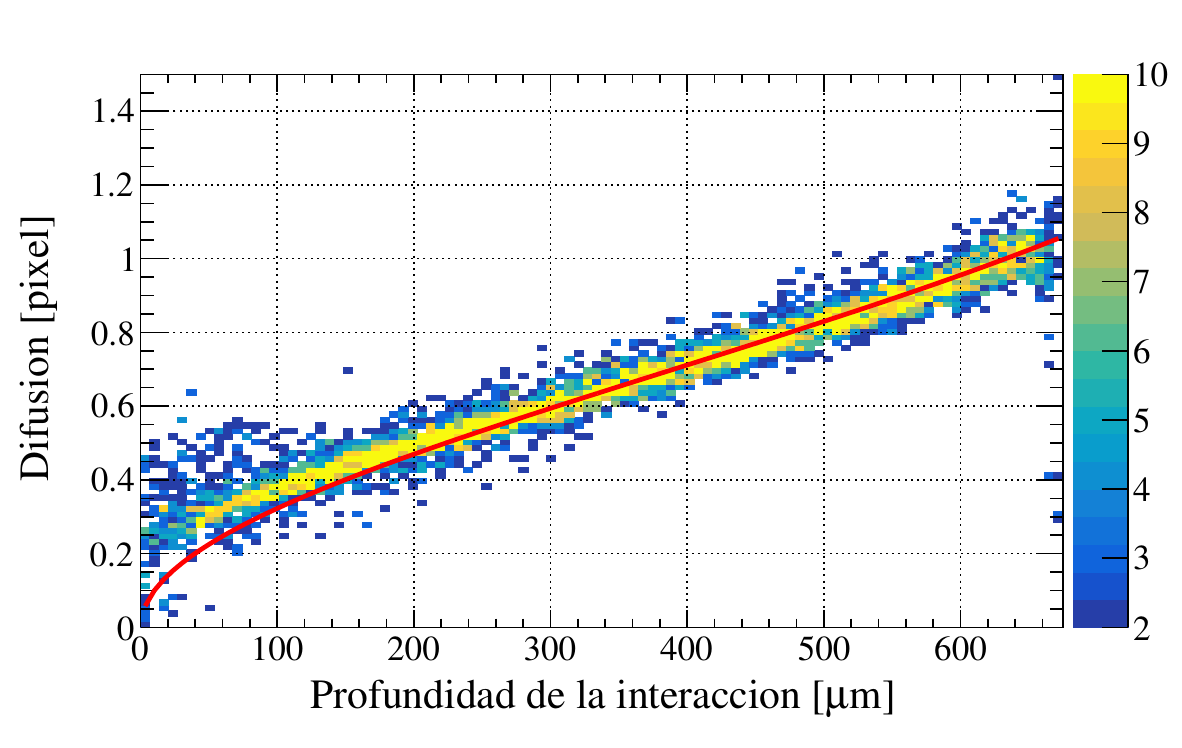}
    \caption{Fit performed using Equation \eqref{eq.diffusion} on the pixel spread of 85 muons extracted from the science images. The color scale shows the number of muons per bin for each interaction depth level. The red line shows the fitted curve. Figure extracted from \cite{SENSEI2020}.}
    \label{fig.SENSEI2020difusion}
\end{figure}

The charges generated in the CCD will diffuse towards the surface due to the applied electric potential, following a two-dimensional Gaussian distribution parallel to the detector's surface. The dispersion of this Gaussian is described by Equation \eqref{eq.diffusion}:

\[
\sigma_{xy}=-A \ln |1-bz|
\]

Where A and b are two parameters to estimate. For this data analysis, these parameters were estimated using muons that pass through the CCD completely, allowing data extraction for all depths. In Figure \ref{fig.SENSEI2020difusion}, the result of overlaying 85 muons and extracting the charge dispersion for them can be seen. The \textit{y}-axis shows the dispersion in pixels, while the \textit{x}-axis shows the estimated depth at which the diffused charges were generated. In red, a fit is performed using Equation \eqref{eq.diffusion}, which fits satisfactorily (by visual inspection) between 300 and 675 $\mu$m and begins to deviate for events that diffuse very close to the surface due to pixel quantization of the dispersion. The fit provides the following parameter estimates: $A=218.715$~$\mu$m$^2$ and $b=1.015\times 10^{-3}/\mu$m.

Using these parameters and Equation \eqref{eq.diffusion}, the diffusion efficiency can be reconstructed for different energies through a Monte Carlo simulation. The code that performs this simulation was co-written in C/C++ with Dr. Guillermo Fernández Moroni. It simulates the interaction position in depth, length, and width (uniform for dark matter) and then uses Equation \eqref{eq.diffusion} to simulate diffusion in the xy plane. The events are saved in a ROOT catalog, similar to the one described in Section \ref{sec:clusterizacion}.

\begin{figure}[h!]
    \centering
    \includegraphics[width=0.7\textwidth]{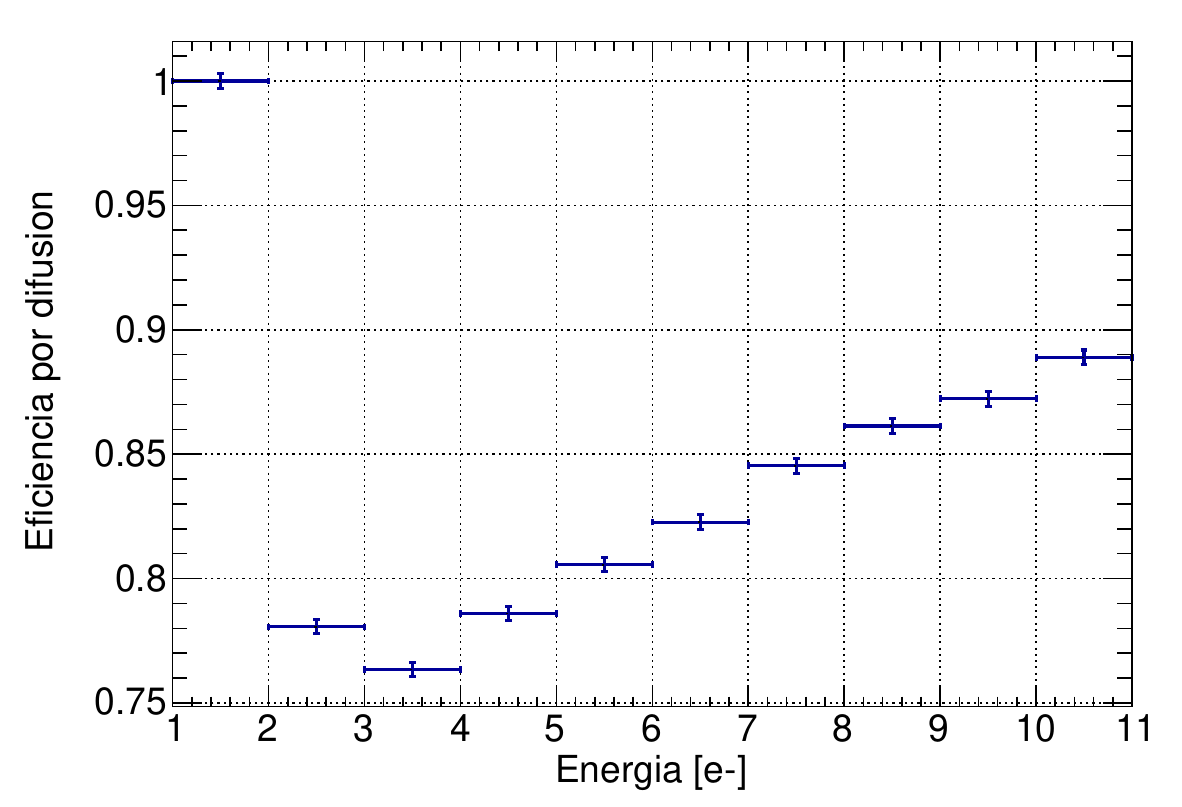}
    \caption{Histogram of the diffusion efficiency obtained using the code described in the text and the parameters obtained from the fit in Figure \ref{fig.SENSEI2020difusion}.}
    \label{fig.diffusion3}
\end{figure}

In this way, it is possible to reconstruct the probability that a dark matter event producing \textit{n} electrons ends up in ${\rm i<=n}$ pixels. Figure \ref{fig.diffusion3} shows the probability that events from 1 to 10 electrons generate continuous clusters, meaning that the pixels that make up these clusters are connected by their vertices or sides. This efficiency monotonically increases from 3 electrons as the energy of the dark matter candidate increases, due to the increasing number of possible configurations becoming available, and remains above 75$\%$ for energies equal to or less than 4 electrons. Additionally, it can be seen trivially that the efficiency for the 1-electron channel is 100$\%$.

Once the diffusion efficiencies have been obtained for each channel (1, 0.228, 0.761, and 0.778, respectively), this number is multiplied by the total exposure to obtain the efficiency-corrected exposure. It should be noted that for the 2-electron channel, the efficiency is 0.228 since only single-pixel events will be used for the analysis of this channel.

\subsubsection{Geometric Efficiency}
\label{sec:2020eficienciageometrica}

The third efficiency, geometric efficiency, results from the different geometric shapes that clusters present for different energies. The simplest case to understand is to consider a SEE and a 4-electron event with 4 pixels in a diagonal arrangement. It is clear that if the SEE is positioned in each of the unmasked pixels, the geometric efficiency will be 100$\%$, as the event will never encounter a masked pixel, by construction. However, the 4-pixel diagonal event will have situations in which at least one of its pixels is unmasked while the other three (or one or two) are masked. We call this geometric efficiency, an efficiency that decreases with energy as the generated cluster configurations become more diverse.

\begin{figure}[h!]
    \centering
    \includegraphics[width=0.7\textwidth]{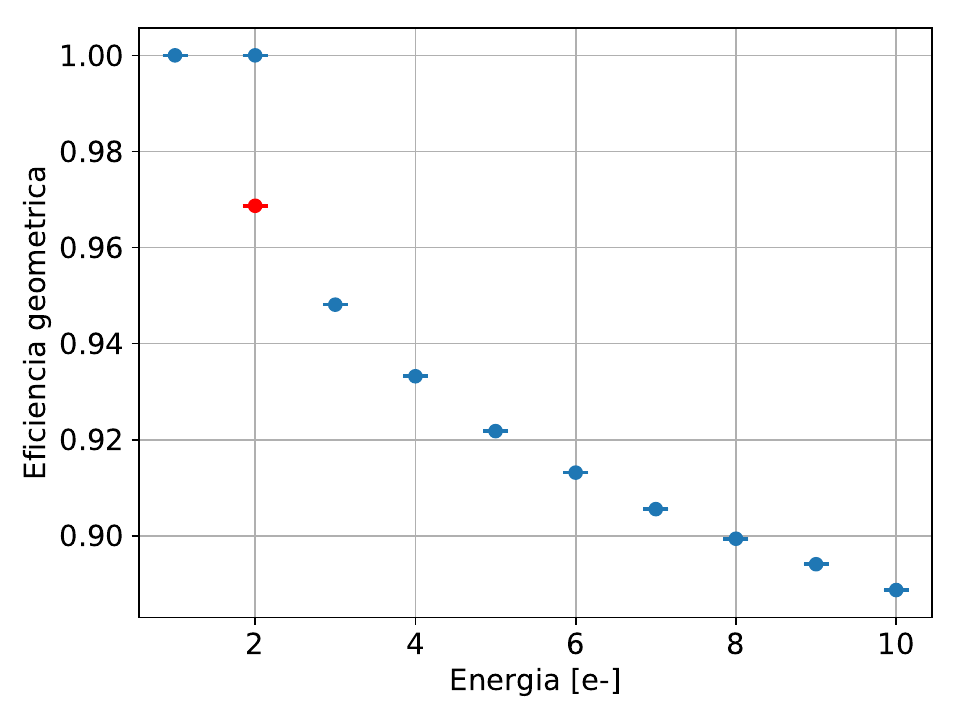}
    \caption{Histogram of geometric efficiency. Note that the geometric efficiency for the 2-electron channel is 100$\%$ as only single-pixel events are used for the mentioned analysis. As an illustrative example, the red point shows the geometric efficiency when considering all possible geometries for the 2-electron channel.}
    \label{fig.geometriceff}
\end{figure}

Figure \ref{fig.geometriceff} shows the estimated geometric efficiency for different energies and continuous cluster configurations. This figure is the product of a code written in C/C++ that calculates the exposure in a set of images, including the efficiency. To calculate the latter, the code injects events into unmasked regions of the images and counts whether the entire event was masked or not. For the 1-electron channel, this algorithm trivially results in 100$\%$ efficiency. However, for the 2-electron channel, for example, a two-pixel event adjacent horizontally may be masked if injected at the edge of a mask in such a way that one pixel is masked, and the other is not \footnote{It is worth noting that our masking algorithm is sensitive to edges, and in the case of finding a multi-pixel cluster where only one pixel is masked (e.g., the edge of a high-energy halo), the entire event is masked.}. The same applies to higher energy levels.

It is worth noting that the geometric efficiency strongly depends on the event selection criteria applied to the set of images. For example, the horizontal streak event mask, which generates horizontal lines in the images, strongly affects vertically elongated pixel configurations, while the hot column mask affects horizontally oriented configurations. In summary, the geometric efficiencies for the studied energy levels are obtained (1, 1, 0.953, and 0.937, respectively), taking into account that only single-pixel events were used for the 2-electron channel, and applied to the total exposure.

\subsection{Results and Dark Matter Exclusion Limits}
\label{sec:2020limits}

\begin{table}[]
\centering
\begin{tabular}{@{}lcccc@{}}
\toprule
                                     & 1 electron    & 2 electrons & 3 electrons & 4 electrons \\ \midrule
Total Exposure {[}gram-days{]}       & 19.93         & 19.93        & 27.82        & 27.82        \\
Geometric Efficiency                & 1             & 1            & 0.953        & 0.937        \\
Diffusion Efficiency                & 1             & 0.228        & 0.761        & 0.778        \\
Cuts Efficiency                     & 0.069         & 0.461        & 0.448        & 0.448        \\
Effective Exposure {[}gram-days{]}   & 1.38          & 2.09         & 9.03         & 9.10         \\
Observed Events                     & 1311.7        & 5            & 0            & 0            \\
90$\%$ CL Upper Limits {[}gram-days$^{-1}${]} & 575.5$^{(*)}$ & 4.449        & 0.255        & 0.253        \\ \bottomrule
\end{tabular}
\caption{Table with exposures, number of events, and efficiencies per channel. The exposure used to calculate the 90$\%$ confidence level upper limits is shown in the row labeled "Effective Exposure," and the results of these limits, along with the "Observed Events," are presented in the last row. (*) For the 1-electron channel, the upper limit was calculated based on the 749 events estimated in Section \ref{sec:2020conteo1e}. \label{tab:SENSEI2020results}}
\end{table}

The Table \ref{tab:SENSEI2020results} presents the results obtained with the mentioned procedure. The count of 1-electron events was carried out as described in Section \ref{sec:2020conteo1e}. Each of the channels shows the 90$\%$ upper limit for the event occurrence rate. In particular, it can be mentioned that, according to what was obtained in the 1-electron channel, fewer 2-electron events in a single pixel were expected. The number of events for such a configuration can be estimated as the SEE rate squared and divided by two, following a Poisson distribution, taking care to apply the same event selection criteria as for the 2-electron channel. In summary, 2.2 events of 2 electrons were expected, and 5 were measured. However, the 90$\%$ upper limit of 2.2 is 5.3, which is within statistical margins.

\begin{figure}[ht!]
    \centering
    \begin{subfigure}[c]{0.49\textwidth}
         \centering
         \caption{}
         \includegraphics[width=\textwidth]{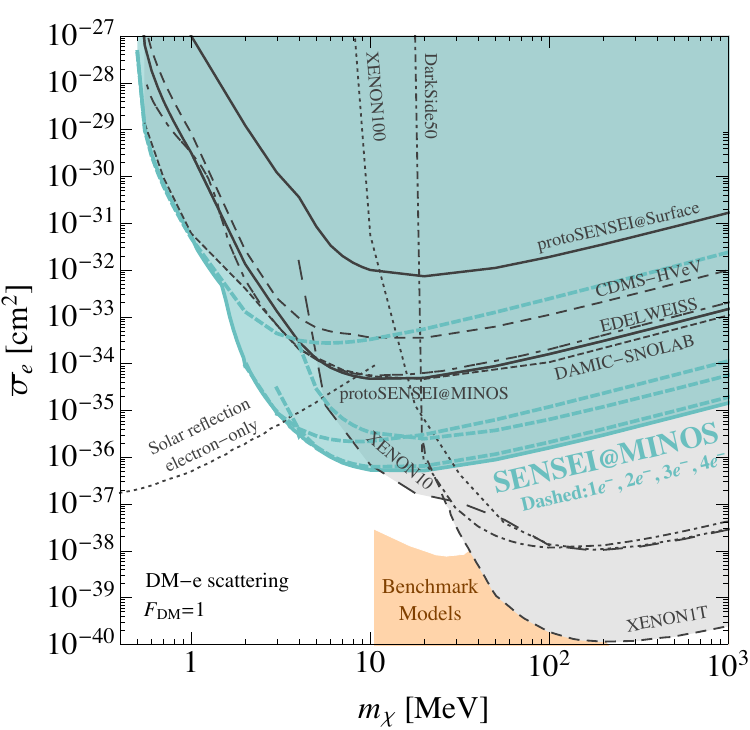}
         \label{fig.2020heavymediator}
     \end{subfigure}
     \hfill
     \begin{subfigure}[c]{0.49\textwidth}
         \centering
         \caption{}
         \includegraphics[width=\textwidth]{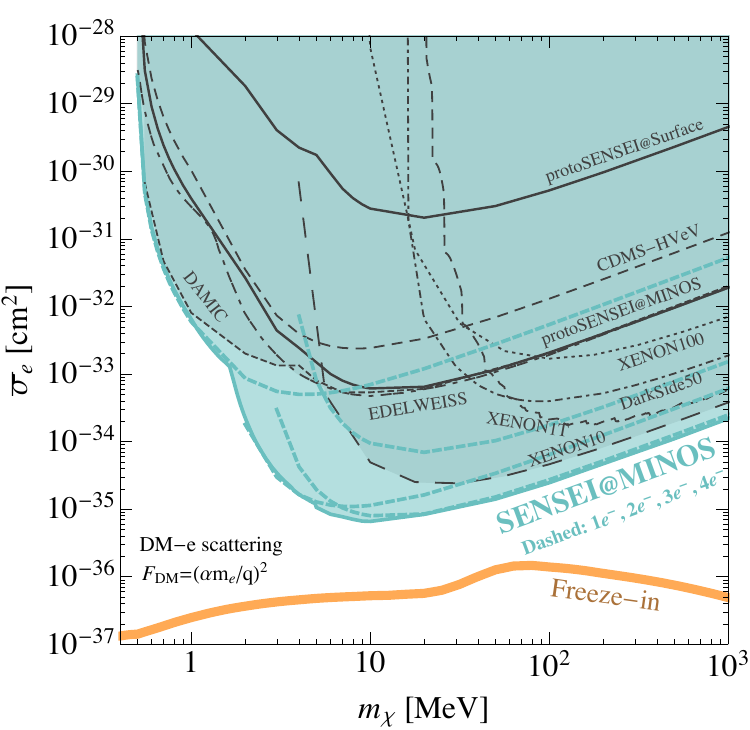}
         \label{fig.2020lightmediator}
     \end{subfigure}
        \caption{90$\%$ exclusion limits for the effective DM-electron scattering cross-section for a heavy mediator (Figure \ref{fig.2020heavymediator}) and a light mediator (Figure \ref{fig.2020lightmediator}). The limits are shown in shaded blue. Additionally, the limits obtained for each of the analyzed energy channels are presented separately (in order of appearance, from left to right, from 1 to 4 electrons), along with previously published limits by the XENON10/100 \cite{XENON10100}, DarkSide-50 \cite{agnes2018constraints}, EDELWEISS \cite{EDELWEISS2020}, CDMS-HVeV \cite{SUPERCDMS2018}, XENON1T \cite{XENON1T2019}, and DAMIC \cite{DAMIC2019} collaborations. Some relevant theoretical models for dark mediator (in orange) and light mediator (in orange) are also presented (\cite{essig2012direct,essig2016direct,boehm2004scalar,lin2012symmetric,PhysRevLett.113.171301,kuflik2017phenomenology,PhysRevLett.124.151801}). Figures extracted from \cite{SENSEI2020}.}
        \label{fig.limitesdeexclusion}
\end{figure}

\begin{figure}[h!]
    \centering
    \includegraphics[width=0.7\textwidth]{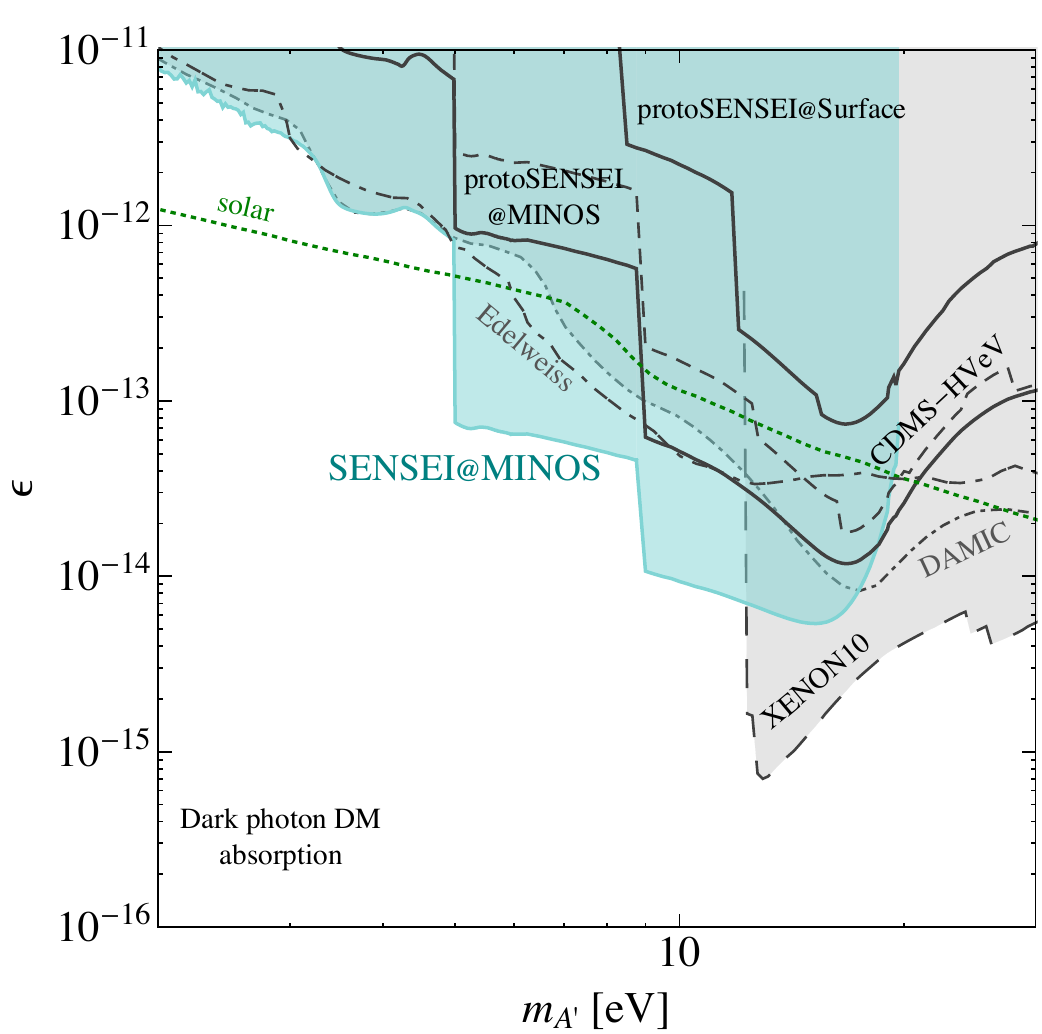}
    \caption{90$\%$ exclusion limits for the kinetic mixing parameter $\epsilon$. Also shown are the limits from DAMIC \cite{DAMIC2017,DAMIC2019}, EDELWEISS \cite{EDELWEISS2020}, CDMSlite \cite{bloch2017searching}, and the Sun \cite{bloch2017searching,an2013new,redondo2013solar}. Figure extracted from \cite{SENSEI2020}.}
    \label{fig.2020absorcion}
\end{figure}

The final step in obtaining exclusion limits is to compare the results obtained with the expected ones according to the dark matter models being tested: scattering by a heavy mediator, scattering by a light mediator, and dark matter absorption. This was detailed in Sections \ref{sec:electronrecoils}, \ref{sec:dmabsorption}, and \ref{sec:limits} of Chapter \ref{cap:1} and involves calculating the expected number of events for different dark matter masses (for each model) and setting the exclusion limit for the lowest possible effective cross-section that is compatible with a 90$\%$ confidence level with the observed signal.

The exclusion limits obtained for the cited mass ranges can be seen in Figures \ref{fig.limitesdeexclusion} and \ref{fig.2020absorcion}. Going from left to right, for electron scattering with dark matter through a heavy mediator (Figure \ref{fig.2020heavymediator}), the most restrictive limits were obtained below 10 MeV. Dashed lines show the limits obtained for each channel separately: from left to right, in order of appearance, the 1, 2, 3, and 4-electron channels. Analyzing these results, it can be seen that the 1-electron channel is the most sensitive to dark matter around 1 MeV, the 2-electron channel between 2 and 3 MeV, and the 3-electron channel above 3 MeV. Additionally, experiments using xenon as the target material (XENON10, XENON100, XENON1T, and DarkSide50) find greater sensitivity for higher masses, mainly due to their heavier detectors. In contrast, for masses below 10 MeV, both SENSEI and other experiments using silicon and germanium as target materials (CDMS, EDELWEISS, and DAMIC) achieve higher sensitivity due to their low detection threshold.

Figure \ref{fig.2020lightmediator} shows the limits if a model is used in which the mediator of the interaction is light, and its form factor goes as $F_{DM}=(\frac{\alpha m_{e}}{q})^{2}$. It can be seen how, with the results mentioned in this thesis, the best exclusion limits were obtained for the LDM's (Light Dark Matter) range of interest. Similarly to what was mentioned for the heavy mediator, the 1-electron channel is the most sensitive for masses around 1 MeV, the 2-electron channel between 2 and 5 MeV, and the 3-electron channel for higher masses.

In the case of dark matter absorption (Figure \ref{fig.2020absorcion}), the best limits are established below $\sim$10~eV for the mass of the dark photon. In this case, the limits are not separately marked for each electronic channel, but they can be understood as the sensitivity increases rapidly when changing the detection channel (at approximately 4.75 eV and 8.5 eV).

\section{Current Status and Future Prospects}
\label{sec:future}

\begin{figure}[h!]
    \centering
    \includegraphics[width=0.7\textwidth]{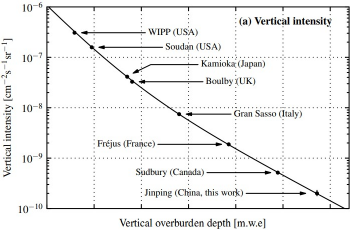}
    \caption{Measurements of vertical muon flux intensity in various underground facilities, including SNOLAB. Figure extracted from \cite{guo2021muon}.}
    \label{fig.muonflux}
\end{figure}

At the time of writing this thesis, the SENSEI experiment has 12 functional quadrants (6.678~grams) in the SNOLAB underground laboratory in Sudbury, Canada, the second deepest underground facility in the world. This laboratory features a Class 2000 cleanroom located at a depth of 2070 meters alongside an active nickel mine. This depth significantly reduces the number of muons reaching the detector compared to what was observed in MINOS (see Figure \ref{fig.muonflux}).

\begin{figure}[h!]
    \centering
    \includegraphics[width=0.98\textwidth]{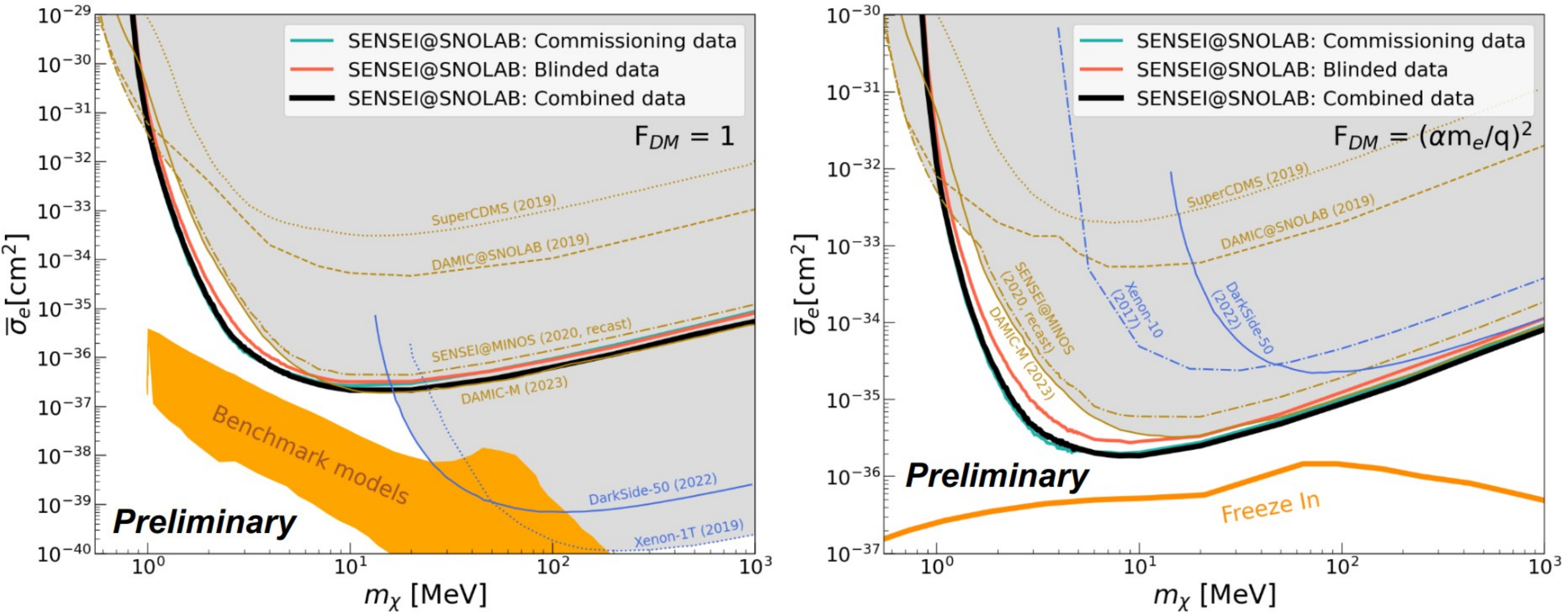}
    \caption{Exclusion limits for light dark matter scattering in the case of heavy mediator (left) and light mediator (right). The shaded gray areas represent the previous limits before the presentation of the new results. In addition to the results mentioned in Figure \ref{fig.limitesdeexclusion}, the results from DAMIC-M \cite{arnquist2023first} and DarkSide-50 \cite{agnes2023search} are added. The limits are preliminary in nature.}
    \label{fig.SENSEI2023}
\end{figure}

The depth allowed for a reduction in the measured high-energy background. Additionally, the use of external lead and polyethylene shielding helped mitigate the environmental radiation background from the mine. This resulted in a significant improvement in the efficiency of the quality cuts used, particularly regarding the high-energy halo.
Recently, preliminary results from a new run using SCCDs installed in SNOLAB were presented \cite{SENSEI2023}, including 45 commissioning images and 37 science images. The results for the electron recoil spectrum can be seen in Figure \ref{fig.SENSEI2023}, which demonstrates the improvement compared to previous results. Currently, the collaboration is in the process of writing a publication that details the new results from the SENSEI experiment at SNOLAB.

In the coming months, new SCCDs will be added to the experiment to achieve the ultimate goal of testing the most well-motivated dark matter models in the parameter space under study.

\section{Conclusions}
\label{cap:2020conclusiones}

In this chapter, we have discussed in detail the key points, steps, and concepts relevant to obtaining the results achieved in 2020, which formed a central part of my doctoral research. 
It is worth highlighting the achievement of limits for dark matter scattering with electrons (and dark photon absorption) that were globally leading across a significant portion of the mass range under study. These results were obtained with a single detector and only 0.08$\%$ of the total exposure set as the collaboration's goal. To emphasize the impact of these results, it took nearly three years for another experiment to surpass these limits.
Furthermore, with respect to the latter point, preliminary results obtained at the SNOLAB laboratory once again positioned SENSEI as a global leader in the search for light dark matter.

Currently, within the SENSEI collaboration, we continue to work on improving the performance of our detectors, as well as achieving a better understanding of the contributions and possible sources of background to the signal of interest under investigation. It is a crucial moment for both the collaboration and the scientific community at large, as current experiments, led by SENSEI, are very close to exploring the parameter space where various well-motivated models reside. The results of these searches will not be long in coming in the coming years.



\cleardoublepage 
\phantomsection  
\renewcommand*{\bibname}{Bibliography}

\addcontentsline{toc}{chapter}{\textbf{Bibliography}}
\printbibliography

@article{hinshaw2013nine,
  title={Nine-year Wilkinson Microwave Anisotropy Probe (WMAP) observations: cosmological parameter results},
  author={Hinshaw, Gary and Larson, D and Komatsu, Eiichiro and Spergel, David N and Bennett, CLaa and Dunkley, Joanna and Nolta, MR and Halpern, M and Hill, RS and Odegard, N and others},
  journal={The Astrophysical Journal Supplement Series},
  volume={208},
  number={2},
  pages={19},
  year={2013},
  publisher={IOP Publishing}
}

@article{spergel2003first,
  title={First-year Wilkinson Microwave Anisotropy Probe (WMAP)* observations: determination of cosmological parameters},
  author={Spergel, David N and Verde, Licia and Peiris, Hiranya V and Komatsu, Eiichiro and Nolta, MR and Bennett, Charles L and Halpern, Mark and Hinshaw, Gary and Jarosik, Norman and Kogut, Alan and others},
  journal={The Astrophysical Journal Supplement Series},
  volume={148},
  number={1},
  pages={175},
  year={2003},
  publisher={IOP Publishing}
}

@article{fixsen1996cosmic,
  title={The cosmic microwave background spectrum from the full cobe* firas data set},
  author={Fixsen, DJ and Cheng, ES and Gales, JM and Mather, John C and Shafer, RA and Wright, EL},
  journal={The Astrophysical Journal},
  volume={473},
  number={2},
  pages={576},
  year={1996},
  publisher={IOP Publishing}
}

@article{mather1990preliminary,
  title={A preliminary measurement of the cosmic microwave background spectrum by the Cosmic Background Explorer (COBE) satellite},
  author={Mather, John C and Cheng, ES and Eplee Jr, RE and Isaacman, RB and Meyer, SS and Shafer, RA and Weiss, R and Wright, EL and Bennett, CL and Boggess, NW and others},
  journal={The Astrophysical Journal},
  volume={354},
  pages={L37--L40},
  year={1990}
}

@ARTICLE{1965ApJ...142..419P,
       author = {Penzias, A.A. and Wilson, R.W.},
        title = {A Measurement of Excess Antenna Temperature at 4080 Mc/s.},
      journal = {The Astrophysical Journal},
         year = 1965,
       volume = {142},
        pages = {419-421},
          doi = {10.1086/148307},
}

@MISC{bulletref,
    title = {Composite image showing the galaxy cluster 1E0657-56, the Bullet cluster},
    year = {2006},
    note = {X-ray: NASA/CXC/CfA/M.Markevitch Optical: NASA/STScI; Magellan/U.Arizona/D.Clowe Lensing Map: NASA/STScI; ESO WFI; Magellan/U.Arizona/D.Clowe},
}

@article{clowe2006direct,
  title={A direct empirical proof of the existence of dark matter},
  author={Clowe, Douglas and Brada{\v{c}}, Maru{\v{s}}a and Gonzalez, Anthony H and Markevitch, Maxim and Randall, Scott W and Jones, Christine and Zaritsky, Dennis},
  journal={The Astrophysical Journal},
  volume={648},
  number={2},
  pages={L109},
  year={2006},
  publisher={IOP Publishing}
}

@book{sanders2010dark,
  title={The dark matter problem: a historical perspective},
  author={Sanders, Robert H},
  year={2010},
  publisher={Cambridge University Press}
}

@article{van1985distribution,
  title={Distribution of dark matter in the spiral galaxy NGC 3198},
  author={van Albada, Tjeerd S and Bahcall, John N and Begeman, K and Sancisi, R},
  journal={The Astrophysical Journal},
  volume={295},
  pages={305--313},
  year={1985}
}

@article{rubin1980rotational,
  title={Rotational properties of 21 SC galaxies with a large range of luminosities and radii, from NGC 4605/R= 4kpc/to UGC 2885/R= 122 kpc},
  author={Rubin, Vera C and Ford Jr, W Kent and Thonnard, Norbert},
  journal={The Astrophysical Journal},
  volume={238},
  pages={471--487},
  year={1980}
}

@article{rubin1983dark,
  title={Dark matter in spiral galaxies},
  author={Rubin, Vera C},
  journal={Scientific American},
  volume={248},
  number={6},
  pages={96--109},
  year={1983},
  publisher={JSTOR}
}

@article{rubin1970rotation,
  title={Rotation of the Andromeda nebula from a spectroscopic survey of emission regions},
  author={Rubin, Vera C and Ford Jr, W Kent},
  journal={The Astrophysical Journal},
  volume={159},
  pages={379},
  year={1970}
}

@article{zwicky1933rotverschiebung,
  title={Die rotverschiebung von extragalaktischen nebeln},
  author={Zwicky, Fritz},
  journal={Helvetica physica acta},
  volume={6},
  pages={110--127},
  year={1933}
}

@article{aghanim2020planck,
  title={Planck 2018 results-VI. Cosmological parameters},
  author={Aghanim, Nabila and Akrami, Yashar and Ashdown, Mark and Aumont, J and Baccigalupi, C and Ballardini, M and Banday, AJ and Barreiro, RB and Bartolo, N and Basak, S and others},
  journal={Astronomy \& Astrophysics},
  volume={641},
  pages={A6},
  year={2020},
  publisher={EDP sciences}
}

@article{Randall_2008,
	doi = {10.1086/587859},
	url = {https://doi.org/10.1086/587859},
	year = 2008,
	month = {jun},
	publisher = {American Astronomical Society},
	volume = {679},
	number = {2},
	pages = {1173--1180},
	author = {Scott W. Randall and Maxim Markevitch and Douglas Clowe and Anthony H. Gonzalez and Marusa Brada{\v{c}}},
	title = {Constraints on the Self-Interaction Cross Section of Dark Matter from Numerical Simulations of the Merging Galaxy Cluster 1E 0657-56},
	journal = {The Astrophysical Journal},
}

@article{audren2014strongest,
  title={Strongest model-independent bound on the lifetime of Dark Matter},
  author={Audren, Benjamin and Lesgourgues, Julien and Mangano, Gianpiero and Serpico, Pasquale Dario and Tram, Thomas},
  journal={Journal of Cosmology and Astroparticle Physics},
  volume={2014},
  number={12},
  pages={028},
  year={2014},
  publisher={IOP Publishing}
}

@article{particle2020review,
  title={Review of particle physics},
  author={Particle Data Group and Zyla, PAea and Barnett, RM and Beringer, J and Dahl, O and Dwyer, DA and Groom, DE and Lin, C-J and Lugovsky, KS and Pianori, E and others},
  journal={Progress of Theoretical and Experimental Physics},
  volume={2020},
  number={8},
  pages={083C01},
  year={2020},
  publisher={Oxford University Press}
}

@article{tremaine1979dynamical,
  title={Dynamical role of light neutral leptons in cosmology},
  author={Tremaine, Scott and Gunn, James E},
  journal={Physical Review Letters},
  volume={42},
  number={6},
  pages={407},
  year={1979},
  publisher={APS}
}

@article{randall2017cores,
  title={Cores in dwarf galaxies from Fermi repulsion},
  author={Randall, Lisa and Scholtz, Jakub and Unwin, James},
  journal={Monthly Notices of the Royal Astronomical Society},
  volume={467},
  number={2},
  pages={1515--1525},
  year={2017},
  publisher={Oxford University Press}
}

@misc{https://doi.org/10.48550/arxiv.1904.07915,
  doi = {10.48550/ARXIV.1904.07915},
  
  url = {https://arxiv.org/abs/1904.07915},
  
  author = {Lin, Tongyan},
  
  title = {TASI lectures on dark matter models and direct detection},
  
  publisher = {arXiv},
  
  year = {2019},
  
  copyright = {arXiv.org perpetual, non-exclusive license}
}

@BOOK{1990eaun.book.....K,
       author = {{Kolb}, Edward W. and {Turner}, Michael S.},
        title = "{The early universe}",
         year = 1990,
       volume = {69},
       adsurl = {https://ui.adsabs.harvard.edu/abs/1990eaun.book.....K}
}

@article{jungman1996supersymmetric,
  title={Supersymmetric dark matter},
  author={Jungman, Gerard and Kamionkowski, Marc and Griest, Kim},
  journal={Physics Reports},
  volume={267},
  number={5-6},
  pages={195--373},
  year={1996},
  publisher={Elsevier}
}

@article{hall2010freeze,
  title={Freeze-in production of FIMP dark matter},
  author={Hall, Lawrence J and Jedamzik, Karsten and March-Russell, John and West, Stephen M},
  journal={Journal of High Energy Physics},
  volume={2010},
  number={3},
  pages={1--33},
  year={2010},
  publisher={Springer}
}

@article{zurek2014asymmetric,
  title={Asymmetric dark matter: theories, signatures, and constraints},
  author={Zurek, Kathryn M},
  journal={Physics Reports},
  volume={537},
  number={3},
  pages={91--121},
  year={2014},
  publisher={Elsevier}
}

@article{bernabei2008first,
  title={First results from DAMA/LIBRA and the combined results with DAMA/NaI},
  author={Bernabei, R and Belli, P and Cappella, F and Cerulli, R and Dai, CJ and d’Angelo, A and He, HL and Incicchitti, A and Kuang, HH and Ma, JM and others},
  journal={The European Physical Journal C},
  volume={56},
  number={3},
  pages={333--355},
  year={2008},
  publisher={Springer}
}

@article{billard2022direct,
  title={Direct detection of dark matter--APPEC committee report},
  author={Billard, Julien and Boulay, Mark and Cebri{\'a}n, Susana and Covi, Laura and Fiorillo, Giuliana and Green, Anne M and Kopp, Joachim and Majorovits, B{\'e}la and Palladino, Kimberly and Petricca, Federica and others},
  journal={Reports on Progress in Physics},
  year={2022},
  publisher={IOP Publishing}
}

@article{dine1981simple,
  title={A simple solution to the strong CP problem with a harmless axion},
  author={Dine, Michael and Fischler, Willy and Srednicki, Mark},
  journal={Physics letters B},
  volume={104},
  number={3},
  pages={199--202},
  year={1981},
  publisher={Elsevier}
}

@article{battaglieri2017us,
  title={US cosmic visions: new ideas in dark matter 2017: community report},
  author={Battaglieri, Marco and Belloni, Alberto and Chou, Aaron and Cushman, Priscilla and Echenard, Bertrand and Essig, Rouven and Estrada, Juan and Feng, Jonathan L and Flaugher, Brenna and Fox, Patrick J and others},
  journal={arXiv preprint arXiv:1707.04591},
  year={2017}
}

@article{bloch2017searching,
  title={Searching for dark absorption with direct detection experiments},
  author={Bloch, Itay M and Essig, Rouven and Tobioka, Kohsaku and Volansky, Tomer and Yu, Tien-Tien},
  journal={Journal of High Energy Physics},
  volume={2017},
  number={6},
  pages={1--21},
  year={2017},
  publisher={Springer}
}

@article{goldberg1986new,
  title={A new candidate for dark matter},
  author={Goldberg, Haim and Hall, Lawrence J},
  journal={Physics Letters B},
  volume={174},
  number={2},
  pages={151--155},
  year={1986},
  publisher={Elsevier}
}

@article{lin2012symmetric,
  title={Symmetric and asymmetric light dark matter},
  author={Lin, Tongyan and Yu, Hai-Bo and Zurek, Kathryn M},
  journal={Physical Review D},
  volume={85},
  number={6},
  pages={063503},
  year={2012},
  publisher={APS}
}

@article{nelson2011dark,
  title={Dark light, dark matter, and the misalignment mechanism},
  author={Nelson, Ann E and Scholtz, Jakub},
  journal={Physical Review D},
  volume={84},
  number={10},
  pages={103501},
  year={2011},
  publisher={APS}
}

@article{essig2012direct,
  title={Direct detection of sub-GeV dark matter},
  author={Essig, Rouven and Mardon, Jeremy and Volansky, Tomer},
  journal={Physical Review D},
  volume={85},
  number={7},
  pages={076007},
  year={2012},
  publisher={APS}
}

@article{bulbul2014detection,
  title={Detection of an unidentified emission line in the stacked X-ray spectrum of galaxy clusters},
  author={Bulbul, Esra and Markevitch, Maxim and Foster, Adam and Smith, Randall K and Loewenstein, Michael and Randall, Scott W},
  journal={The Astrophysical Journal},
  volume={789},
  number={1},
  pages={13},
  year={2014},
  publisher={IOP Publishing}
}

@article{essig2016direct,
  title={Direct detection of sub-GeV dark matter with semiconductor targets},
  author={Essig, Rouven and Fernandez-Serra, Marivi and Mardon, Jeremy and Soto, Adrian and Volansky, Tomer and Yu, Tien-Tien},
  journal={Journal of High Energy Physics},
  volume={2016},
  number={5},
  pages={1--54},
  year={2016},
  publisher={Springer}
}

@article{zeplin,
title = {Measurement of single electron emission in two-phase xenon},
journal = {Astroparticle Physics},
volume = {30},
number = {2},
pages = {54-57},
year = {2008},
issn = {0927-6505},
doi = {https://doi.org/10.1016/j.astropartphys.2008.06.006},
url = {https://www.sciencedirect.com/science/article/pii/S0927650508000789},
author = {B. Edwards and H.M. Araújo and V. Chepel and D. Cline and T. Durkin and J. Gao and C. Ghag and E.V. Korolkova and V.N. Lebedenko and A. Lindote and M.I. Lopes and R. Lüscher and A.St.J. Murphy and F. Neves and W. Ooi and J. Pinto {da Cunha} and R.M. Preece and G. Salinas and C. Silva and V.N. Solovov and N.J.T. Smith and P.F. Smith and T.J. Sumner and C. Thorne and R.J. Walker and H. Wang and J.T. White and F.L.H. Wolfs},
}

@article{xenon10,
  title = {Search for Light Dark Matter in XENON10 Data},
  author = {Angle, J. and Aprile, E. and Arneodo, F. and Baudis, L. and Bernstein, A. and Bolozdynya, A. I. and Coelho, L. C. C. and Dahl, C. E. and DeViveiros, L. and Ferella, A. D. and Fernandes, L. M. P. and Fiorucci, S. and Gaitskell, R. J. and Giboni, K. L. and Gomez, R. and Hasty, R. and Kastens, L. and Kwong, J. and Lopes, J. A. M. and Madden, N. and Manalaysay, A. and Manzur, A. and McKinsey, D. N. and Monzani, M. E. and Ni, K. and Oberlack, U. and Orboeck, J. and Plante, G. and Santorelli, R. and dos Santos, J. M. F. and Schulte, S. and Shagin, P. and Shutt, T. and Sorensen, P. and Winant, C. and Yamashita, M.},
  collaboration = {XENON10 Collaboration},
  journal = {Phys. Rev. Lett.},
  volume = {107},
  issue = {5},
  pages = {051301},
  numpages = {5},
  year = {2011},
  month = {Jul},
  publisher = {American Physical Society},
  doi = {10.1103/PhysRevLett.107.051301},
  url = {https://link.aps.org/doi/10.1103/PhysRevLett.107.051301}
}

@article{hochberg2017,
  title = {Absorption of light dark matter in semiconductors},
  author = {Hochberg, Yonit and Lin, Tongyan and Zurek, Kathryn M.},
  journal = {Phys. Rev. D},
  volume = {95},
  issue = {2},
  pages = {023013},
  numpages = {8},
  year = {2017},
  month = {Jan},
  publisher = {American Physical Society},
  doi = {10.1103/PhysRevD.95.023013},
  url = {https://link.aps.org/doi/10.1103/PhysRevD.95.023013}
}

@article{boylenobel,
  title = {Nobel Lecture: The invention and early history of the CCD},
  author = {Smith, George E.},
  journal = {Rev. Mod. Phys.},
  volume = {82},
  issue = {3},
  pages = {2307--2312},
  numpages = {0},
  year = {2010},
  month = {Aug},
  publisher = {American Physical Society},
  doi = {10.1103/RevModPhys.82.2307},
  url = {https://link.aps.org/doi/10.1103/RevModPhys.82.2307}
}

@article{janesick,
author = {Janesick, James},
year = {2001},
month = {01},
pages = {},
title = {Scientific Charge-Couple Devices},
volume = {83},
journal = {Scientific charge-coupled devices, Bellingham, WA: SPIE Optical Engineering Press, 2001, xvi, 906 p. SPIE Press monograph, PM 83. ISBN 0819436984},
doi = {10.1117/12.7974139}
}

@article{SENSEI2022,
  title = {SENSEI: Characterization of Single-Electron Events Using a Skipper Charge-Coupled Device},
  author = {Barak, Liron and Bloch, Itay M. and Botti, Ana and Cababie, Mariano and Cancelo, Gustavo and Chaplinsky, Luke and Chierchie, Fernando and Crisler, Michael and Drlica-Wagner, Alex and Essig, Rouven and Estrada, Juan and Etzion, Erez and Fernandez Moroni, Guillermo and Gift, Daniel and Holland, Stephen E. and Munagavalasa, Sravan and Orly, Aviv and Rodrigues, Dario and Singal, Aman and Haro, Miguel Sofo and Stefanazzi, Leandro and Tiffenberg, Javier and Uemura, Sho and Volansky, Tomer and Yu, Tien-Tien},
  collaboration = {SENSEI Collaboration},
  journal = {Phys. Rev. Applied},
  volume = {17},
  issue = {1},
  pages = {014022},
  numpages = {9},
  year = {2022},
  month = {Jan},
  publisher = {American Physical Society},
  doi = {10.1103/PhysRevApplied.17.014022},
  url = {https://link.aps.org/doi/10.1103/PhysRevApplied.17.014022}
}

@article{holland2003fully,
  title={Fully depleted, back-illuminated charge-coupled devices fabricated on high-resistivity silicon},
  author={Holland, Stephen E and Groom, Donald E and Palaio, Nick P and Stover, Richard J and Wei, Mingzhi},
  journal={IEEE Transactions on Electron Devices},
  volume={50},
  number={1},
  pages={225--238},
  year={2003},
  publisher={IEEE}
}

@misc{holland2001fully,
  title={Fully depleted back illuminated CCD},
  author={Holland, Stephen Edward},
  year={2001},
  month=jul # "~10",
  publisher={Google Patents},
  note={US Patent 6,259,085}
}

@article{bebek2017status,
  title={Status of the CCD development for the Dark Energy Spectroscopic Instrument},
  author={Bebek, CJ and Emes, JH and Groom, DE and Haque, S and Holland, SE and Jelinsky, PN and Karcher, A and Kolbe, WF and Lee, JS and Palaio, NP and others},
  journal={Journal of Instrumentation},
  volume={12},
  number={04},
  pages={C04018},
  year={2017},
  publisher={IOP Publishing}
}

@inproceedings{janesick1990new,
  title={New advancements in charge-coupled device technology: subelectron noise and 4096 x 4096 pixel CCDs},
  author={Janesick, James R and Elliott, Tom S and Dingiziam, Arsham and Bredthauer, Richard A and Chandler, Charles E and Westphal, James A and Gunn, James E},
  booktitle={Charge-Coupled Devices and Solid State Optical Sensors},
  volume={1242},
  pages={223--237},
  year={1990},
  organization={SPIE}
}

@article{SENSEI2018,
  title={SENSEI: first direct-detection constraints on sub-GeV dark matter from a surface run},
  author={Crisler, Michael and Essig, Rouven and Estrada, Juan and Fernandez, Guillermo and Tiffenberg, Javier and Haro, Miguel Sofo and Volansky, Tomer and Yu, Tien-Tien and Sensei Collaboration and others},
  journal={Physical review letters},
  volume={121},
  number={6},
  pages={061803},
  year={2018},
  publisher={APS}
}

@article{SENSEI2019,
  title={SENSEI: direct-detection constraints on sub-GeV dark matter from a shallow underground run using a prototype skipper CCD},
  author={Abramoff, Orr and Barak, Liron and Bloch, Itay M and Chaplinsky, Luke and Crisler, Michael and Drlica-Wagner, Alex and Essig, Rouven and Estrada, Juan and Etzion, Erez and Fernandez, Guillermo and others},
  journal={Physical review letters},
  volume={122},
  number={16},
  pages={161801},
  year={2019},
  publisher={APS}
}

@article{tiffenberg2017,
  title={Single-electron and single-photon sensitivity with a silicon Skipper CCD},
  author={Tiffenberg, Javier and Sofo-Haro, Miguel and Drlica-Wagner, Alex and Essig, Rouven and Guardincerri, Yann and Holland, Steve and Volansky, Tomer and Yu, Tien-Tien},
  journal={Physical review letters},
  volume={119},
  number={13},
  pages={131802},
  year={2017},
  publisher={APS}
}

@article{haro2020studies,
  title={Studies on small charge packet transport in high-resistivity fully depleted CCDs},
  author={Haro, Miguel Sofo and Moroni, Guillermo Fernandez and Tiffenberg, Javier},
  journal={IEEE Transactions on Electron Devices},
  volume={67},
  number={5},
  pages={1993--2000},
  year={2020},
  publisher={IEEE}
}

@article{DAMIC2016,
  title={Search for low-mass WIMPs in a 0.6 kg day exposure of the DAMIC experiment at SNOLAB},
  author={Aguilar-Arevalo, A and Amidei, D and Bertou, X and Butner, M and Cancelo, G and V{\'a}zquez, A Casta{\~n}eda and Vergara, BA Cervantes and Chavarria, AE and Chavez, CR and de Mello Neto, JRT and others},
  journal={Physical Review D},
  volume={94},
  number={8},
  pages={082006},
  year={2016},
  publisher={APS}
}

@article{CONNIE2016,
  title={Results of the engineering run of the Coherent Neutrino Nucleus Interaction Experiment (CONNIE)},
  author={Aguilar-Arevalo, A and Bertou, X and Bonifazi, C and Butner, M and Cancelo, G and V{\'a}zquez, A Casta{\~n}eda and Vergara, B Cervantes and Chavez, CR and Da Motta, H and D'Olivo, JC and others},
  journal={Journal of Instrumentation},
  volume={11},
  number={07},
  pages={P07024},
  year={2016},
  publisher={IOP Publishing}
}

@article{ramanathan,
  title = {Ionization yield in silicon for eV-scale electron-recoil processes},
  author = {Ramanathan, K. and Kurinsky, N.},
  journal = {Phys. Rev. D},
  volume = {102},
  issue = {6},
  pages = {063026},
  numpages = {14},
  year = {2020},
  month = {Sep},
  publisher = {American Physical Society},
  doi = {10.1103/PhysRevD.102.063026},
  url = {https://link.aps.org/doi/10.1103/PhysRevD.102.063026}
}

@misc{ccdinventionholland,
 title={Fully depleted back illuminated CCD},
 author={Stephen E. Holland},
 year={U.S. Patent 6 259 085, 2001}
 }

@inbook{ashcroft28and29,
  title="Solid State Physics",
  author="Ashcroft, N.W. and Mermin, N.D",
  year="1976",
  chapter="28-29",
}

@inbook{ashcroft12,
  title="Solid State Physics",
  author="Ashcroft, N.W. and Mermin, N.D",
  year="1976",
  chapter="12",
}

@article{wen1974design,
  title={Design and operation of a floating gate amplifier},
  author={Wen, David D},
  journal={IEEE Journal of Solid-State Circuits},
  volume={9},
  number={6},
  pages={410--414},
  year={1974},
  publisher={IEEE}
}

@article{moroni2012,
  title={Sub-electron readout noise in a Skipper CCD fabricated on high resistivity silicon},
  author={Fern{\'a}ndez Moroni, Guillermo and Estrada, Juan and Cancelo, Gustavo and Holland, Stephen E and Paolini, Eduardo E and Diehl, H Thomas},
  journal={Experimental Astronomy},
  volume={34},
  number={1},
  pages={43--64},
  year={2012},
  publisher={Springer}
}

@phdthesis{miguelthesis,
    title    = {Sensores multipixel CCD de ultra bajo ruido de lectura para detección de partículas},
    school   = {Instituto Balseiro, Universidad Nacional de Cuyo.},
    author   = {Sofo Haro, Miguel},
    year     = {2017}, %other attributes omitted
}

@article{chavarria2015damic,
  title={Damic at snolab},
  author={Chavarria, Alvaro E and Tiffenberg, Javier and Aguilar-Arevalo, Alexis and Amidei, Dan and Bertou, Xavier and Cancelo, Gustavo and D’Olivo, Juan Carlos and Estrada, Juan and Moroni, Guillermo Fernandez and Izraelevitch, Federico and others},
  journal={Physics Procedia},
  volume={61},
  pages={21--33},
  year={2015},
  publisher={Elsevier}
}

@article{lindhard1963integral,
  title={Integral equations governing radiation effects},
  author={Lindhard, Jens and Nielsen, V and Scharff, M and Thomsen, PV},
  journal={Mat. Fys. Medd. Dan. Vid. Selsk},
  volume={33},
  number={10},
  pages={1--42},
  year={1963}
}

@article{freedman1974,
  title = {Coherent effects of a weak neutral current},
  author = {Freedman, Daniel Z.},
  journal = {Phys. Rev. D},
  volume = {9},
  issue = {5},
  pages = {1389--1392},
  numpages = {0},
  year = {1974},
  month = {Mar},
  publisher = {American Physical Society},
  doi = {10.1103/PhysRevD.9.1389},
  url = {https://link.aps.org/doi/10.1103/PhysRevD.9.1389}
}

@article{CONNIE2015,
  title = {Charge coupled devices for detection of coherent neutrino-nucleus scattering},
  author = {Fernandez Moroni, Guillermo and Estrada, Juan and Paolini, Eduardo E. and Cancelo, Gustavo and Tiffenberg, Javier and Molina, Jorge},
  journal = {Phys. Rev. D},
  volume = {91},
  issue = {7},
  pages = {072001},
  numpages = {9},
  year = {2015},
  month = {Apr},
  publisher = {American Physical Society},
  doi = {10.1103/PhysRevD.91.072001},
  url = {https://link.aps.org/doi/10.1103/PhysRevD.91.072001}
}

@article{drlica2020,
  title={Characterization of skipper CCDs for cosmological applications},
  author={Drlica-Wagner, Alex and Villalpando, Edgar Marrufo and O'Neil, Judah and Estrada, Juan and Holland, Stephen and Kurinsky, Noah and Li, Ting and Moroni, Guillermo Fernandez and Tiffenberg, Javier and Uemura, Sho},
  booktitle={X-Ray, Optical, and Infrared Detectors for Astronomy IX},
  volume={11454},
  pages={210--223},
  year={2020},
  organization={SPIE}
}

@inproceedings{widenhorn2002temperature,
  title={Temperature dependence of dark current in a CCD},
  author={Widenhorn, Ralf and Blouke, Morley M and Weber, Alexander and Rest, Armin and Bodegom, Erik},
  booktitle={Sensors and Camera Systems for Scientific, Industrial, and Digital Photography Applications III},
  volume={4669},
  pages={193--201},
  year={2002},
  organization={SPIE}
}

@ARTICLE{burke1991,  author={Burke, B.E. and Gajar, S.A.},  journal={IEEE Transactions on Electron Devices},   title={Dynamic suppression of interface-state dark current in buried-channel CCDs},   year={1991},  volume={38},  number={2},  pages={285-290},  doi={10.1109/16.69907}}

@article{SENSEI2020,
  title={Sensei: Direct-detection results on sub-gev dark matter from a new skipper ccd},
  author={Barak, Liron and Bloch, Itay M and Cababie, Mariano and Cancelo, Gustavo and Chaplinsky, Luke and Chierchie, Fernando and Crisler, Michael and Drlica-Wagner, Alex and Essig, Rouven and Estrada, Juan and others},
  journal={Physical Review Letters},
  volume={125},
  number={17},
  pages={171802},
  year={2020},
  publisher={APS}
}

@inproceedings{moroni2019low,
  title={Low threshold acquisition controller for skipper charge coupled devices},
  author={Moroni, G Fernandez and Chierchie, F and Haro, M Sofo and Stefanazzi, L and Soto, A and Paolini, EE and Cancelo, G and Treptow, K and Wilcer, N and Zmuda, T and others},
  booktitle={2019 Argentine Conference on Electronics (CAE)},
  pages={86--91},
  year={2019},
  organization={IEEE}
}

@article{cancelo2021,
author = {Gustavo I. Cancelo and Claudio Chavez and Fernando Chierchie and Juan Estrada and Guillermo Fernandez-Moroni and Eduardo E. Paolini and Miguel Sofo Haro and Angel Soto and Leandro Stefanazzi and Javier Tiffenberg and Ken Treptow and Neal Wilcer and Ted J. Zmuda},
title = {{Low threshold acquisition controller for Skipper charge-coupled devices}},
volume = {7},
journal = {Journal of Astronomical Telescopes, Instruments, and Systems},
number = {1},
publisher = {SPIE},
pages = {015001},
year = {2021},
doi = {10.1117/1.JATIS.7.1.015001},
URL = {https://doi.org/10.1117/1.JATIS.7.1.015001}
}

@INPROCEEDINGS{8214366,
  author={Haro, Miguel Sofo and Soto, Angel and Moroni, Guillermo Fernandez and Chierchie, Fernando and Stefanazzi, Leandro and Chavez, Rodrigo and Castaneda, Alejandro and Hernandez, Karen and Zmuda, Ted and Wilser, Neal and Paolini, Eduardo and Oliva, Alejandro and Cancelo, Gustavo},
  booktitle={2017 XVII Workshop on Information Processing and Control (RPIC)}, 
  title={A low noise digital readout system for scientific charge coupled devices}, 
  year={2017},
  volume={},
  number={},
  pages={1-5},
  doi={10.23919/RPIC.2017.8214366}}

@article{du2022sources,
  title={Sources of low-energy events in low-threshold dark-matter and neutrino detectors},
  author={Du, Peizhi and Egana-Ugrinovic, Daniel and Essig, Rouven and Sholapurkar, Mukul},
  journal={Physical Review X},
  volume={12},
  number={1},
  pages={011009},
  year={2022},
  publisher={APS}
}

@article{DAMIC2019,
  title = {Constraints on Light Dark Matter Particles Interacting with Electrons from DAMIC at SNOLAB},
  author = {Aguilar-Arevalo, A. and Amidei, D. and Baxter, D. and Cancelo, G. and Cervantes Vergara, B. A. and Chavarria, A. E. and Darragh-Ford, E. and de Mello Neto, J. R. T. and D'Olivo, J. C. and Estrada, J. and Ga\"{\i}or, R. and Guardincerri, Y. and Hossbach, T. W. and Kilminster, B. and Lawson, I. and Lee, S. J. and Letessier-Selvon, A. and Matalon, A. and Mello, V. B. B. and Mitra, P. and Molina, J. and Paul, S. and Piers, A. and Privitera, P. and Ramanathan, K. and Da Rocha, J. and Sarkis, Y. and Settimo, M. and Smida, R. and Thomas, R. and Tiffenberg, J. and Torres Machado, D. and Vilar, R. and Virto, A. L.},
  collaboration = {DAMIC Collaboration},
  journal = {Phys. Rev. Lett.},
  volume = {123},
  issue = {18},
  pages = {181802},
  numpages = {6},
  year = {2019},
  month = {Oct},
  publisher = {American Physical Society},
  doi = {10.1103/PhysRevLett.123.181802},
  url = {https://link.aps.org/doi/10.1103/PhysRevLett.123.181802}
}

@article{ramanujanlost,
  title={The lost notebook and other unpublished papers, 1988},
  author={Ramanujan, S},
  journal={Narosa, New Delhi}
}

@article{tsang1997,
  title={Picosecond hot electron light emission from submicron complementary metal--oxide--semiconductor circuits},
  author={Tsang, JC and Kash, JA},
  journal={Applied Physics Letters},
  volume={70},
  number={7},
  pages={889--891},
  year={1997},
  publisher={American Institute of Physics}
}

@article{bartelink1963,
  title={Hot-electron emission from shallow p- n junctions is silicon},
  author={Bartelink, Dirk Jan and Moll, JL and Meyer, NI},
  journal={Physical Review},
  volume={130},
  number={3},
  pages={972},
  year={1963},
  publisher={APS}
}

@article{toriumi1987,
  title={A study of photon emission from n-channel MOSFET's},
  author={Toriumi, Akira and Yoshimi, Makoto and Iwase, Masao and Akiyama, Yutaka and Taniguchi, Kenji},
  journal={IEEE Transactions on Electron Devices},
  volume={34},
  number={7},
  pages={1501--1508},
  year={1987},
  publisher={IEEE}
}

@article{lanzoni1991,
  title={Extended (1.1-2.9 eV) hot-carrier-induced photon emission in n-channel Si MOSFETs},
  author={Lanzoni, Massimo and Sangiorgi, Enrico and Fiegna, Claudio and Manfredi, Manfredo and Ricco, Bruno},
  journal={IEEE electron device letters},
  volume={12},
  number={6},
  pages={341--343},
  year={1991},
  publisher={IEEE}
}

@article{bude1992,
  title={Hot-carrier luminescence in Si},
  author={Bude, Jeff and Sano, Nobuyuki and Yoshii, Akira},
  journal={Physical Review B},
  volume={45},
  number={11},
  pages={5848},
  year={1992},
  publisher={APS}
}

@article{newman1955visible,
  title={Visible light from a silicon p- n junction},
  author={Newman, Roger},
  journal={Physical review},
  volume={100},
  number={2},
  pages={700},
  year={1955},
  publisher={APS}
}

@article{herzog1989electromagnetic,
  title={Electromagnetic radiation from hot carriers in FET-devices},
  author={Herzog, M and Schels, M and Koch, F and Moglestue, C and Rosenzweig, J},
  journal={Solid-state electronics},
  volume={32},
  number={12},
  pages={1765--1769},
  year={1989},
  publisher={Elsevier}
}

@article{das1990luminescence,
  title={Luminescence spectra of an n-channel metal-oxide-semiconductor field-effect transistor at breakdown},
  author={Das, NC and Arora, BM},
  journal={Applied physics letters},
  volume={56},
  number={12},
  pages={1152--1153},
  year={1990},
  publisher={American Institute of Physics}
}

@phdthesis{venter2013reach,
  title={Reach Through Hot Carrier Silicon Electroluminescence In Standard Cmos},
  author={Venter, Petrus Johannes and others},
  year={2013},
  school={University of Pretoria}
}

@article{bernstein2017instrumental,
  title={Instrumental response model and detrending for the Dark Energy Camera},
  author={Bernstein, GM and Abbott, TMC and Desai, Shantanu and Gruen, D and Gruendl, RA and Johnson, MD and Lin, H and Menanteau, F and Morganson, E and Neilsen, E and others},
  journal={Publications of the Astronomical Society of the Pacific},
  volume={129},
  number={981},
  pages={114502},
  year={2017},
  publisher={IOP Publishing}
}

@article{agnes2018constraints,
  title={Constraints on sub-GeV dark-matter--electron scattering from the DarkSide-50 experiment},
  author={Agnes, Paolo and Albuquerque, Ivone Freire da Mota and Alexander, T and Alton, AK and Araujo, GR and Asner, DM and Ave, M and Back, HO and Baldin, B and Batignani, G and others},
  journal={Physical review letters},
  volume={121},
  number={11},
  pages={111303},
  year={2018},
  publisher={APS}
}

@article{arnquist2023first,
  title={First Constraints from DAMIC-M on Sub-GeV Dark-Matter Particles Interacting with Electrons},
  author={Arnquist, I and Avalos, N and Baxter, D and Bertou, X and Castello-Mor, N and Chavarria, AE and Cuevas-Zepeda, J and Gutierrez, J Cortabitarte and Duarte-Campderros, J and Dastgheibi-Fard, A and others},
  journal={arXiv preprint arXiv:2302.02372},
  year={2023}
}

@article{XENON10100,
title = {New constraints and prospects for sub-GeV dark matter scattering off electrons in xenon},
author = {Essig, Rouven and Volansky, Tomer and Yu, Tien-Tien},
journal = {Phys. Rev. D},
volume = {96},
issue = {4},
pages = {043017},
numpages = {9},
year = {2017},
month = {Aug},
publisher = {American Physical Society},
doi = {10.1103/PhysRevD.96.043017},
url = {https://link.aps.org/doi/10.1103/PhysRevD.96.043017}
}

@article{EDELWEISS2020,
title = {First Germanium-Based Constraints on Sub-MeV Dark Matter with the EDELWEISS Experiment},
author = {The EDELWEISS collaboration},
collaboration = {EDELWEISS Collaboration},
journal = {Phys. Rev. Lett.},
volume = {125},
issue = {14},
pages = {141301},
numpages = {6},
year = {2020},
month = {Oct},
publisher = {American Physical Society},
doi = {10.1103/PhysRevLett.125.141301},
url = {https://link.aps.org/doi/10.1103/PhysRevLett.125.141301}
}

@article{SUPERCDMS2018,
title = {First Dark Matter Constraints from a SuperCDMS Single-Charge Sensitive Detector},
author = {The Super-CDMS collaboration},
journal = {Phys. Rev. Lett.},
volume = {121},
issue = {5},
pages = {051301},
numpages = {7},
year = {2018},
month = {Aug},
publisher = {American Physical Society},
doi = {10.1103/PhysRevLett.121.051301},
url = {https://link.aps.org/doi/10.1103/PhysRevLett.121.051301}
}

@article{XENON1T2019,
title = {Light Dark Matter Search with Ionization Signals in XENON1T},
author = {The XENON-1T collaboration.},
collaboration = {XENON Collaboration},
journal = {Phys. Rev. Lett.},
volume = {123},
issue = {25},
pages = {251801},
numpages = {8},
year = {2019},
month = {Dec},
publisher = {American Physical Society},
doi = {10.1103/PhysRevLett.123.251801},
url = {https://link.aps.org/doi/10.1103/PhysRevLett.123.251801}
}

@article{solarreflection,
title = {Directly Detecting MeV-Scale Dark Matter Via Solar Reflection},
author = {An, Haipeng and Pospelov, Maxim and Pradler, Josef and Ritz, Adam},
journal = {Phys. Rev. Lett.},
volume = {120},
issue = {14},
pages = {141801},
numpages = {6},
year = {2018},
month = {Apr},
publisher = {American Physical Society},
doi = {10.1103/PhysRevLett.120.141801},
url = {https://link.aps.org/doi/10.1103/PhysRevLett.120.141801}
}

@article{DAMIC2017,
title = {First Direct-Detection Constraints on eV-Scale Hidden-Photon Dark Matter with DAMIC at SNOLAB},
author = {The DAMIC collaboration.},
collaboration = {DAMIC Collaboration},
journal = {Phys. Rev. Lett.},
volume = {118},
issue = {14},
pages = {141803},
numpages = {5},
year = {2017},
month = {Apr},
publisher = {American Physical Society},
doi = {10.1103/PhysRevLett.118.141803},
url = {https://link.aps.org/doi/10.1103/PhysRevLett.118.141803}
}

@article{an2013new,
title={New stellar constraints on dark photons},
author={An, Haipeng and Pospelov, Maxim and Pradler, Josef},
journal={Physics Letters B},
volume={725},
number={4-5},
pages={190--195},
year={2013},
publisher={Elsevier}
}

@article{redondo2013solar,
title={Solar constraints on hidden photons re-visited},
author={Redondo, Javier and Raffelt, Georg},
journal={Journal of Cosmology and Astroparticle Physics},
volume={2013},
number={08},
pages={034},
year={2013},
publisher={IOP Publishing}
}

@article{boehm2004scalar,
title={Scalar dark matter candidates},
author={Boehm, C and Fayet, Pierre},
journal={Nuclear Physics B},
volume={683},
number={1-2},
pages={219--263},
year={2004},
publisher={Elsevier}
}

@article{PhysRevLett.113.171301,
title = {Mechanism for Thermal Relic Dark Matter of Strongly Interacting Massive Particles},
author = {Hochberg, Yonit and Kuflik, Eric and Volansky, Tomer and Wacker, Jay G.},
journal = {Phys. Rev. Lett.},
volume = {113},
issue = {17},
pages = {171301},
numpages = {5},
year = {2014},
month = {Oct},
publisher = {American Physical Society},
doi = {10.1103/PhysRevLett.113.171301},
url = {https://link.aps.org/doi/10.1103/PhysRevLett.113.171301}
}

@article{kuflik2017phenomenology,
title={Phenomenology of ELDER dark matter},
author={Kuflik, Eric and Perelstein, Maxim and Lorier, Nicolas Rey-Le and Tsai, Yu-Dai},
journal={Journal of High Energy Physics},
volume={2017},
number={8},
pages={1--30},
year={2017},
publisher={Springer}
}

@article{PhysRevLett.124.151801,
title = {Thermal Relic Targets with Exponentially Small Couplings},
author = {D'Agnolo, Raffaele Tito and Pappadopulo, Duccio and Ruderman, Joshua T. and Wang, Po-Jen},
journal = {Phys. Rev. Lett.},
volume = {124},
issue = {15},
pages = {151801},
numpages = {6},
year = {2020},
month = {Apr},
publisher = {American Physical Society},
doi = {10.1103/PhysRevLett.124.151801},
url = {https://link.aps.org/doi/10.1103/PhysRevLett.124.151801}
}

@article{chu2012four,
title={The four basic ways of creating dark matter through a portal},
author={Chu, Xiaoyong and Hambye, Thomas and Tytgat, Michel HG},
journal={Journal of Cosmology and Astroparticle Physics},
volume={2012},
number={05},
pages={034},
year={2012},
publisher={IOP Publishing}
}

@article{dvorkin2019making,
title={Making dark matter out of light: freeze-in from plasma effects},
author={Dvorkin, Cora and Lin, Tongyan and Schutz, Katelin},
journal={Physical Review D},
volume={99},
number={11},
pages={115009},
year={2019},
publisher={APS}
}

@article{tulin2018dark,
  title={Dark matter self-interactions and small scale structure},
  author={Tulin, Sean and Yu, Hai-Bo},
  journal={Physics Reports},
  volume={730},
  pages={1--57},
  year={2018},
  publisher={Elsevier}
}

@article{chavarria2016measurement,
  title={Measurement of the ionization produced by sub-keV silicon nuclear recoils in a CCD dark matter detector},
  author={Chavarria, AE and Collar, JI and Pena, JR and Privitera, P and Robinson, AE and Scholz, B and Sengul, C and Zhou, J and Estrada, J and Izraelevitch, F and others},
  journal={Physical Review D},
  volume={94},
  number={8},
  pages={082007},
  year={2016},
  publisher={APS}
}

@article{FANO,
title = {Absolute measurement of the Fano factor using a Skipper-CCD},
journal = {Nuclear Instruments and Methods in Physics Research Section A: Accelerators, Spectrometers, Detectors and Associated Equipment},
volume = {1010},
pages = {165511},
year = {2021},
issn = {0168-9002},
doi = {https://doi.org/10.1016/j.nima.2021.165511},
url = {https://www.sciencedirect.com/science/article/pii/S0168900221004964},
author = {D. Rodrigues and K.Andersson and M. Cababie and A. Donadon and A. Botti and G. Cancelo and J. Estrada and G. Fernandez-Moroni and R.Piegaia and M. Senger and M. Sofo Haro and L. Stefanazzi and J. Tiffenberg and S. Uemura}
}

@misc{rodrigues2023unraveling,
      title={Unraveling Fano noise and partial charge collection effect in X-ray spectra below 1 keV}, 
      author={Dario Rodrigues and Mariano Cababie and Ignacio Gomez Florenciano and Ana Botti and Juan Estrada and Guillermo Fernandez-Moroni and Javier Tiffenberg and Sho Uemura},
      year={2023},
      eprint={2305.09005},
      archivePrefix={arXiv},
      primaryClass={physics.ins-det}
}

@article{moroni2022skipper,
  title={Skipper charge-coupled device for low-energy-threshold particle experiments above ground},
  author={Moroni, Guillermo Fernandez and Chierchie, Fernando and Tiffenberg, Javier and Botti, Ana and Cababie, Mariano and Cancelo, Gustavo and Depaoli, Eliana L and Estrada, Juan and Holland, Stephen E and Rodrigues, Dario and others},
  journal={Physical Review Applied},
  volume={17},
  number={4},
  pages={044050},
  year={2022},
  publisher={APS}
}

@article{comptonbotti,
  title = {Constraints on the electron-hole pair creation energy and Fano factor below 150 eV from Compton scattering in a skipper CCD},
  author = {Botti, A. M. and Uemura, S. and Moroni, G. Fernandez and Barak, L. and Cababie, M. and Essig, R. and Etzion, E. and Rodrigues, D. and Saffold, N. and Sofo Haro, M. and Tiffenberg, J. and Volansky, T.},
  journal = {Phys. Rev. D},
  volume = {106},
  issue = {7},
  pages = {072005},
  numpages = {8},
  year = {2022},
  month = {Oct},
  publisher = {American Physical Society},
  doi = {10.1103/PhysRevD.106.072005},
  url = {https://link.aps.org/doi/10.1103/PhysRevD.106.072005}
}

@article{PhysRevD.96.042002,
  title = {Measurement of low energy ionization signals from Compton scattering in a charge-coupled device dark matter detector},
  author = {Ramanathan, K. and Kavner, A. and Chavarria, A. E. and Privitera, P. and Amidei, D. and Chou, T.-L. and Matalon, A. and Thomas, R. and Estrada, J. and Tiffenberg, J. and Molina, J.},
  journal = {Phys. Rev. D},
  volume = {96},
  issue = {4},
  pages = {042002},
  numpages = {8},
  year = {2017},
  month = {Aug},
  publisher = {American Physical Society},
  doi = {10.1103/PhysRevD.96.042002},
  url = {https://link.aps.org/doi/10.1103/PhysRevD.96.042002}
}

@ARTICLE{199363,
  author={Lacaita, A.L. and Zappa, F. and Bigliardi, S. and Manfredi, M.},
  journal={IEEE Transactions on Electron Devices}, 
  title={On the bremsstrahlung origin of hot-carrier-induced photons in silicon devices}, 
  year={1993},
  volume={40},
  number={3},
  pages={577-582},
  doi={10.1109/16.199363}}

@INPROCEEDINGS{499199,
author={Selmi, L. and Mastrapasqua, M. and Boulin, D.M. and Bude, J.D. and Manfredi, M. and Sangiorgi, E.S. and Pinto, M.R.},
booktitle={Proceedings of International Electron Devices Meeting}, 
title={Characterization and modeling of hot-carrier luminescence in silicon n/sup +//n/n/sup -/ devices}, 
year={1995},
volume={},
number={},
pages={293-296},
doi={10.1109/IEDM.1995.499199}}

@article{PhysRevB.52.10993,
  title = {Photon emission from hot electrons in silicon},
  author = {Villa, S. and Lacaita, A. L. and Pacelli, A.},
  journal = {Phys. Rev. B},
  volume = {52},
  issue = {15},
  pages = {10993--10999},
  numpages = {0},
  year = {1995},
  month = {Oct},
  publisher = {American Physical Society},
  doi = {10.1103/PhysRevB.52.10993},
  url = {https://link.aps.org/doi/10.1103/PhysRevB.52.10993}
}

@article{obeidat1997model,
  title={A model for visible photon emission from reverse-biased silicon pn junctions},
  author={Obeidat, Amjad T and Kalayjian, Zaven and Andreou, Andreas G and Khurgin, Jacob B},
  journal={Applied physics letters},
  volume={70},
  number={4},
  pages={470--471},
  year={1997},
  publisher={American Institute of Physics}
}

@article{feldman1998unified,
  title={Unified approach to the classical statistical analysis of small signals},
  author={Feldman, Gary J and Cousins, Robert D},
  journal={Physical review D},
  volume={57},
  number={7},
  pages={3873},
  year={1998},
  publisher={APS}
}

@article{guo2021muon,
  title={Muon flux measurement at china jinping underground laboratory},
  author={Guo, Zi-yi and Bathe-Peters, Lars and Chen, Shao-min and Chouaki, Mourad and Dou, Wei and Guo, Lei and Hussain, Ghulam and Li, Jin-jing and Liu, Qian and Luo, Guang and others},
  journal={Chinese Physics C},
  volume={45},
  number={2},
  pages={025001},
  year={2021},
  publisher={IOP Publishing}
}

@article{agnes2023search,
  title={Search for dark matter particle interactions with electron final states with DarkSide-50},
  author={Agnes, P and Albuquerque, IFM and Alexander, T and Alton, AK and Ave, M and Back, HO and Batignani, G and Biery, K and Bocci, V and Bonivento, WM and others},
  journal={Physical Review Letters},
  volume={130},
  number={10},
  pages={101002},
  year={2023},
  publisher={APS}
}

@Conference{SENSEI2023,
  author = "Nate Saffold on behalf of the SENSEI collaboration",
  title  = "Sub-GeV dark matter searches with SENSEI",
  booktitle = "APS April Meeting",
  year = "2023"
}

@article{PANDA2021,
  title={Search for light dark matter--electron scattering in the PandaX-II Experiment},
  author={Cheng, Chen and Xie, Pengwei and Abdukerim, Abdusalam and Chen, Wei and Chen, Xun and Chen, Yunhua and Cui, Xiangyi and Fan, Yingjie and Fang, Deqing and Fu, Changbo and others},
  journal={Physical Review Letters},
  volume={126},
  number={21},
  pages={211803},
  year={2021},
  publisher={APS}
}

@article{XENON1T2022,
  title={Emission of single and few electrons in XENON1T and limits on light dark matter},
  author={Aprile, Elena and Abe, K and Agostini, F and Maouloud, S Ahmed and Alfonsi, M and Althueser, L and Angelino, E and Angevaare, JR and Antochi, Vasile C and Martin, D Ant{\'o}n and others},
  journal={Physical Review D},
  volume={106},
  number={2},
  pages={022001},
  year={2022},
  publisher={APS}
}

\nocite{*}


\addcontentsline{toc}{chapter}{Publications}
\chapter*{List of publications}

\begin{longtable}{r|p{11cm}}

\textsc{}{2023}

& \textsc{}{Dario Rodrigues, Mariano Cababie, Ignacio Gomez Florenciano, Ana Botti, Juan Estrada, Guillermo Fernandez Moroni, Javier Tiffenberg, Sho Uemura, Miguel Sofo Haro. ``Unraveling Fano noise and partial charge collection effect in X-ray spectra below 1 keV". 2023, 2305.09005 (arXiv pre-print)} \\
\\

& \textsc{}{The SENSEI collaboration. ``SENSEI: Search for Millicharged Particles produced in the NuMI Beam". 2023, 2305.04964 (arXiv pre-print).}\\

\\
\textsc{}{2022}

& \textsc{}{The Oscura collaboration. ``The Oscura Experiment". 2022, 2202.10518 (arXiv pre-print).} \\

\\

& \textsc{}{A.M. Botti, S. Uemura, G. Fernandez Moroni, L. Barak, M. Cababie, R. Essig, J. Estrada, E. Etzion, D. Rodrigues, N. Saffold, M. Sofo Haro, J. Tiffenberg, T. Volansky. ``Constraints on the electron-hole pair creation energy and Fano factor below 150 eV from Compton scattering in a Skipper-CCD". 2022, Phys. Rev. D 106, 072005.}\\

\\

& \textsc{}{The SENSEI collaboration. ``Sensei: Characterization of single-electron events using a skipper-ccd." 2022, Phys.Rev.Applied 17 (2022) 1, 014022.} \\

\\

\textsc{}{2021}

& \textsc{}{Moroni, Guillermo Fernandez, Fernando Chierchie, Javier Tiffenberg, Ana Botti, Mariano Cababie, Gustavo Cancelo, Eliana L. Depaoli et al. ``The Skipper CCD for low-energy threshold particle experiments above ground." Phys. Rev. Applied 17, 044050 (2021).} \\ 

\\

& \textsc{}{Dario Rodrigues, Kevin Andersson, Mariano Cababie, Andre Donadon, Ana Botti, Gustavo Cancelo, Juan Estrada et al. ``Absolute measurement of the Fano factor using a Skipper-CCD." Nuclear Instruments and Methods in Physics Research Section A: Accelerators, Spectrometers, Detectors and Associated Equipment (2021): 165511.} \\

\\

\textsc{}{2020}

& \textsc{}{The SENSEI collaboration. ``SENSEI: direct-detection results on sub-GeV dark matter from a new Skipper CCD." Physical Review Letters 125, no. 17 (2020): 171802.} \\
\end{longtable}


\appendix
\addcontentsline{toc}{chapter}{APPENDIXES}
\chapter{Glossary of Datasets Used}
\label{cap:glossaryofdatasets}

In this thesis, results from the following datasets were presented\footnote{Note that the datasets used in Section \ref{sec:antecedentes} are not included since they were not analyzed in this thesis.} (see Table \ref{tab:glosario}):

\newcommand{\specialcell}[2][c]{%
  \begin{tabular}[#1]{@{}c@{}}#2\end{tabular}}

\renewcommand{\arraystretch}{1.7}
\begin{longtable}{cp{12cm}}
\centering
\textbf{\specialcell[t]{Setup \\ Images}} & 7 images taken to study the performance of \textit{SENSEI-SCCD} in \textit{MINOS} and later used to establish the event selection criteria in science images (see Section \ref{sec:2020}). This dataset was used repeatedly throughout the thesis as an example of a particular phenomenon, procedure, or effect. \\
\textbf{\specialcell[t]{Science \\ Images}} & 22 images used to set the dark matter exclusion limits presented in Section \ref{sec:2020}. \\
\textbf{A} & 8 images, each with a different exposure time $t_{\rm EXP}$ but the same readout time $t_{\rm RO}$. They were used to determine $\lambda_{\rm DC}$ in Section \ref{sec:resultados2022}. \\
\textbf{B} & 9 datasets, each consisting of six zero-exposure images; each dataset has a different voltage ${\rm V_{DD}}$ applied to the M1 transistor (see Figure \ref{fig.vdd_pra_v3}), ranging from $-17$~V to $-25$~V in 1 V steps. They were used to study the amplifier light contribution in Section \ref{sec:resultados2022}. \\
\textbf{C} & 7 datasets, each composed of two zero-exposure images and different readout times $t_{\rm RO}$. ${\rm V_{DD}}$ was set to $-22$~V. The experimental setup was used in \textit{MINOS} without the external shielding specified in Chapter \ref{cap:3}. These images were used to measure $\lambda_{\rm AL}$ and $\mu_{\rm SC}$ in Section \ref{sec:resultados2022}. \\
\textbf{D} & Dataset identical to C, except that ${\rm V_{DD}}$ was set to $-21$~V, and the additional external shielding was added. These images were also used to measure $\lambda_{\rm AL}$ and $\mu_{\rm SC}$ in Section \ref{sec:resultados2022}. \\
\textbf{SC} & Images extracted from a dataset measured in \textit{SiDet} and used to study the performance of SCCDs before their installation in \textit{MINOS} or another underground facility. Two images were taken for each measured voltage, as shown in Appendix \ref{appendix:cti}. The name \textbf{SC} comes from the vacuum chamber used, called the \textit{Supercube}.
\end{longtable}

\newpage

\setlength{\tabcolsep}{3pt}
\begin{sidewaystable}
  \centering
\begin{tabular}{cccccccc}
\hline\hline
\hline
& \textbf{A}  &  \textbf{B} &  \textbf{C} &  \textbf{D} &  \textbf{SC} & \textbf{Setup}  & \textbf{Science} \\
\hline
\textit{SENSEI-SCCD} & Yes & Yes & Yes & Yes & Yes(*) & Yes & Yes  \\
\hline
Location & \textit{MINOS} & \textit{MINOS} & \textit{MINOS} & \textit{MINOS} & \textit{SiDet} & \textit{MINOS} & \textit{MINOS}  \\
\hline
Skipper Samples & 250 & 250 & 200-950 & 200-950 & 400 & 300 & 300  \\
\hline
${\rm V_{DD}}$ (-V) & 22 & 17-25 & 22 & 21 & 21 & 21 & 21 \\
\hline
Additional Shielding & No & Yes & Yes & Yes & No & Yes & Yes  \\
\hline
Exposure Time (hours) & 0-8 & 0 & 0 & 0 & 0(**) & 20 & 20  \\
\hline
Readout Time (h:m:s) & 4:57:00 & 0:55:43 & Variable & Variable & 0:16:00 & 5:09:11 & 5:09:11  \\
\hline\hline
\end{tabular}
\caption{Glossary of datasets used. (*) The SCCDs used, as mentioned in Appendix \ref{appendix:cti}, are identical in their fabrication to the \textit{SENSEI-SCCDs} used in this thesis but differ in their dimensions: they are 512 rows by 3072 columns. (**) The exposure time for these images was 30 seconds.}
\label{tab:glosario}
\end{sidewaystable}



\chapter{Charge Transfer Inefficiency Study}
\label{appendix:cti}

As mentioned in Section \ref{sec:ctivssc}, the results obtained for Figure \ref{fig.CTIvsSC} are the outcome of a quality test to which the \textit{SENSEI-SCCDs} are subjected. This study was conducted at \textit{SiDet} and involved a complete temperature cycle, starting from room temperature up to 135K, followed by a return to the initial temperature. Throughout this cycle, a series of measurements were recorded to characterize dark current, spurious charge, CTI (Charge Transfer Inefficiency), and the uniformity of collected charges, among other observables. The purpose of these tests was to certify the quality of these devices for dark matter searches and to investigate any operational deficiencies.

While the complete study's results are not included in this manuscript, we will detail the tests conducted to estimate spurious charge and the CTI generated in the horizontal register, as they are relevant to understanding these effects.

The devices used are identical to the \textit{SENSEI-SCCDs} described in Chapter \ref{cap:3}, with the exception that their active area is slightly larger (6291456 as opposed to 5443584, a 15$\%$ increase). Additionally, while the previously described \textit{SENSEI-SCCDs} have a "short" horizontal register of 443 pixels per quadrant, these SCCDs have 3072 pixels.

The horizontal spurious charge study involved capturing an image of 40 rows by 800 columns with binning factors of 16 and 512 pixels, respectively, and without any prior exposure. We refer to the term "binning" as the simultaneous readout of multiple physical pixels, so that 16 rows and 512 columns are represented as a single "super-pixel" in the recorded image. This procedure amplifies the effect of spurious charge (512 times for horizontal transfers, 16 times for vertical transfers), making the signal more easily measurable. Additionally, surface measurements are challenging due to the large number of high-energy events reaching the device, and binning significantly reduces reading times. The rate of Single Event Effects (SEEs) per super-pixel is then calculated as usual during the manuscript's development.

The test conducted to measure horizontal CTI began by exposing the SCCD to a LED light source within the vacuum chamber for a few seconds. Subsequently, an image of 640 rows by 61 columns was read, skipping the reading in the first 3067 rows. This was done to obtain the charge collection profile in the last rows of the active area and the first rows of the overscan region. The skip reading involves transferring the charge collected in the pixels to the drain voltage, where they are discarded without being read.

\begin{figure}[h!]
    \centering
    \includegraphics[width=0.8\textwidth]{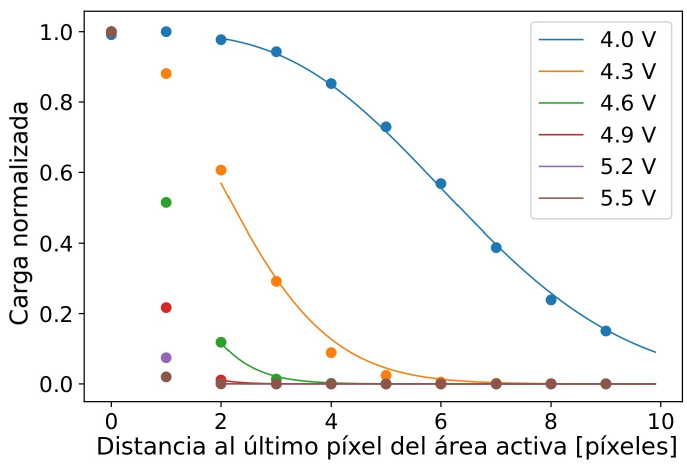}
    \caption{Normalized collected charge as a function of the distance from the last pixel of the active area. The normalization was performed with respect to the charge in the first pixel of the overscan region.}
    \label{fig.HCTI}
\end{figure}

Once the image is obtained, the charge profile per pixel as a function of the column number is determined for a certain range of rows, specifically for the columns at the edge of the active area. To simplify the analysis, this profile, as illustrated in Figure \ref{fig.HCTI}, is normalized by the charge value from the pixels in the last column of the active area. The Charge Transfer Inefficiency (CTI) is then calculated using Equation \ref{eq.cti}. However, there are certain considerations to keep in mind when performing this adjustment. 

Firstly, the charge accumulated in, let's say, the first pixel of the overscan region does not originate solely from the last pixel of the active area but also from the preceding ones. To address this, a sum of the charge accumulated by CTI in the last 20 pixels was performed, which yielded satisfactory results for the analyzed quadrant. 

Secondly, Equation \ref{eq.cti} has the disadvantage of containing a factorial term that is impossible to compute for certain values of the distance from the active area. Additionally, it does not provide a continuous fit but is partially discrete since the factorial operation can only be applied to integers. Therefore, an approximation of the factorial term developed by Ramanujan in \cite{ramanujanlost} was used. This approximation is highly accurate (to within hundredths of a percent) for values greater than or equal to 2. Consequently, values not meeting this condition were excluded from the fit.

\printglossary
\cleardoublepage
\phantomsection		

\end{document}